\documentclass[12pt,oneside,a4paper,onecolumn]{elsarticle}
\usepackage{times}
\usepackage{fullpage}

\usepackage{gensymb}
\usepackage{tikz}
\usetikzlibrary{matrix,shapes,arrows,positioning,chains,shapes.geometric,calc,patterns,decorations.pathmorphing,decorations.markings}
\usepackage{alltt}
\usepackage{upquote}
\usepackage{float}
\usepackage{array}
\usepackage{pgfplots}
\usepackage{graphicx}
\usepackage[toc,page]{appendix}
\usepackage{amsmath,mathtools}
\usetikzlibrary{decorations.markings}
\usepackage{amsthm}
\usepackage{amsfonts}
\usepackage{enumitem}
\usepackage{subcaption}
\usepackage{caption}
\usepackage{xcolor}
\usepackage{scalerel}
\usepackage[export]{adjustbox}
\usepackage[absolute,overlay]{textpos}
	\usepackage{mwe}

\newcommand{\omegab}{{\boldsymbol \omega}}

\renewcommand{\u}{\boldsymbol{u}^\mathrm{f}}
\newcommand{\un}{{u}^\mathrm{f}}

\newcolumntype{M}[1]{>{\centering\arraybackslash}m{#1}}
\newcolumntype{N}{@{}m{0pt}@{}}

\tikzset{>=latex}
\graphicspath{ {1D_results/}{2D_results/SQ/}{2D_results/UQ/}{2D_results/AQ/}{2D_results/UAQ/}{Fourier_analysis/}}
\def\@author#1{\g@addto@macro\elsauthors{\normalsize%
    \def\baselinestretch{1}%
    \upshape\authorsep#1\unskip\textsuperscript{%
      \ifx\@fnmark\@empty\else\unskip\sep\@fnmark\let\sep=,\fi
      \ifx\@corref\@empty\else\unskip\sep\@corref\let\sep=,\fi
      }%
    \def\authorsep{\unskip,\space}%
    \global\let\@fnmark\@empty
    \global\let\@corref\@empty
    \global\let\sep\@empty}%
    \@eadauthor={#1}
}
\begin{document}
\begin{frontmatter}
\title{Transverse flow-induced vibrations of a sphere in the proximity of a free surface: A numerical study}
\author[UBC]{Amir Chizfahm}

\author[UBC]{Vaibhav Joshi}

\author[UBC]{Rajeev Jaiman\corref{cor1}}
\ead{rjaiman@mech.ubc.ca}
\cortext[cor1]{Corresponding author}
\address[UBC]{Department of Mechanical Engineering, University of British Columbia, Vancouver, Canada}

\begin{abstract}
\noindent 
In this paper, we present a numerical study on the transverse flow-induced vibration (FIV) of an elastically mounted sphere in the vicinity of a free surface at subcritical Reynolds numbers. We assess the interaction dynamics and the vibration characteristics of fully submerged and piercing spheres that are free to vibrate in the transverse direction. We employ the recently developed three-dimensional two-phase flow-structure interaction solver to investigate fully and partially submerged configurations of an elastically mounted sphere. To begin, we examine the vortex-induced vibration (VIV) phenomenon and the vortex-shedding modes of a fully-submerged sphere vibrating freely in all three spatial directions. We systematically verify and analyze the mode transitions and the motion trajectories in the three degrees-of-freedom (3-DOF) for the Reynolds number up to $30\,000$. We next simulate the transversely vibrating (1-DOF) full-submerged sphere over a wide range of reduced velocities $3 \leq U^* \leq 20$, whereby the reduced velocity is adjusted by changing the freestream Reynolds number. The VIV response amplitude and the topology of the wake structure are compared with the measurements for the mode I and mode II response branches.
We further look into the effect of the free surface on the FIV response of a transversely vibrating sphere in the proximity of a free surface. The response dynamics of the sphere is studied for three representative values of normalized immersion ratio ($h^*=h/D$, where $h$ is the distance from the top of the sphere to undisturbed free-surface level and $D$ is the sphere diameter), at $h^*=1$ (fully submerged sphere with no free-surface effect), $h^*=0$ (where the top of the sphere touches the free surface) and $h^*=-0.25$ (where the sphere pierces the free surface). At the lock-in range, we observe that the amplitude response of the sphere at $h^*=0$ is decreased significantly compared to the case at $h^*=1$. It is found that the vorticity flux is diffused due to the free-surface boundary and the free surface acts as a sink of energy that leads to a reduction in the transverse force and amplitude response. When the sphere pierces the free surface at $h^*=-0.25$, the amplitude response at the lock-in state is found to be greater than all the submerged cases studied with the maximum peak-to-peak amplitude of $\sim2D$.
We find that the interaction of the piercing sphere with the air-water interface causes a relatively large surface deformation and has a significant impact on the synchronization of the vortex shedding and the vibration frequency. The streamwise vorticity contours and pressure distribution are employed to understand the VIV characteristics and wake dynamics. Increased streamwise vorticity gives rise to a relatively larger transverse force to the piercing sphere at $h^*=-0.25$, resulting in greater positive energy transfer per cycle to sustain the large-amplitude vibration. Lasty, we study the sensitivity of large-amplitude vibration on the mass ratio, $m^*$, and, Froude number, $Fr$, at the lock-in state. 
\end{abstract}
\end{frontmatter}

\section{Introduction}
Fluid-structure interaction (FSI) of spherical bodies is omnipresent and has numerous applications in marine and offshore engineering. For example, flow-induced vibrations (FIV) of an elastically mounted or tethered spherical configuration can be useful for power generation and wave energy harvesting while such vibrations are undesirable on spherical marine/offshore structures such as low-aspect-ratio escort tugboats connected with ships \cite{Tugboat}. Physical understanding of these problems poses serious challenges due to the richness and complexity of nonlinear coupled FIV phenomenon together with vorticity/free-surface interactions. A prototypical geometry of the sphere (i.e., axisymmetric bluff body) close to the free surface or piercing the free surface can be considered as an idealized model, which serves as a generic problem to examine the coupled fluid-structure and free-surface interactions. A freely vibrating sphere in the vicinity of free surface can exhibit complex spatial-temporal dynamics and synchronization as functions of physical and geometric parameters.
Synchronization or lock-in is a general nonlinear physical phenomenon in fluid-structure systems whereby the coupled system has an intrinsic ability to lock at a preferred frequency and amplitude. The phenomenon of lock-in and vortex-induced vibrations are extensively reviewed for cylindrical structures in \cite{sarpkaya2004,williamson2004,bearman2011}. When the natural frequency of an elastically-mounted configuration of a spherical body approaches to the frequency of the unsteady vortex shedding, the sphere can undergo vortex-induced vibration similar to two-dimensional bluff bodies \cite{williamson1997dynamics,williamson2004, govardhan2005vortex}. In contrast to two-dimensional circular cylinder wakes, the vortex topology and shedding process are significantly different for a three-dimensional configuration of an elastically mounted sphere. Furthermore, due to the vorticity/free-surface interactions, the vortex-induced vibration of the sphere can be significantly altered as reported recently in \cite{sareen2018}. Through high-fidelity numerical simulations, the central intent of this paper is to explore the effect of free surface on the vortex-induced vibration of an elastically mounted sphere at subcritical Reynolds numbers  (based on the freestream velocity and the sphere diameter).
%

Many have studied the vortex dynamics of circular cylinders in the close proximity of a free surface. The free surface deforms to satisfy the fluid-fluid (i.e., air-water) interface conditions such that  the tangential stress maintains zero stress condition and the normal stress remains constant. The vorticity generation process at a free surface is different in contrast to a no-slip solid surface, whereby the flow velocity  at the surface can slip freely to satisfy the zero stress condition and the surface can deform or distort. Owing to complex dynamical interactions between the deformed free surface and the flow field, free-surface boundaries can act as sources or sinks for the vorticity. To characterize the free-surface dynamics, the ratio of the inertial force to the gravitational force and surface deformation is defined by Froude number $Fr=U/\sqrt{gD}$, where $U$ is the freestream velocity, $D$ is the characteristic length and $g$ is the gravitational acceleration. The experimental work on flow past a cylinder close to the free surface in \cite{sheridan1997flow}, investigated the effect of the Froude number at high range of $Fr\in[0.47,0.72]$.  They reported that for high Froude numbers, the wake of a cylinder close to the free surface is fundamentally different than the wake of a cylinder far beneath the free surface. The generation of a vorticity layer from the free surface was observed in their experiments. The authors found that for low Froude number $(Fr\leq0.3)$ with small free-surface deformation, the problem is analogous to the flow past a cylinder close to the no-slip wall. 
The free surface was found to act like a rigid free-slip boundary at low Froude number. However, for higher Froude numbers, the surface can 
deform significantly, giving rise to a larger surface vorticity that can defuse or convect into the main flow and modify the wake dynamics. 
The numerical study on the two-dimensional flow past a cylinder close to the free surface at $Re=180$ was performed in \cite{reichl2005flow}. It was found that the free-surface curvature can lead to relatively larger diffusion of vorticity. A series of numerical studies considered the flow past a cylinder piercing the free surface \cite{kawamura2002large, yu2008large}. The study in \cite{yu2008large} on the piercing cylinder, considered the effect of the free surface on the vortex pattern in the near wake for the Froude number up to $Fr=3$. They found that the free surface prevents the vortex generation in the near wake, and therefore reduces the vorticity and vortex shedding. At $Fr=0.8$, 2-D vortex structures were spotted in the deep wake while in the proximity of the free surface, the vortex structures showed strong 3-D features. At higher Froude number $Fr=2$, the effect of the free surface propagated throughout the wake and prevented the regular vortex shedding and vortices with less intensity dominated the region below the free surface. 

While there are plethora of publications on the FIV of a cylindrical body, a few studies focusing on a freely vibrating sphere are available in the literature. The first study on the FIV of sphere was performed in \cite{govardhan1997vortex, williamson1997dynamics}. The authors conducted an experimental study on a fully submerged tethered sphere in a steady fluid flow. Their experimental observation on the tethered sphere, uncovered the vigorous vibration of the sphere similar to the cylinders when exposed to the uniform flow stream. The authors experimentally observed that a tethered sphere will undergo a large peak-to-peak amplitude of about two diameters of the sphere over a wide range of flow velocities. The motion trajectory of the sphere was found to form a ``figure-eight" trajectory and by the increase in the mass ratio ($m^*=m/m_d$, where $m$ is the mass of the structural body and $m_d$ is the mass of the displaced fluid), the motion trajectory transformed to the ``crescent shape". Further investigation of the effect of the mass ratio on the amplitude response for the tethered and elastically mounted sphere configurations was studied in \cite{govardhan2005vortex}. They showed the maximum peak amplitude as a function of the mass-damping parameter $(m^*\zeta)$ on the Griffin plot. The plot exhibited a good collapse of data where a saturation maximum amplitude of around $(0.9D)$ was recorded for all cases. 
The existence of multiple modes of vibration was reported for the tethered sphere configuration in the experimental studies performed in \cite{govardhan1997vortex, williamson1997dynamics, govardhan2005vortex, jauvtis2001multiple}. These modes were identified based on the amplitude response curve ($A^*$-$U^*$), where $A^*$ is the non-dimensional amplitude defined as $A^*=A/D$, and $U^*$ is the reduced velocity defined as $U^*=U/f_nD$ ($f_n$ is the natural frequency of the system in vacuum). For the range of reduced velocities $U^*\in[5\sim10]$, the authors identified mode I and mode II of vibrations that correspond to the lock-in region where the sphere vibration synchronized with the shedding frequency and the natural frequency of the system \cite{govardhan1997vortex, williamson1997dynamics}. Mode III of vibration was found to exist at a higher reduced velocity range $U^*\in[20\sim40]$, where the shedding frequency is three to eight times higher than the sphere vibration frequency \cite{govardhan2005vortex}. The existence of mode IV (intermittent mode) at much higher reduced velocities $(U^*\geq 100)$ for a tethered sphere was reported in \cite{jauvtis2001multiple}, where non-periodic large amplitude response was observed. The experimental and numerical studies in \cite{lee2013vortex, rajamuni2020vortex} further explained the existence of non-stationary chaotic dynamic response similar to the mode IV in \cite{jauvtis2001multiple}, at lower reduced velocity range for low inertia systems.

Unlike a vast amount of literature available on two-dimensional geometry of elastically-mounted circular cylinder, there are a handful of numerical studies on a three-dimensional geometry of a sphere undergoing flow-induced vibration. A numerical study on the VIV of a freely oscillating sphere in all three spatial directions has been performed in \cite{behara2011vortex} at $Re=300$ and the reduced velocity range of $U^*\in[4,9]$. The authors observed two distinct VIV response and wake modes, termed as the hairpin mode and the spiral mode, at the same reduced velocity. It was found that the motion trajectory corresponding to the hairpin mode followed a linear path, while for the spiral mode, the sphere shifted to a circular orbit \cite{behara2011vortex, behara2016}. In another recent numerical study of \cite{rajamuni2019} on 3-DOF elastically mounted sphere, the authors reported the instability of the hairpin mode for the range of $Re\in[300,2\,000]$. They observed that the hairpin mode was always followed by the spiral mode during the transformation from a transient state to the stationary state for the Reynolds number range studied. The VIV response of elastically mounted sphere restricted to move in the transverse direction was carried out experimentally and numerically in \cite{rajamuni2018transverse}. The authors provided additional insight into the experimental study of the sphere transverse motion \cite{sareen2018vortex}, and explored some distinctions from the numerical and experimental studies with specific constraints such as the tethered configuration \cite{williamson1997dynamics} or 3-DOF elastically mounted \cite{behara2011vortex}.

All the aforementioned studies on the FIV of the sphere were performed for the flow past a fully submerged structural body with no free-surface effect. However, the VIV response of the elastically mounted sphere close to the free surface or piercing it could be very different due to the complications of vorticity/free-surface interactions. The experimental study in \cite{mirauda2014dynamic}, was conducted for a tethered sphere in shallow water. The authors reported the reduction of the amplitude response due to the presence of the free surface when the sphere is fully submerged. In the recent experimental investigation of \cite{sareen2018}, the authors systematically studied the effect of the free surface on the vortex-induced vibration of fully and semi-submerged elastically mounted sphere for the range of Reynolds number $Re\in [5\,000,30\,000]$, the reduced velocity $U^*\in [3,20]$, and the immersion depth ratio $h^*=h/D\in[-0.75,2.5]$, where $h$ denotes the distance from the top of the sphere to the free surface.  When the sphere came closer to the free surface for a range of immersion ratio $(0.185\leq h^*\leq 1)$, 
the authors observed the reduction in the peak amplitude response and the transverse fluctuating force acting on the sphere. However, by further decreasing the immersion ratio in the range $(0\leq h^*\leq 0.185)$ for the submerged sphere and in the range $(-0.375\leq h^*\leq 0)$ for the piercing sphere, the peak amplitude response of the oscillations was found to increase significantly. Several modes of vibrations were categorized in their study for the VIV response of fully and partially submerged cases based on force measurement, the total phase (phase difference between the fluid force and the body displacement) and the vortex phase (phase difference between the vortex force and the body displacement). Due to the limitation on the PIV imaging set-up in their experiments, capturing the vorticity formation close to the sphere was found to be challenging. The reasons for the large-amplitude response for the piercing sphere cases could not be fully explained. The effect of the free surface on the wake dynamics, the vortex forcing and the unsymmetrical geometry of the semi-submerged immersed body once part of it lies above the waterline added to the complexity of the problem.

\begin{figure}[!h]
	\centering
	\adjincludegraphics[scale=0.43,trim={0.1\width} {0.11\width} {0.1\width} {0.12\width},clip]{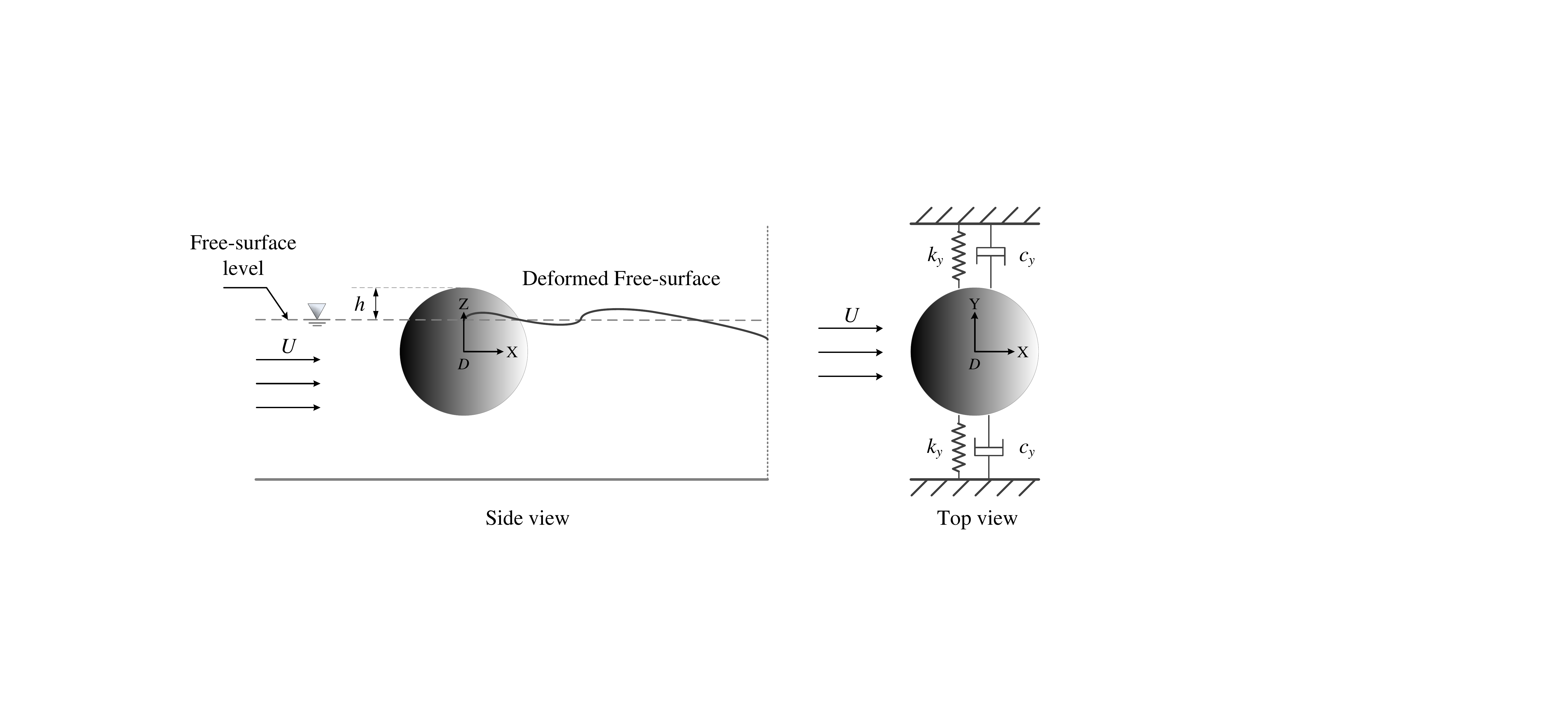}
	\caption{Schematic illustrating a side view of sphere piercing a viscous free surface (left) and a top view of  1-DOF elastically mounted sphere in steady incident flow (right). Here $D$ denotes the sphere diameter, $U$ is the freestream velocity, $k$ is the spring stiffness and $c$ is the structural damping. $h$ measures the distance from top of the sphere to the free-surface level. }
	\label{domain_FS}		
	
\end{figure}

In the present study, the role of streamwise vorticity/free-surface interaction on the VIV response of a freely vibrating sphere is analyzed numerically as functions of immersion ratio, the mass ratio and Froude number. The physical setup for the elastically-mounted sphere along the free surface is similar to the experimental configuration of \cite{sareen2018}. Fig \ref{domain_FS} shows a schematic of the problem setup for a piercing sphere case together with the geometric and flow parameters. Building upon the experimental investigations, we attempt to understand the origin of the large amplitude vibration when the sphere pierces the free surface. We consider a representative $h^*=-0.25$ to investigate the large-amplitude excitation mechanism via numerical computations. We employ a first-principle based fully-coupled continuum mechanics formulation for solving a multiphase fluid-structure interaction at sub-critical 
Reynolds number \cite{Joshi_IJNME}. The free-surface effects are taken into account by modeling the air-water interface with the aid of the phase-field Allen-Cahn equation and the turbulence is modeled via dynamic large eddy simulation (LES) \cite{NIFC_2}. Successful validation of the 1-DOF vibrating sphere by considering the effect of the free surface is established through detailed quantitative and qualitative comparisons with the experiments. We systematically examine the effect of Reynolds number on the mode transitions at VIV regime for a freely vibrating sphere, and its substantial effect on the coupled dynamical behavior and the motion trajectories. 

The central intent of this work is to perform a numerical investigation of the unsteady flow fields and the vibrational characteristics of the elastically mounted sphere subjected to the vorticity/free-surface interactions. The insight gained is used for identifying the wake modes and the coupled dynamical interactions that lead to vortex-induced vibration with a large amplitude response for the sphere configuration piercing the free surface. Coupled dynamics of unsteady wake-sphere interaction, the force and amplitude characteristics and the vorticity and pressure distributions are investigated during the oscillation. We deduce that the extra vorticity generation at the free surface for the piercing sphere has a significant impact on the synchronization of the vortex shedding and the vibration frequency, and could be the major cause of the large-amplitude VIV response. 
Such physical insight on the VIV and free-surface interactions may guide to develop effective active or passive suppression devices.
 In Section \ref{numerical_framework}, we summarize the two-phase FSI framework which is followed by brief implementation details in Section \ref{Implitation_detail}. While Section \ref{Conv_Val} presents the validation of the FSI solver,  Section \ref{results} focuses on the FIV response of a fully and partially submerged sphere with a particular emphasis on the origin of large amplitude response of the piercing sphere. The major conclusions of this work are reported in Section \ref{Conclusion}. 

\section{Numerical methodology}
\label{numerical_framework}
A brief description about the coupled two-phase fluid-structure solver based on the spatially filtered two-phase Navier-Stokes equations in the moving boundary arbitrary Lagrangian-Eulerian (ALE) framework and six-degrees of freedom structural equation is presented in this section along with its variational form.

\subsection{Two-phase flow modeling with moving boundary}
The spatially filtered Navier-Stokes equations in an ALE framework for an incompressible flow are given as
\begin{align} \label{NS}
\rho^\mathrm{f}\frac{\partial \bar{\boldsymbol{u}}^\mathrm{f}}{\partial t}\bigg|_{\hat{{x}}^\mathrm{f}} + \rho^\mathrm{f}(\bar{\boldsymbol{u}}^\mathrm{f} - {\boldsymbol{u}^\mathrm{m}})\cdot\nabla\bar{\boldsymbol{u}}^\mathrm{f} &= \nabla\cdot \bar{\boldsymbol{\sigma}}^\mathrm{f} + \nabla\cdot{\boldsymbol{\sigma}}^\mathrm{sgs} + \boldsymbol{b}^\mathrm{f}\ \ \ \mathrm{on\ \ \Omega^\mathrm{f}(t)},\\
\nabla\cdot\bar{\boldsymbol{u}}^\mathrm{f} &= 0\ \ \ \mathrm{on\ \ \Omega^\mathrm{f}(t)},
\end{align}
where $\bar{\boldsymbol{u}}^\mathrm{f} = \bar{\boldsymbol{u}}^\mathrm{f}(\boldsymbol{x}^\mathrm{f},t)$ and $\boldsymbol{u}^\mathrm{m}=\boldsymbol{u}^\mathrm{m}(\boldsymbol{x}^\mathrm{f},t)$ represent the fluid and mesh velocities defined for each spatial point $\boldsymbol{x}^\mathrm{f} \in \Omega^\mathrm{f}(t)$ respectively. The fluid density is denoted by $\rho^\mathrm{f}$ and $\boldsymbol{b}^\mathrm{f}$ represents the body force acting on the fluid and $\bar{\boldsymbol{\sigma}}^\mathrm{f}$ is the Cauchy stress tensor for a Newtonian fluid which is given as
\begin{align}
\bar{\boldsymbol{\sigma}}^\mathrm{f} = -\bar{p}\boldsymbol{I} + \mu^\mathrm{f}( \nabla\bar{\boldsymbol{u}}^\mathrm{f} + (\nabla\bar{\boldsymbol{u}}^\mathrm{f})^T),
\end{align}
where $\bar{p}$ is the filtered fluid pressure, $\boldsymbol{I}$ denotes the second-order identity tensor, $\mu^\mathrm{f}$ represents the dynamic viscosity of the fluid, and $\boldsymbol{\sigma}^\mathrm{sgs}$ is the extra turbulent stress term based on subgrid filtering procedure for large eddy simulation. Details about the dynamic subgrid model utilized in the present formulation can be found in \cite{NIFC_2}. The partial derivative with respect to the ALE referential coordinate $\hat{x}^\mathrm{f}$ is kept fixed in Eq. (\ref{NS}). The density $\rho^\mathrm{f}$ and viscosity $\mu^\mathrm{f}$ for two-phase flows in Eq.~(\ref{NS}) depend on the phase-indicator $\phi$ as
\begin{align}
\rho^\mathrm{f}(\phi) = \frac{1+\phi}{2}\rho^\mathrm{f}_1 + \frac{1-\phi}{2}\rho^\mathrm{f}_2,\\
\mu^\mathrm{f}(\phi) = \frac{1+\phi}{2}\mu^\mathrm{f}_1 + \frac{1-\phi}{2}\mu^\mathrm{f}_2,
\end{align} 
where $\rho^\mathrm{f}_1$, $\mu^\mathrm{f}_1$ and $\rho^\mathrm{f}_2$, $\mu^\mathrm{f}_2$ represent the densities and viscosities of the two phases respectively.

In contrast to the traditional approaches to evolve the fluid-fluid interface of the two-phase flow like volume-of-fluid (VOF) and level-set which require some kind of geometric manipulation which can be computationally expensive in three-dimensions \cite{Joshi_OMAE_2018}, we utilize the diffuse interface description  which originate from thermodynamically consistent theories of phase transitions and avoid any kind of geometric manipulations. The diffused interface has a finite thickness ($\mathcal{O}(\varepsilon)$) and is evolved by the minimization of the Ginzburg-Landau energy functional,
\begin{align} \label{energy_functional}
\mathcal{E}(\phi) = \int_{\Omega^\mathrm{f}(t)} \bigg(\frac{\varepsilon^2}{2}|\nabla\phi|^2 + F(\phi)\bigg) \mathrm{d}\Omega,
\end{align}
where $\varepsilon$ is the representative length scale of the finite thickness of the fluid-fluid interface.  It represents a balance between the interfacial energy and the bulk energy of the two-phase system. 
The convective form of the conventional Allen-Cahn equation \cite{Allen_Cahn} is considered in the present formulation with a Lagrange multiplier for mass conservation.  
The order parameter $\phi$ distinguishes between the two phases, being $1$ in water and $-1$ in air.
The order parameter $\phi$ changes continuously but steeply across the phase interface from one phase to the other in the interface thickness  $\varepsilon$. 
 The conservative Allen-Cahn equation is written on $\Omega^\mathrm{f}(t)$ as:
\begin{align}
\label{CAC}
\frac{\partial\phi}{\partial t}\bigg|_{\hat{{x}}^\mathrm{f}} + (\bar{\boldsymbol{u}}^\mathrm{f}-\boldsymbol{u}^\mathrm{m})\cdot\nabla\phi - \gamma\big(\varepsilon^2\nabla^2\phi - F'(\phi) + \beta(t)\sqrt{F(\phi)}\big) = 0,
\end{align}
where $\bar{\boldsymbol{u}}^\mathrm{f}$ is the convection velocity which is coupled with the Navier-Stokes equations and $\gamma$ is a relaxation factor with units of $[T^{-1}]$ which is selected as $1$ for the present study. The term $F'(\phi)$ denotes the derivative of $F(\phi)$ with respect to $\phi$, with $F(\phi)=(1/4)(\phi^2-1)^2$ being a double well potential function which has minima at $\phi=\pm 1$ indicating the two stable phases. The parameter $\beta(t)$ is the time dependent part of the Lagrange multiplier given as
\begin{align} \label{eqn:beta}
\beta(t) = \frac{\int_{\Omega^\mathrm{f}(t)} F'(\phi)\mathrm{d}\Omega}{\int_{\Omega^\mathrm{f}(t)} \sqrt{F(\phi)}\mathrm{d}\Omega}.
\end{align}

The convection-diffusion-reaction form of the convective Allen-Cahn equation (Eq. \ref{CAC}) can be written as,
\begin{align}
\partial_t{\phi} + \hat{\boldsymbol{u}}\cdot\nabla\phi - \nabla\cdot(\hat{k}\nabla\phi) + \hat{s}\phi - \hat{f} = 0,
\end{align}
where $\hat{\boldsymbol{u}}=(\bar{\boldsymbol{u}}^\mathrm{f}-\boldsymbol{u}^\mathrm{m})$, $\hat{k}$, $\hat{s}$ and $\hat{f}$ are the modified convection velocity, diffusion coefficient, reaction coefficient and source terms respectively, the expressions of which are given in detail in \cite{PPV_JCP, AC_JCP}.

\subsection{Structural modeling}
The structure is modeled as a rigid body having six degrees of freedom of motion. Consider a mapping function $\boldsymbol{\varphi}^\mathrm{s}(\boldsymbol{x}^\mathrm{s},t)$ which denotes the position vector and maps the reference configuration of the rigid body $\boldsymbol{x}^\mathrm{s}$ at $t=0$ to its position at the deformed state $\Omega^\mathrm{s}(t)$. If $\boldsymbol{\eta}^\mathrm{s}(\boldsymbol{x}^\mathrm{s},t)$ is the displacement of the rigid body, the position vector is given by $\boldsymbol{\varphi}(\boldsymbol{x}^\mathrm{s},t) = \boldsymbol{\eta}^\mathrm{s}(\boldsymbol{x}^\mathrm{s},t) + \boldsymbol{x}^\mathrm{s}$.
Let the center of mass of the body in the reference configuration $\boldsymbol{x}^\mathrm{s}$ and the current configuration $\boldsymbol{\varphi}^\mathrm{s}$ be $\boldsymbol{x}^\mathrm{s}_{0}$ and $\boldsymbol{\varphi}^\mathrm{s}_{0}$ respectively and $\boldsymbol{\eta}^\mathrm{s}_{0}$ denote the displacement of the center of mass due to translation of the body. Therefore, the rigid body kinematics is given by
\begin{align} \label{RBD_eqn}
\boldsymbol{\varphi}^\mathrm{s} = \boldsymbol{Q}(\boldsymbol{x}^\mathrm{s}-\boldsymbol{x}^\mathrm{s}_{0}) + \boldsymbol{\varphi}^\mathrm{s}_{0} = \boldsymbol{Q}(\boldsymbol{x}^\mathrm{s}-\boldsymbol{x}^\mathrm{s}_{0}) + \boldsymbol{x}^\mathrm{s}_{0} + \boldsymbol{\eta}^\mathrm{s}_{0},
\end{align}
where $\boldsymbol{Q}$ is a rotation matrix. The displacement can be expressed as, 
\begin{align}
\boldsymbol{\eta}^\mathrm{s} = (\boldsymbol{Q}-\boldsymbol{I})(\boldsymbol{x}^\mathrm{s}-\boldsymbol{x}^\mathrm{s}_{0}) + \boldsymbol{\eta}^\mathrm{s}_{0}, \label{etaeqn}\\
\frac{\partial\boldsymbol{\eta}^\mathrm{s}}{\partial t} = \frac{\partial\boldsymbol{Q}}{\partial t}(\boldsymbol{x}^\mathrm{s}-\boldsymbol{x}^\mathrm{s}_{0}) + \frac{\partial\boldsymbol{\eta}^\mathrm{s}_{0}}{\partial t}, \label{detadteqn}
\end{align}
where $\boldsymbol{I}$ is the identity matrix and Eq.~(\ref{detadteqn}) is obtained by differentiating Eq.~(\ref{etaeqn}) with respect to time. Suppose the rotational degrees of freedom for the body are given by $\boldsymbol{\theta}^\mathrm{s}$. Equation (\ref{detadteqn}) can be restructured in terms of the angular velocity of the body denoted by $\boldsymbol{\omega}^\mathrm{s} = \partial\boldsymbol{\theta}^\mathrm{s}/dt$ as
\begin{align}
\frac{\partial\boldsymbol{\eta}^\mathrm{s}}{\partial t} = \boldsymbol{\omega}^\mathrm{s}\times(\boldsymbol{\varphi}^\mathrm{s}-\boldsymbol{\varphi}^\mathrm{s}_{0}) + \frac{\partial\boldsymbol{\eta}^\mathrm{s}_{0}}{\partial t}
\end{align}
Therefore, the  six degrees-of-freedom rigid body motion is governed by the matrix form,
\begin{align}
\boldsymbol{M}^\mathrm{s}\frac{\partial^2\boldsymbol{\eta}^\mathrm{s}_{0}}{\partial t^2} + \boldsymbol{C}_{\eta}\frac{\partial\boldsymbol{\eta}^\mathrm{s}_{0}}{\partial t} + \boldsymbol{K}_{\eta}\boldsymbol{\eta}^\mathrm{s}_{0} &= \boldsymbol{f}^\mathrm{s},\ \mathrm{on}\ \Omega^\mathrm{s},\\
\boldsymbol{I}^\mathrm{s}\frac{\partial^2\boldsymbol{\theta}^\mathrm{s}}{\partial t^2} + \boldsymbol{C}_{\theta}\frac{\partial\boldsymbol{\theta}^\mathrm{s}}{\partial t} + \boldsymbol{K}_{\theta}\boldsymbol{\theta}^\mathrm{s} &= \boldsymbol{\tau}^\mathrm{s},\ \mathrm{on}\ \Omega^\mathrm{s},	
\end{align}
where $\boldsymbol{M}^\mathrm{s}$, $\boldsymbol{C}_\eta$ and $\boldsymbol{K}_\eta$ denote the mass, damping and stiffness matrices for the translational degrees of freedom respectively, $\boldsymbol{I}^\mathrm{s}$, $\boldsymbol{C}_\theta$ and $\boldsymbol{K}_\theta$ represent the moment of inertia, damping and stiffness matrices for the rotational degrees of freedom respectively, and $\boldsymbol{f}^\mathrm{s}$ and $\boldsymbol{\tau}^\mathrm{s}$ denote the forces and the moments applied on the body respectively.  

\subsection{Fluid-structure interface}
It is imperative for a fluid-structure interaction problem that the kinematic and dynamic equilibrium are satisfied at the fluid-structure interface $\Gamma^\mathrm{fs}$. Mathematically, these relations can be written as
\begin{align}
{\bar{\boldsymbol{u}}}^\mathrm{f}(\boldsymbol{\varphi}^\mathrm{s}(\boldsymbol{x}^\mathrm{s},t),t) &= \boldsymbol{u}^\mathrm{s}(\boldsymbol{x}^\mathrm{s},t),\\
\int_{\Gamma^\mathrm{fs}} \bar{\boldsymbol{\sigma}}^\mathrm{f}(\boldsymbol{x}^\mathrm{f},t)\cdot \boldsymbol{n} \mathrm{d\Gamma} +  \boldsymbol{f}^\mathrm{s}  &= 0
\end{align}
where $\boldsymbol{\varphi}^\mathrm{s}$ is the position vector mapping the initial position $\boldsymbol{x}^\mathrm{s}$ of the rigid body to its position at time $t$, $\boldsymbol{f}^\mathrm{s}$ is the fluid force acting on the body and $\boldsymbol{u}^\mathrm{s}$ is the structural velocity at time $t$ given by $\boldsymbol{u}^\mathrm{s} = \partial\boldsymbol{\varphi}^\mathrm{s}/\partial t$. Here, $\boldsymbol{n}$ is the outer normal to the fluid-body interface $\Gamma^\mathrm{fs}$ in the reference configuration. 
Owing to the body-fitted ALE formulation, the fluid velocity is exactly equal to the velocity of the body along the interface. 
The motion of the immersed body is governed by the fluid forces which include the integration of pressure and shear stress effects on the body surface.
The coupling algorithm between the two-phase fluid and the rigid-body structural equations is based on the nonlinear iterative force correction (NIFC) scheme presented in \cite{Joshi_IJNME}.

\section{Implementation details}
\label{Implitation_detail}
The continuum equations with their variational form presented in the previous section are coupled in a nonlinear partitioned iterative manner. 
The movement of the internal ALE nodes is evaluated by considering a continuum hyperelastic model for the fluid mesh such that the mesh quality does not deteriorate
as the displacement of the body increases during wake-induced vibration. For fluid-structure interaction with strong added
mass effects ($m^* \approx \mathcal{O}(1)$), a partitioned iterative scheme based on nonlinear iterative force correction has been employed \cite{NIFC_1}. The temporal discretization of both the fluid and the structural equations is embedded by energy
conservative implementation of the generalized-$\alpha$ framework \cite{Gen_alpha}.

The equations are linearized via the Newton-Raphson technique and are then solved in a predictor-corrector format. Further details about the coupling procedure for the two-phase fluid-structure interaction problems can be found in \cite{Joshi_IJNME}. While the displacement after solving the structure equations forms a predictor step, the transfer of corrected forces via the NIFC algorithm is a corrector step in a particular nonlinear iteration. The increments of the velocity and pressure fields in the linearized system of equation are then evaluated by the Generalized Minimal Residual (GMRES) algorithm \cite{saad1986}. The left-hand side matrix is not constructed explicitly for this procedure, but we perform matrix-vector products of each block matrix for the GMRES algorithm. On the other hand, the mesh equation is solved by conjugate gradient method owing to the symmetric property of the left-hand side matrix. The NIFC scheme \cite{NIFC_2,NIFC_1} increases the stability of the fluid-structure coupling for low structure-to-fluid mass ratio regimes. All the variables are interpolated using first-order Lagrange polynomials.

For the parallel computing, the solver relies on a standard master-slave strategy for distributed memory clusters by message passing interface (MPI) which depends on domain decomposition strategy of the computational domain \cite{mpi}. The partition of the mesh is generated by the master process into different subgrids with the help of an automatic graph partitioner \cite{metis}. The master process performs the operations at the root subgrid and all other subgrids behave as the slave processes. The matrices at the element level are computed by the slave processes and then the system of equations is then solved across different compute nodes.
The adopted fluid-structure interaction solver has been extensively validated for a wide range of fluid–structure interaction problems at subcritical Reynolds number 
without free-surface effects \cite{jaiman2016partitioned,li2016vortex,mysa2017,law2018passive,miyanawala2018square}.
A systematic validation of the free-surface interaction with floating objects is provided in \cite{Joshi_IJNME}.

\section{Convergence study and validation}
\label{Conv_Val}
Before we proceed to a detailed analysis of sphere VIV with free-surface effects, we first perform a convergence study and validate the solver by comparing with the experimental and available numerical data.  
The definitions of some relevant important non-dimensional parameters are summarized in Table \ref{Parameters}. 
The non-dimensional amplitude response $A^*$ is defined as ${A^*=A/D}$ and $f^*$ denotes the normalized frequency and $f_n=\frac{1}{2\pi}\sqrt{\frac{k}{m}}$ is the natural frequency of the spring-mass system in vacuum, where $m$ is the mass of the sphere and $k$ is the spring stiffness.  The mass ratio is given by ${m}^*={m}/{m_d}$, where $m$ is the mass of sphere and $m_d$ is the mass of displaced fluid.  For the cases of partially submerged bodies in the flow field, the mass ratio would be increased  as the mass of the displaced fluid is reduced. 
In our numerical analysis based on the coupling of incompressible Navier–Stokes and rigid body equations, we use the natural frequency $f_n$ in vacuum for the purpose of non-dimensionalization. During this fluid-structure coupling cycle, the added mass effect is implicitly accounted in the coupled formulation and the response results are appropriately adjusted to match the experimental conditions \cite{sareen2018}. 
%
%
%
%
The normalized horizontal and transverse forces are evaluated from the fluid traction, acting on the structural body, where $C_x$ is the normalized horizontal force, $C_y$ and $C_z$ are the normalized transverse forces in $y$ and $z$ directions, respectively. The normalized force coefficients are evaluated as follows
\begin{align}
C_x=\frac{1}{\frac{1}{2}\rho U^2 S}\int_{\Gamma} (\bar{\boldsymbol{\sigma}}^\mathrm{f}\cdot \boldsymbol{n})\cdot \boldsymbol{n_x} \mathrm{d\Gamma} \\
C_y=\frac{1}{\frac{1}{2}\rho U^2 S}\int_{\Gamma} (\bar{\boldsymbol{\sigma}}^\mathrm{f}\cdot \boldsymbol{n})\cdot \boldsymbol{n_y} \mathrm{d\Gamma} \\
C_z=\frac{1}{\frac{1}{2}\rho U^2 S}\int_{\Gamma} (\bar{\boldsymbol{\sigma}}^\mathrm{f}\cdot \boldsymbol{n})\cdot \boldsymbol{n_z} \mathrm{d\Gamma} 
\end{align}
where $\boldsymbol{n_x}$, $\boldsymbol{n_y}$ and $\boldsymbol{n_z}$ are the Cartesian components of the unit normal $\boldsymbol{n}$ to the sphere surface, and $S$ is the relevant surface area which is defined as $S=\pi D^2/4$. 


The transfer of energy between the flow and oscillating sphere can be characterized by means of the normalized transverse force in phase with the sphere velocity. Thus the non-dimensional time-averaged quantity of fluid-structure energy transfer $E$ over a period $T$ of motion can be expressed as 
\begin{align}
E=\int_{t}^{t+T} C_y(t)\left(\frac{u_y}{U} \right)\ dt 
\end{align}
where $u_y$ is the transverse component of the sphere velocity. To identify the frequencies and the direction of the energy transfer, we make use of the time-dependent energy coefficient ($C_E$), which can be given as
\begin{align}
C_E(t) = C_y(t)\bar{u}_y,\ \ \ \ \ \  \bar{u}_y=\left(\frac{u_y}{U} \right)  
\end{align}
The sign of $C_E$ demonstrates the relative direction between the transverse force and velocity, hence the direction of energy transfer between the rigid body and the fluid flow. While the positive value ($C_E > 0$) represents the supply of energy from the flow to the structure, the flow damps the body oscillations for the negative value ($C_E < 0$). 

\renewcommand{\arraystretch}{0.5}
\begin{table}[htbp!]
	\caption{Definition of the non-dimensional parameters and post-processing quantities }
	\centering
	\begin{tabular}{  M{6cm}  M{6cm}  N }
		\hline
		\raggedright
		Parameter & \raggedright Definition & \\[10pt]
		\hline
		\raggedright Reynolds number  & \raggedright $Re={\rho ^\mathrm{f} U D}/{\mu ^\mathrm{f}}$   & \\[10pt]
		\raggedright Reduced velocity  & \raggedright $U^*={U}/{f_n D}$   & \\[10pt]
		\raggedright Mass ratio  & \raggedright ${m}^*={m}/{m_d}$   & \\[10pt]
		\raggedright Damping ratio  & \raggedright $\zeta=c/2\sqrt{mk}$   & \\[10pt]
		\raggedright Froude Number  & \raggedright $Fr=U/\sqrt{gD}$   & \\[10pt]
		\raggedright Non-dimensional amplitude  & \raggedright ${A}^*_\mathrm{rms}=\sqrt{2}{A_\mathrm{rms}/D}$   & \\[10pt]
		\raggedright Immersion ratio  & \raggedright $h^*=h/D$   & \\[10pt]
		\raggedright Normalized horizontal force  & \raggedright $C_x={f}^{\mathrm{s}}_x/(\frac{1}{2}\rho U^2 S)$   & \\[10pt]
		\raggedright Normalized transverse force  & \raggedright $C_y={f}^{\mathrm{s}}_y/(\frac{1}{2}\rho U^2 S)$   & \\[10pt]
		\raggedright Normalized vertical force  & \raggedright $C_z={f}^{\mathrm{s}}_z/(\frac{1}{2}\rho U^2 S)$   & \\[10pt]
		\raggedright Normalized frequency  & \raggedright $f^*=f/f_n$   & \\[10pt]
		
		\hline
		
	\end{tabular}
	\label{Parameters}
\end{table}

\subsection{VIV of fully submerged freely vibrating elastically mounted sphere}
At very low Reynolds numbers $(Re \leq 200 )$, the flow past a stationary sphere is steady and axisymmetric
but it loses the axisymmetry first and then the steadiness with increasing of Reynolds number \cite{johnson1999sphere,mittal1999sphere}.
While a pair of streamwise vortices are formed behind the sphere without shedding $(210 \leq \mathrm{Re} \leq 270)$,
hairpin-shaped vortices are periodically shed with the same strength in a fixed orientation for the unsteady planar-symmetric flow $(280 \leq \mathrm{Re}<375)$.
As the Reynolds number is increased further, the strength and shedding orientation of the hairpin vortices vary in time and thus the flow becomes asymmetric $(375 \leq \mathrm{Re}<800)$ \cite{tomboulides2000sphere}.
In the case of the subcritical flow over a stationary sphere (non-lock-in condition), there exists the high-frequency mode associated with the small-scale instability of the separating shear layer and the low-frequency mode related to the large-scale instability of the wake due to the vortex shedding \cite{taneda1978}.

\begin{figure}[!h]
	\centering
	\adjincludegraphics[scale=0.3,trim={0\width} {0\width} {0\width} {0.0\width},clip]{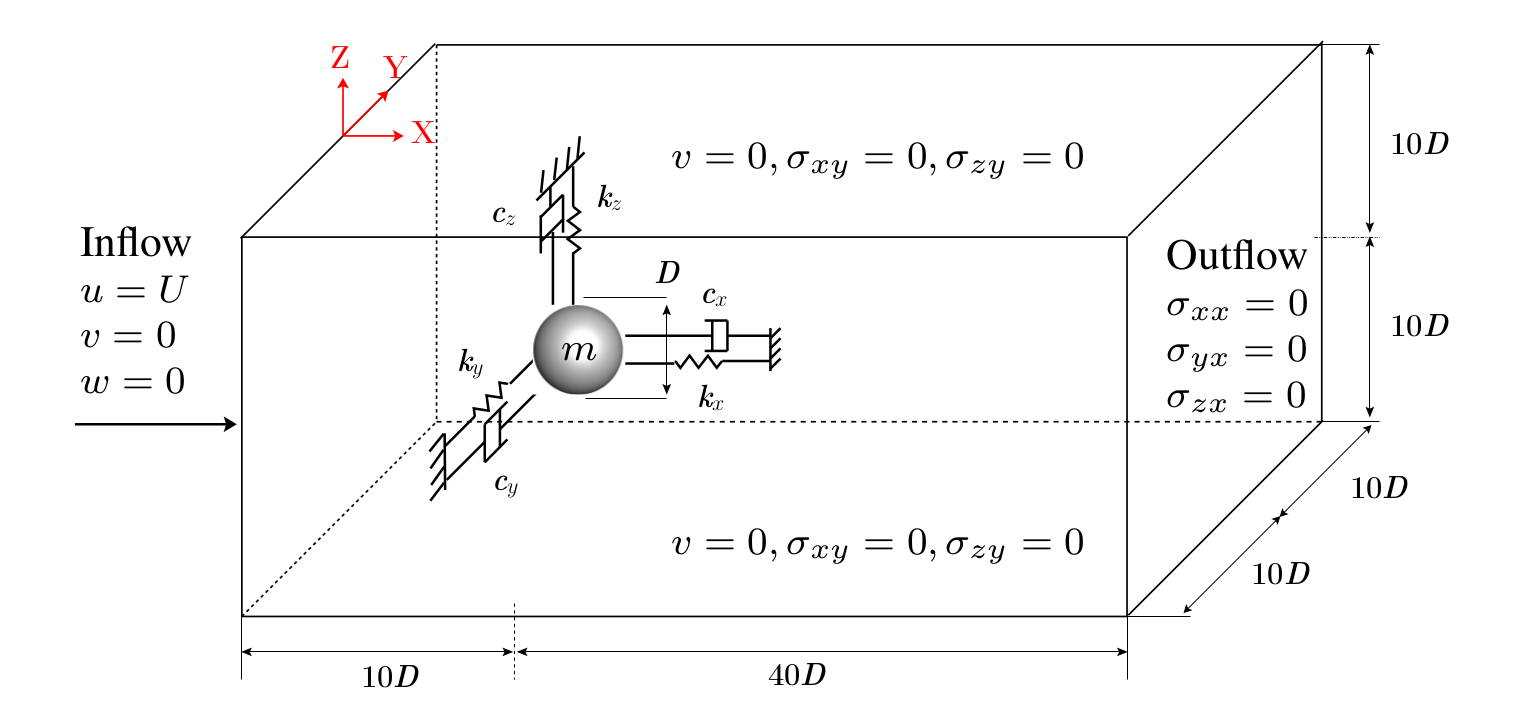}
	\caption{Schematic and associated boundary conditions of the fluid flow past a fully submerged elastically mounted sphere with 3-DOF.
	}
	\label{domain_3d}		
	
\end{figure}

For the present VIV study, a representative case of a fully submerged sphere that is free to translate in all spatial directions is considered for the grid convergence study.
Fig. \ref{domain_3d} depicts a three-dimensional computational domain of the size ($50 \times 20 \times 20$)$D$ set up with a sphere of diameter $D$ placed at an offset of $10D$ from the inflow surface. The origin of the coordinate system is fixed at the center of the sphere.
We consider the $x$-axis as streamwise flow direction, the $y$-axis in horizontal and perpendicular to the flow direction, and the $z$-axis is the vertical direction.
While the streamwise motion corresponds to the freestream ($x$-direction), the transverse motion is parallel to the $y$-direction. A uniform freestream flow with velocity $U$ is along the $x$-axis. At the inlet boundary, a stream of water enters into the domain at a horizontal velocity $(u, v, w)=(U, 0, 0)$ where $u$, $v$ and $w$ denote the streamwise, transverse and vertical velocities in $x, y$ and $z$ directions, respectively. The sphere is elastically mounted on springs with a stiffness value of $k$ and linear dampers with a damping value of $c$ in all three spatial directions. We have considered the slip-wall boundary condition along the top, bottom and side surfaces, in addition to the Dirichlet and traction free Neumann boundary conditions along the inflow and outflow boundaries, respectively.

\begin{figure}[!h]
	\begin{subfigure}[b]{0.65\textwidth}
		\centering
		\adjincludegraphics[scale=0.17,trim={0\width} {0\width} {0\width} {0.0\width}]{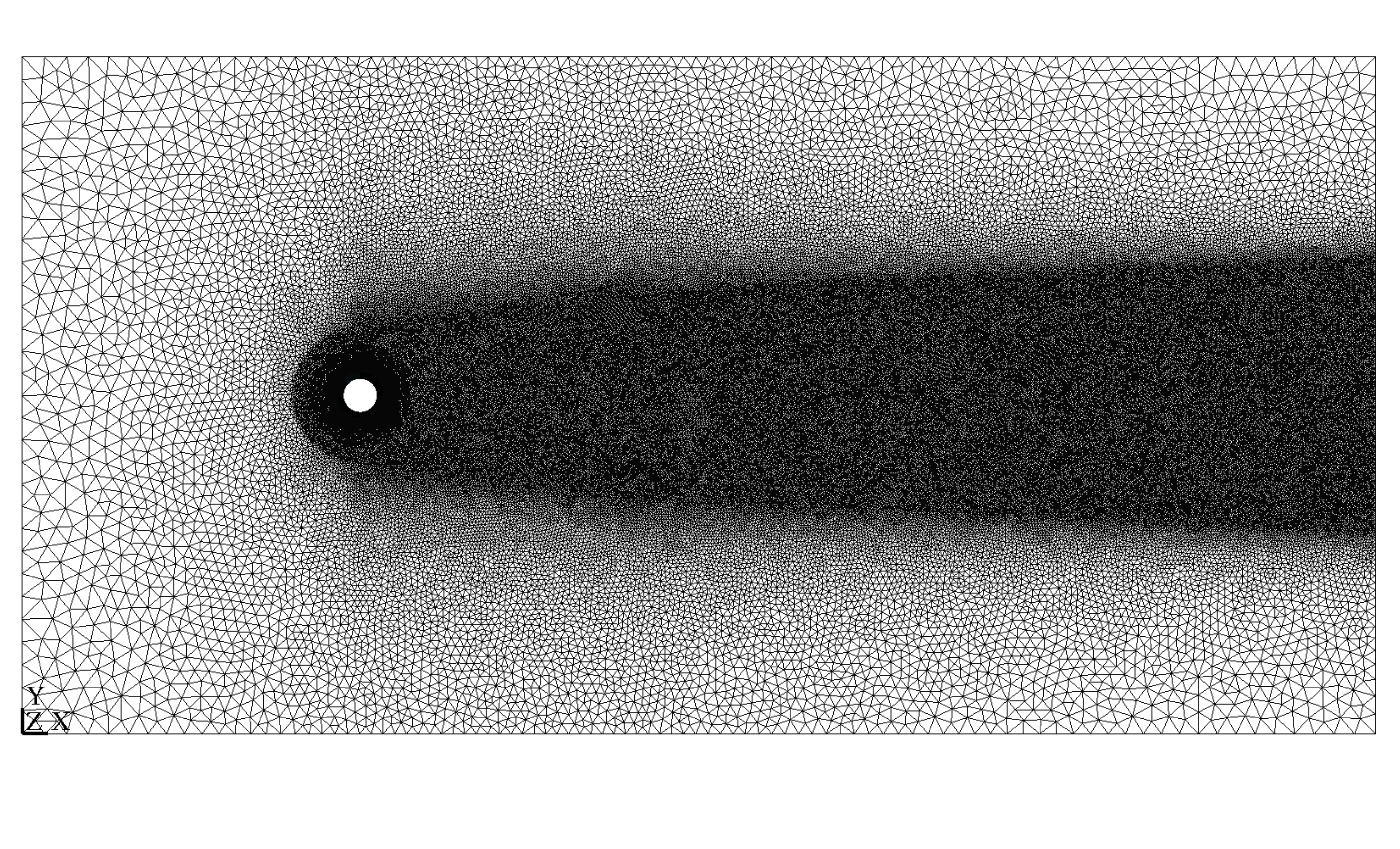}
		\caption{}
	\end{subfigure}%
	\begin{subfigure}[b]{0.35\textwidth}
		\centering
		\adjincludegraphics[scale=0.135,trim={{0.15\width} {0.0\width} {0.22\width} {0.0\width}},clip]{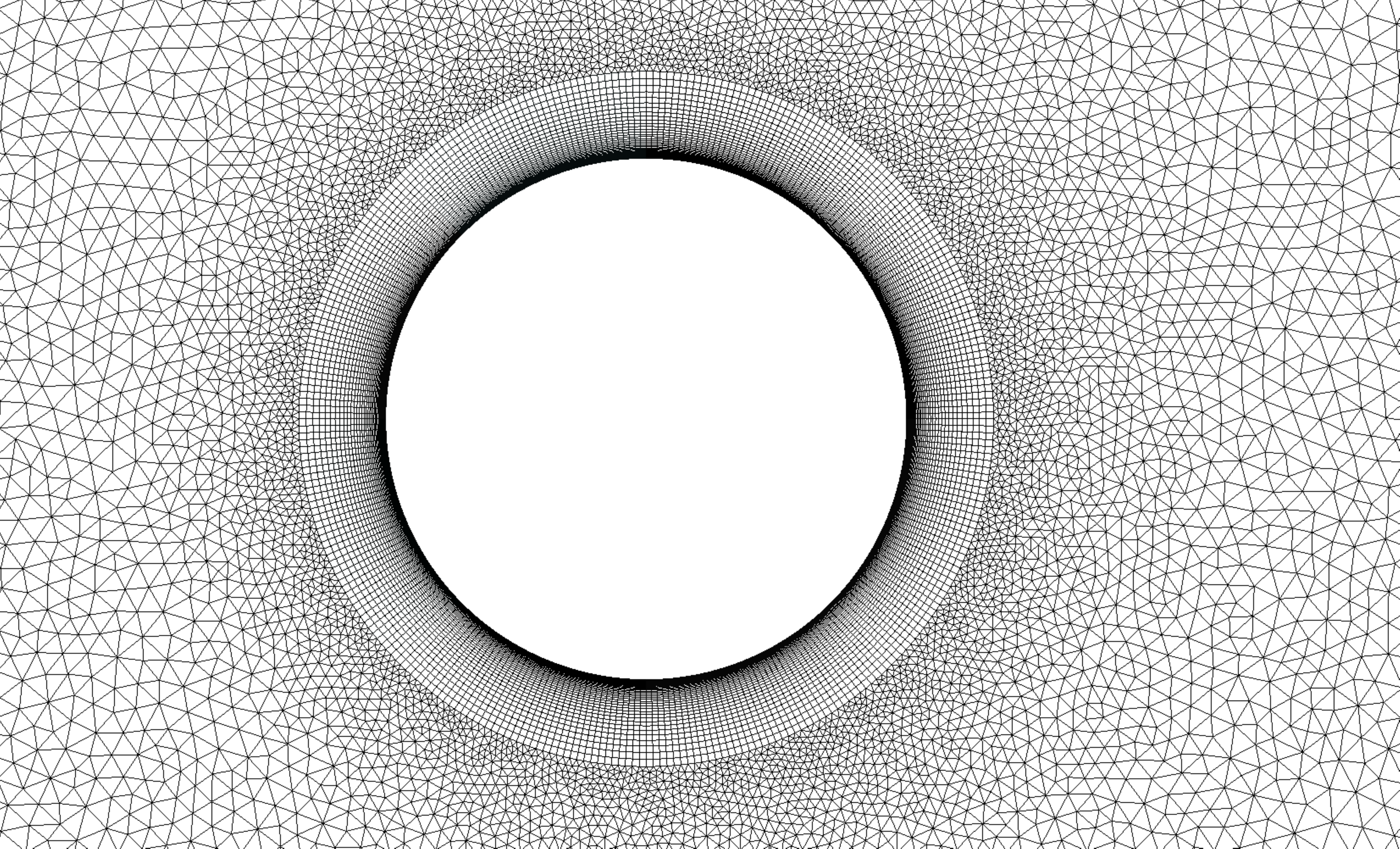}
		\vspace{8.5mm}
		\caption{}
	\end{subfigure}%
	\caption{A representative computational mesh employed for an elastically mounted sphere in a uniform steady flow: (a) two-dimensional slice of the mesh for the entire domain along the $X$-$Y$ plane, 
and (b) zoomed view of mesh in the vicinity of the sphere.}
	\label{Msh_Sphere}		
\end{figure}

\renewcommand{\arraystretch}{0.5}
\begin{table}[!h]
	\caption{Mesh statistics and convergence study of the flow parameters for VIV of a 3-DOF elastically mounted sphere with the mass ratio $m^*=3.82$ at $Re=2\,000$ and $U^*=9$. The error deviation is evaluated based on the corresponding value of M4 mesh. The r.m.s. value of the sphere dimensionless amplitude response in $y$-direction, $A^*_{y,\mathrm{rms}}$, the r.m.s. value of the normalized transverse force in $y$ direction, $C_{y,\mathrm{rms}}$ and normalized mean horizontal force, $\overline{C}_{x}$ are also recorded.}
	
	\centering
	\begin{tabular}{  M{3.5cm}  M{2.5cm}  M{2.5cm}   M{2.5cm} M{2.5cm} N }
		\hline
		\centering
		Mesh & Nodes ($\times 10^6 $) & $A^*_{y,\mathrm{rms}} $ & $C_{{y,\mathrm{rms}}} $ & $\overline{C}_{x} $ &\\[10pt]
		\hline
		\centering
		M1  & 0.637   &  0.758 (13.6\%)   & 0.029 (25.6\%)   & 0.947 (9.4\%) &\\[10pt]
		\centering
		M2  & 1.330   &  0.681 (2.0\%)   &  0.0383 (1.7\%) & 0.885 (2.3\%) &\\[10pt]
		\centering
		M3  & 2.963   &  0.682 (0.7\%)    &  0.0392 (0.6\%)  & 0.876 (1.2\%) &\\[10pt]
		\centering
		M4  & 5.947   &   0.667    &  0.0390   &  0.865 &\\[10pt]
		\hline
		\centering
		Rajamuni et al. \cite{rajamuni2019}& 1.2$\times 10^6$ Cells &\hspace{2 mm}0.61&0.041&0.77&\\[10pt]
		\hline
		
	\end{tabular}
	\label{Statistics}
\end{table}

Fig. \ref{Msh_Sphere} shows a representative computational mesh for the simulations which contains structured prismatic six-node wedge elements at boundary layer region and unstructured four-node tetrahedral elements elsewhere. For the spatial convergence, to maintain the accuracy of the boundary layer dynamics, the refinement is kept such that the non-dimensional wall unit $y^+$ remains less than $1$. In this study, we have performed a grid convergence study for a freely vibrating sphere in all spatial directions. The maximum amplitude response ($A^*$) of the 3-DOF fully submerged vibrating sphere at VIV regime is observed at $U^* \approx 9$ for $m^*=3.82$ \cite{rajamuni2019}. The grid convergence study is carried out at $U^*=9$ and $Re=2\,000$ for four different grid size domains. The spatial convergence with the mesh details are given in Table \ref{Statistics}. By considering M4 as the reference, the differences between the amplitude and force resulted from M1, M2, and M3 and those from M4 are thus calculated and noted in the corresponding brackets. The differences between the results are approximately within 1\% for the two finer meshes, M3 and M4. Considering computational efficiency, M3 mesh is selected for the validation with the experimental results and all the simulations in this study. Here we aim to examine the VIV of the 3-DOF fully submerged sphere for a higher range of Reynolds number up to $30\,000$. In particular, we verify and examine the effect of Reynolds number on the mode transition.

\begin{figure}[htbp!]
	\centering
	\includegraphics[scale=0.35]{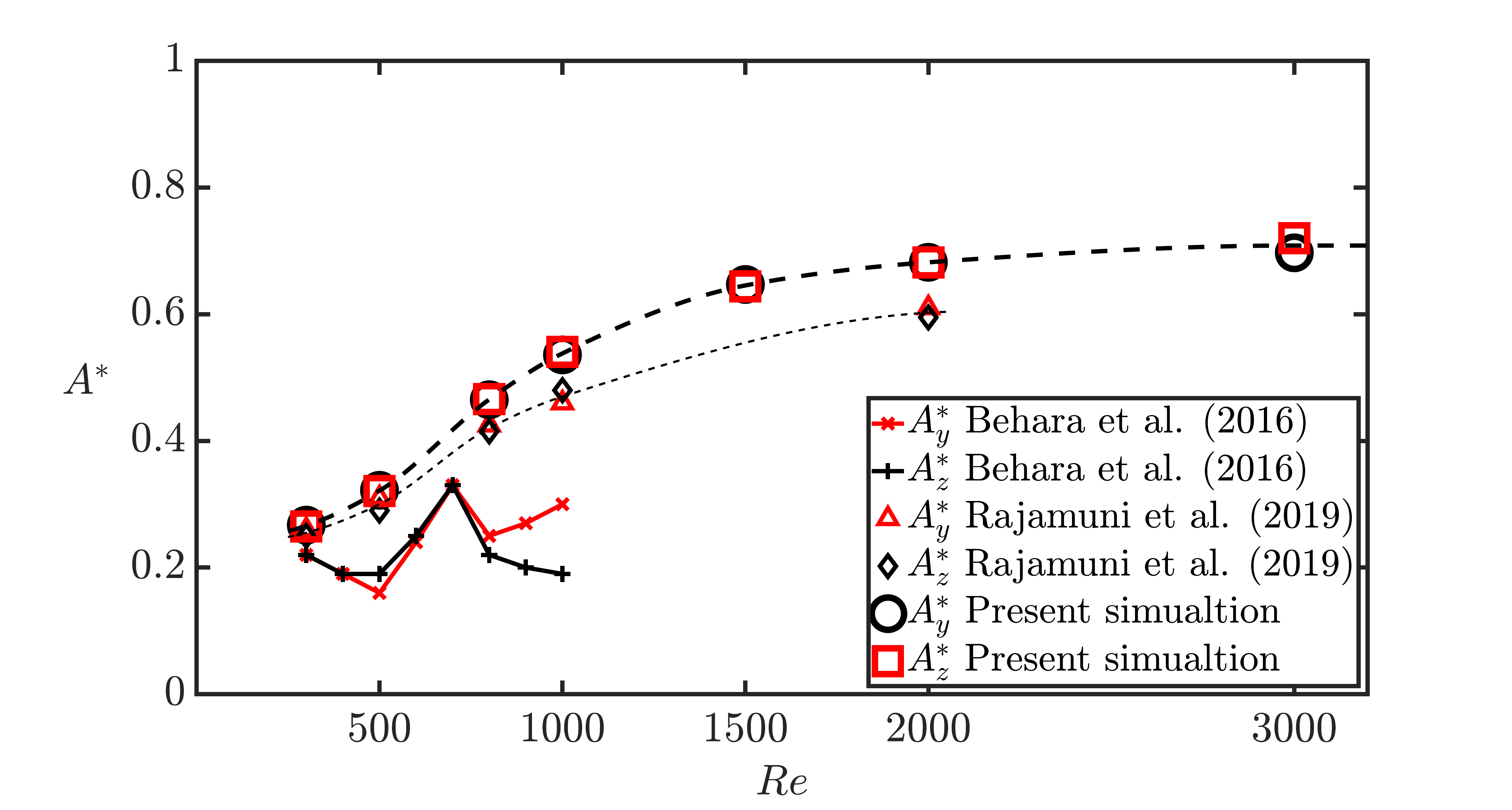} 
	\caption{Dependence of transverse ($A^*_y$) and lateral ($A^*_z$) amplitudes on Reynolds number for elastically mounted 3-DOF sphere at reduced velocity $U^*=9$ and mass ratio $m^*=3.82$. The response amplitudes are contrasted with the numerical results of \cite{behara2016} and \cite{rajamuni2019}. }
	\label{3d_Trend_Re}
\end{figure}

\begin{figure}[htbp!]
	\centering
	
	\begin{subfigure}[b]{0.65\textwidth}
		\adjincludegraphics[scale=0.2,trim={{0\width} 0 {0\width} 0},clip]{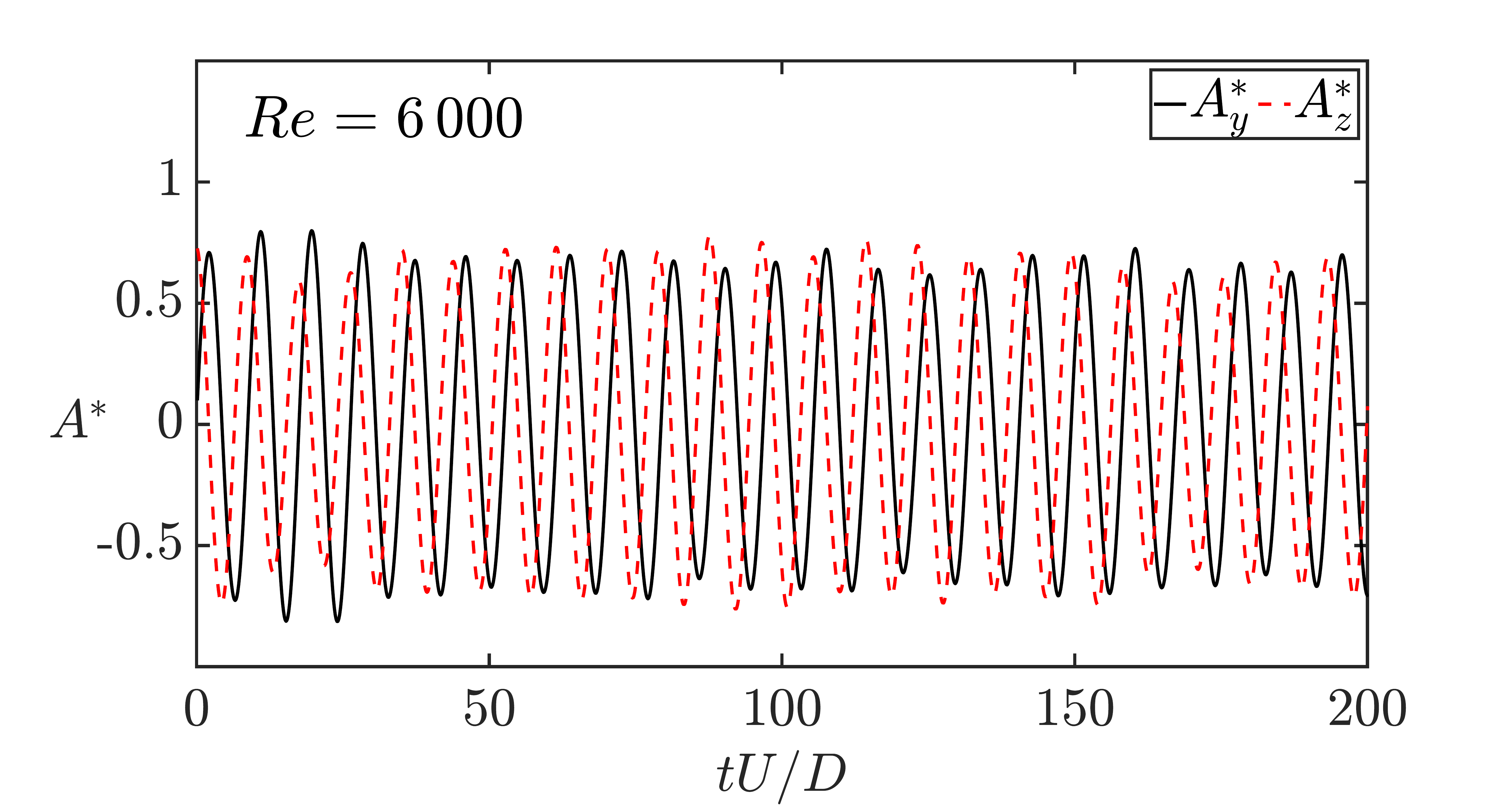}
		\caption*{$(a$ - $1)$}
	\end{subfigure}%
	\begin{subfigure}[b]{0.4\textwidth}
		\adjincludegraphics[scale=0.2,trim={0.2\width} {0\width} {0.18\width} {0.0\width},clip]{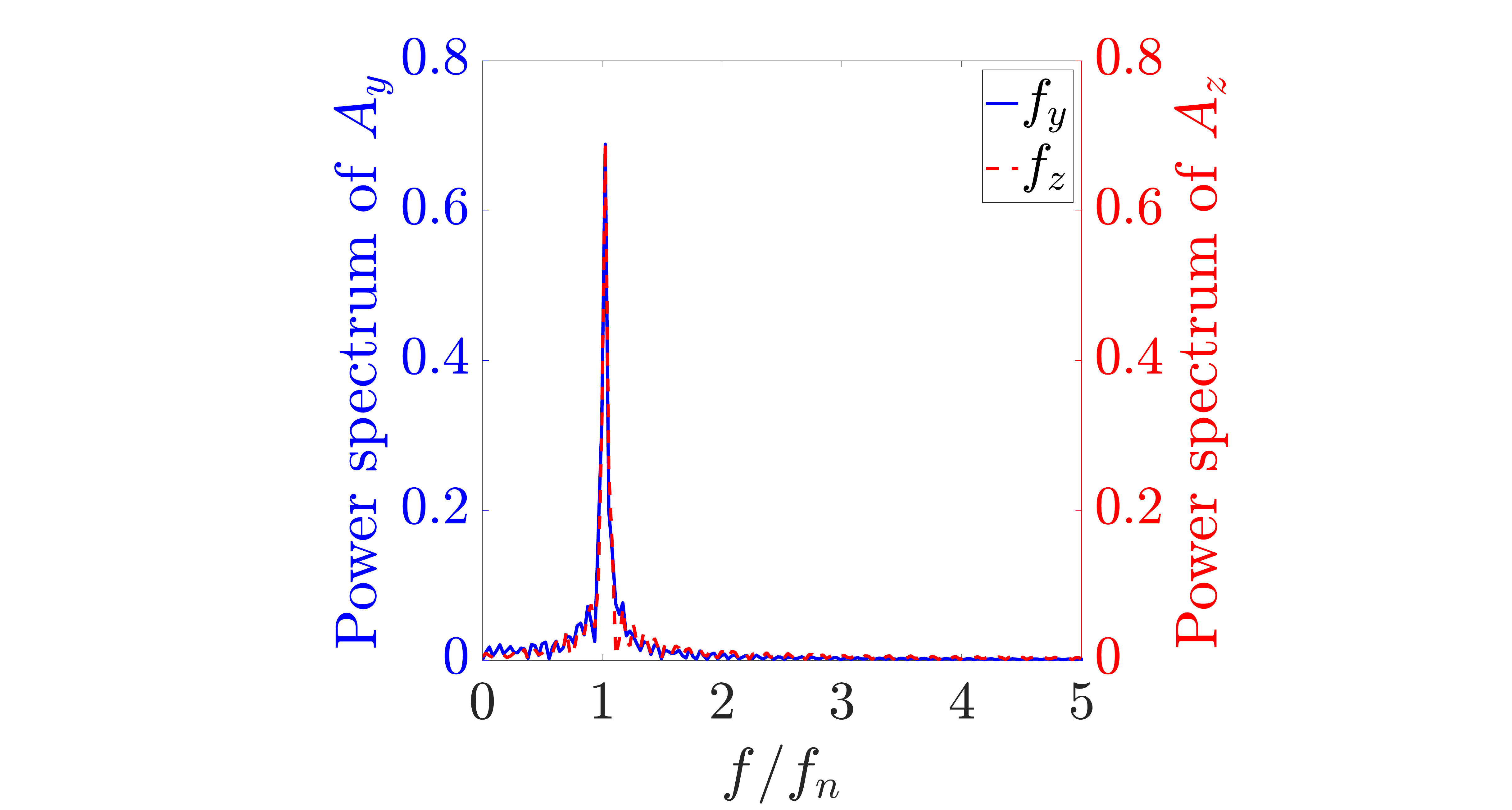}
		\caption*{$(a$ - $2)$}
	\end{subfigure}
	
	\begin{subfigure}[b]{0.65\textwidth}
		\adjincludegraphics[scale=0.2,trim={{0\width} 0 {0\width} 0},clip]{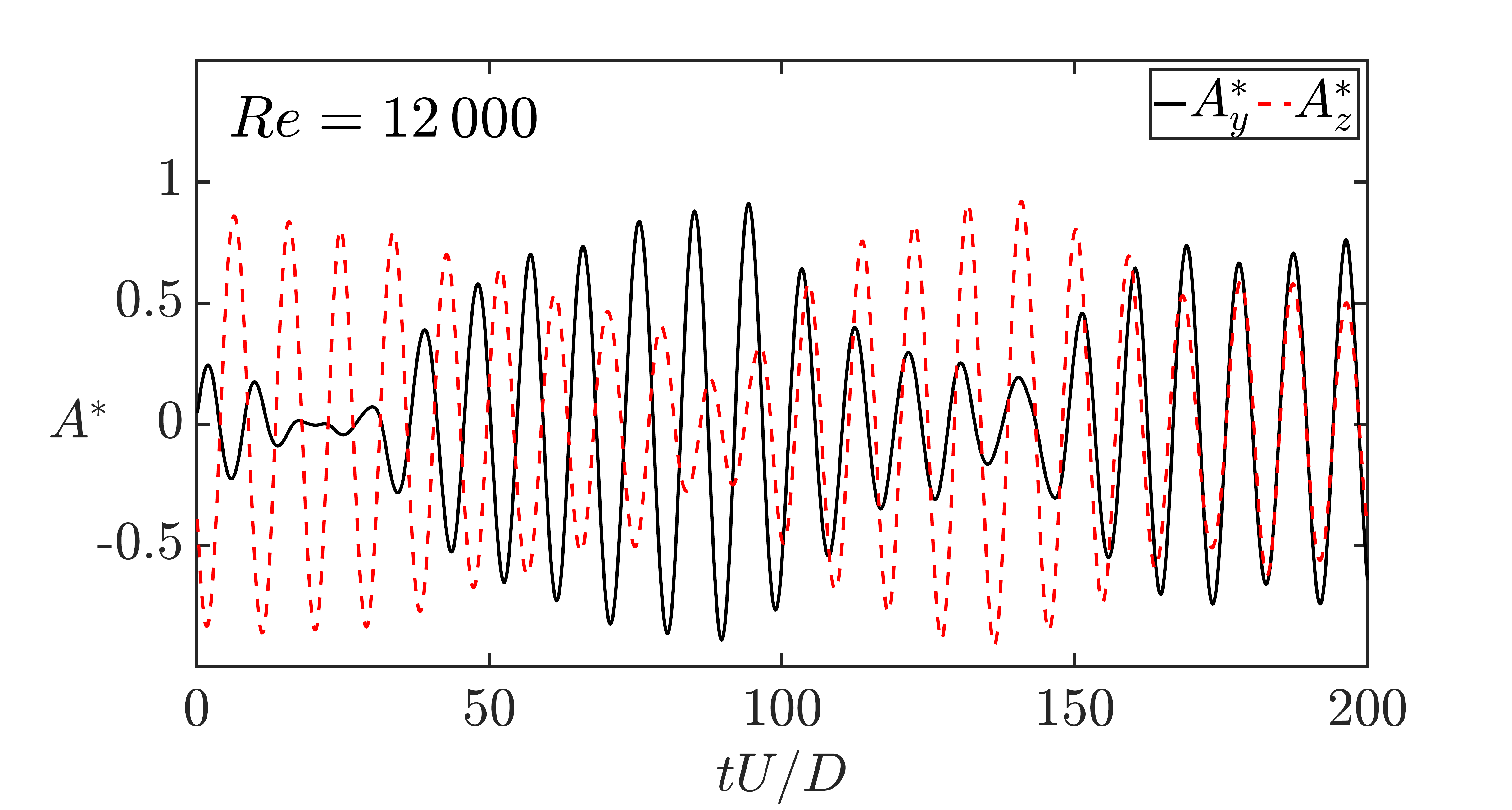}
		\caption*{$(b$ - $1)$}
	\end{subfigure}%
	\begin{subfigure}[b]{0.4\textwidth}
		\adjincludegraphics[scale=0.2,trim={0.2\width} {0\width} {0.18\width} {0.0\width},clip]{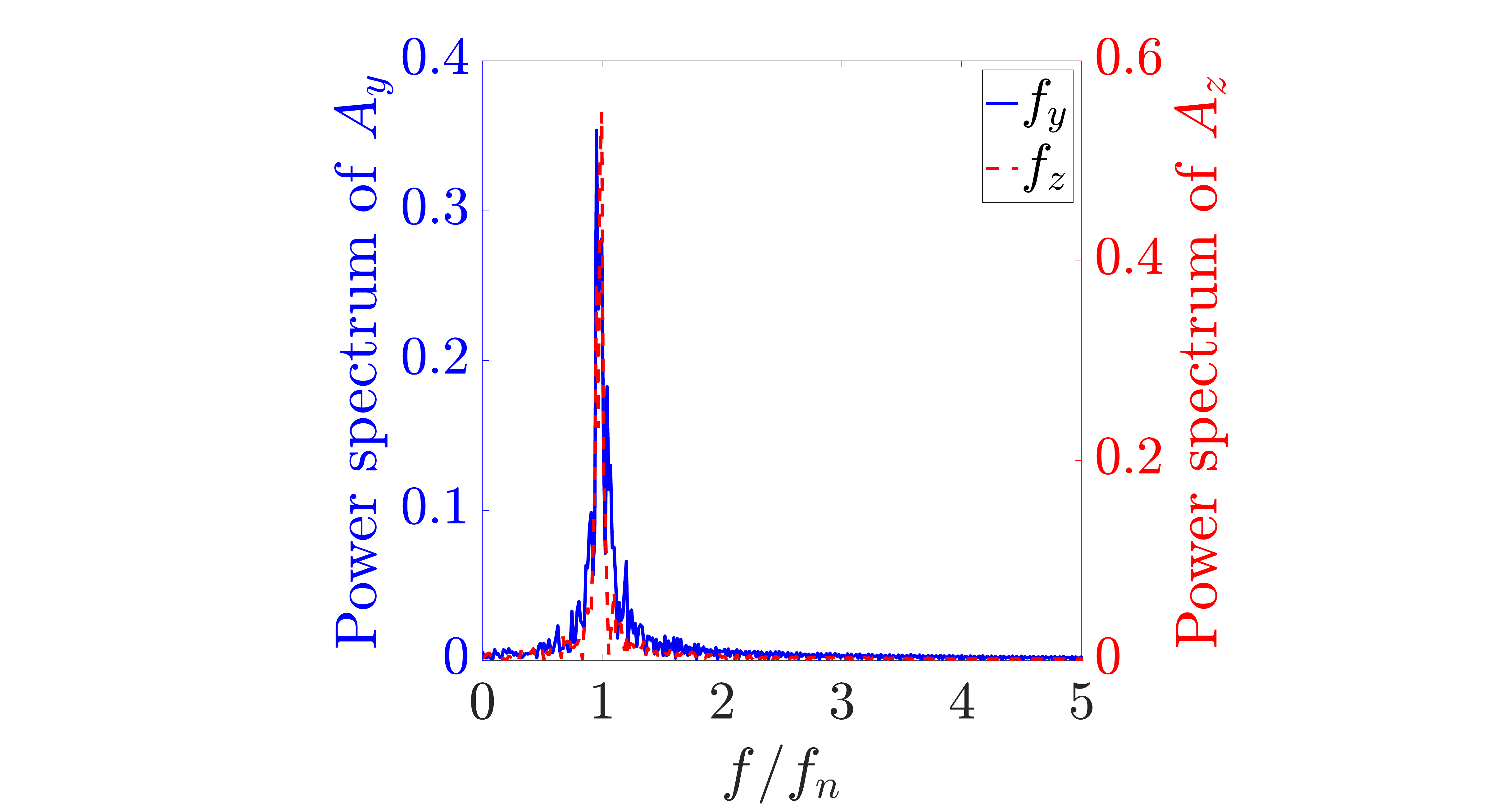}
		\caption*{$(b$ - $2)$}
	\end{subfigure}
	
	\begin{subfigure}[b]{0.65\textwidth}
		\adjincludegraphics[scale=0.2,trim={{0\width} 0 {0\width} 0},clip]{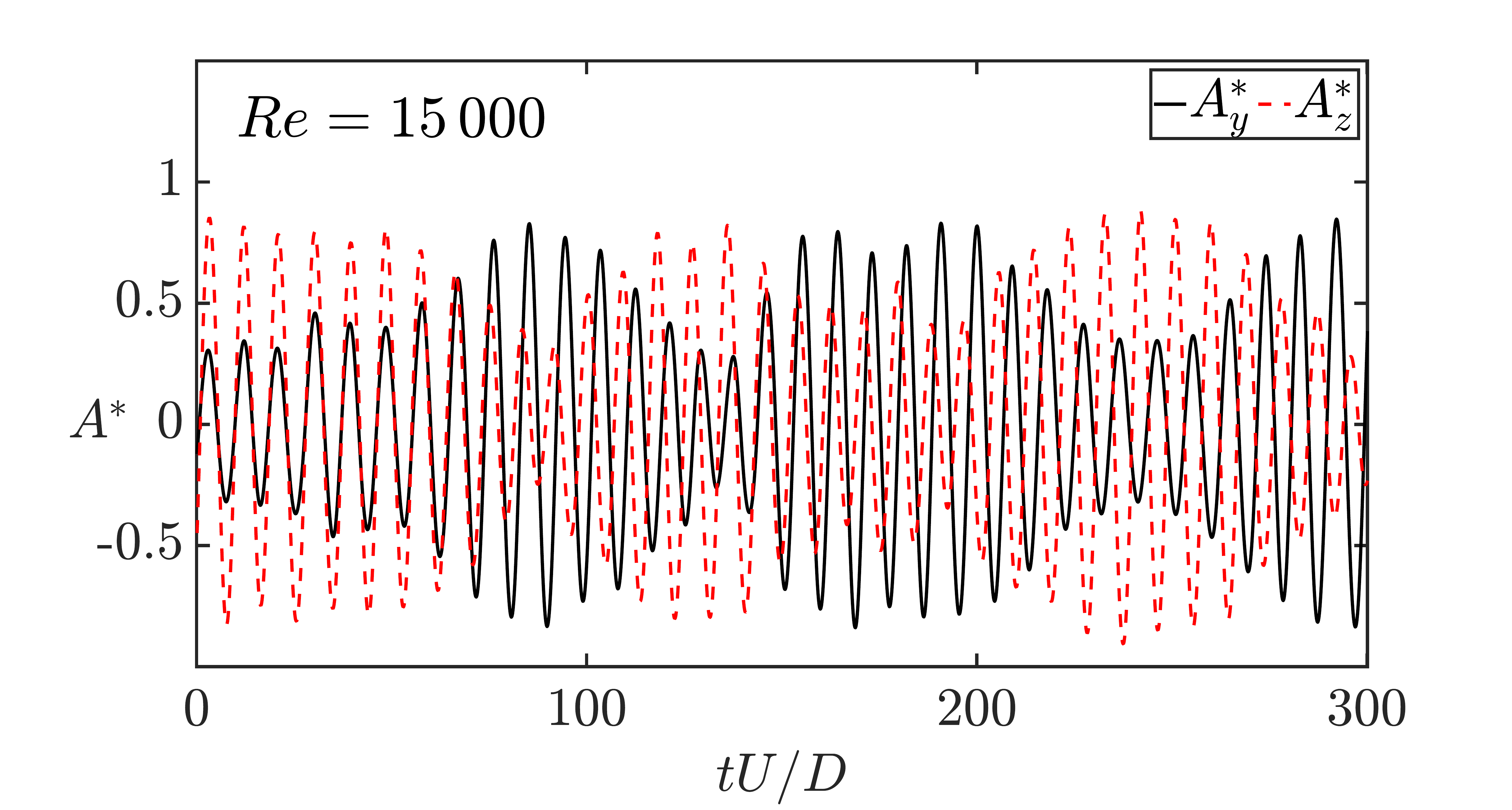}
		\caption*{$(c$ - $1)$}
	\end{subfigure}%
	\begin{subfigure}[b]{0.4\textwidth}
		\adjincludegraphics[scale=0.2,trim={0.2\width} {0\width} {0.18\width} {0.0\width},clip]{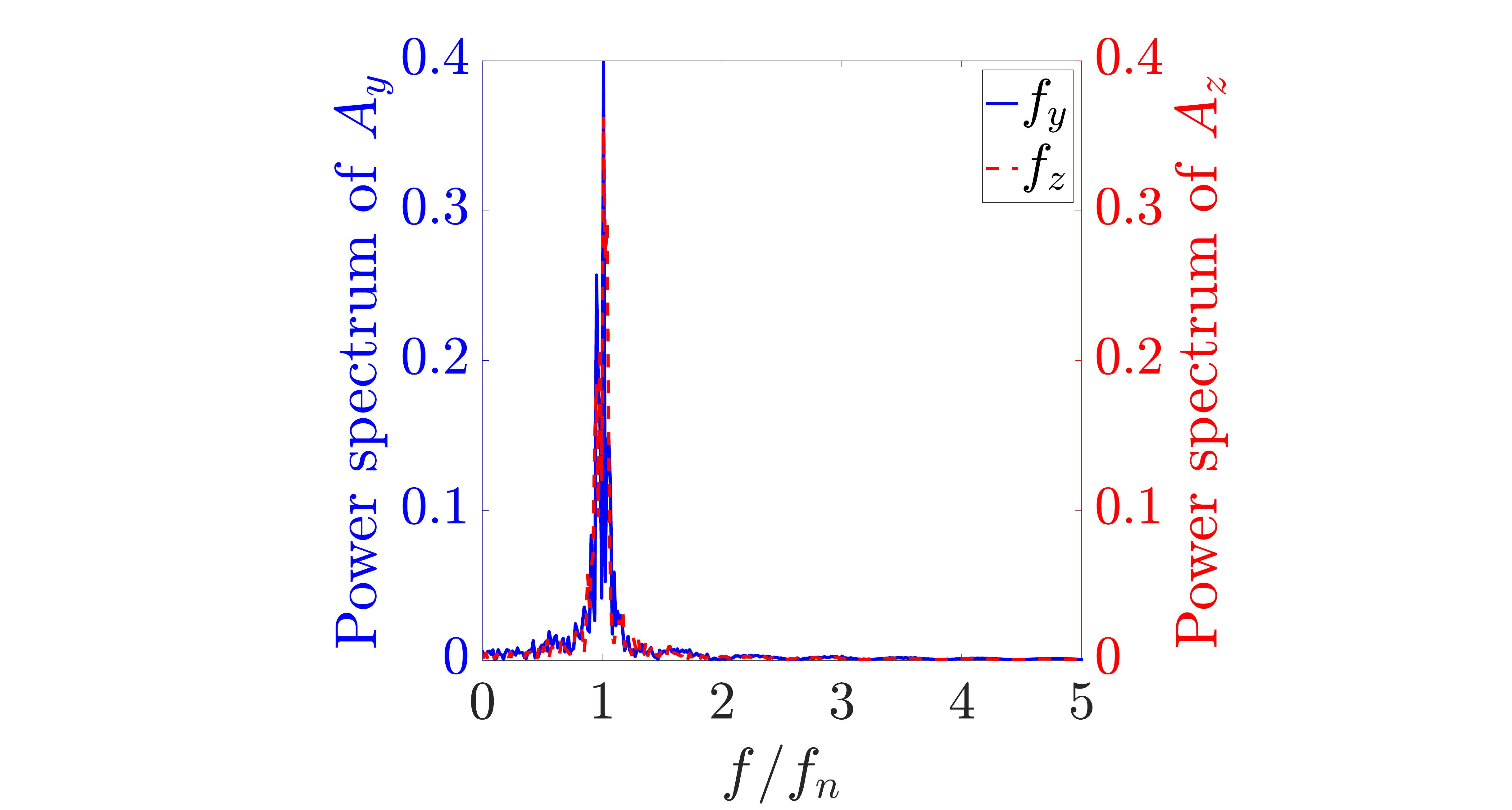}
		\caption*{$(c$ - $2)$}
	\end{subfigure}
	\caption{Time histories of the stationary-state amplitude response of the sphere versus non-dimensional time at $U^*=9$, and the corresponding frequency spectrum of the oscillations in transverse direction ($f_{A_y}$) and vertical direction ($f_{A_z}$) at $Re=6\,000, 12\,000$ and $15\,000$. }
	\label{Time_history_3dof}
\end{figure}

\begin{figure}[htbp!]
	\centering

	\begin{subfigure}[b]{0.25\textwidth}
		\adjincludegraphics[scale=0.15,trim={{0.2\width} {0\width} {0.25\width} 0.2},clip]{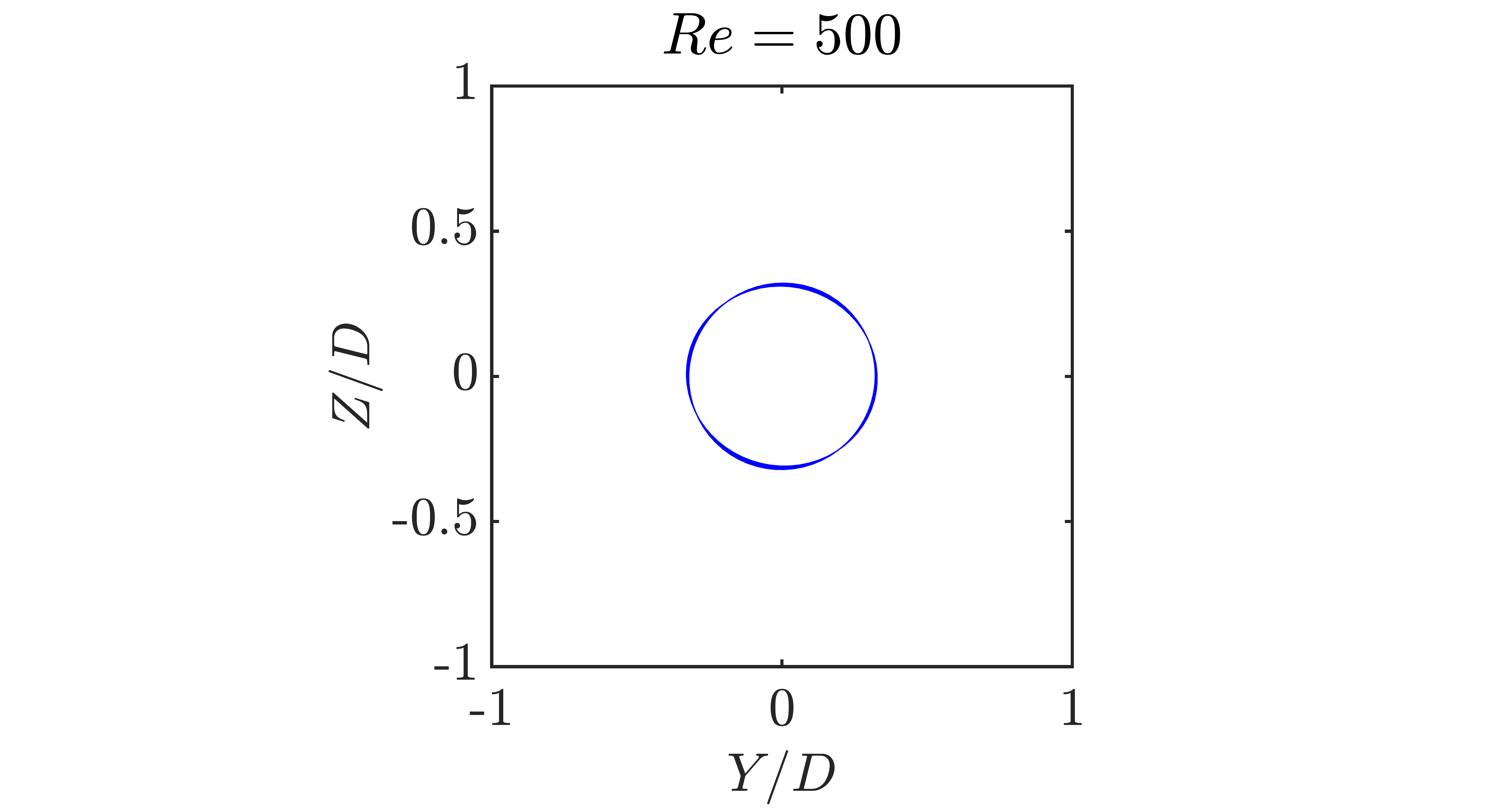} 
	\end{subfigure}%
	\begin{subfigure}[b]{0.25\textwidth}
		\adjincludegraphics[scale=0.15,trim={{0.2\width} {0\width} {0.25\width} 0.2},clip]{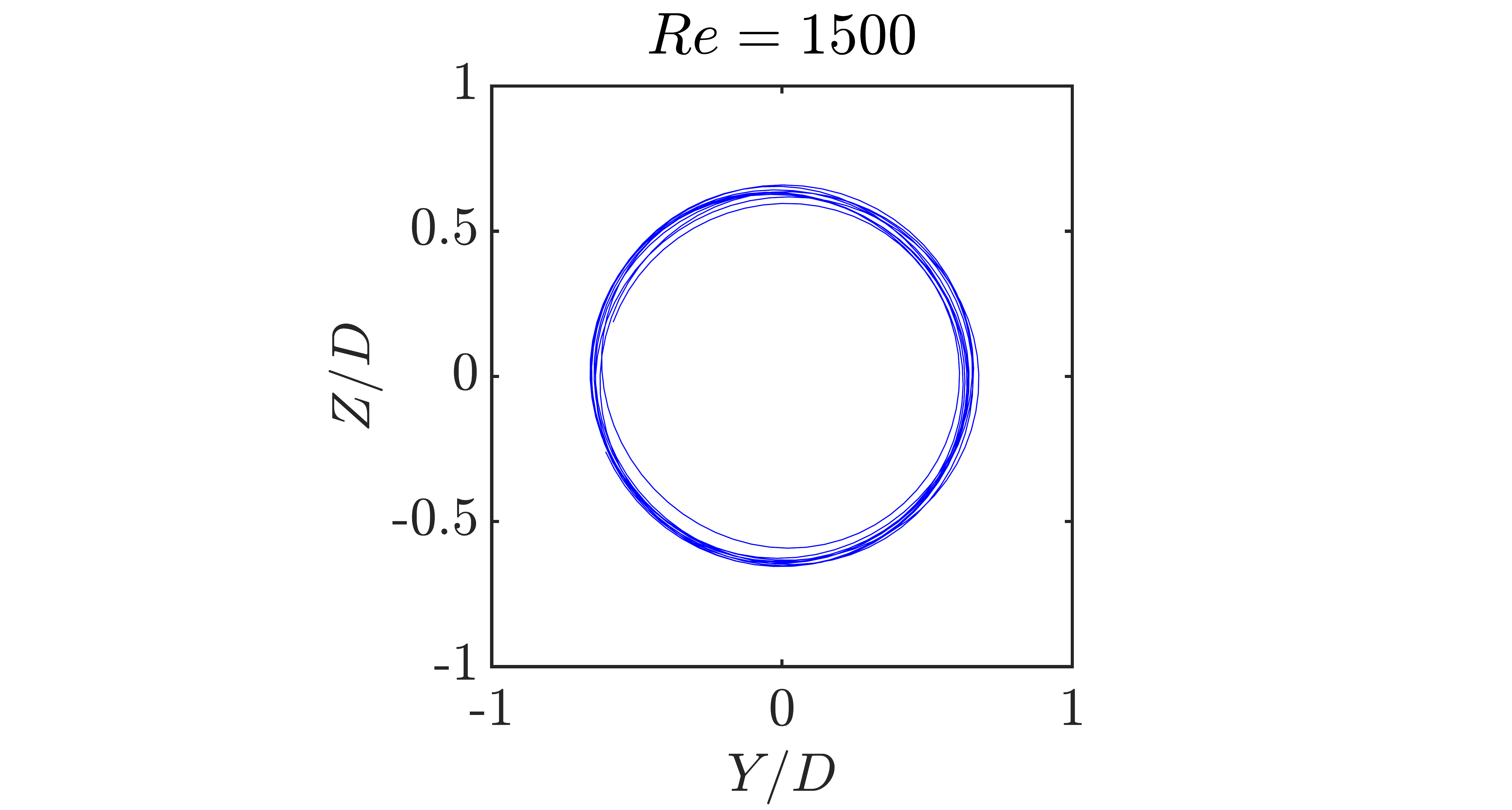} 
	\end{subfigure}%
	\begin{subfigure}[b]{0.25\textwidth}
		\adjincludegraphics[scale=0.15,trim={{0.2\width} {0\width} {0.25\width} 0.2},clip]{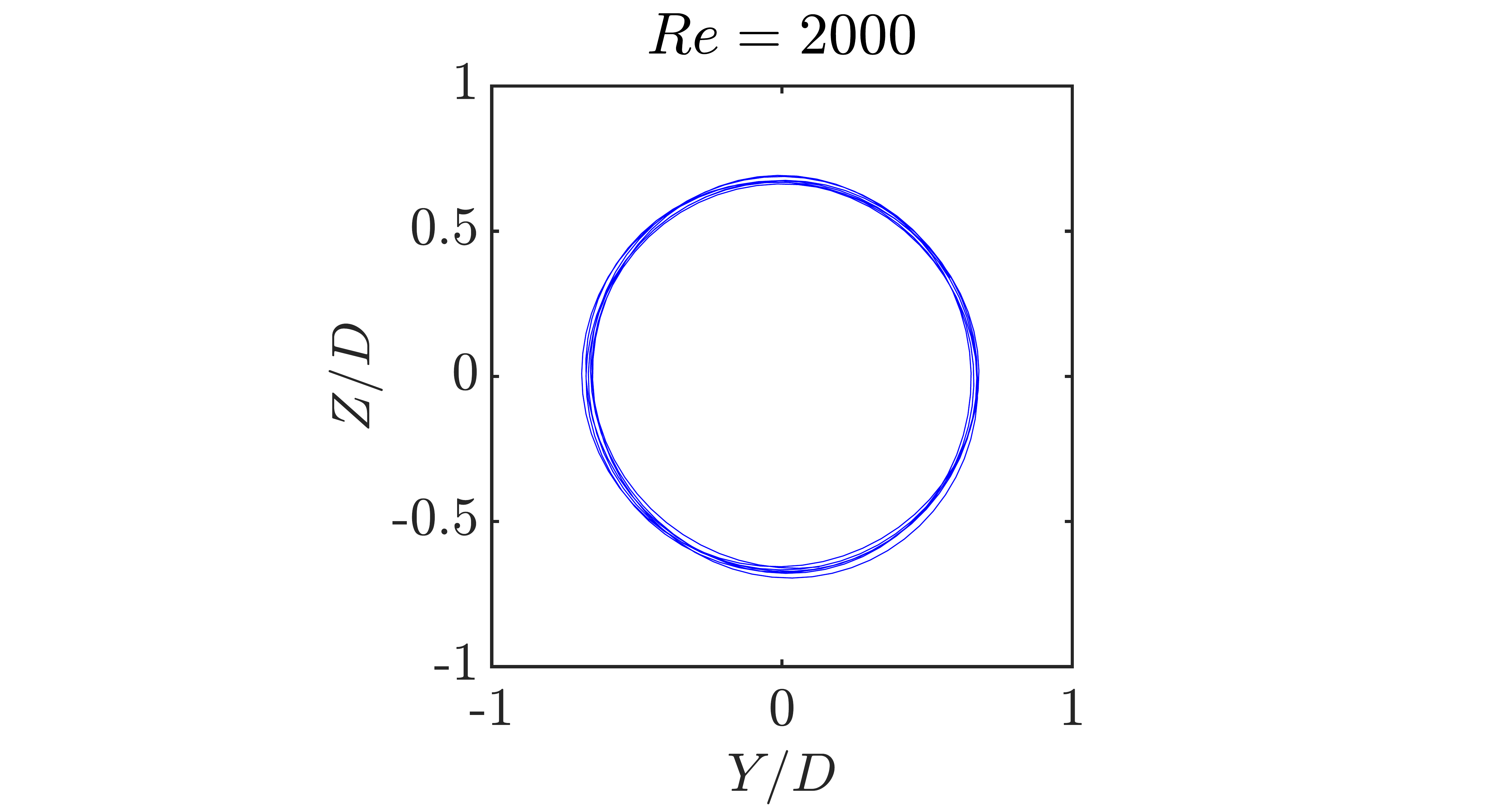} 
	\end{subfigure}%
	\begin{subfigure}[b]{0.25\textwidth}
		\adjincludegraphics[scale=0.15,trim={{0.2\width} {0\width} {0.25\width} 0.2},clip]{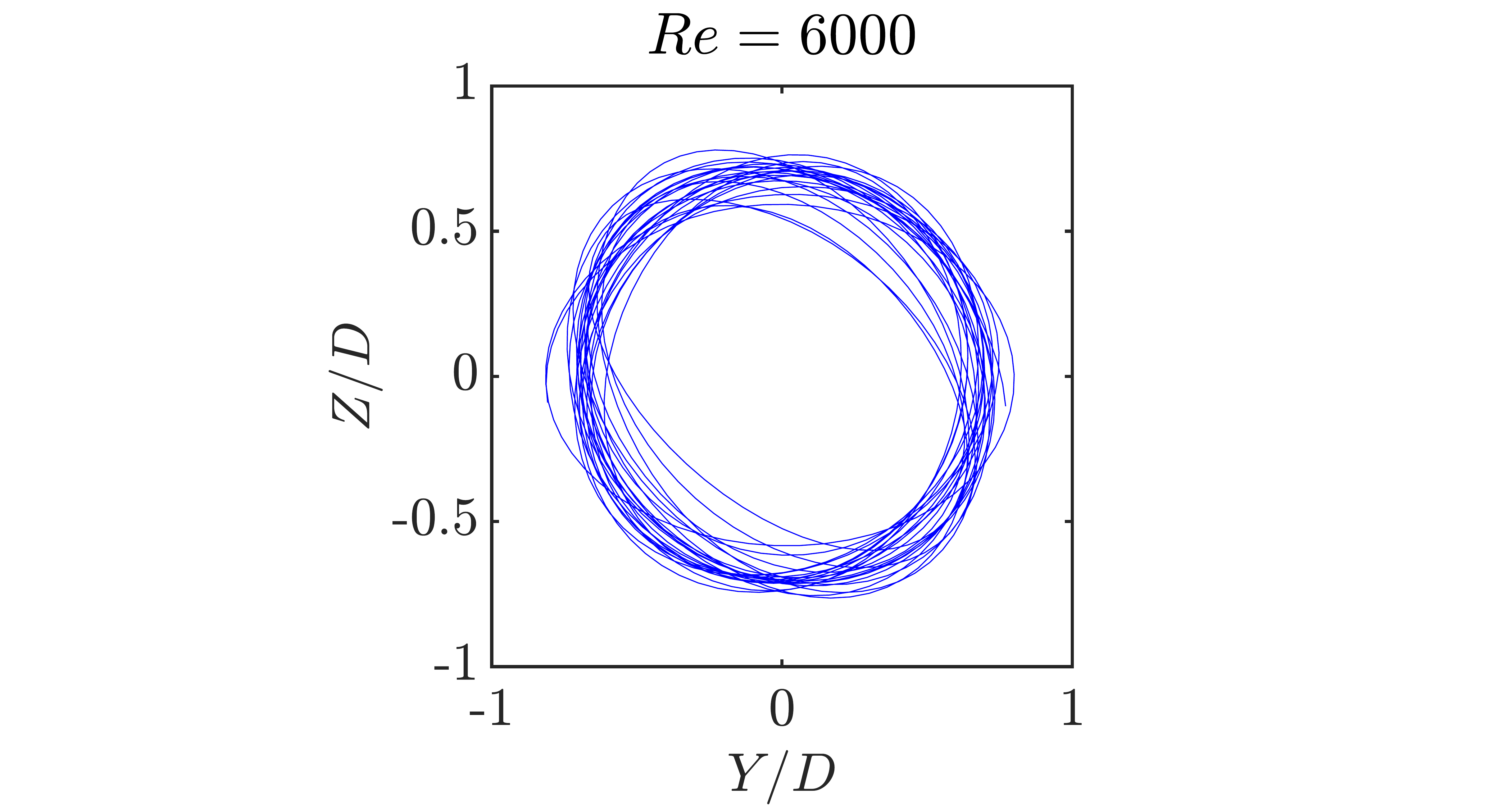} 
	\end{subfigure}
	\begin{subfigure}[b]{0.25\textwidth}
		\adjincludegraphics[scale=0.15,trim={{0.2\width} 0 {0.25\width} 0.2},clip]{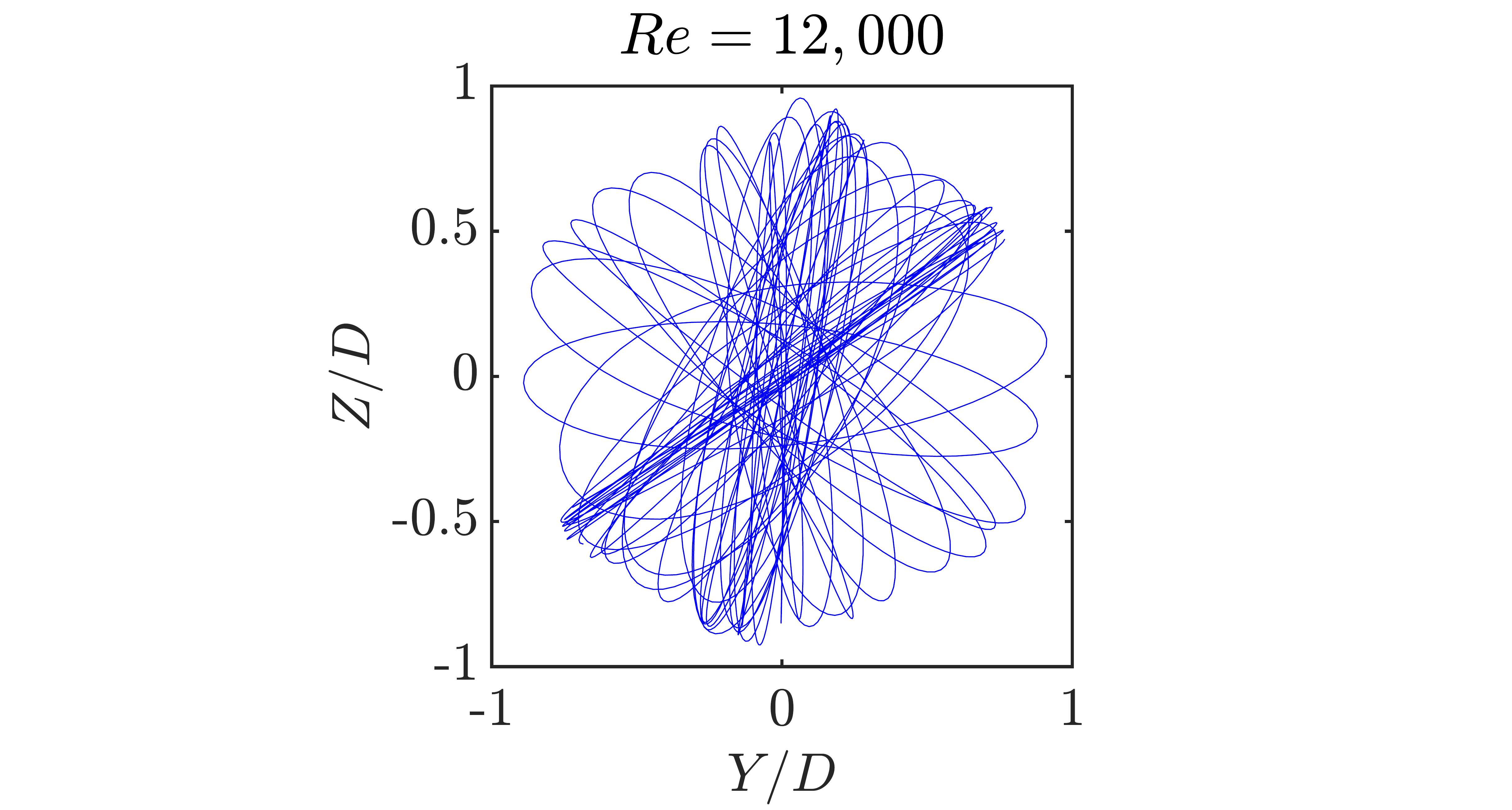} 
	\end{subfigure}%
		\begin{subfigure}[b]{0.25\textwidth}
		\adjincludegraphics[scale=0.15,trim={{0.2\width} 0 {0.25\width} 0.2},clip]{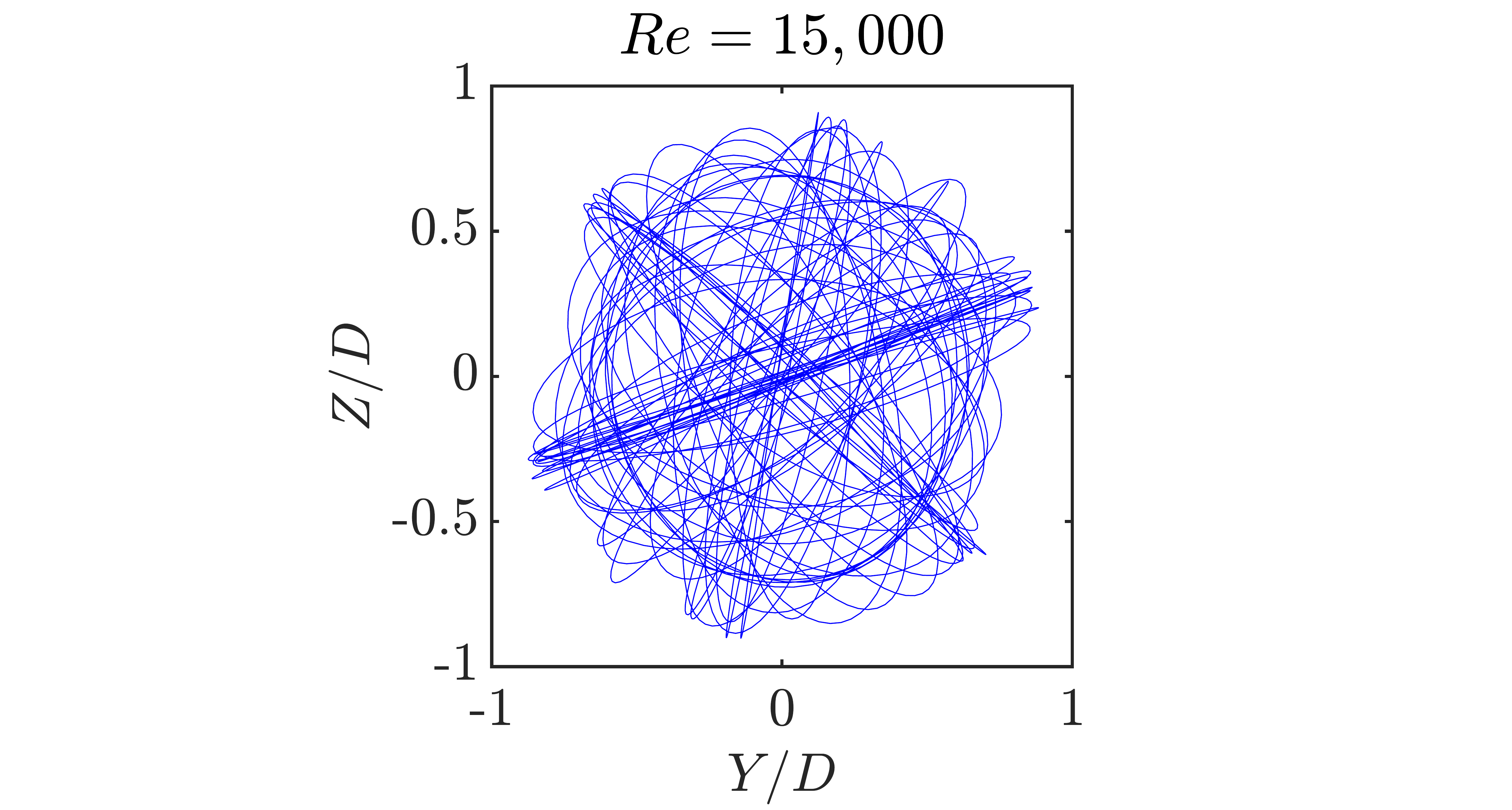}
	\end{subfigure}%
	\begin{subfigure}[b]{0.25\textwidth}
		\adjincludegraphics[scale=0.15,trim={{0.2\width} 0 {0.25\width} 0.2},clip]{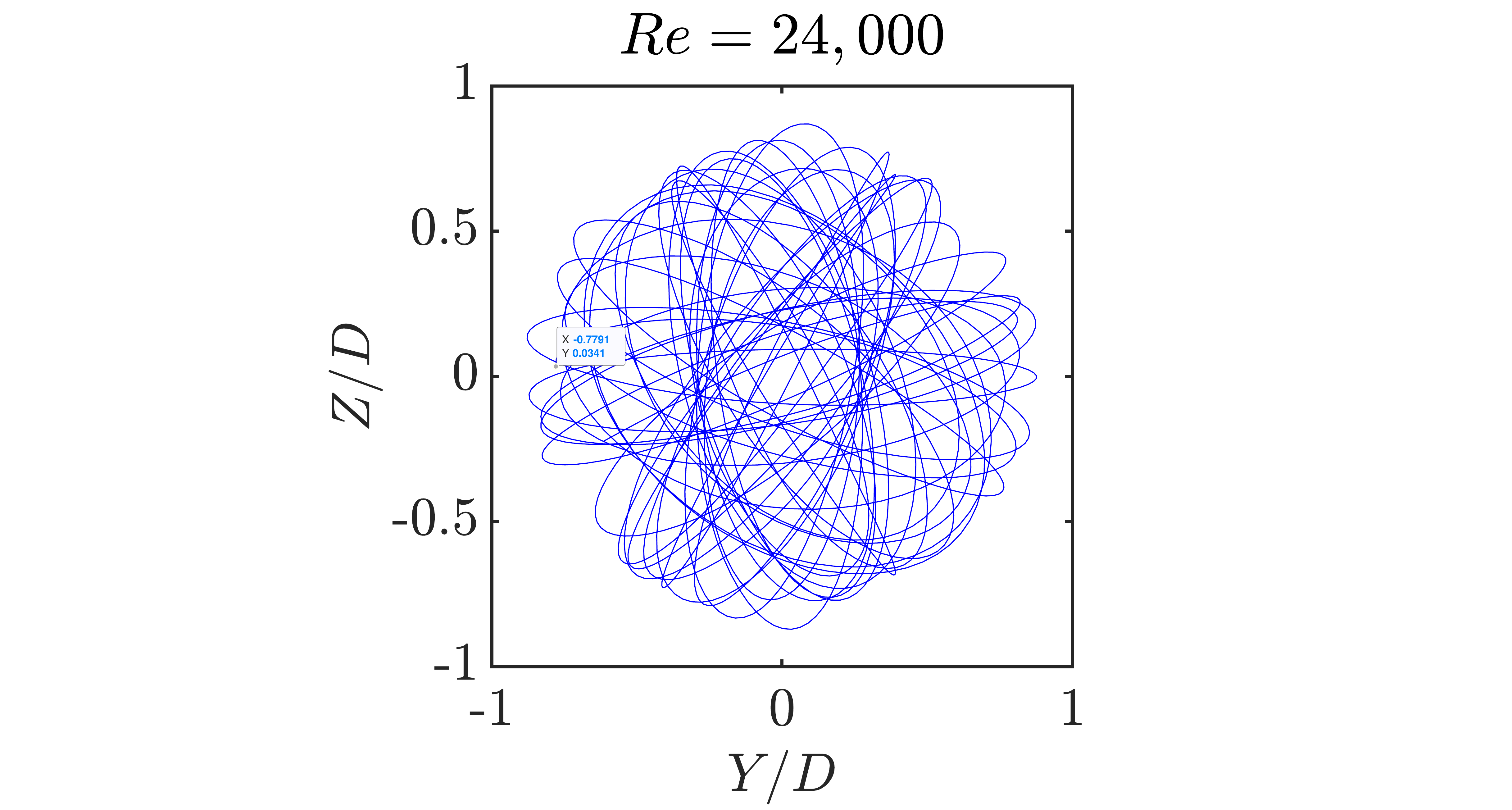}
	\end{subfigure}%
	\begin{subfigure}[b]{0.25\textwidth}
		\adjincludegraphics[scale=0.15,trim={{0.2\width} 0 {0.25\width} 0.2},clip]{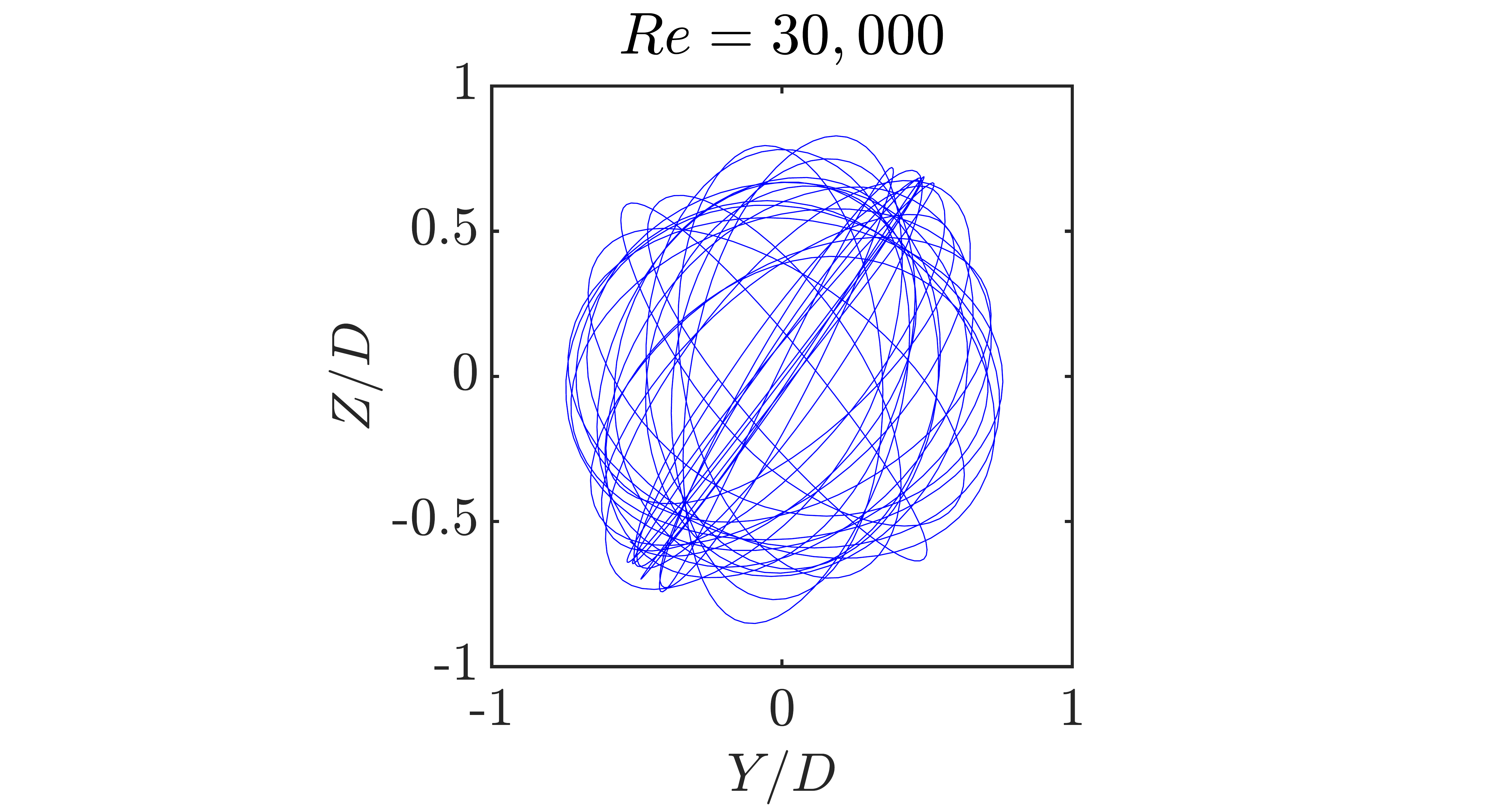}
	\end{subfigure}%
	\caption{The sphere trajectories in the $Y$-$Z$ plane at $U^*=9$ and $m^*=3.82$ for a range of $Re\in[500,30\,000]$. }
	\label{Trajectories_3d}
\end{figure}

Fig. \ref{3d_Trend_Re} compares the results obtained for the sphere r.m.s. amplitude response as a function of Reynolds number in the transverse direction ($A^*_y $) and the vertical direction ($A^*_z $) with the numerical results of \cite{behara2016} and \cite{rajamuni2019} for the Reynolds number range of $Re\in[300,2\,000]$. Our results show a similar trend with the results of \cite{rajamuni2019}, and the response amplitudes increase as a function of Reynolds number. As elucidated by \cite{rajamuni2019}, the simulations of \cite{behara2016} did not reach the asymptotic state and the response amplitudes are
not comparable with our numerical prediction. Fig. \ref{Time_history_3dof} shows time histories of the amplitude response in the transverse ($A^*_y$) and vertical ($A^*_z$) directions at $Re=6\,000, 12\,000$ and $Re=15\,000$ at the periodic state. A noticeable difference in the amplitude response is observed for the case at $Re=6\,000$ compared to the cases at higher Reynolds number at $Re=12\,000$ and 15\,000. At $Re=6\,000$ the variation of the peak amplitude oscillations is small (Fig. \ref{Time_history_3dof} (a - 1)) and the frequency of the transverse oscillation ($f_{A_y}$) and vertical oscillation ($f_{A_z}$) is matched (Fig. \ref{Time_history_3dof} (a - 2)), therefore, the phase difference between ($A^*_y$) and ($A^*_z$) does not change with time ($\phi \approx \pi/2$). This represents the circular type motion in the transverse plane similar to the cases in the range of Reynolds number $Re\in[300,2\,000]$. While at higher Reynolds number ($Re=12\,000$ and $Re=15\,000$), the amplitude response in both transverse and vertical directions ($A^*_y$ and $A^*_z$), have significant variation with time, similar to beating-type behaviour (Fig. \ref{Time_history_3dof} (b - 1) and (c - 1)). The phase difference between the transverse and vertical motion is found to change with time due to the difference in the oscillation frequencies of $f_{A_y}$ and $f_{A_z}$, Fig. \ref{Time_history_3dof} (b - 2) and (c - 2).
Fig. \ref{Trajectories_3d} shows the trajectory response of the sphere motion exposed to the unsteady flow field at the lock-in regime ($U^*=9$) for eight cases in the Reynolds number range of $Re\in[300,30\,000]$ at their periodic states. As it can be seen, the motion trajectories for the Reynolds number range of $Re\in[2\,000,6\,000]$ show circular-type motion. However, at higher Reynolds numbers $Re\in[12\,000,30\,000]$ the behavior of the motion trajectories is found to be chaotic, and consists of a combination of linear and circular-type motions.

\begin{figure}[htbp!]
	\centering
	
	\begin{subfigure}[b]{0.25\textwidth}
		\adjincludegraphics[scale=0.15,trim={{0.2\width} 0 {0.25\width} 0.2},clip]{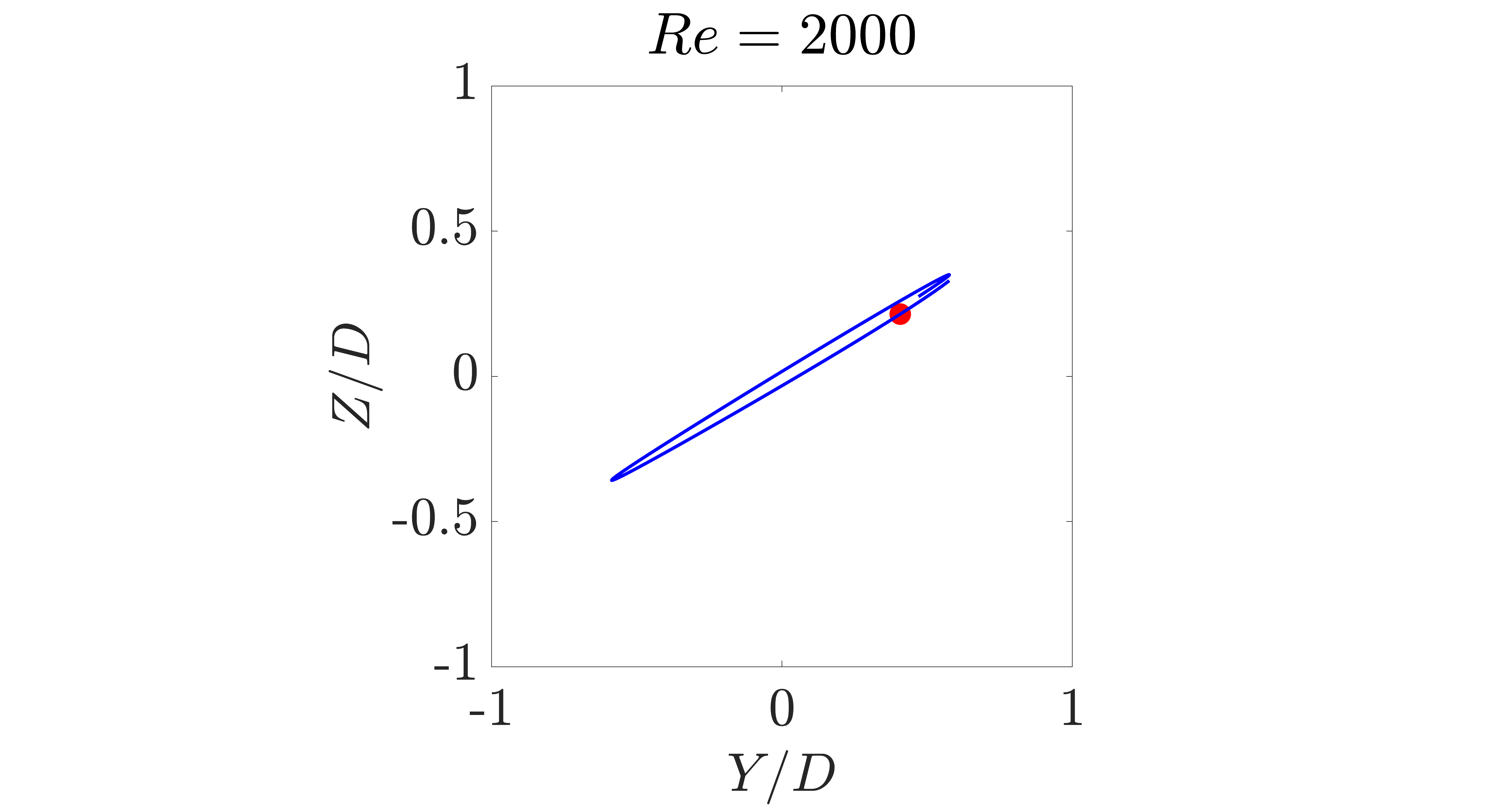}
	\end{subfigure}%
	\begin{subfigure}[b]{0.75\textwidth}
		\adjincludegraphics[scale=0.4,trim={{0.07\width} {0.32\width} {0.05\width} {0.3\width}},clip]{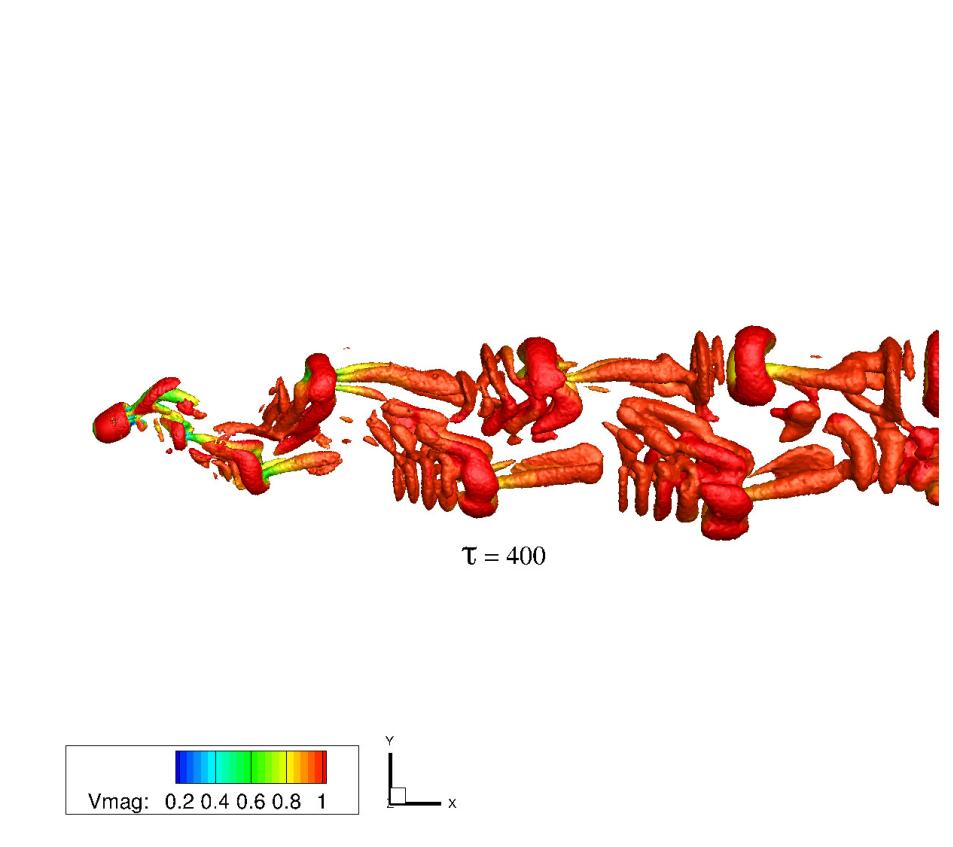} 
		\caption{Initial state}	
	\end{subfigure}
	\begin{subfigure}[b]{0.25\textwidth}
		\adjincludegraphics[scale=0.15,trim={{0.2\width} 0 {0.25\width} 0.2},clip]{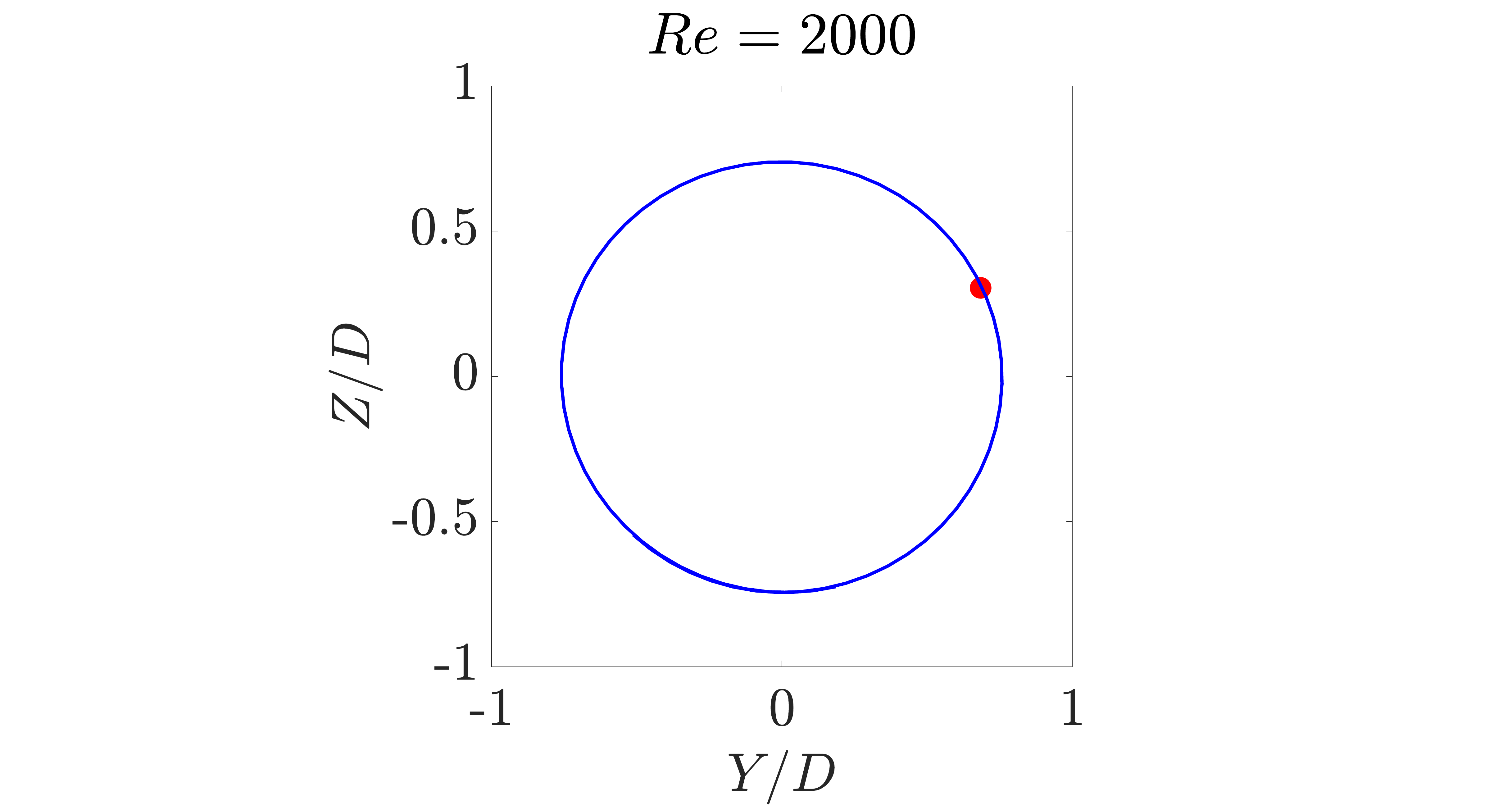}
	\end{subfigure}%
	\begin{subfigure}[b]{0.75\textwidth}
		\adjincludegraphics[scale=0.4,trim={{0.07\width} {0.32\width} {0.05\width} {0.3\width}},clip]{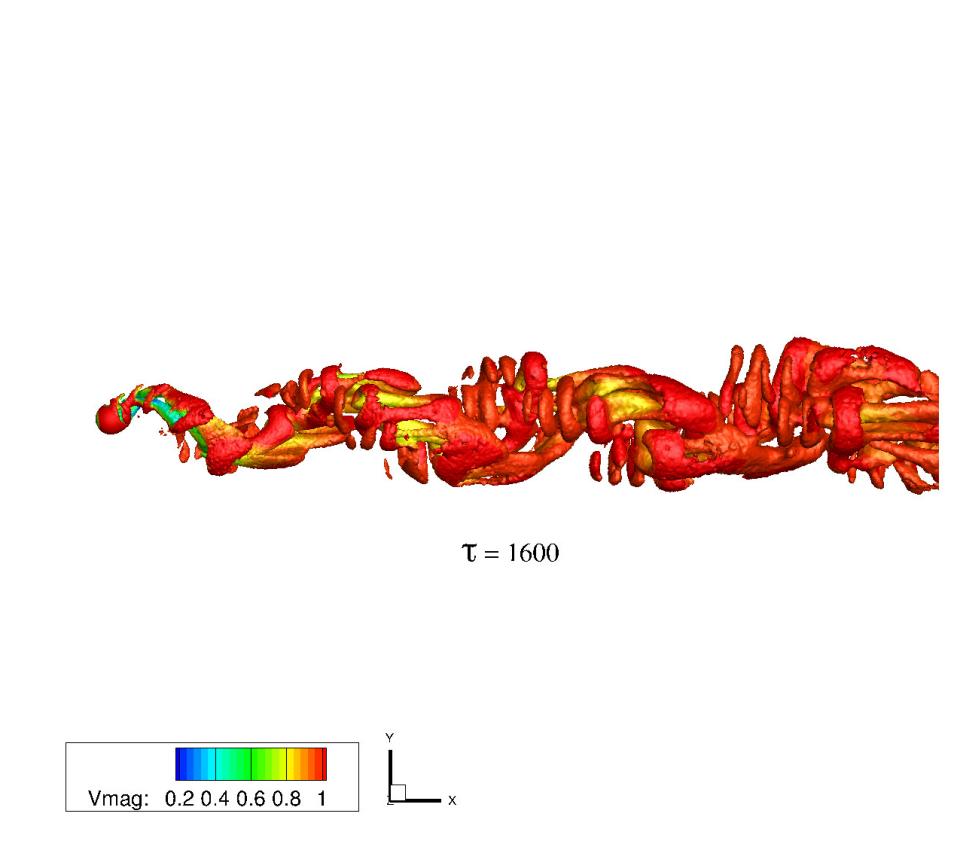} 
		\caption{Final state}
	\end{subfigure}%
	
	\caption{Iso-surface of three-dimensional wake structures formed behind the 3-DOF sphere at $U^*=9$ and $Re=2\,000$: (a) initial state with linear path, and (b) final stationary state with circular motion. Iso-surfaces are plotted by the Q-criterion ($Q=0.001$).} 
	\label{Q_Criterion_high_Resolution_3d_Sphere}
\end{figure}

To further look into the vortex formation and the wake structure,  we employ the Q-criterion \cite{hunt} which is given as
\begin{align}
Q=\frac{1}{2}\left(\|\boldsymbol{\Omega}\|^{2}-\|\boldsymbol{S}\|^{2}\right)
\end{align}
where $\|\boldsymbol{\Omega}\|^{2}=\left[\operatorname{tr}\left(\boldsymbol{\Omega} \boldsymbol{\Omega}^{t}\right)\right]$ and $\|\boldsymbol{S}\|^{2}=\left[\operatorname{tr}\left(\boldsymbol{S} \boldsymbol{S}^{t}\right)\right]$, $\boldsymbol{\Omega}$ and $\boldsymbol{S}$ represents the antisymmetric and symmetric components of the velocity gradient tensor $\nabla \u$ and $\operatorname{tr}(\cdot)$ denotes the trace operator. When $Q > 0$, vorticity prevails over strain and the strength of rotation dominates the strain.
In Fig. \ref{Q_Criterion_high_Resolution_3d_Sphere}, we observe that at each Reynolds number, hairpin vortex loops from the opposite sides of the sphere forms in the wake at the initial state. The sphere initially begins to vibrate in a linear path as shown in Fig. \ref{Q_Criterion_high_Resolution_3d_Sphere} (a). For the range of Reynolds number $Re \in [2\,000,6\,000]$, the hairpin mode is found to be unstable and the wake mode transforms to spiraling vortical structure behind the sphere at the final state. The sphere motion merges to circular trajectory orthogonal to the flow as shown in Fig. \ref{Q_Criterion_high_Resolution_3d_Sphere} (b). The wake mode transition for the higher Reynolds number range $Re\in[12\,000,30\,000]$ is found to be quite different, where both the hairpin mode and the spiral mode were identified as unstable states. For this higher range of Reynolds number, it is found that the vortical structures transform frequently from the hairpin mode to the spiral mode and \textit{vice versa}.
%

%

\subsection{VIV of submerged elastically mounted sphere close to the free surface} 
\begin{figure}[htbp!]
	\centering
	
	\begin{subfigure}[b]{0.5\textwidth}
		\adjincludegraphics[scale=0.17,trim={{0\width} 0 {0\width} {0.05\width}},clip]{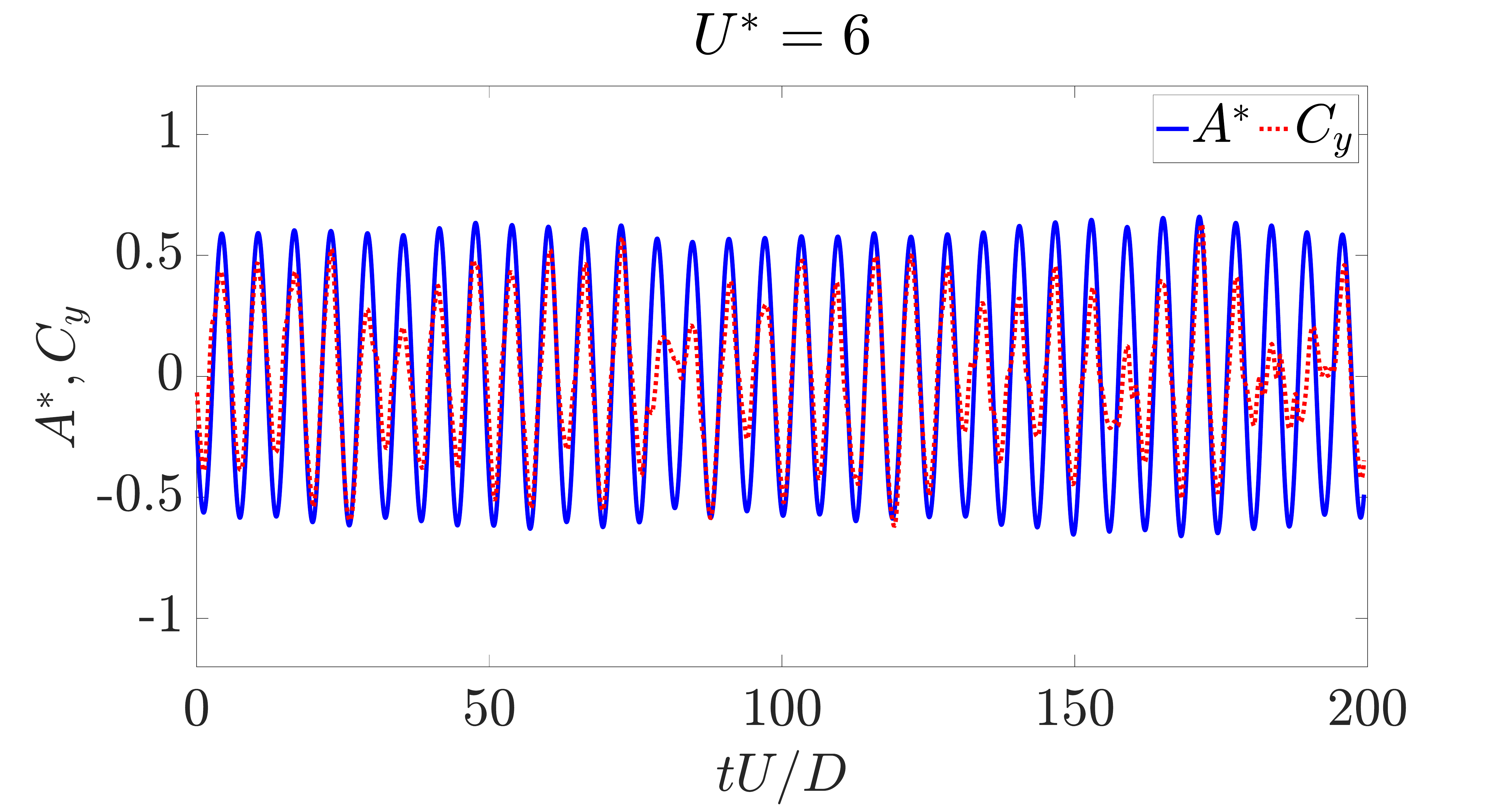} 
		\caption{$U^*=6$}
	\end{subfigure}%
	\begin{subfigure}[b]{0.5\textwidth}
		\adjincludegraphics[scale=0.17,trim={{0\width} 0 {0\width} {0.05\width}},clip]{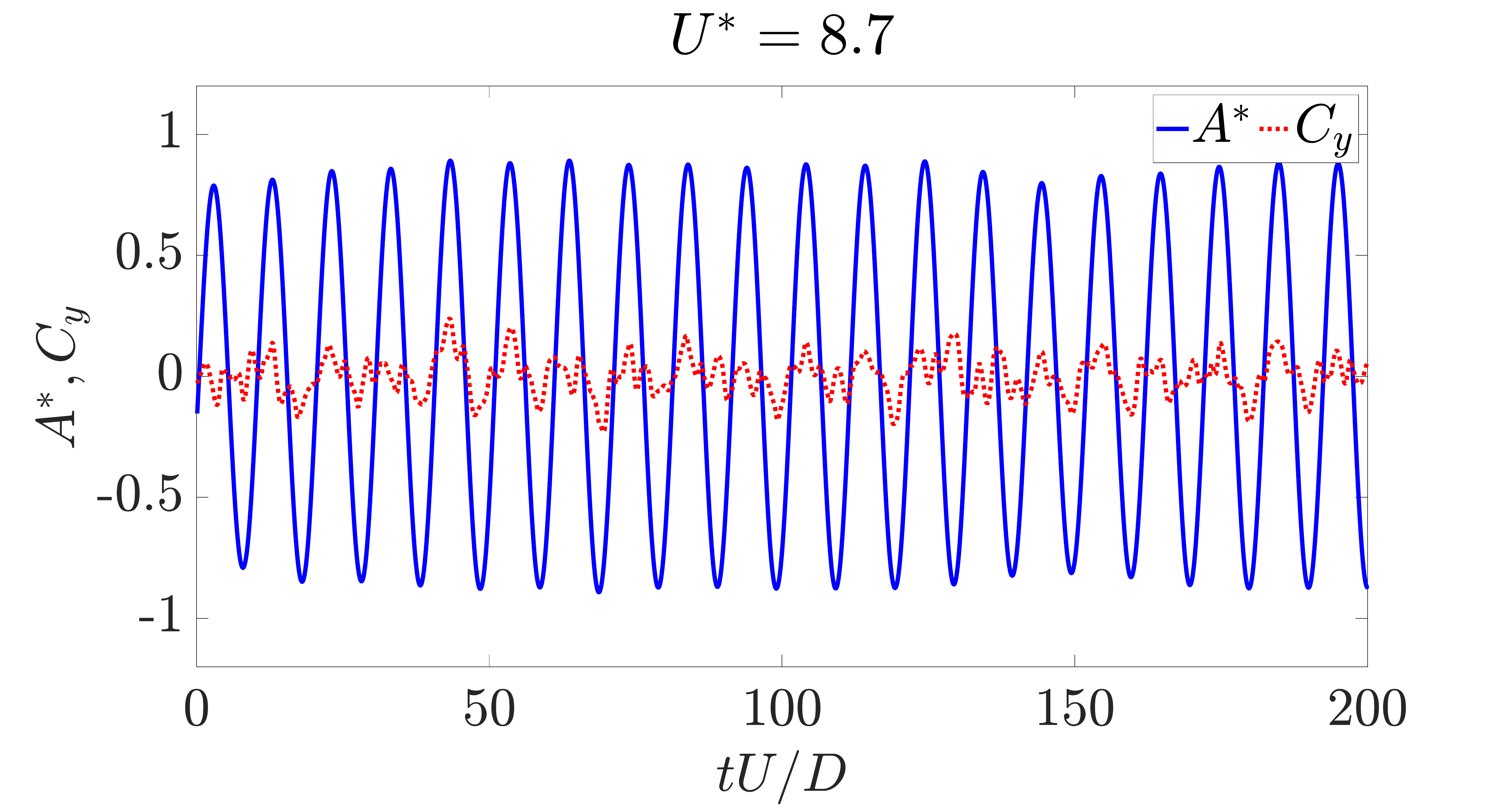} 
		\caption{$U^*=8.7$}
	\end{subfigure}

\vspace{0.4cm}

	\begin{subfigure}[b]{0.5\textwidth}
		\adjincludegraphics[scale=0.17,trim={{0\width} 0 {0\width} {0.05\width}},clip]{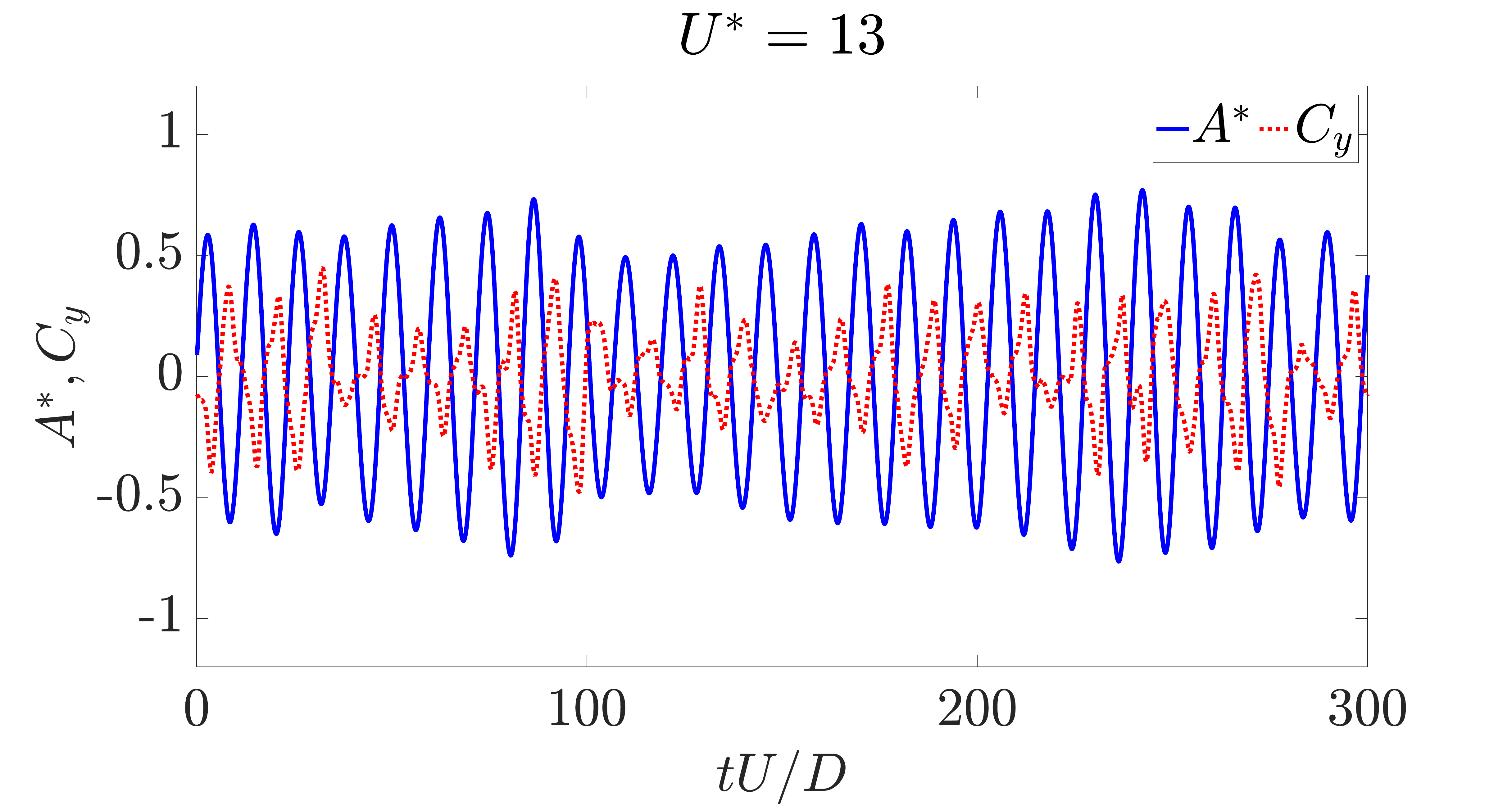} 
		\caption{$U^*=13$}
	\end{subfigure}%
	\begin{subfigure}[b]{0.5\textwidth}
		\adjincludegraphics[scale=0.17,trim={{0\width} 0 {0\width} {0.05\width}},clip]{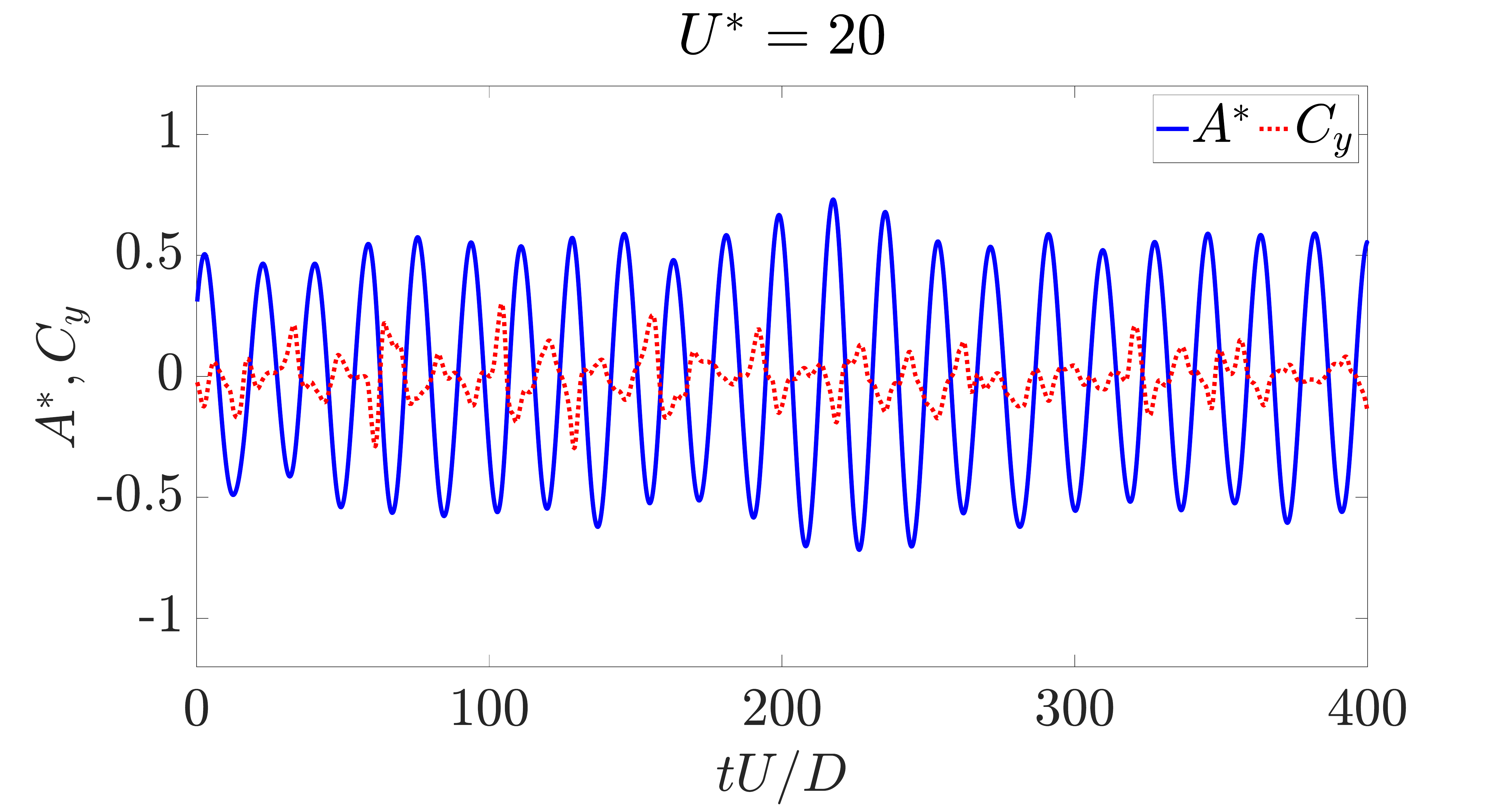} 
		\caption{$U^*=20$}
	\end{subfigure}%
	
	\caption{Time histories of the amplitude response ($A^*$) and the normalized transverse force ($C_y$) with non-dimensional time for 1-DOF sphere at $h^*=1$ at different reduced velocities. }
	\label{FS_Validation_TH}
\end{figure}

To validate the accuracy of our two-phase FSI solver, we next consider a fully submerged elastically mounted sphere restricted to move in the transverse $y$-direction. Our simulation results are compared against the measurement data of \cite{sareen2018} at the immersion ratio of $h^*=1$.
Consistent with the experimental set-up, a low mass ratio of $m^{*} = 7.8$ and a damping ratio of $\zeta = 0.002$ are considered. The 3D VIV simulations are performed for the Reynolds number $Re$ in the range $5\,000 \le Re \le 30\,000$, which corresponds to the reduced velocity $U^*$ range of $3 \le U^* \le 20$. The goal of this validation study is to establish the predictive capability of our solver in the two regimes (mode I and mode II) of sphere VIV.
The maximum amplitude is extracted from 10 oscillation cycles when the system reaches a steady-state. Fig. \ref{FS_Validation_TH} shows the time histories of the amplitude response for several selected $U^*$ values at mode I ($U^*=6$), transition mode ($\sim U^*=8.7$) and mode II ($U^*=13$, $U^*=20$) response at $h^*=1$.

\begin{figure}[htbp!]

	\begin{subfigure}[b]{1\textwidth}
		\hspace{-1cm}
		\adjincludegraphics[scale=0.36,trim={0\width} {0\width} {0\width} {0\width},clip]{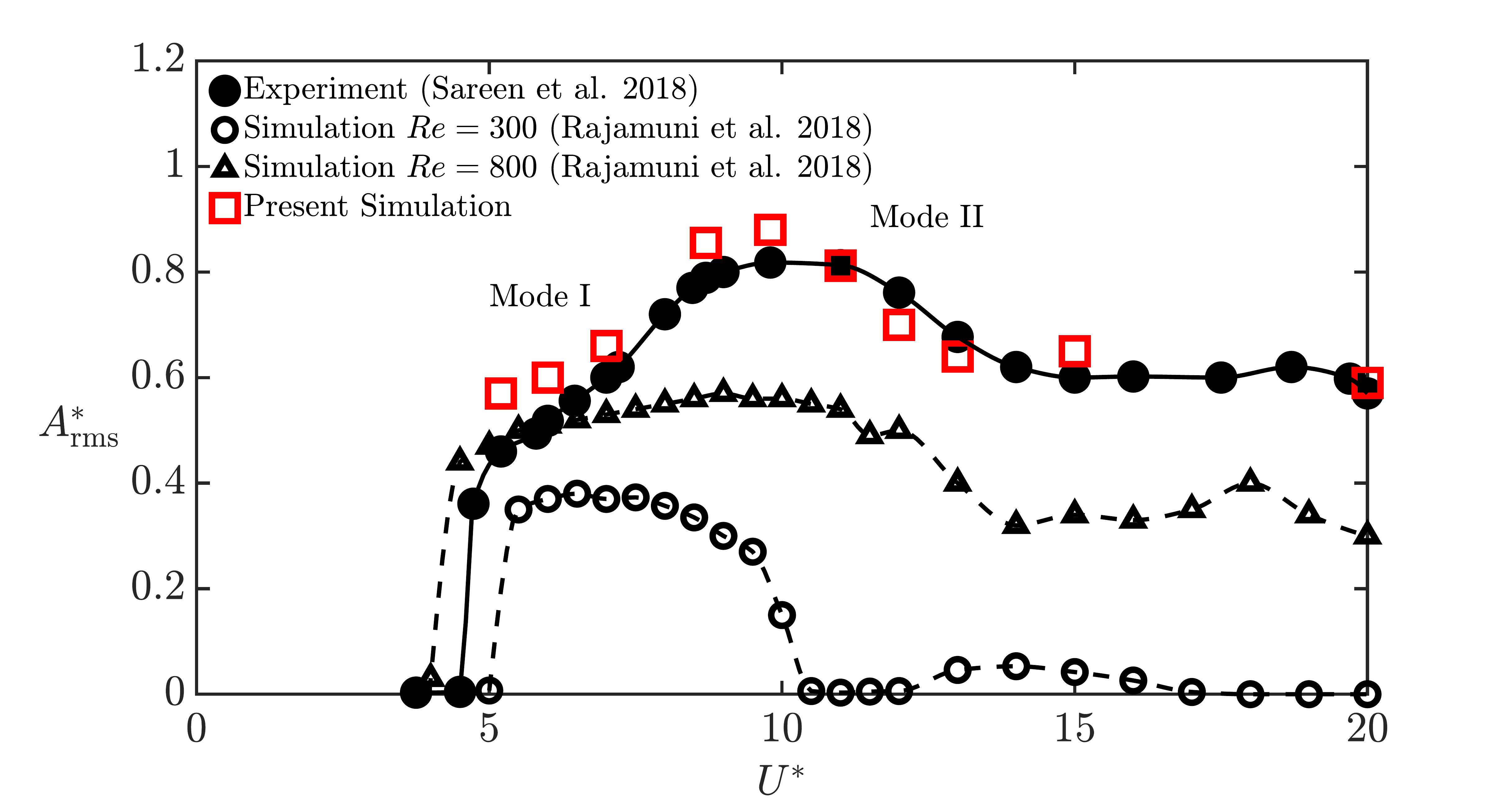}%
		\caption{}
	\end{subfigure}

	\begin{subfigure}[b]{0.5\textwidth}
		\hspace{0.3cm}
		\adjincludegraphics[scale=0.16,trim={0\width} {0\width} {0\width} {0\width},clip]{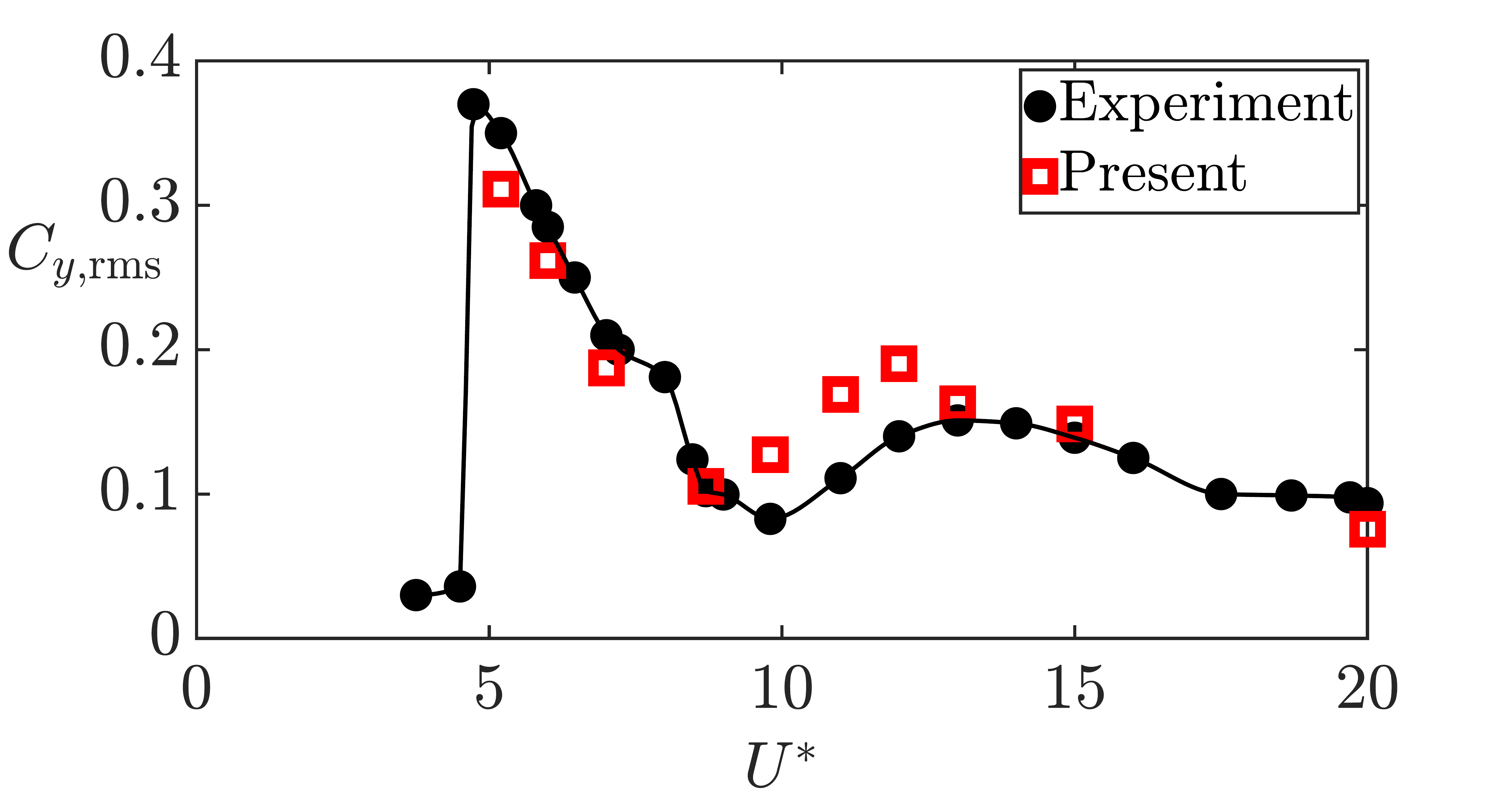}%
		\caption{}
	\end{subfigure}%
	\begin{subfigure}[b]{0.5\textwidth}
		\hspace{0.1cm}
		\adjincludegraphics[scale=0.16,trim={0\width} {0\width} {0\width} {0\width},clip]{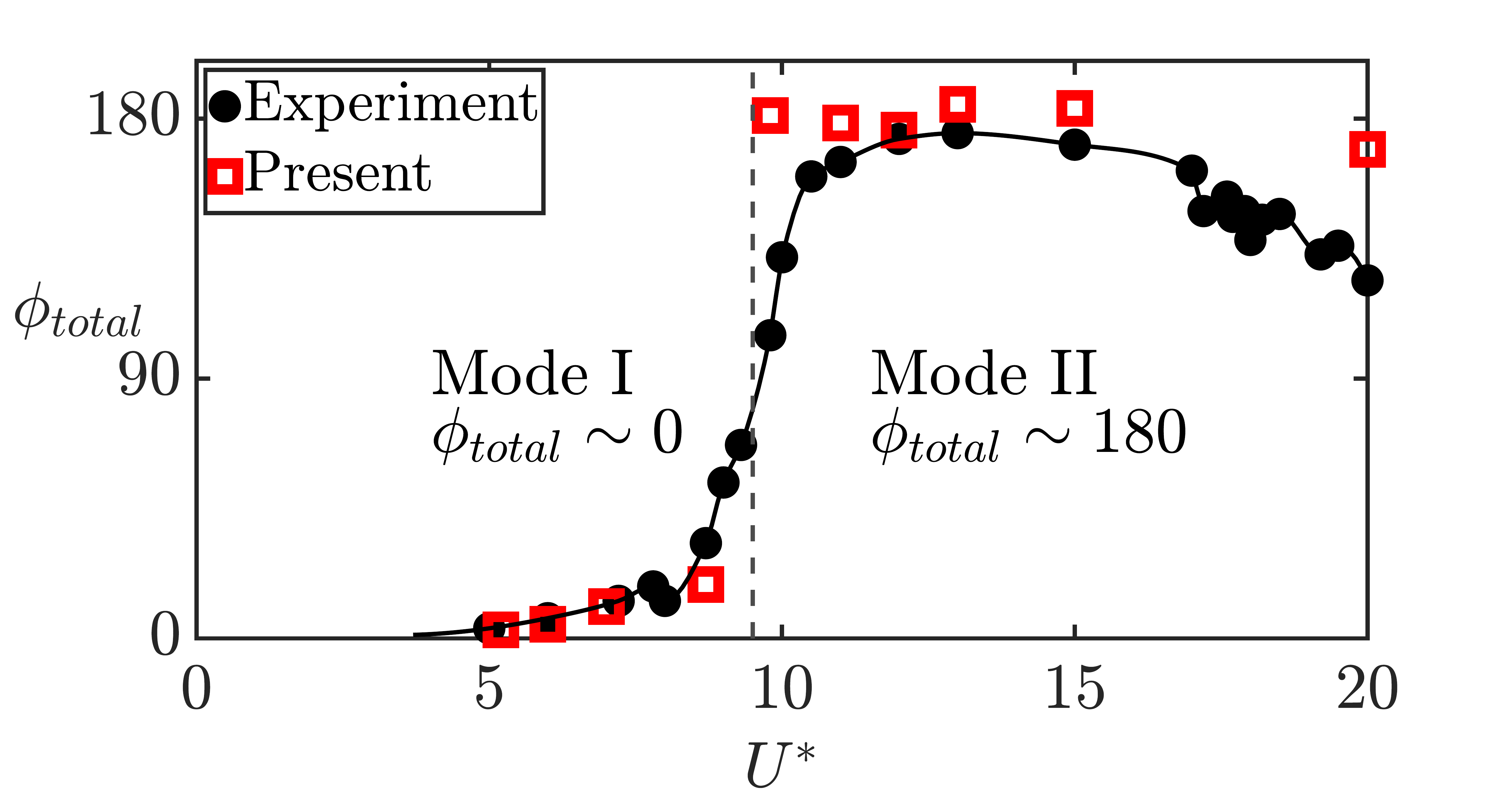}
		\caption{}
	\end{subfigure}%
	\caption{Variation of VIV response parameters as a function of the reduced velocity $U^*$ at $h^*=1$, $m^*=7.8$ and $\zeta=0.002$: (a) r.m.s. amplitude response ($A^*_{\mathrm{rms}}$), (b) r.m.s. normalized transverse fluid force ($C_{y,\mathrm{rms}}$) and, (c)  total phase difference ($\phi_{total}$). The results are compared with the experimental data of \cite{sareen2018}. } 
	\label{Validation_Trend}		
\end{figure} 

%
%

A comparison of the r.m.s. values of the amplitude response, the normalized transverse ($y$-direction) force and the total phase difference (the phase difference between the sphere vibration frequency and the transverse force frequency) with that of the experiment data in \cite{sareen2018} and the numerical data in \cite{rajamuni2018transverse} is shown in Fig. \ref{Validation_Trend}. 
Our results show a good agreement with the experimental study in \cite{sareen2018} and follow a similar trend. Through our simulation, it can be inferred that the noticeable difference between the numerical simulations in \cite{rajamuni2018transverse} at fixed Reynolds number ($Re=300$ and $Re=800$) with that of the experiment is due to significant effect of Reynolds number on the VIV response. 
%

%
%

\begin{figure}[htbp!]
	\centering
	\hspace{-1cm}
	\adjincludegraphics[scale=0.5,trim={0.5\width} {0.07\width} {0.1\width} {0.1\width},clip]{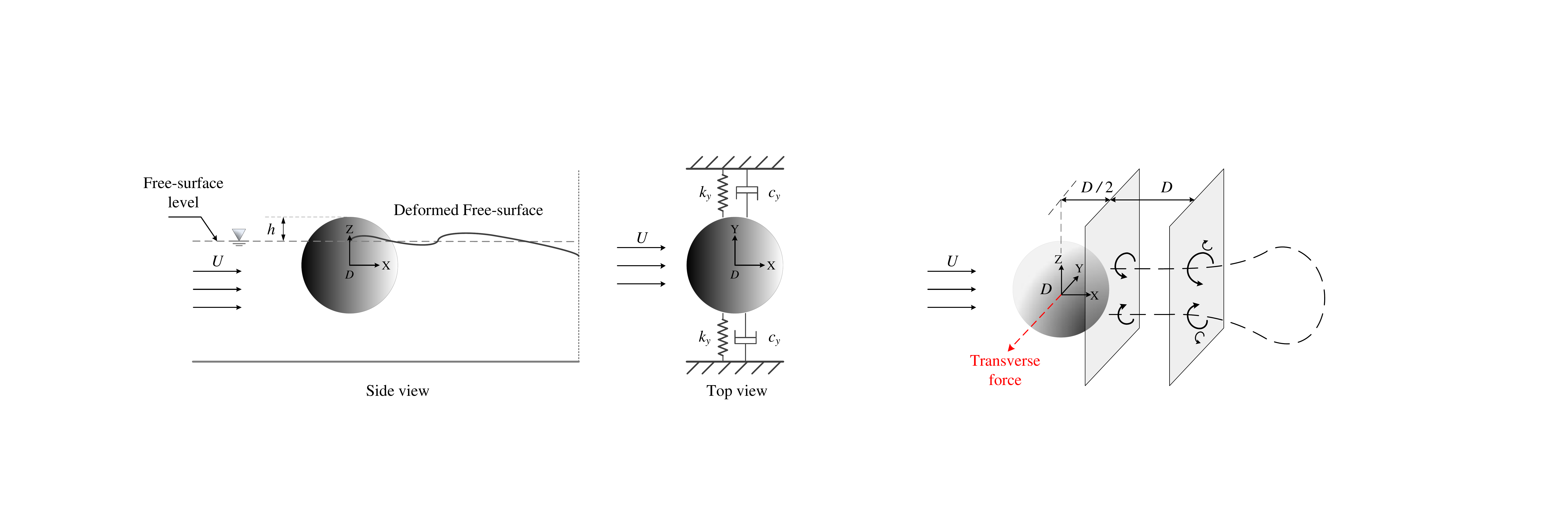}
	
	\caption{Sketch illustrating a sphere in steady flow with streamwise vortex pairs and the transverse force on the sphere.  Two representative $Y$-$Z$ planes in the sphere wake at $0.5D$ and $1.5D$ from the centre are shown for the plotting of streamwise $x$-vorticity contours. Dashed line shows the formation of hairpin vortex loop.} 
	\label{Normal_Plane} 
\end{figure}

\begin{figure}[htbp!]
	\centering
	
	\begin{subfigure}[b]{0.5\textwidth}
		\centering
		\hspace{-0.7cm}
		\adjincludegraphics[scale=0.17,trim={0\width} {0.0\width} {0\width} {0.0\width},clip]{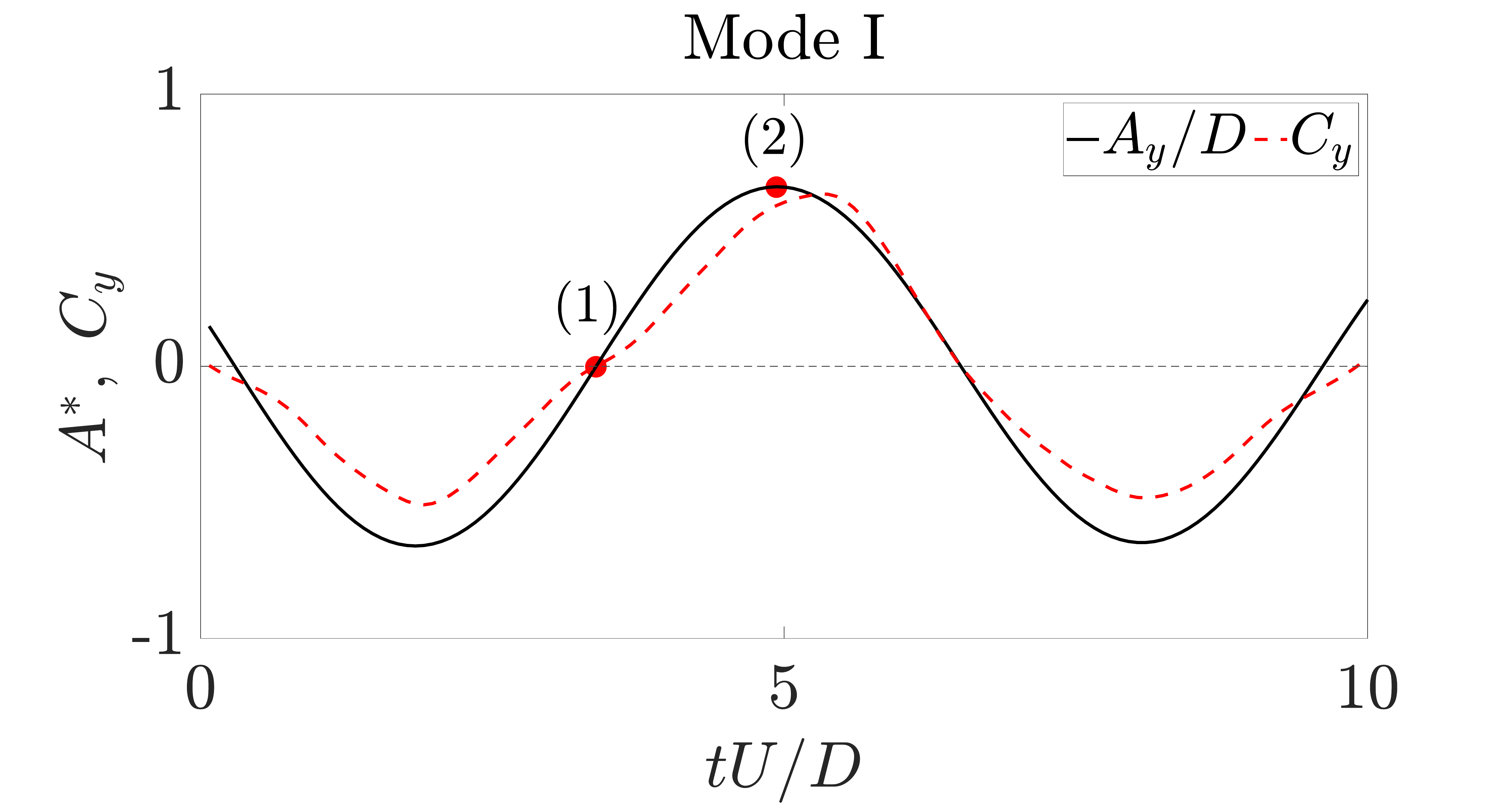}
	\end{subfigure}%
	\begin{subfigure}[b]{0.5\textwidth}
		\centering
		\hspace{-0.7cm}
		\adjincludegraphics[scale=0.17,trim={0\width} {0.0\width} {0\width} {0.0\width},clip]{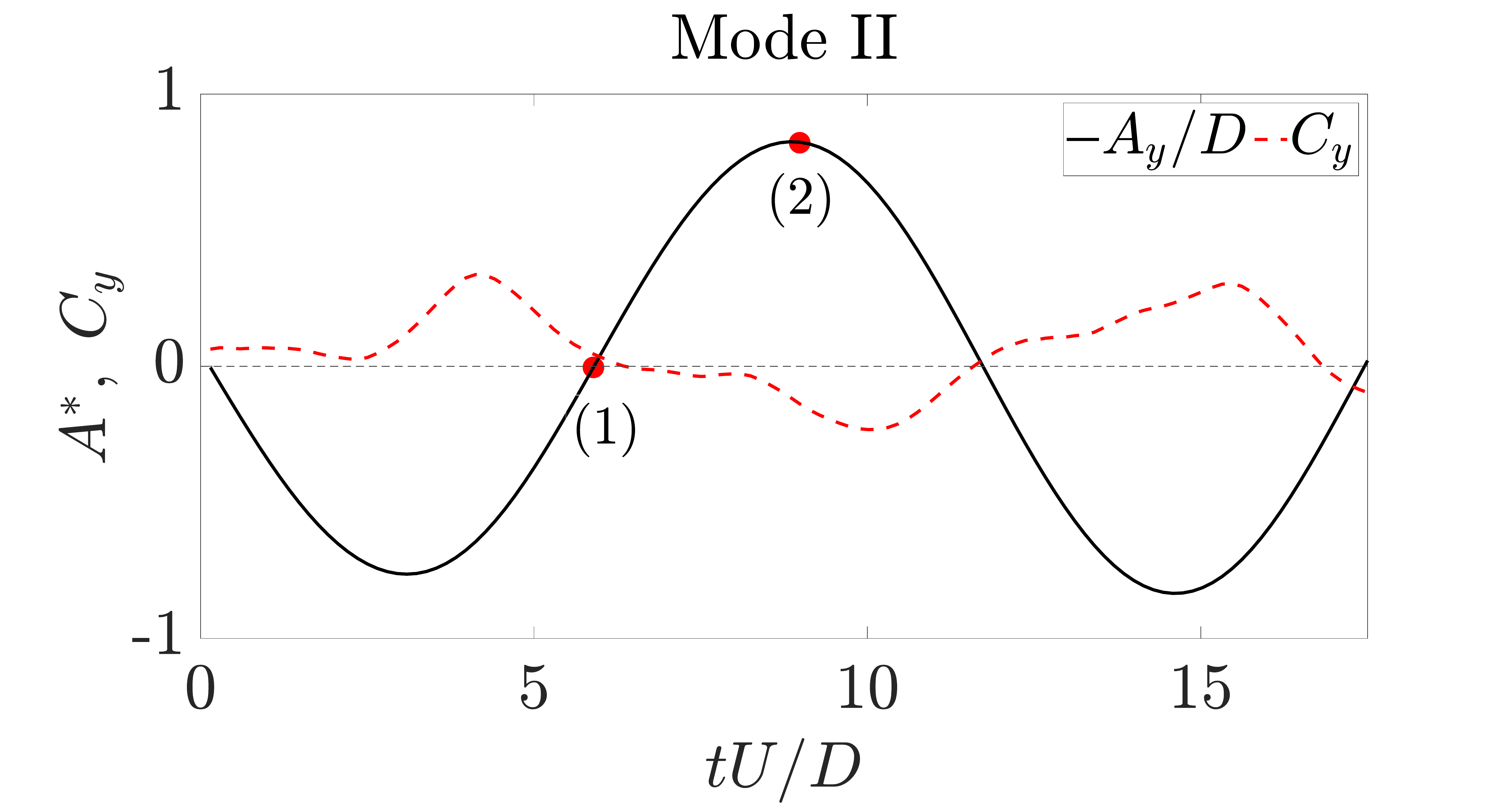}
	\end{subfigure}
	
	\vspace{0.4cm}
	
	\begin{subfigure}[b]{0.5\textwidth}
		\centering
		\adjincludegraphics[scale=0.32,trim={0.05\width} {0.25\width} {0.01\width} {0.25\width},clip]{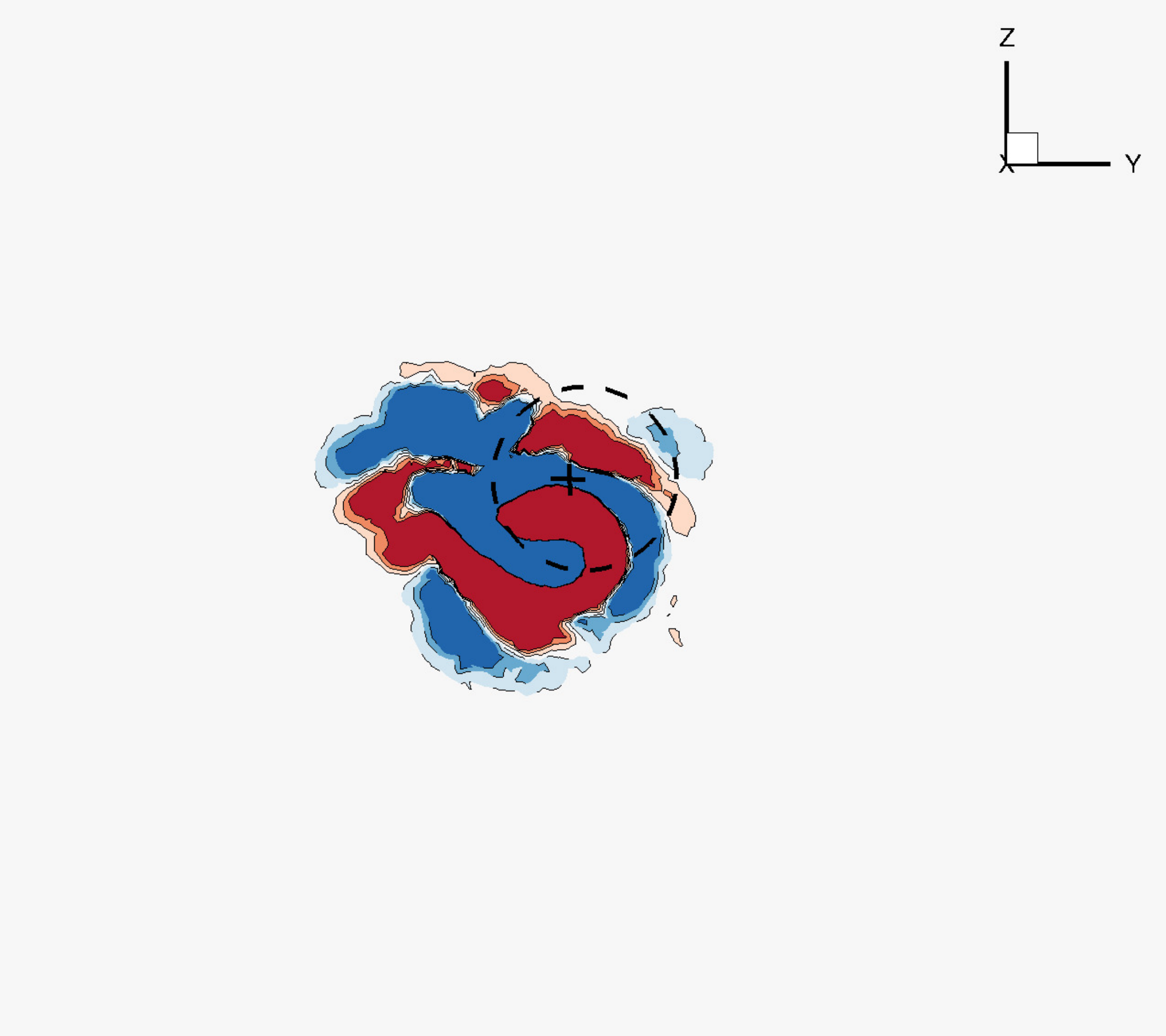}
		\caption*{$(a$ - $1)$}
	\end{subfigure}%
	\begin{subfigure}[b]{0.5\textwidth}
		\centering
		\adjincludegraphics[scale=0.32,trim={0.05\width} {0.25\width} {0.01\width} {0.25\width},clip]{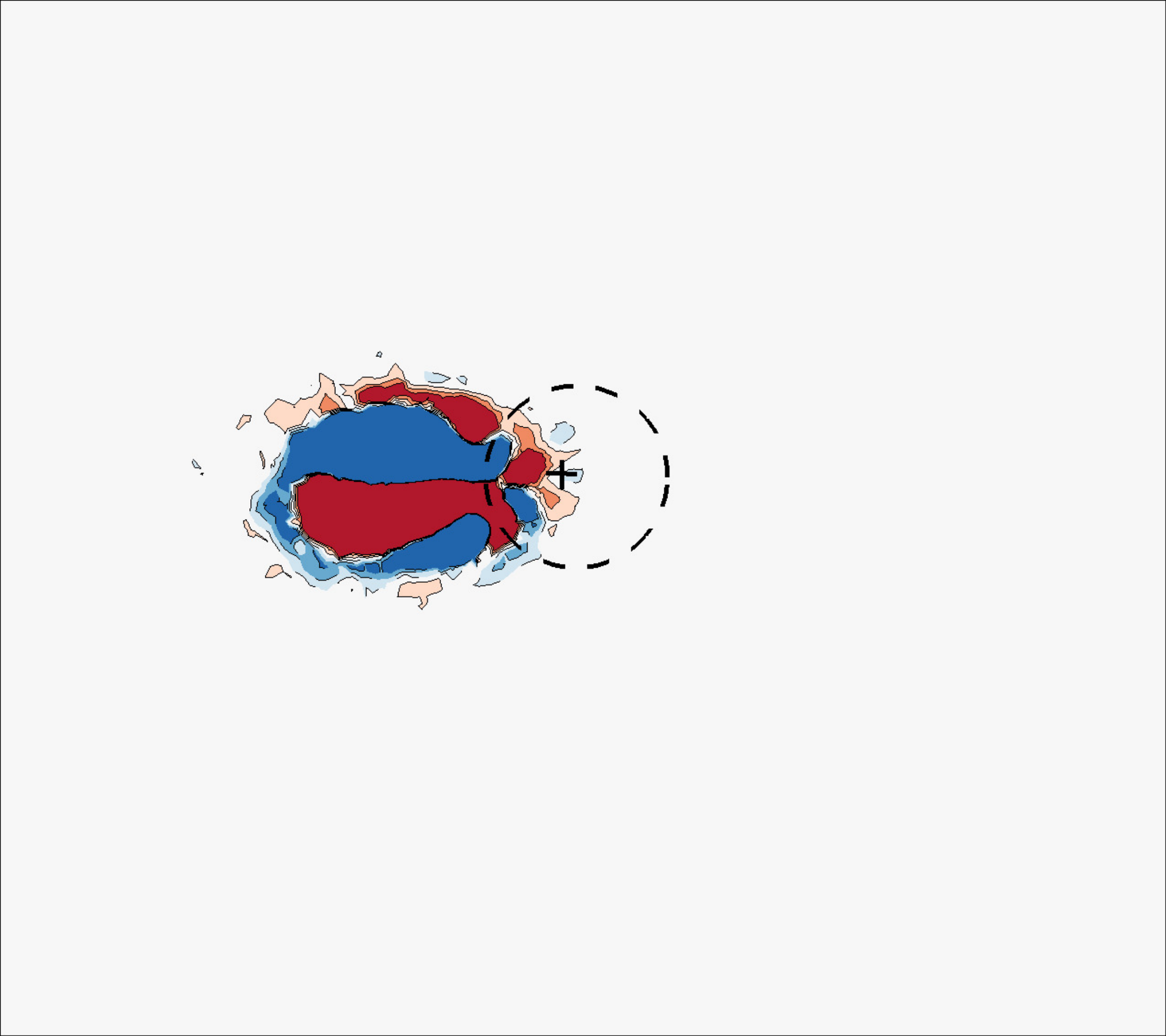}
		\caption*{$(b$ - $1)$}
	\end{subfigure}
	
	\begin{subfigure}[b]{0.5\textwidth}
		\centering
		\adjincludegraphics[scale=0.32,trim={0.05\width} {0.25\width} {0.01\width} {0.25\width},clip]{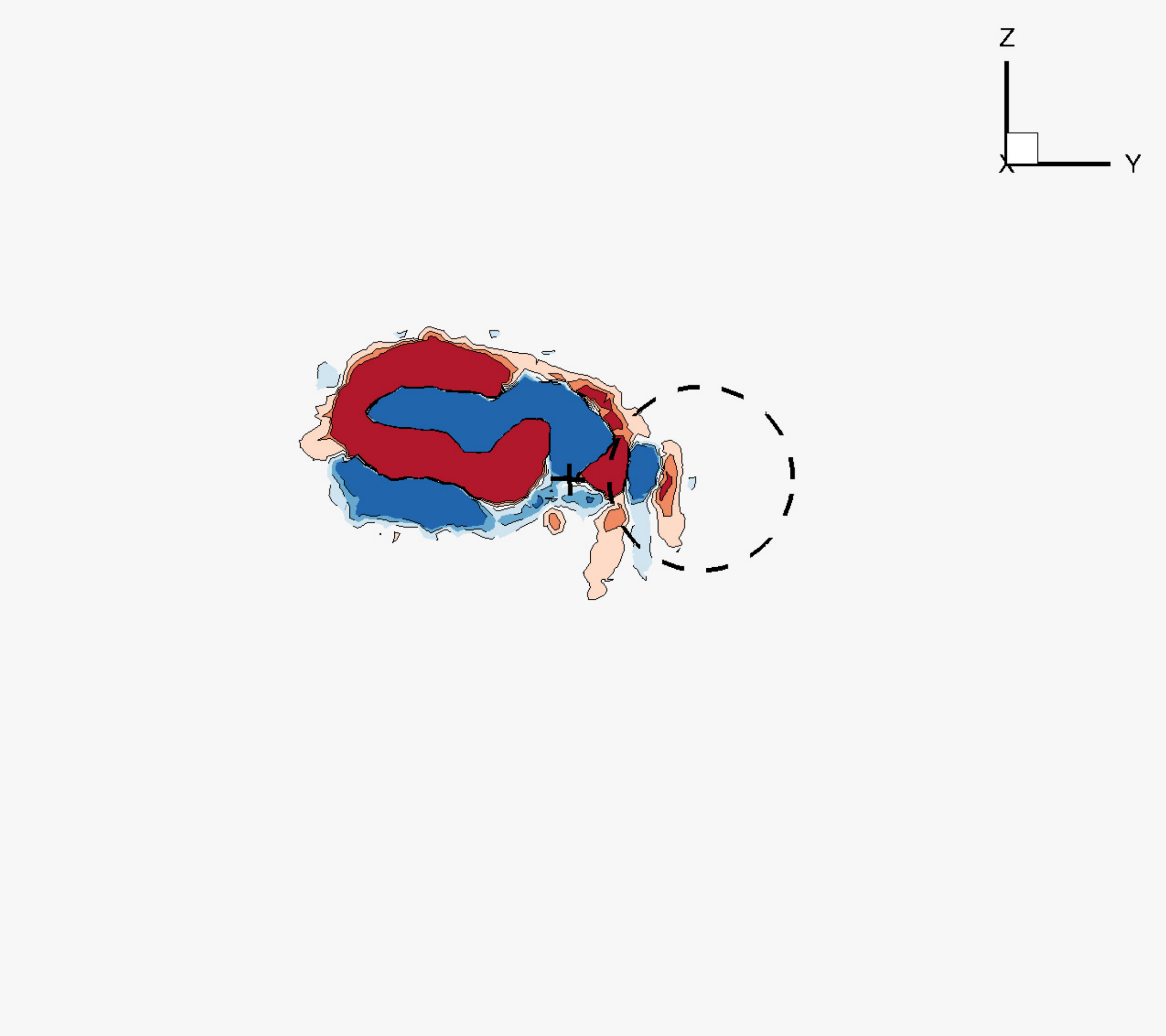}
		\caption*{$(a$ - $2)$}
	\end{subfigure}%
	\begin{subfigure}[b]{0.5\textwidth}
		\centering
		\adjincludegraphics[scale=0.32,trim={0.05\width} {0.25\width} {0.01\width} {0.25\width},clip]{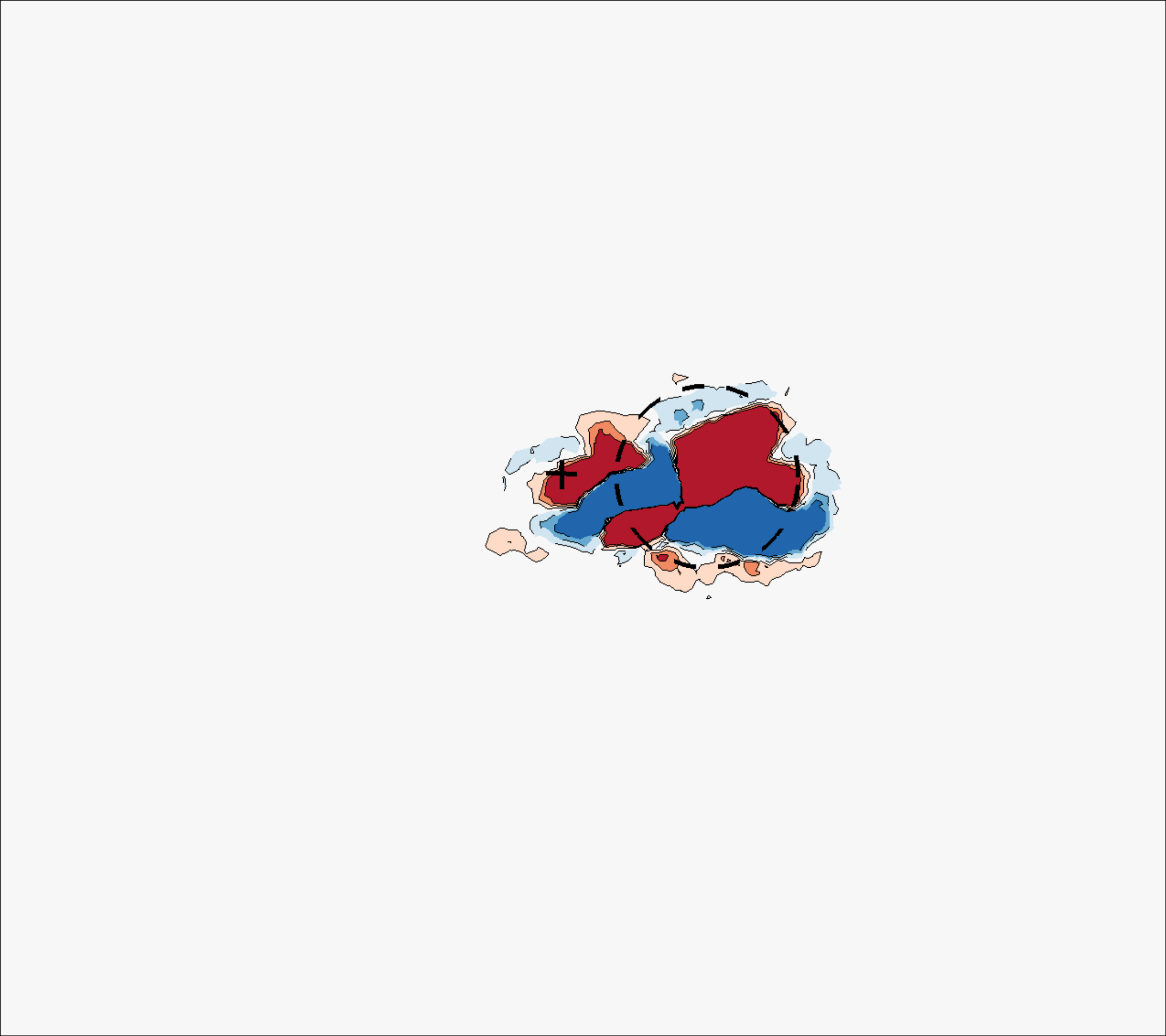}
		\caption*{$(b$ - $2)$}
	\end{subfigure}
	
	\begin{subfigure}[b]{1\textwidth}
		\centering
		\adjincludegraphics[scale=0.4,trim={0.05\width} {0.7\width} {0.01\width} {0.0\width},clip]{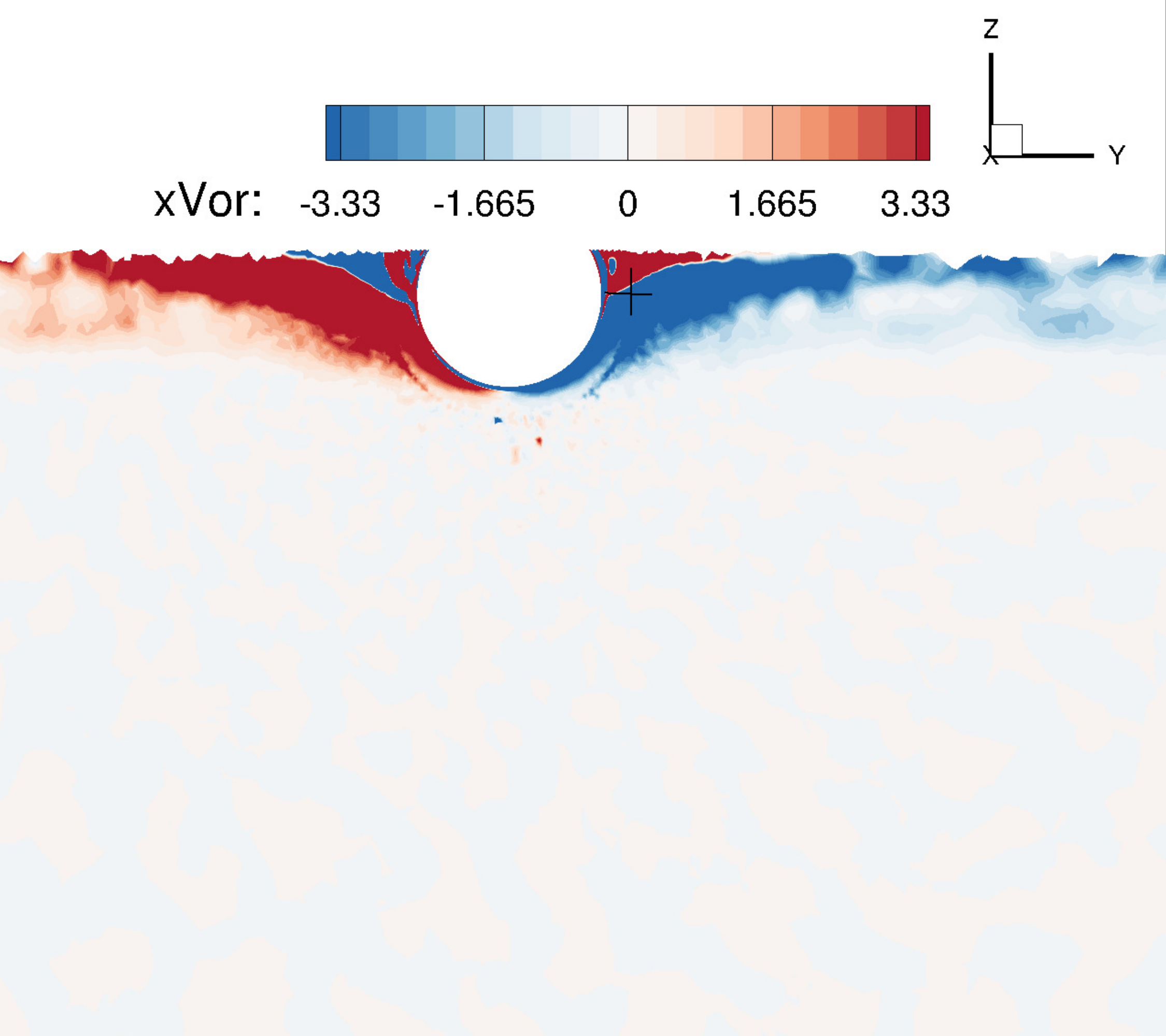}
		\caption*{}
	\end{subfigure}
	
	\caption{Streamwise $x$-vorticity $(\omega_{x}D/U\in[-3.3,3.3])$ contours showing the dominant counter-rotating vortex pair for mode I and mode II at two representative time locations: (a) mode I at $U^*=6$, $Re=9\,400$, and (b) mode II at $U^*=12$, $Re=18\,800$. Blue and red contours show clockwise and anti-clockwise vorticity, respectively. $x$-vorticity contours are plotted on $Y$-$Z$ plane at $1.5D$ downstream from the center of the sphere. } 
	\label{Mode_Compare} 
\end{figure}

We next briefly study the streamwise vortex dynamics which plays a crucial role to sustain the vibration amplitudes through the work done by the transverse force. A sketch to illustrate the formation of the streamwise vortex pairs and the visualization planes is shown in Fig. \ref{Normal_Plane}. Fig. \ref{Mode_Compare} shows the streamwise $x$-vorticity in a plane normal to the flow at $1.5D$ downstream of the sphere center, which enables us to measure the dominant counter-rotating streamwise pairs for both mode I and mode II. The distinct differences in the timing of vortex pair formation for modes I and II in Fig. \ref{Mode_Compare}, is consistent with the differences in the total phase $\phi_{total}$ between the modes, which is quantified in Fig. \ref{Validation_Trend}(c). Our numerical results are qualitatively comparable with the experimental observation in \cite{govardhan2005vortex} for the transversely oscillating sphere at lower Reynolds number $Re=3\,000$. Detailed investigation of vortex wake modes and VIV characteristics is beyond the focus of the present study. The present validation study deems sufficient to serve as the reference to examine the VIV characteristics with the free-surface effect.

\section{RESULTS AND DISCUSSION}
\label{results}
The complexity of the coupled physical phenomena involved in a freely vibrating sphere close to the free surface is enhanced by the wake dynamics and sphere/free-surface interactions. The unsteady wake of the sphere interacting with the free-surface makes the coupled response of the piercing case configuration fundamentally different from the fully submerged sphere counterpart. To understand the coupled dynamics of free-surface VIV, we investigate the effects of immersion ratio $h^*=h/D$ for the piercing case at $h^*=-0.25$ and contrast the VIV behavior with the submerged-sphere counterpart at $h^*=0$ and $h^*=1$. We explore the vibration response and the wake dynamics through the range of immersion ratios. We then proceed to study the sensitivity of large-amplitude oscillation as functions of the mass ratio $m^*$ and Froude number $Fr$ at the lock-in range.

\subsection{VIV of elastically mounted sphere piercing the free surface}
\begin{figure}
	\centering
	\begin{subfigure}[b]{0.55\textwidth}
		\adjincludegraphics[scale=0.25,trim={{0.08\width} {0.018\width} {0.0\width} {0.0\width}},clip]{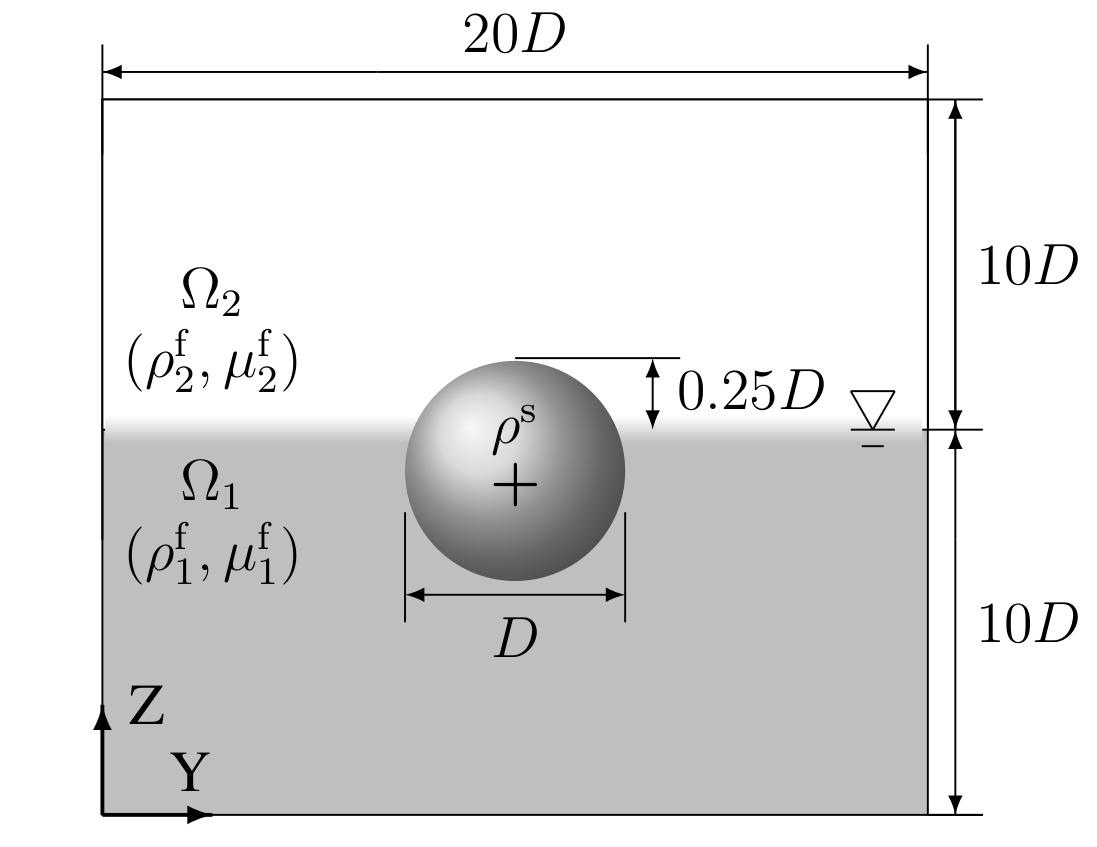} 
		\caption{}
	\end{subfigure}%
	\begin{subfigure}[b]{0.45\textwidth}
		\adjincludegraphics[scale=0.23,trim={{0.15\width} {0.02\width} {0.1\width} {0.0\width}},clip]{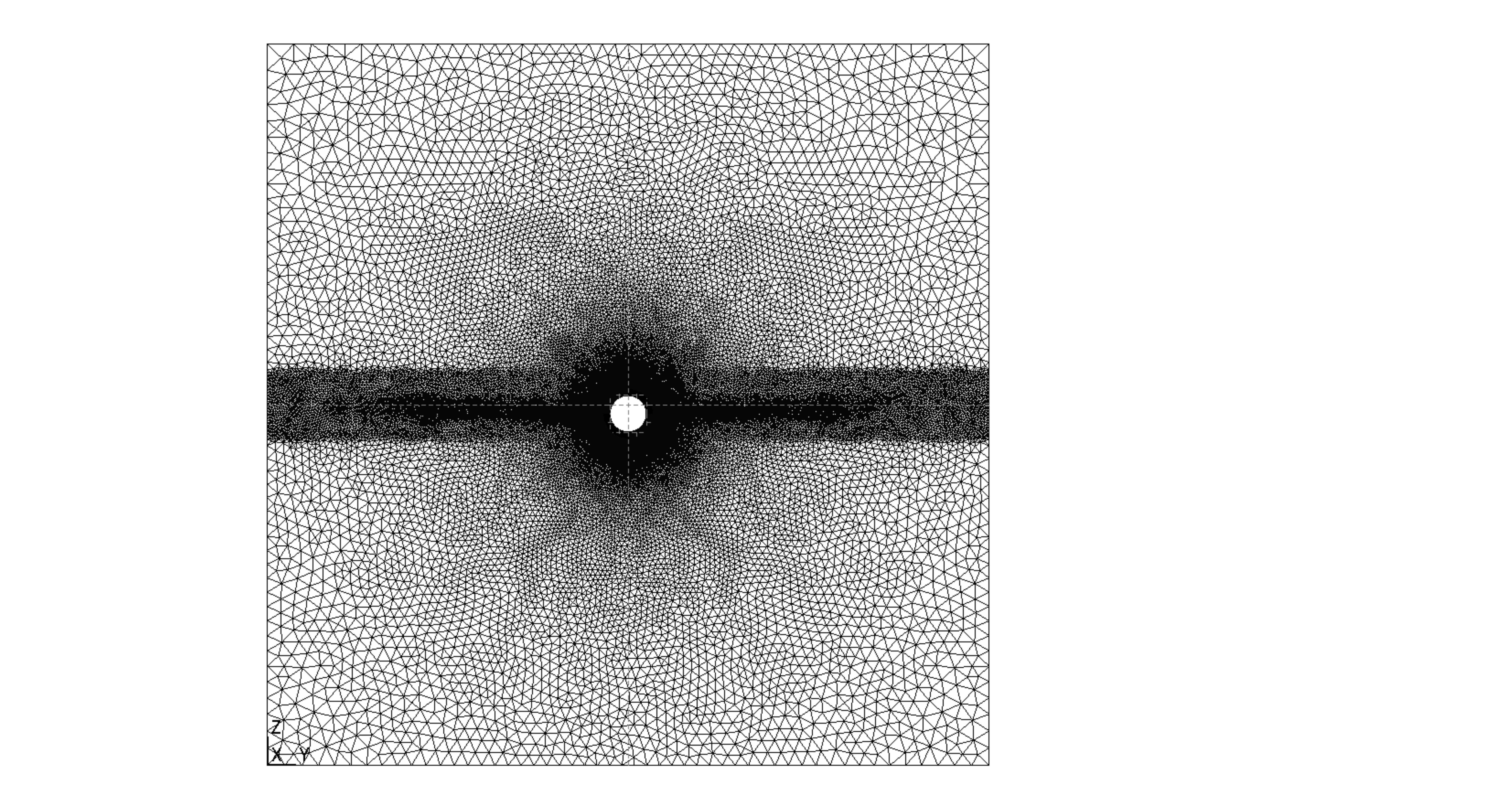} 
		\caption{}
	\end{subfigure}%
	\begin{subfigure}[b]{0.1\textwidth}
		\hspace{-3.1cm}
		\vspace{4.7cm}
		\adjincludegraphics[scale=0.1,trim={{0.27\width} {0.05\width} {0.28\width} {0.06\width}},clip]{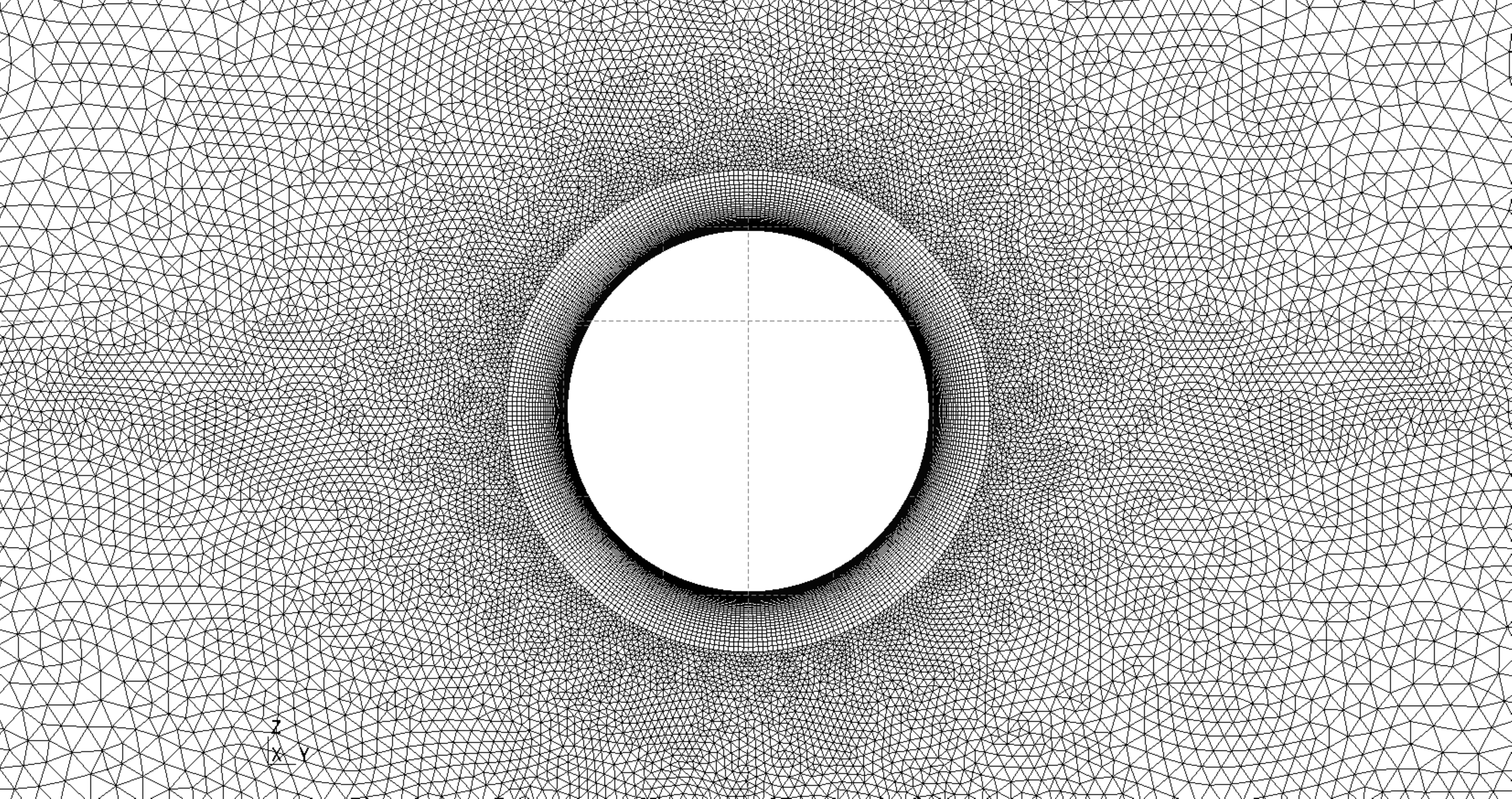} 
	\end{subfigure}
	
	\caption{Problem setup of a piercing sphere at a free surface: (a) sketch showing a cross-section view for the 1-DOF sphere, restricted to move in the transverse direction ($Y$), while piercing the free surface at $h^*=-0.25$,  (b) unstructured finite element mesh in the cross-sectional plane with a close-up view of the boundary layer. The grid size is refined along the air-water interface to capture the free-surface deformation. The flow is in the normal direction ($X$).} 
	\label{Schematic_SPFS}
\end{figure}

\begin{figure}[htbp!]
	\centering
	
	\begin{subfigure}[b]{0.5\textwidth}
		\centering
		\adjincludegraphics[scale=0.2,trim={0\width} {0\width} {0\width} {0.0\width},clip]{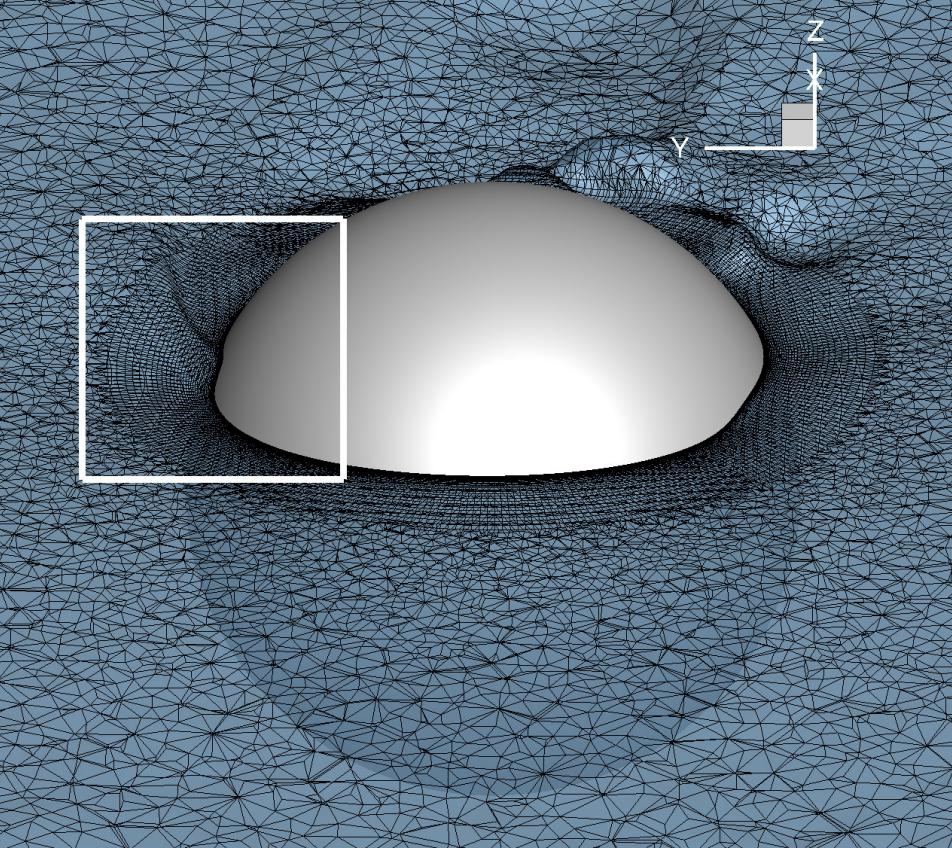}
		\caption{}
	\end{subfigure}%
	\begin{subfigure}[b]{0.5\textwidth}
		\centering
		\adjincludegraphics[scale=0.2,trim={0\width} {0\width} {0\width} {0.0\width},clip]{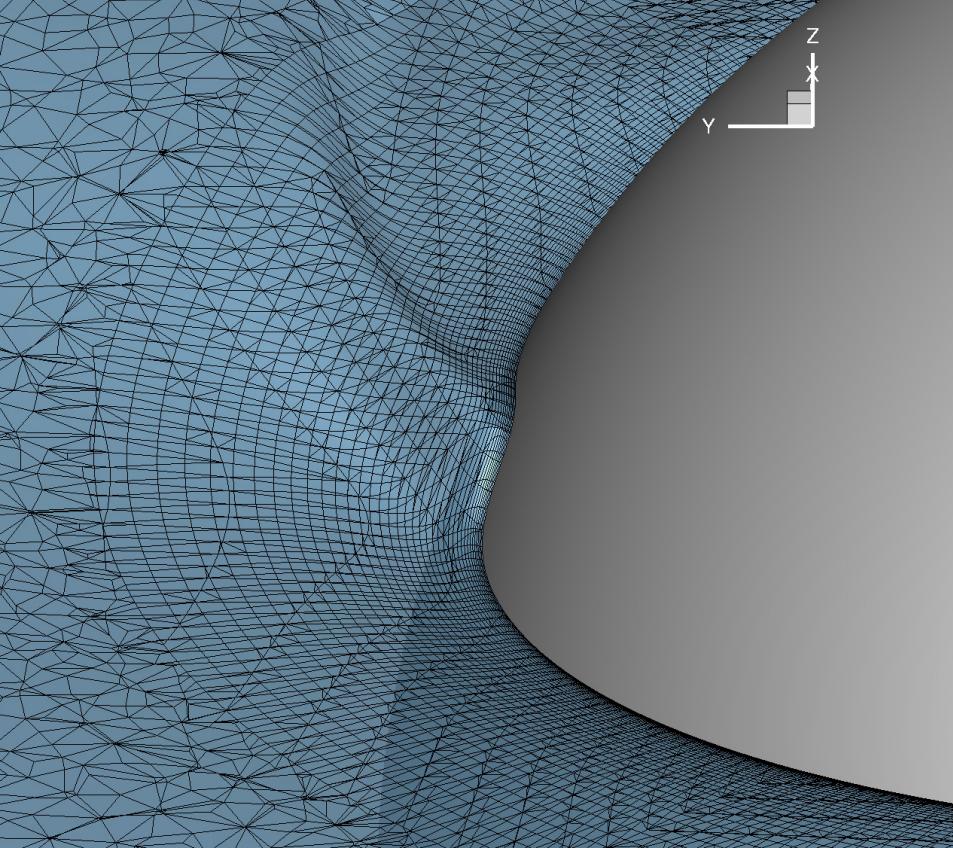}
		\caption{}
	\end{subfigure}%
	\caption{Illustration of mesh motion at the free surface for the hybrid ALE/phase-field formulation: (a) free-surface deformation of 1-DOF elastically mounted sphere piercing the free surface at $h^*=-0.25$ at the lock-in state, and (b) zoomed view from a side of the sphere. }
	\label{Surf_deformation}		
\end{figure}

A schematic of the setup is provided in Fig. \ref{Schematic_SPFS}. To be consistent with the literature, identical parameters are used for the current simulations with the experimental work carried out in \cite{sareen2018} at subcritical Reynolds numbers. A sphere of diameter $D=0.080$ $m$ is placed initially at an offset of $(-0.25D)$ from the free surface in a computational domain $\Omega \in [0,50D]\times[0,20D]\times[0,20D]$. The physical properties of the two phases are $\rho^\mathrm{f}_1=1\,000$, $\rho^\mathrm{f}_2=1.225$, $\mu^\mathrm{f}_1=1\times 10^{-3}$ and $\mu^\mathrm{f}_2=1.983\times 10^{-5}$ and the mass ratio considering the submerged volume of the sphere is $m^{*} = 9.2$ at $h^*=-0.25$. Fig. \ref{Surf_deformation} shows the mesh motion and the free-surface deformation for the case of the piercing sphere at $h^*=-0.25$. Noticeable standing wave structures are formed from sides of the sphere as it pierces the free surface, similar to the observation in \cite{sareen2018}.

\begin{figure}[htbp!]
	\centering
	
	\adjincludegraphics[scale=0.3,trim={0\width} {0\width} {0\width} {0.0\width},clip]{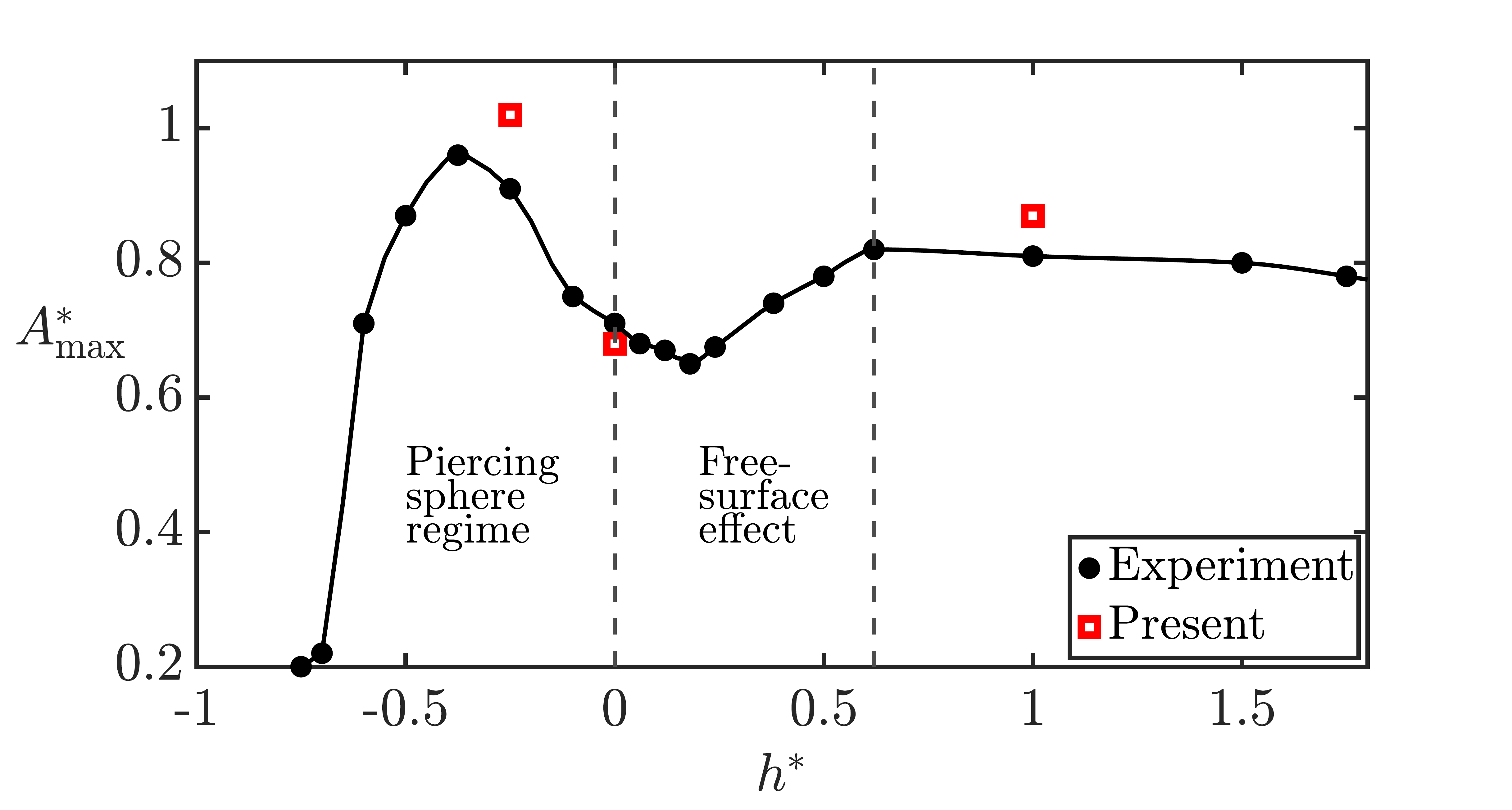}
	
	\caption{The variation of the maximum amplitude response ($A^*_\mathrm{max}$) as a function of the immersion ratio $(h^*)$.}
	\label{hS_Trend}		
\end{figure}

Fig. \ref{hS_Trend} shows the variation of maximum peak amplitude, $A^*_\mathrm{max}$, with immersion ratio $h^*$, for the experiments performed in \cite{sareen2018}. The maximum VIV response in this study for $h^*=0$ and $h^*=-0.25$, is reported for the reduced velocity range $U^*\in[7,15]$. Different regimes were identified, where specific features were dominant \cite{sareen2018}. Table \ref{Comparison_1} compares the amplitude response, the normalized transverse force in $y$-direction and the total phase of the present simulations with experiments in \cite{sareen2018} at the lock-in regime $(U^*=10)$ for the fully and partially submerged cases. It is quantified that the amplitude response for the piercing case is greater than all submerged cases in both experiments and our numerical results at the lock-in state.
Through our results, similar to the observation in \cite{sareen2018}, we find that the peak amplitude response of the submerged sphere at $h^*=0$ is decreased by almost $30\%$ compared to the case at $h^*=1$. It is found that as the sphere pierces the free surface at $h^*=-0.25$, the amplitude response increases substantially with the maximum peak-to-peak amplitude $\sim 2D$. Based on the total phase difference in Table \ref{Comparison_1}, the amplitude response for the submerged cases at $U^*=10$ corresponds to mode II. For the piercing case, the maximum amplitude response at $U^*=10$ corresponds to mode I of vibration. It can be deduced that the lock-in region is shifted toward higher reduced velocities for the piercing sphere case. 

\begin{table}[htbp!]
	\centering
	\caption{Comparison of the r.m.s. amplitude response, $A^*_\mathrm{rms}$, the r.m.s. of the normalized transverse force, $C_{y,\mathrm{rms}}$, and the total phase difference, $\phi_{total}$ for the piercing sphere case with $m^*=9.2$, and fully submerged cases with $m^*=7.8$ at $U^*=10$, $m^*\zeta=0.017$ and $Fr=0.22$ with the experimental data in \cite{sareen2018}. }
	\centering
	\begin{tabular}{  M{2cm}  M{2.5cm}  M{3cm}   M{2.5cm} N }
		\hline
		
		\hline
		
		\centering
		Parameters & Case  & Sareen et al. \cite{sareen2018} & Present study & \\[10pt]
		\hline
		
		\hline
		
		& $h^*=1$   &  0.81   & 0.87   &\\[10pt]
		\large$A^*_{\mathrm{rms}}$ & $h^*\sim0$   &  0.65   &  0.68 &\\[10pt]
		& $h^*=-0.25$   &  0.88   &  1.02 &\\[10pt]
		
		\hline
		& $h^*=1$   &  0.08    &  0.12  &\\[10pt]
		\large$C_{y,\mathrm{rms}}$  & $h^*\sim0$   &   -   &  0.14   &\\[10pt]
		& $h^*=-0.25$   &  -   &  0.10 &\\[10pt]
		
		\hline
		& $h^*=1$   &  $\sim149$    &  $\sim179$  &\\[10pt]
		\large$\phi_{total}$  & $h^*\sim0$   &   $\sim175$   &  $\sim178$   &\\[10pt]
		& $h^*=-0.25$   &  $\sim4$   &  $\sim1$ &\\[10pt]
		\hline
		
		\hline
	\end{tabular}
	\label{Comparison_1}
\end{table}

\begin{figure}[htbp!]
	\centering
	\begin{subfigure}[b]{0.5\textwidth}
		\adjincludegraphics[scale=0.25,trim={0.2\width} {0\width} {0.2\width} {0.0\width},clip]{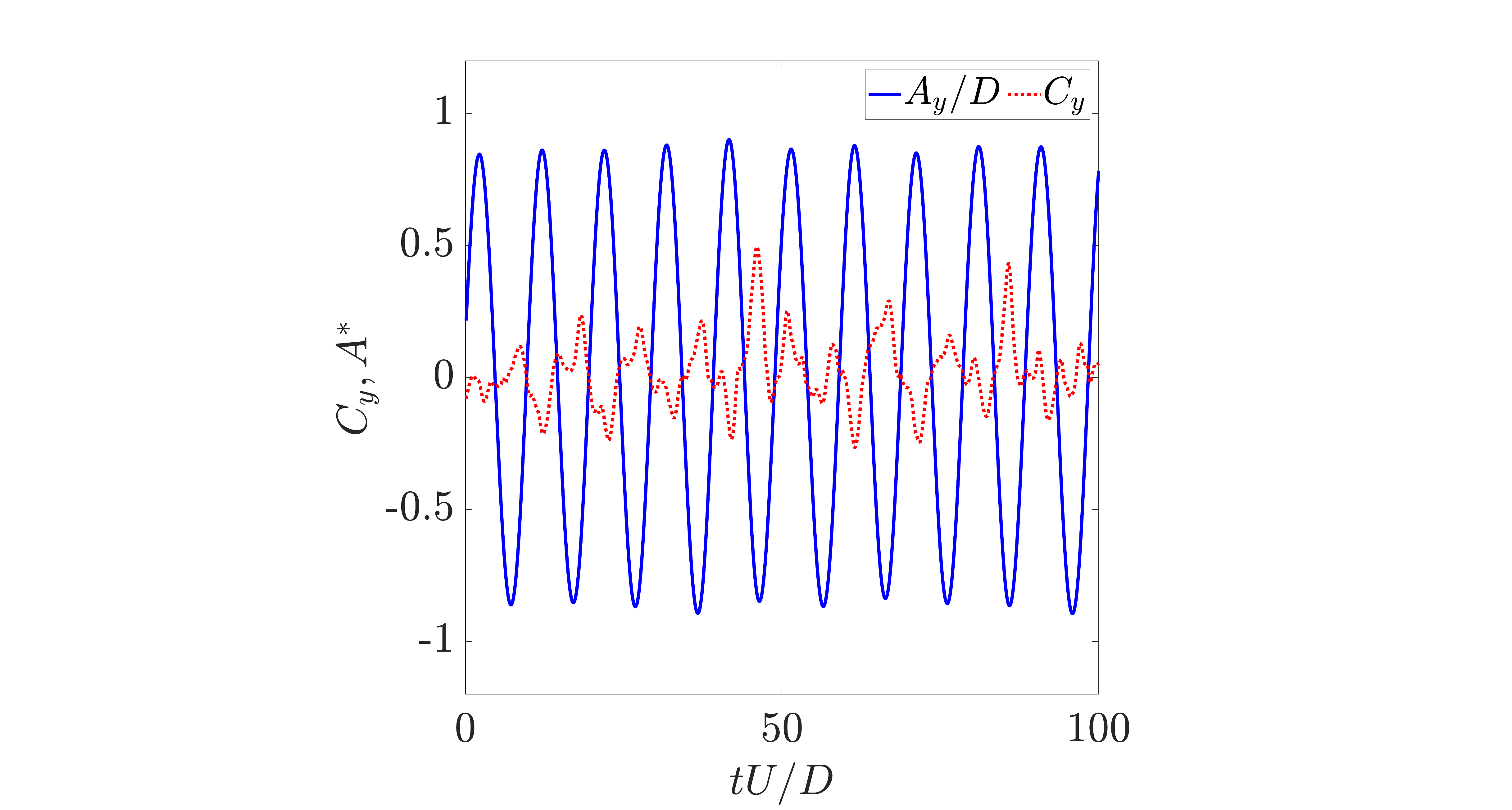}
		\caption*{($a$ - $1$)}
	\end{subfigure}%
	\begin{subfigure}[b]{0.5\textwidth}
		\adjincludegraphics[scale=0.25,trim={0.2\width} {0\width} {0.1\width} {0.0\width},clip]{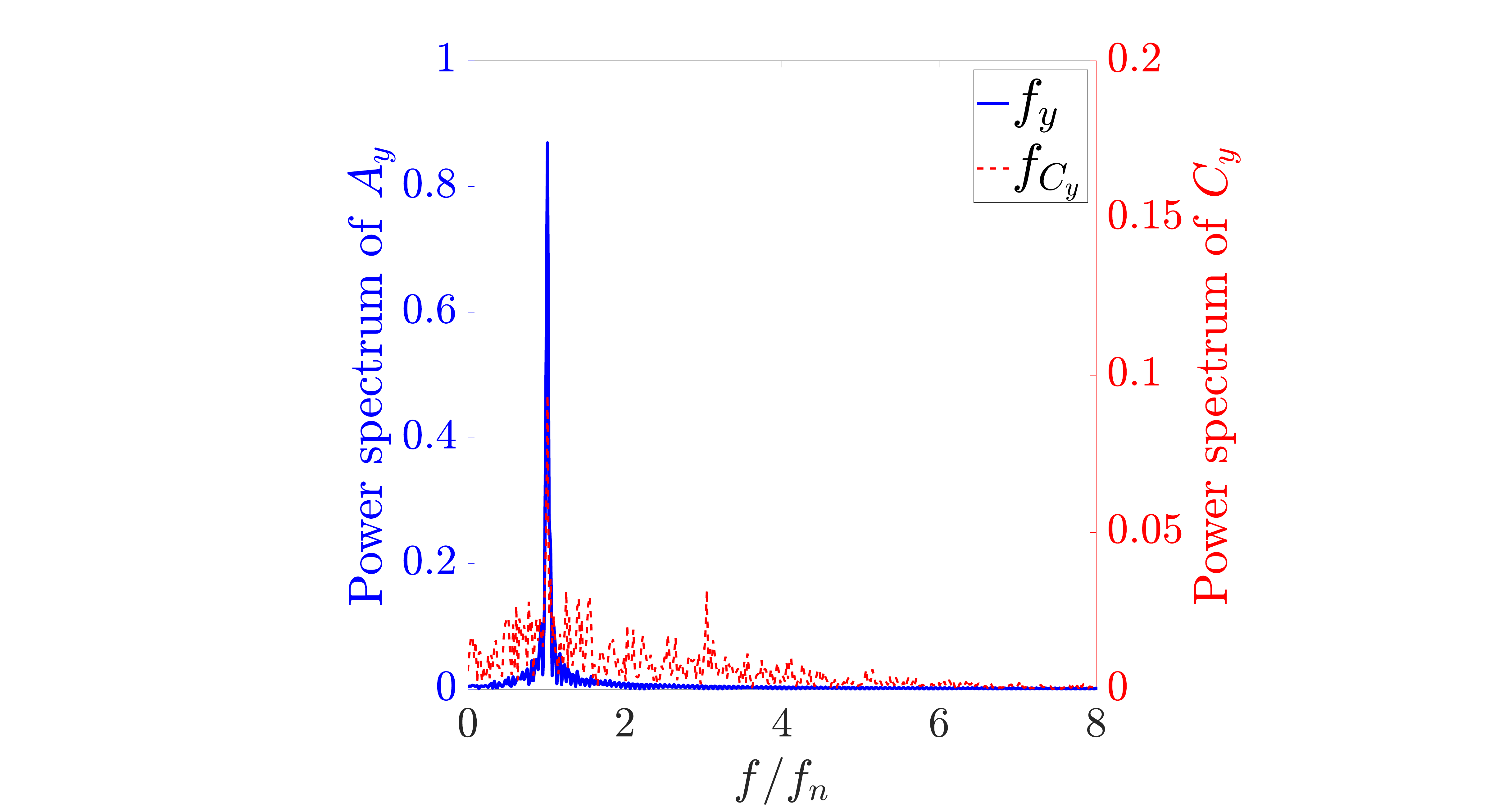}
		\caption*{($a$ - $2$)}
	\end{subfigure}
	\begin{subfigure}[b]{0.5\textwidth}
		\adjincludegraphics[scale=0.25,trim={0.2\width} {0\width} {0.2\width} {0.0\width},clip]{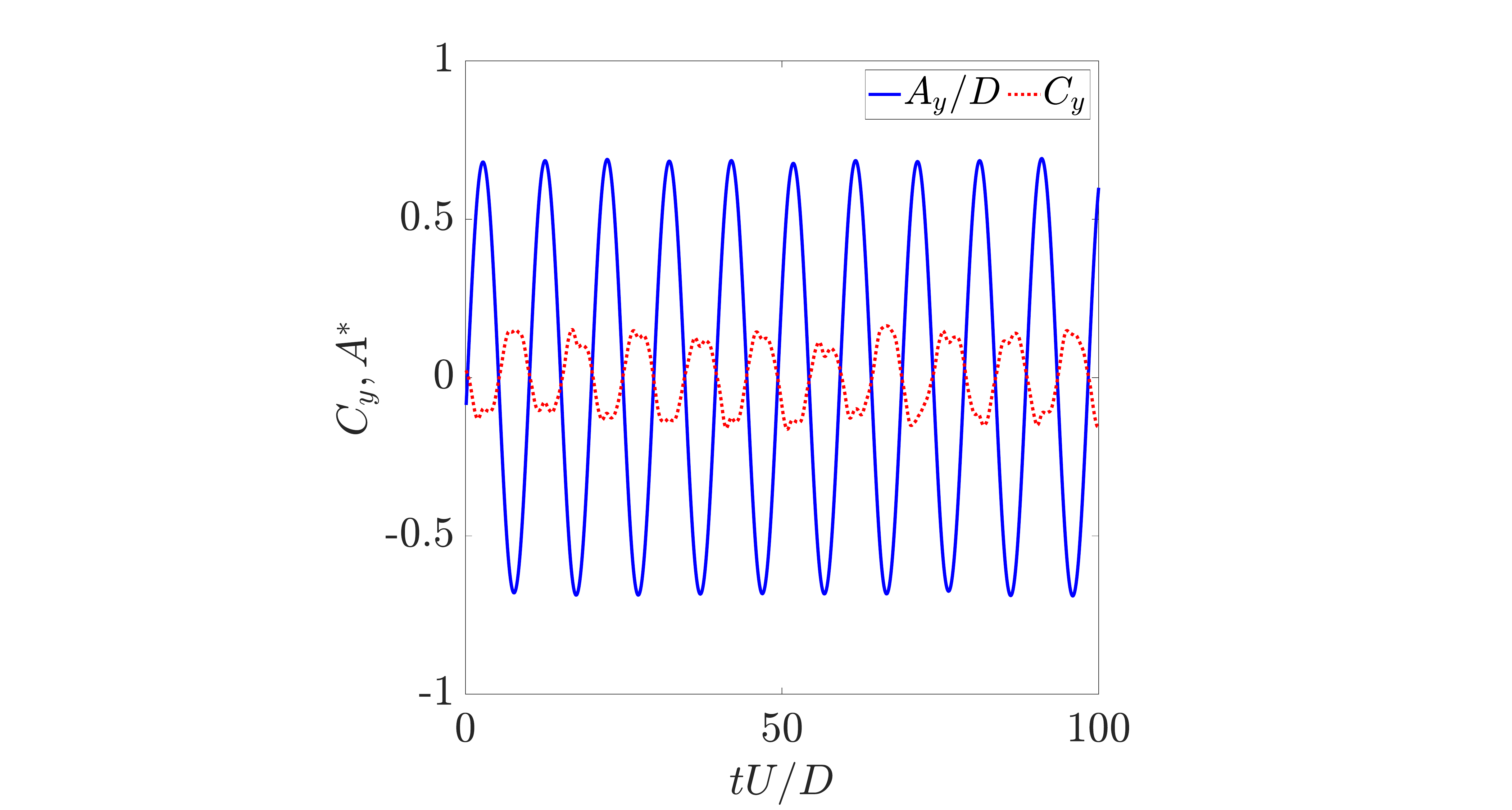}
		\caption*{($b$ - $1$)}
	\end{subfigure}%
	\begin{subfigure}[b]{0.5\textwidth}
		\adjincludegraphics[scale=0.25,trim={0.2\width} {0\width} {0.1\width} {0.0\width},clip]{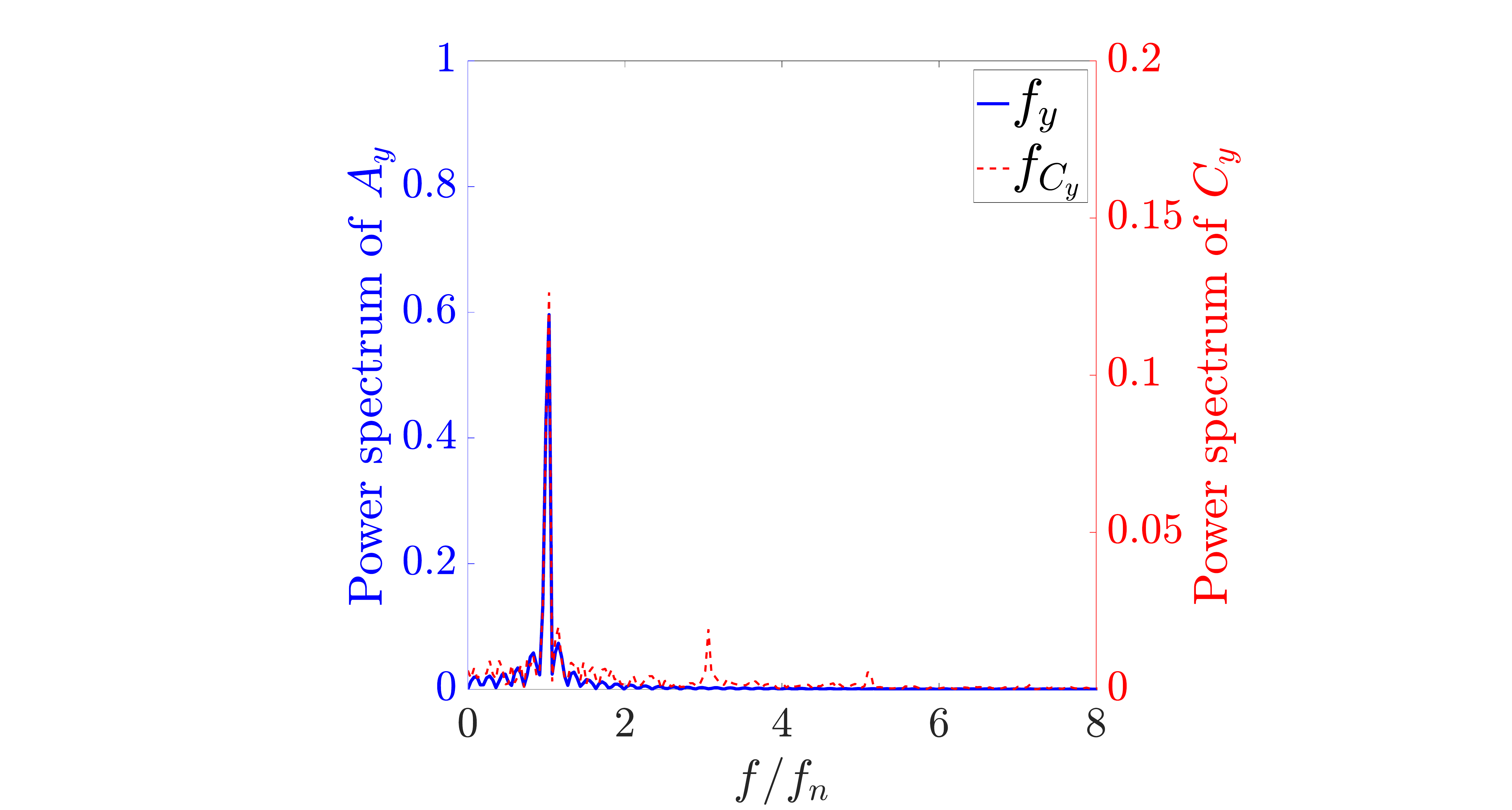}
		\caption*{($b$ - $2$)}
	\end{subfigure}
	\begin{subfigure}[b]{0.5\textwidth}
		\adjincludegraphics[scale=0.25,trim={0.2\width} {0\width} {0.2\width} {0.0\width},clip]{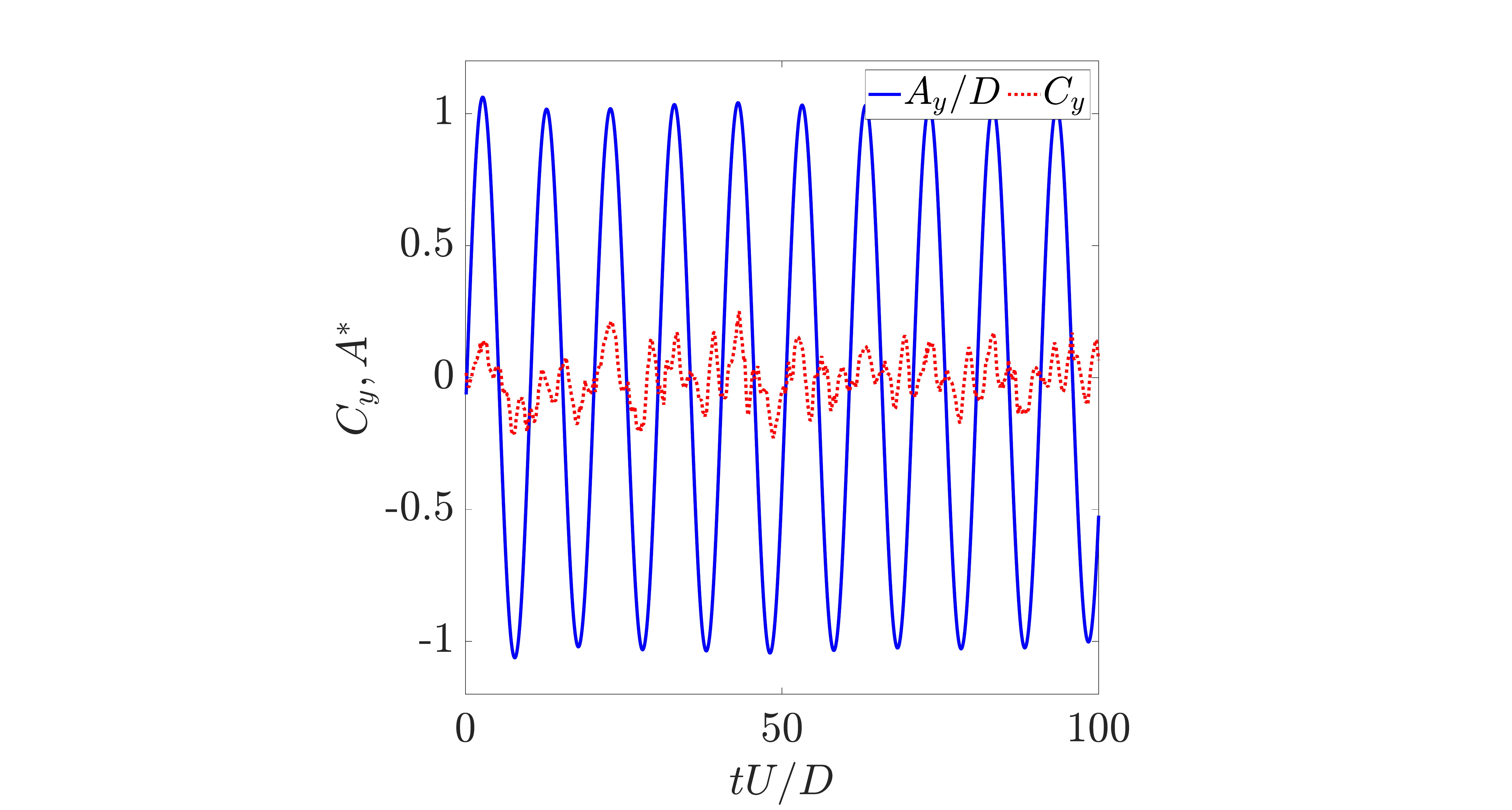}
		\caption*{($c$ - $1$)}
	\end{subfigure}%
	\begin{subfigure}[b]{0.5\textwidth}
		\adjincludegraphics[scale=0.25,trim={0.2\width} {0\width} {0.1\width} {0.0\width},clip]{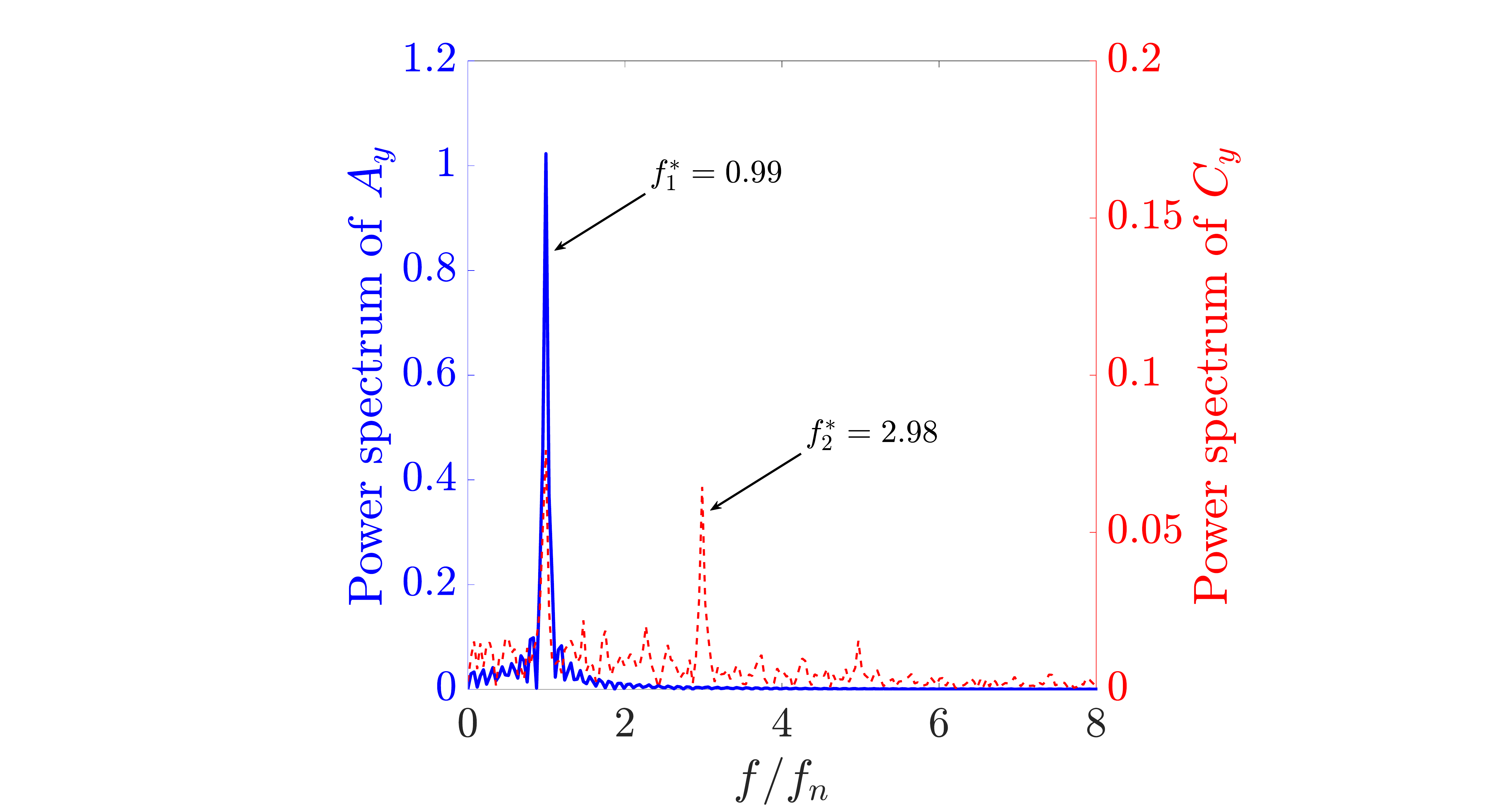}
		\caption*{($c$ - $2$)}
	\end{subfigure}
	\caption{Time histories of the amplitude response ($A^*$) and the normalized transverse force (${C_y}$) with non-dimensional time and their corresponding frequency spectrum with normalized frequency ($f^*$), at $U^*=10$ for 1-DOF sphere at (a) $h^*=1$, (b) $h^*=0$ and (c) $h^*=-0.25$. The mass ratio for the fully submerged cases ($h^*=1$ and $h^*=0$) is $m^*=7.8$ and for partially submerged case ($h^*=-0.25$) is $m^*=9.2$. }	
	\label{FS_TH}	
\end{figure}

\begin{figure}[htbp!]
	\centering
	\begin{subfigure}[b]{0.5\textwidth}
		\adjincludegraphics[scale=0.28,trim={0.2\width} {0\width} {0.2\width} {0.0\width},clip]{Photos/49.pdf}
		\caption*{($a$ - $1$)}
	\end{subfigure}%
	\begin{subfigure}[b]{0.5\textwidth}
		\adjincludegraphics[scale=0.28,trim={0.2\width} {0\width} {0.1\width} {0.0\width},clip]{Photos/50.pdf}
		\caption*{($a$ - $2$)}
	\end{subfigure}
	\begin{subfigure}[b]{0.5\textwidth}
		\adjincludegraphics[scale=0.28,trim={0.2\width} {0\width} {0.2\width} {0.0\width},clip]{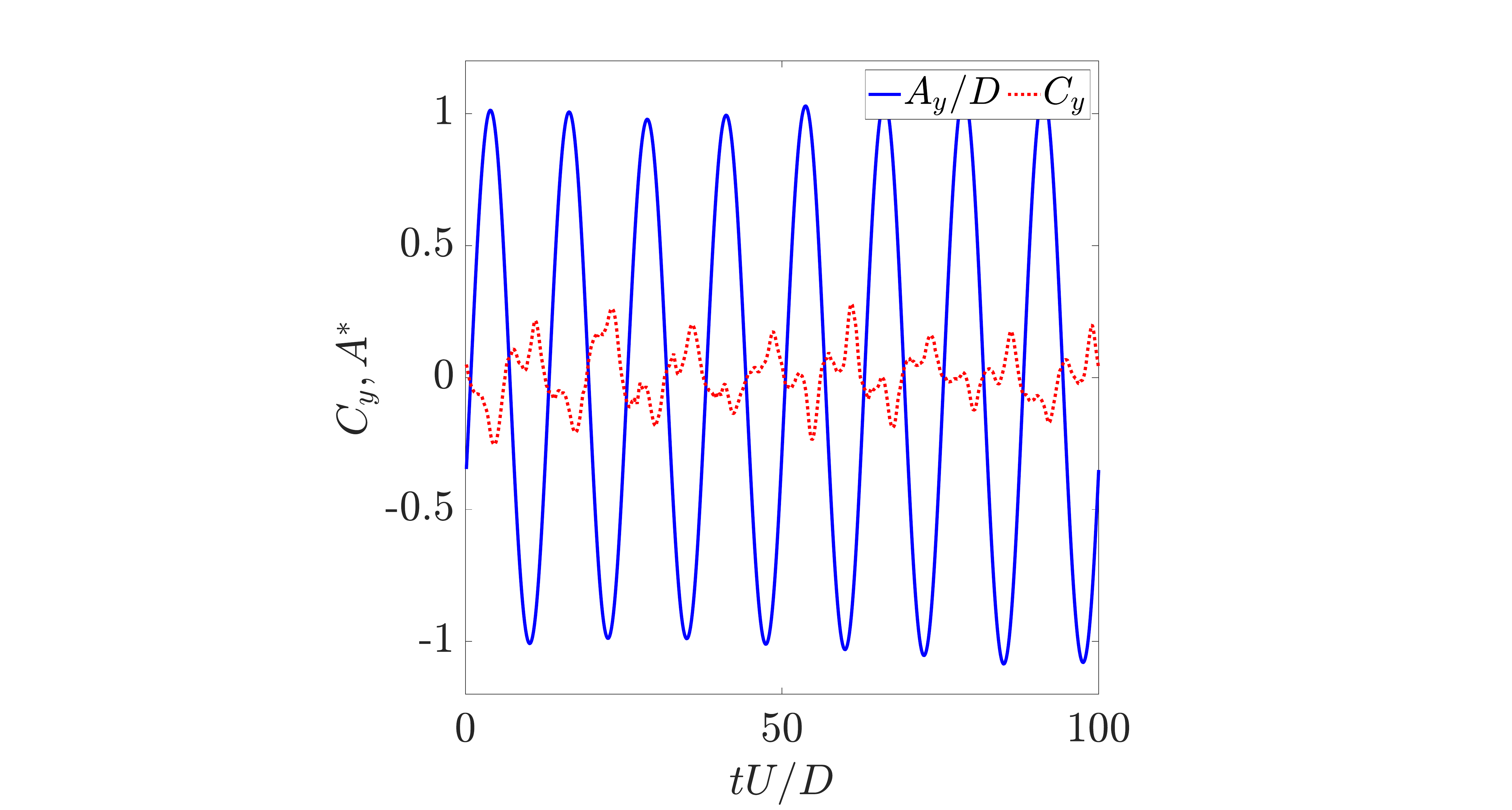}
		\caption*{($b$ - $1$)}
	\end{subfigure}%
	\begin{subfigure}[b]{0.5\textwidth}
		\adjincludegraphics[scale=0.28,trim={0.2\width} {0\width} {0.1\width} {0.0\width},clip]{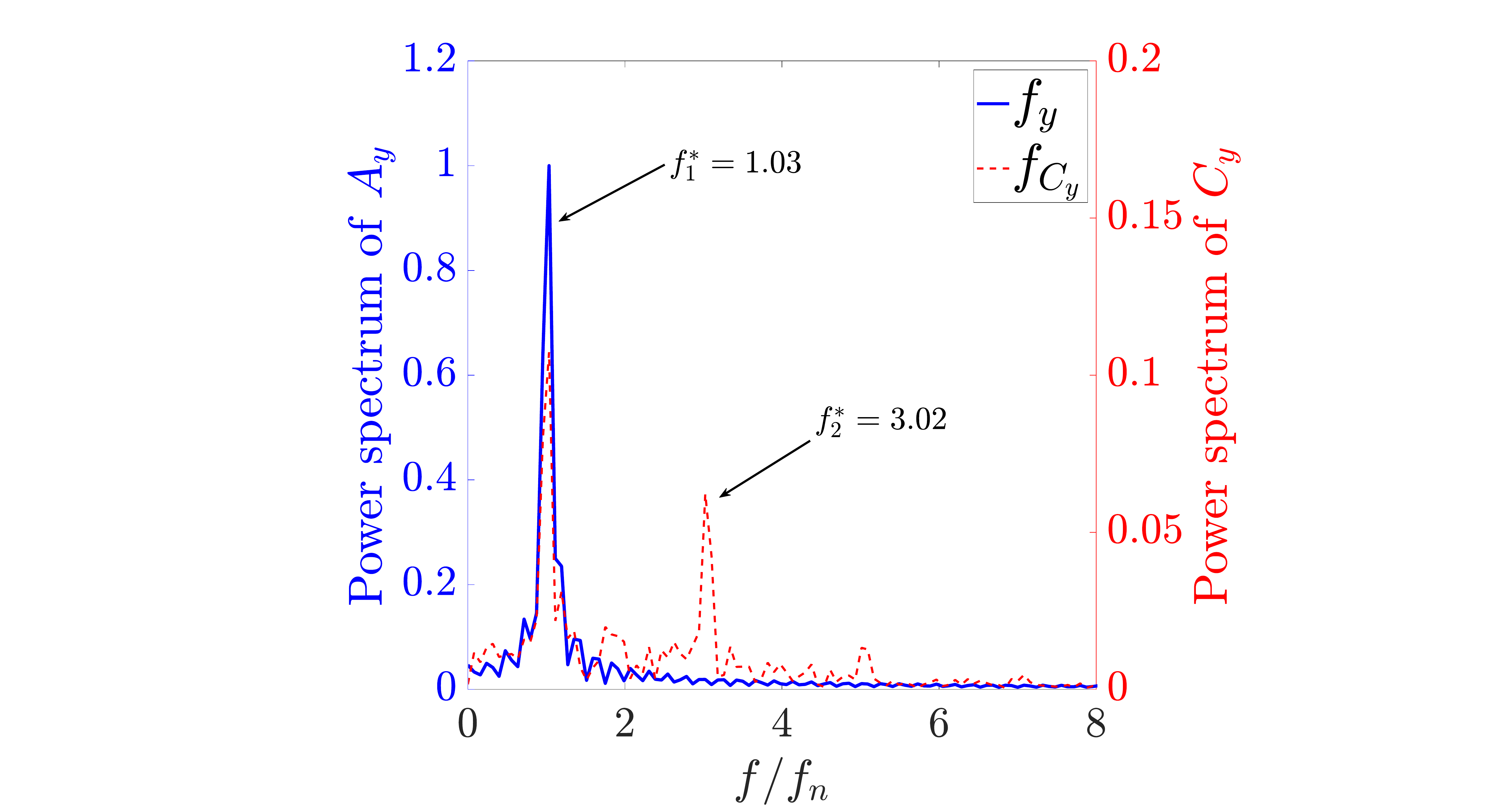}
		\caption*{($b$ - $2$)}
	\end{subfigure}
	\caption{Time histories of the amplitude response ($A^*$) and the normalized transverse force (${C_y}$) and their corresponding frequency spectra  for 1-DOF sphere piercing the free surface at $h^*=-0.25$ at two reduced velocities: (a) $U^*=10$ and, (b) $U^*=12.7$. }	
	\label{FS25_TH_Ur}	
\end{figure}

Fig. \ref{FS_TH} (a) shows the time histories of the amplitude response and the normalized transverse force for the sphere at $h^*=-0.25,0,1$ at $U^*=10$.
For the fully submerged cases, the amplitude response decreases by changing the immersion ratio from $h^*=1$ to $h^*=0$ at the lock-in state. The amplitude response is increased significantly as the sphere pierces the free surface at $h^*=-0.25$. The maximum peak-to-peak amplitude response of $\sim2D$ is observed at the stationary state which is larger than all the submerged cases studied (Fig. \ref{Validation_Trend}).
Further analysis of the frequency spectrum is shown in Fig. \ref{FS_TH}. The only dominant vortex shedding frequency for the fully submerged cases ($h^*=1$ and $h^*=0$) is at $f^*=1$ corresponding to the VIV response. However, two dominant frequencies of the vortex shedding at $f^*_1=0.99$ and $f^*_2=2.98$ are found for the piercing sphere case. The existence of a third-harmonic behavior for the piercing sphere case is related to the vorticity/free-surface interaction. Fig. \ref{FS25_TH_Ur} shows the time traces and the corresponding frequency spectrum of the amplitude response and the normalized transverse force at $h^*=-0.25$ for two different reduced velocities at mode I ($U^*=10$) and mode II ($U^*=12.7$). The third harmonic behavior is observed for both modes of vibration at the VIV regime. As a baseline for our study, where the large-amplitude oscillation is found around $h^*\sim-0.25$, we aim to focus on the VIV response and the wake dynamics for the piercing sphere case and compare with the submerged cases in the next subsection.



\subsection{Vorticity dynamics with free-surface deformation}
In this section, the effects of free surface on the vibration response and the wake dynamics of a fully and partially submerged sphere are investigated. 
On a free surface, the primary driving mechanism of vorticity creation is the balance between the shear stress (measured by tangent vorticity) and the tangent components of the surface-deformation stress. For an incompressible viscous Newtonian fluid, an analytical relationship between the tangential stress and 
the surface vorticity at a free surface was derived in \cite{wu1995}.
To analyze the parallel surface vorticity on a curved free interface $S$, a vector and its derivatives, including gradient operator $\nabla$ can be decomposed into components tangent $\pi$ and normal $n$ to $S$, e.g.
 $\u=\u_{\pi}+\mathbf{n} \un_{n}$, and 
$\nabla(\cdot)=\nabla_{\pi}(\cdot)+\mathbf{n} \frac{\partial(\cdot)}{\partial n}$, 
where  $\un_{n}$ denotes the normal velocity component given by $\un_{n}=\mathbf{n} \cdot \u$  and
 $\u_{\pi}$ is the tangential velocity vector as $\u_{\pi}=\mathbf{n} \times(\u \times \mathbf{n})$.
As derived in \cite{wu1995}, the tangent vorticity $\omegab_{\pi}$ right on the free surface is solely balanced by the tangent components of the surface stress as follows: $\omegab_{\pi}=-2 \mathbf{n} \times\left(\nabla_{\pi} \u_{n}+ \u_{\pi} \cdot \mathbf{K}\right)$,
where $\mathbf{K} \equiv-\nabla_{\pi} \mathbf{n}$ is the surface curvature tensor. This analytical expression clearly underlines the relationship between the surface parallel vorticity and the curvature of a deformed free surface. The vorticity of different signs is created in the flow field whenever there is a curvature in the free surface.
The free surface deforms as it interacts with the vorticity field and \textit{vice versa}. The interaction of the initial vorticity field along the free surface may lead to the generation of additional vorticity by the deformation of the free surface. In the present study, the free surface deforms as it interacts with the spherical body and there is a complex nonlinear interaction between the sphere wake and the vorticity flux at the free surface.  We attempt to explore the complex vorticity interactions with the free surface in the context of piercing sphere VIV response.

\begin{figure}[htbp!]
	\centering
	\begin{subfigure}[b]{0.33\textwidth}
		\adjincludegraphics[scale=0.14,trim={0\width} {0\width} {0\width} {0.0\width},clip]{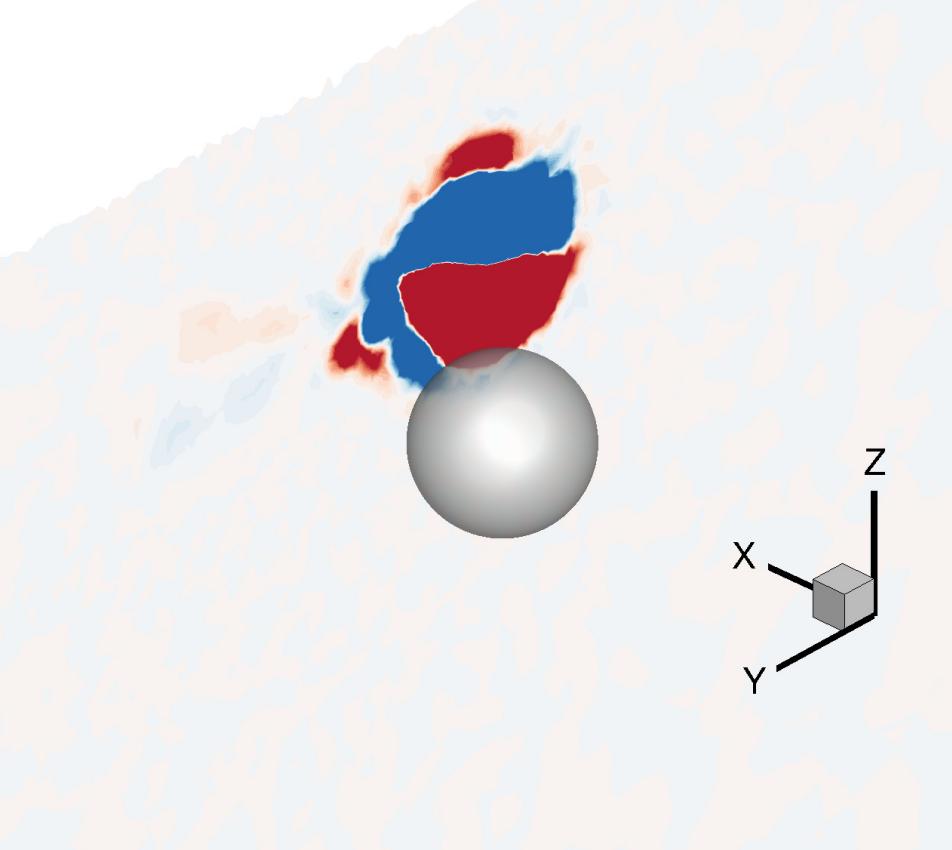}
		\caption{}
	\end{subfigure}%
	\begin{subfigure}[b]{0.33\textwidth}
		\adjincludegraphics[scale=0.14,trim={0\width} {0\width} {0\width} {0.0\width},clip]{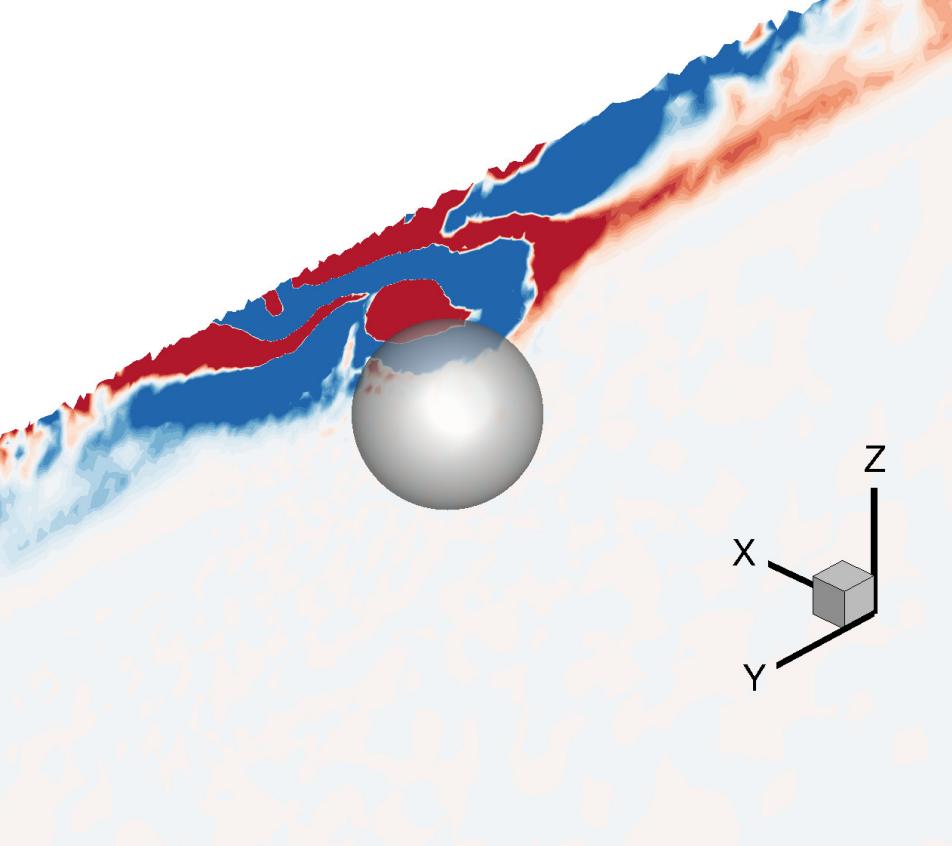}
		\caption{}
	\end{subfigure}%
	\begin{subfigure}[b]{0.33\textwidth}
		\adjincludegraphics[scale=0.14,trim={0\width} {0\width} {0\width} {0.0\width},clip]{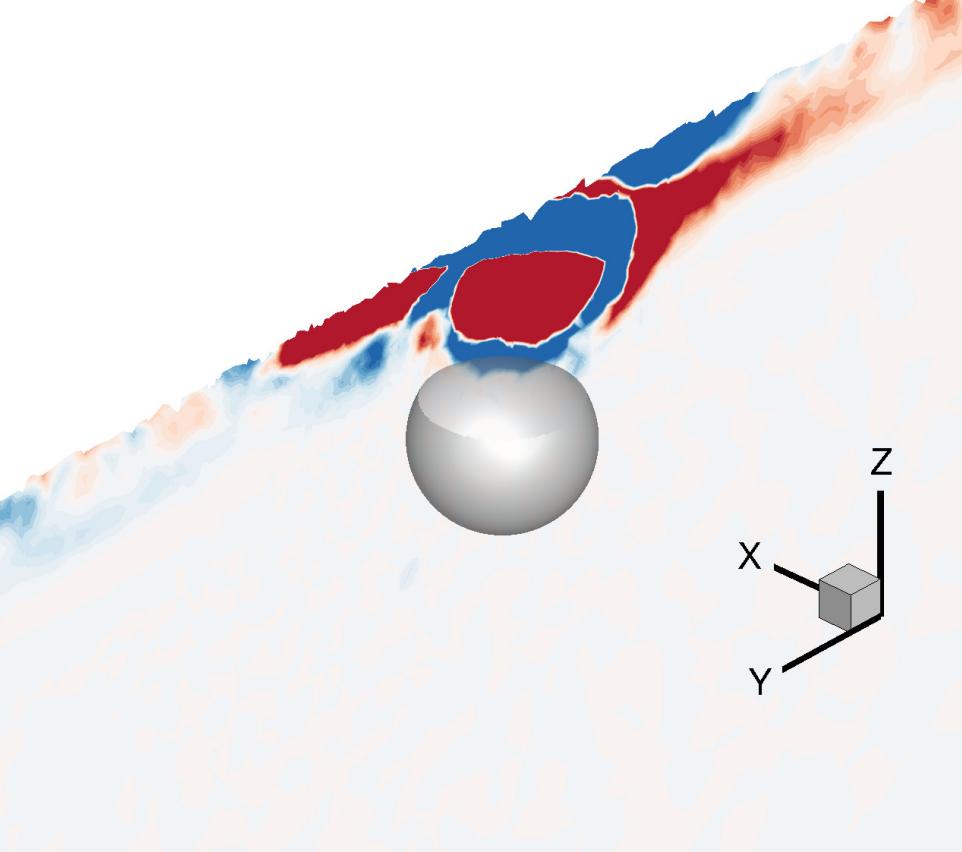}
		\caption{}
	\end{subfigure}
	
		\begin{subfigure}[b]{0.5\textwidth}
			\centering
						\hspace{-2.0cm}
			\adjincludegraphics[scale=0.4,trim={0.05\width} {0.7\width} {0.18\width} {0.07\width},clip]{Photos/38.pdf}
			\caption*{}
		\end{subfigure}
	
	\caption{Streamwise $x$-vorticity $(\omega_{x}D/U\in[-3.3,3.3])$ contour on $Y$-$Z$ plane at $1.5D$ downstream from the centre of the sphere for: (a) the fully submerged sphere at $h^*=1$, (b) $h^*=0$ and (c) the piercing sphere case at $h^*=-0.25$ at mean amplitude position.}
	\label{xVor_FS_Ur10_Compare} 
\end{figure}

Through some quantitative and qualitative comparison with the experiment in \cite{sareen2018}, we study the VIV wake dynamics of the sphere at $h^*=1$ (fully submerged case), $h^*=0$ (when the top of the sphere touches the free surface) and $h^*=-0.25$ (when the sphere is piercing the free surface). The temporal evolution of streamwise vorticity in a plane normal to the flow can provide important insight into wake dynamics for the sphere as the streamwise vortex loops pass through the cross-plane. Hence, in the current study, we have measured the streamwise vorticity in a cross-plane and will compare with the experiments in the literature \cite{govardhan2005vortex, sareen2018}.
Fig. \ref{xVor_FS_Ur10_Compare} shows the $x$-vorticity contour plots for the submerged sphere at $h^*=1$, $h^*=0$ and the piercing sphere at $h^*=-0.25$ at $1.5D$ downstream, while the sphere is at the end of its stroke. The plots correspond to the reduced velocity of $U^*=10$, where the peak amplitude is obtained at the VIV regime. The top boundary in the plots represents the free-surface boundary. The streamwise vorticity for the submerged case at $h^*=1$, Fig. \ref{xVor_FS_Ur10_Compare} (a), consists of two dominant opposite sign vortex pair that is symmetric across the horizontal plane. These vortex loops formation is consistent with the observation in \cite{govardhan2005vortex, sareen2018}.
Fig. \ref{xVor_FS_Ur10_Compare} (b, c) shows the change in the formation of the vortex pairs when the sphere moves closer to the free surface at $h^*=0$ and when it pierces the free surface at $h^*=-0.25$. As it is evident from the plots, the vortex structures change significantly due to the effect of the free surface and the horizontal plane through the sphere center can also no longer act as a plane of symmetry. Stretched vorticity formation (compared to Fig. \ref{xVor_FS_Ur10_Compare} (a)) is observed due to the effect of the free surface.

\begin{figure}[H]
	\centering
	\begin{subfigure}[b]{0.7\textwidth}
			\hspace{0cm}
		\adjincludegraphics[scale=0.2,trim={0\width} {0.0\width} {0\width} {0.0\width},clip]{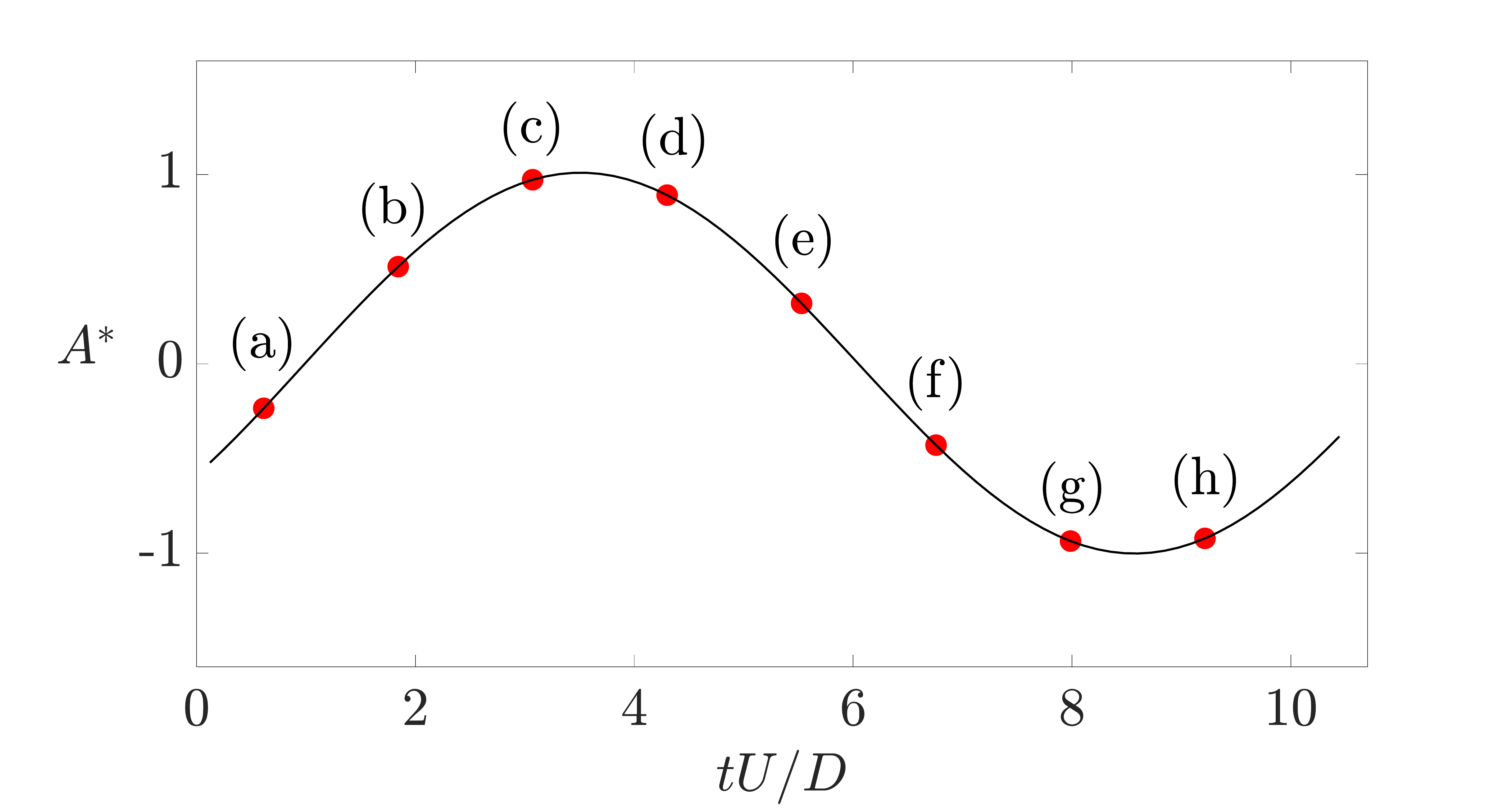}
	\end{subfigure}

	\begin{subfigure}[b]{0.5\textwidth}
		\centering
		\adjincludegraphics[scale=0.4,trim={0.05\width} {0.7\width} {0.01\width} {0.0\width},clip]{Photos/38.pdf}
		\caption*{}
	\end{subfigure}

	\begin{subfigure}[b]{0.25\textwidth}
		\adjincludegraphics[scale=0.1,trim={0\width} {0.2\width} {0\width} {0.0\width},clip]{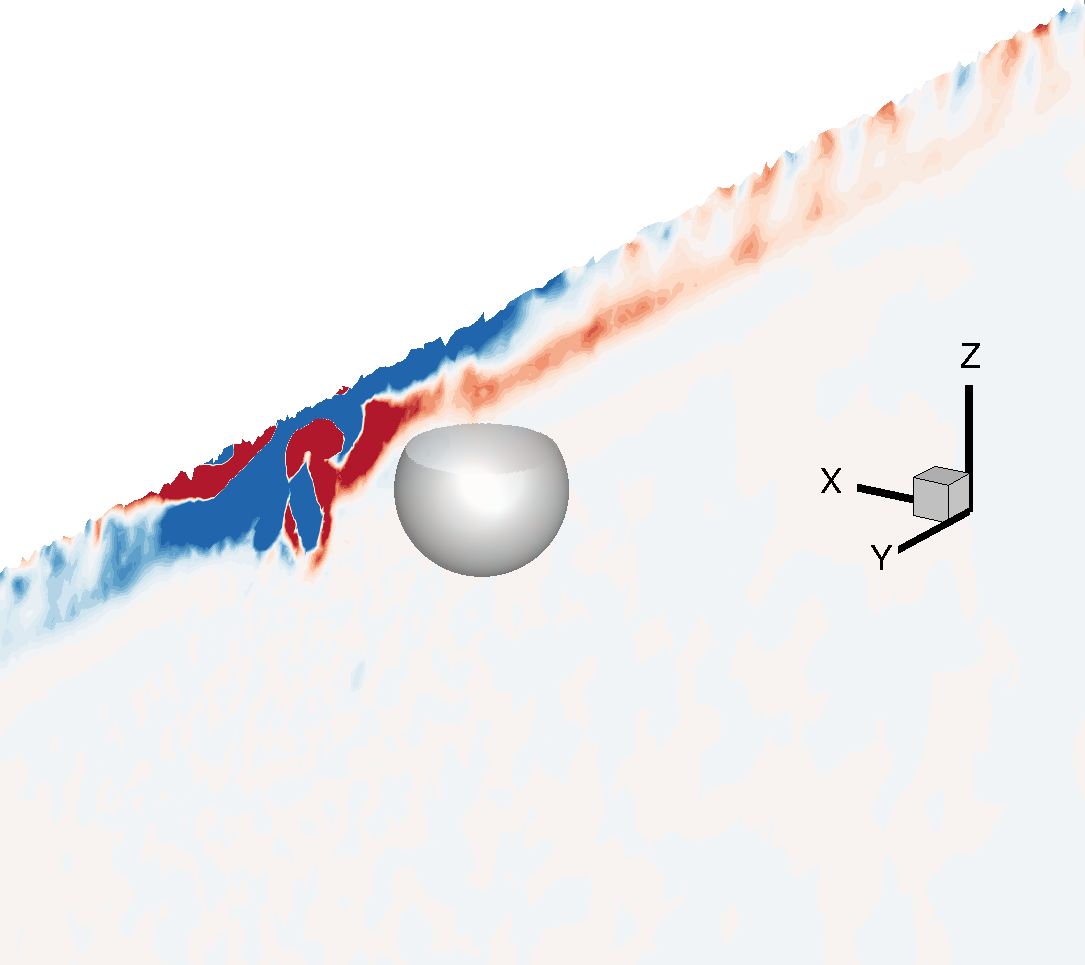}
		\caption*{$(a)$}
	\end{subfigure}%
	\begin{subfigure}[b]{0.25\textwidth}
		\adjincludegraphics[scale=0.1,trim={0\width} {0.2\width} {0\width} {0.0\width},clip]{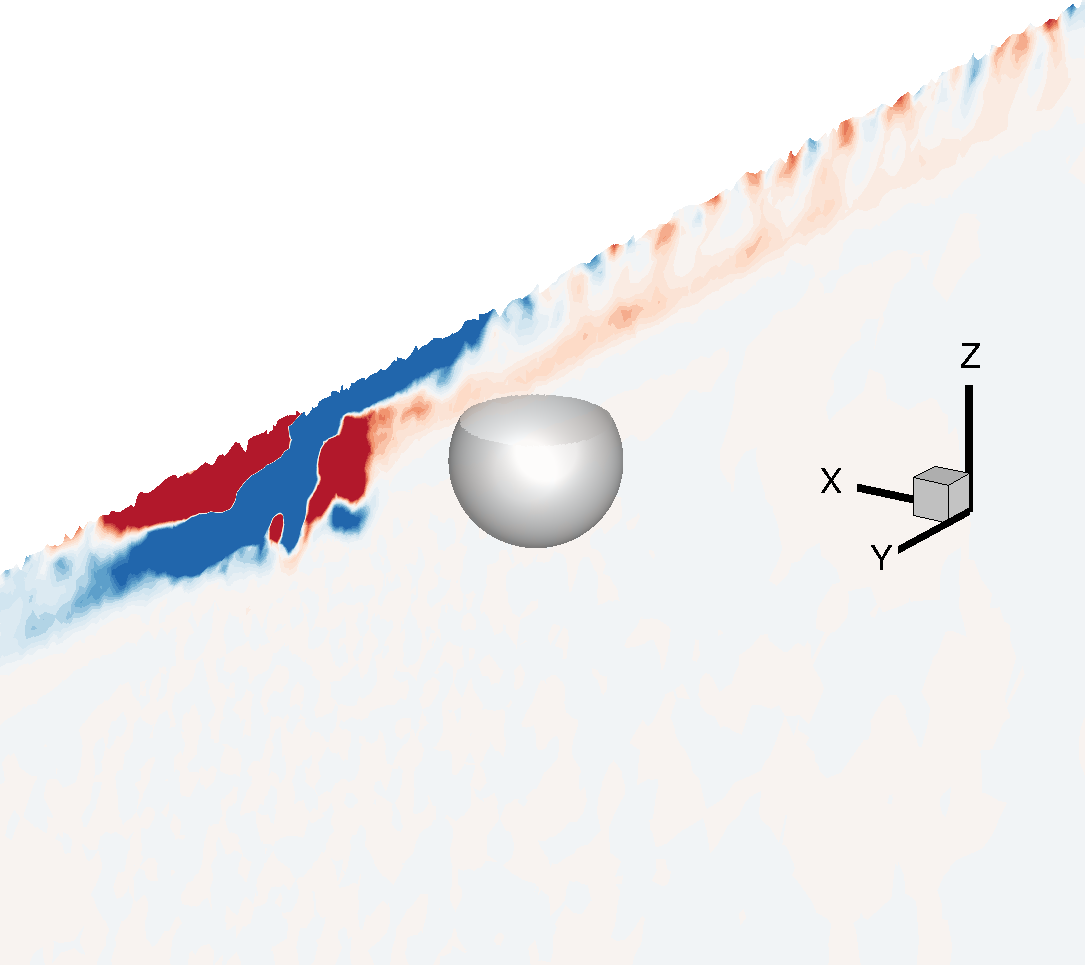}
		\caption*{$(b)$}
	\end{subfigure}%
	\begin{subfigure}[b]{0.25\textwidth}
		\adjincludegraphics[scale=0.1,trim={0\width} {0.2\width} {0\width} {0.0\width},clip]{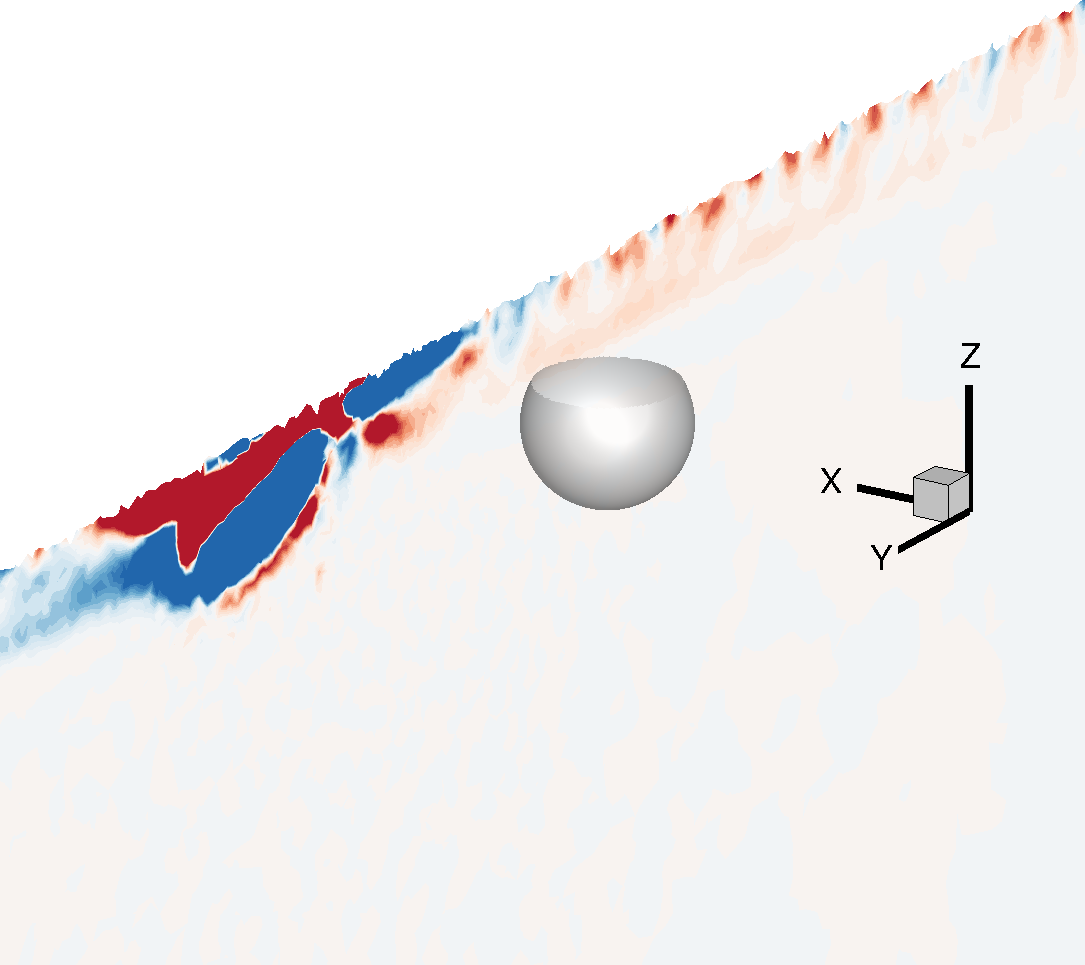}
		\caption*{$(c)$}
	\end{subfigure}%
	\begin{subfigure}[b]{0.24\textwidth}
		\adjincludegraphics[scale=0.1,trim={0\width} {0.2\width} {0\width} {0.0\width},clip]{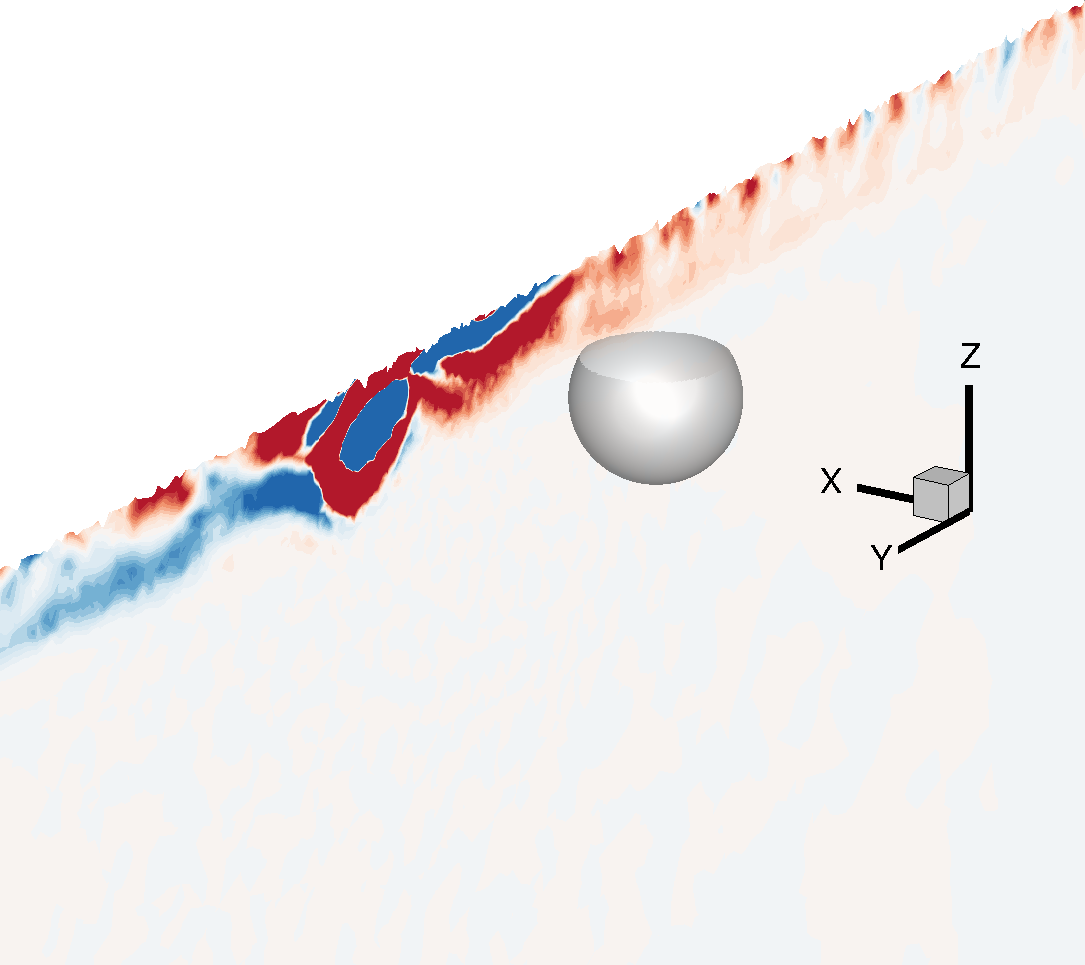}
		\caption*{$(d)$}
	\end{subfigure}
	\begin{subfigure}[b]{0.25\textwidth}
		\adjincludegraphics[scale=0.1,trim={0\width} {0.2\width} {0\width} {0.0\width},clip]{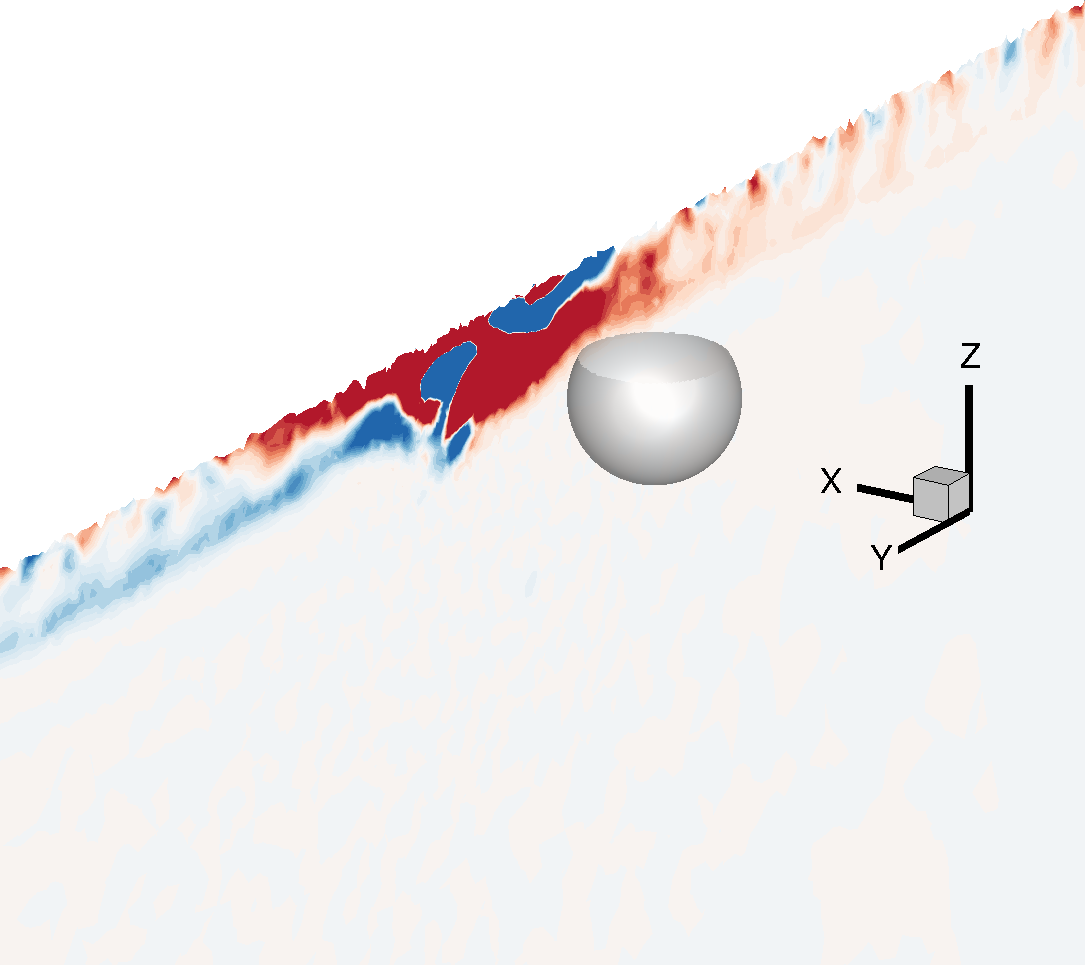}
		\caption*{$(e)$}
	\end{subfigure}%
	\begin{subfigure}[b]{0.25\textwidth}
		\adjincludegraphics[scale=0.1,trim={0\width} {0.2\width} {0\width} {0.0\width},clip]{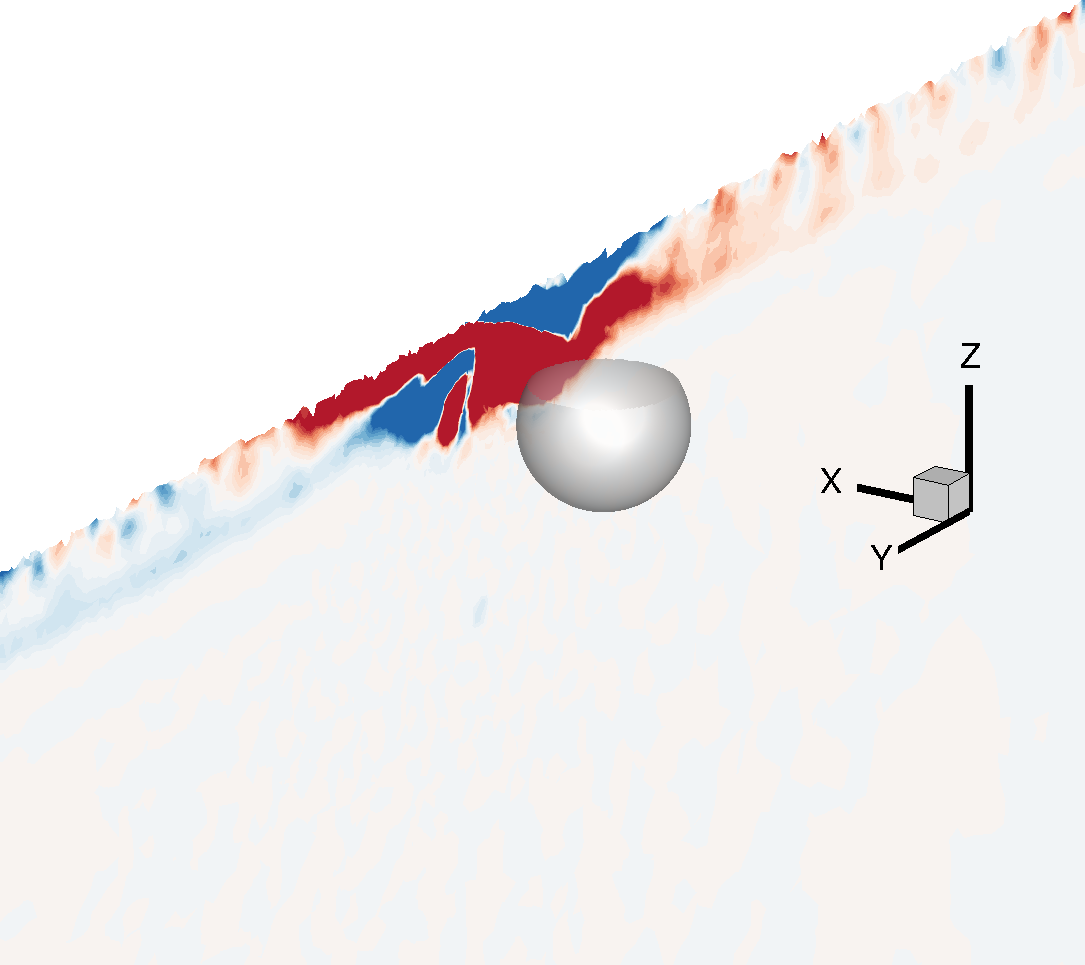}
		\caption*{$(f)$}
	\end{subfigure}%
	\begin{subfigure}[b]{0.25\textwidth}
		\adjincludegraphics[scale=0.1,trim={0\width} {0.2\width} {0\width} {0.0\width},clip]{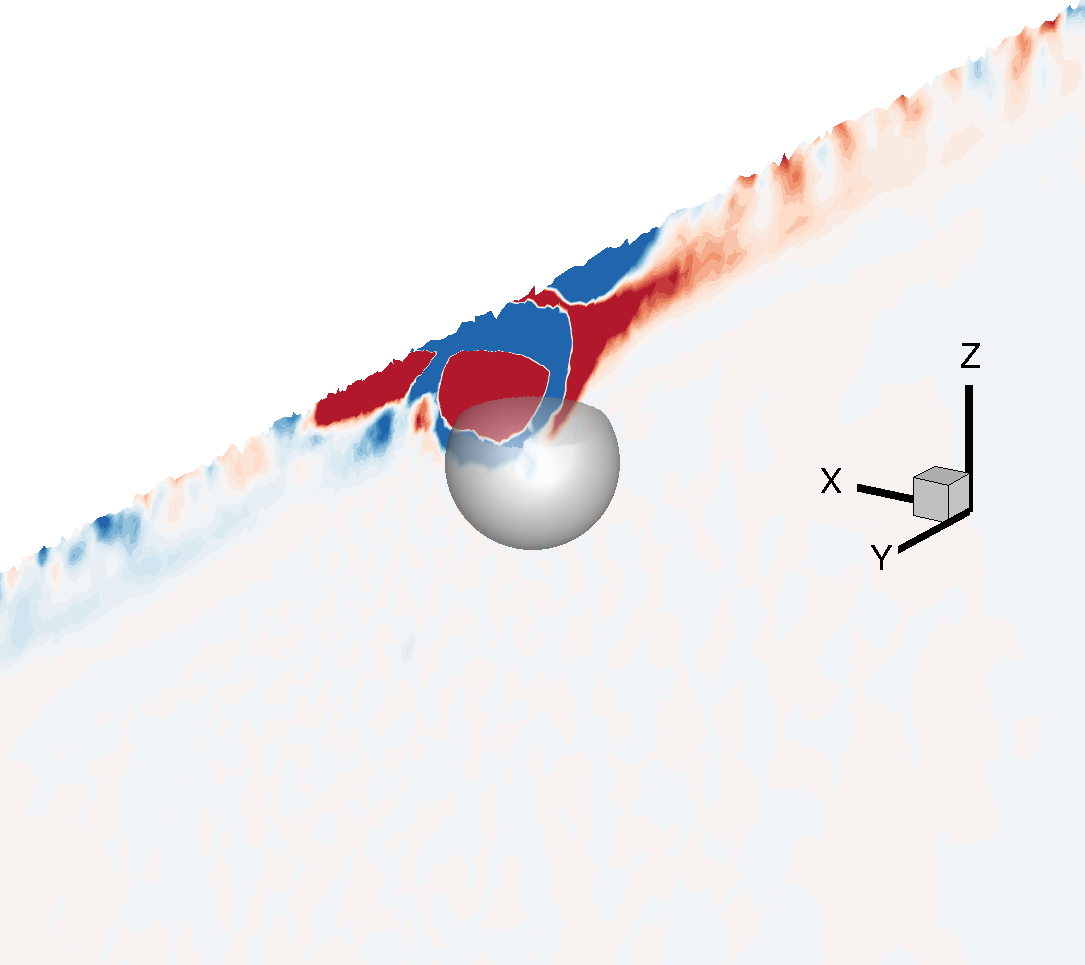}
		\caption*{$(g)$}
	\end{subfigure}%
	\begin{subfigure}[b]{0.25\textwidth}
		\adjincludegraphics[scale=0.1,trim={0\width} {0.2\width} {0\width} {0.0\width},clip]{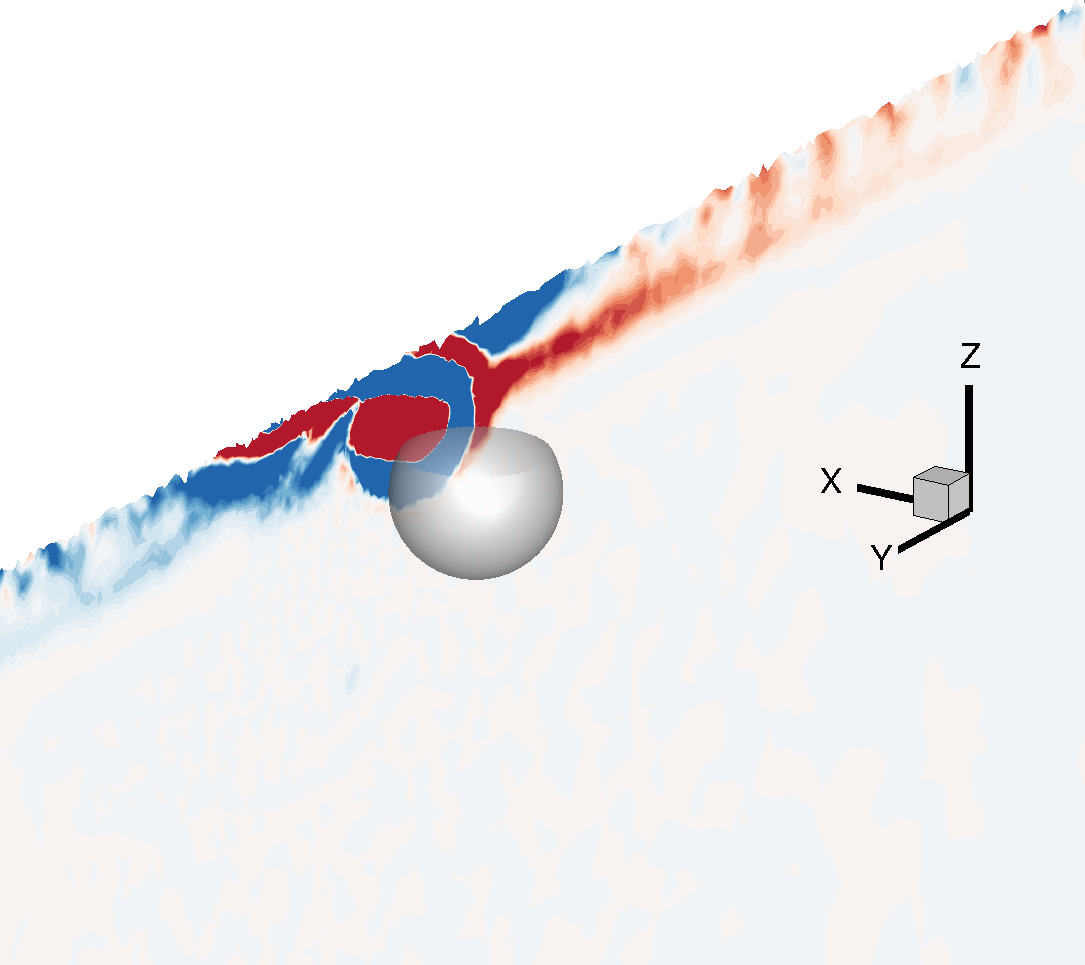}
		\caption*{$(h)$}
	\end{subfigure}
	
	\caption{Streamwise $x$-vorticity contour ($\omega_{x}D/U\in[-3.3,3.3]$) for the piercing sphere at $h^*=-0.25$ and $U^*=10$ for one complete oscillation period. Vorticity plot is taken at $1.5D$ downstream on the $Y$-$Z$ plane.}
	\label{xVor_FS_Ur10_Cycle} 
\end{figure}

\begin{figure}[htbp!]
	\centering
	\begin{subfigure}[b]{0.3\textwidth}
		\adjincludegraphics[scale=0.14,trim={0.0\width} {0.15\width} {0.02\width} {0.1\width},clip]{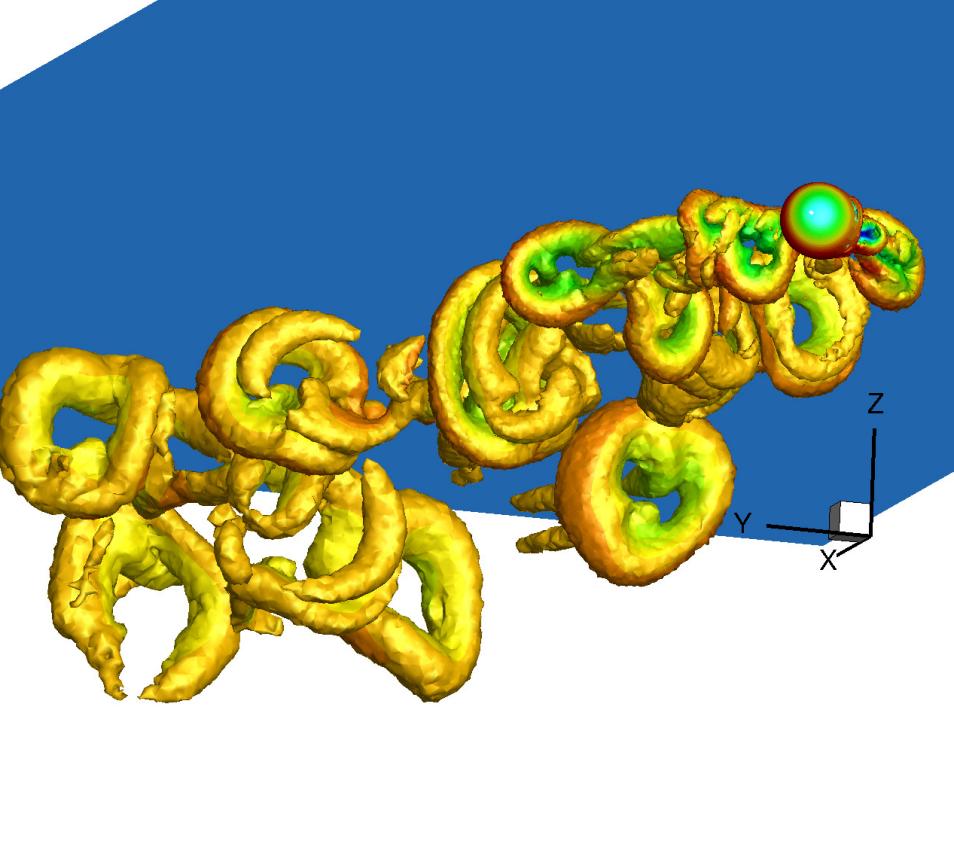}
		\caption{}
	\end{subfigure}%
	\begin{subfigure}[b]{0.3\textwidth}
		\adjincludegraphics[scale=0.14,trim={0.0\width} {0.15\width} {0.02\width} {0.1\width},clip]{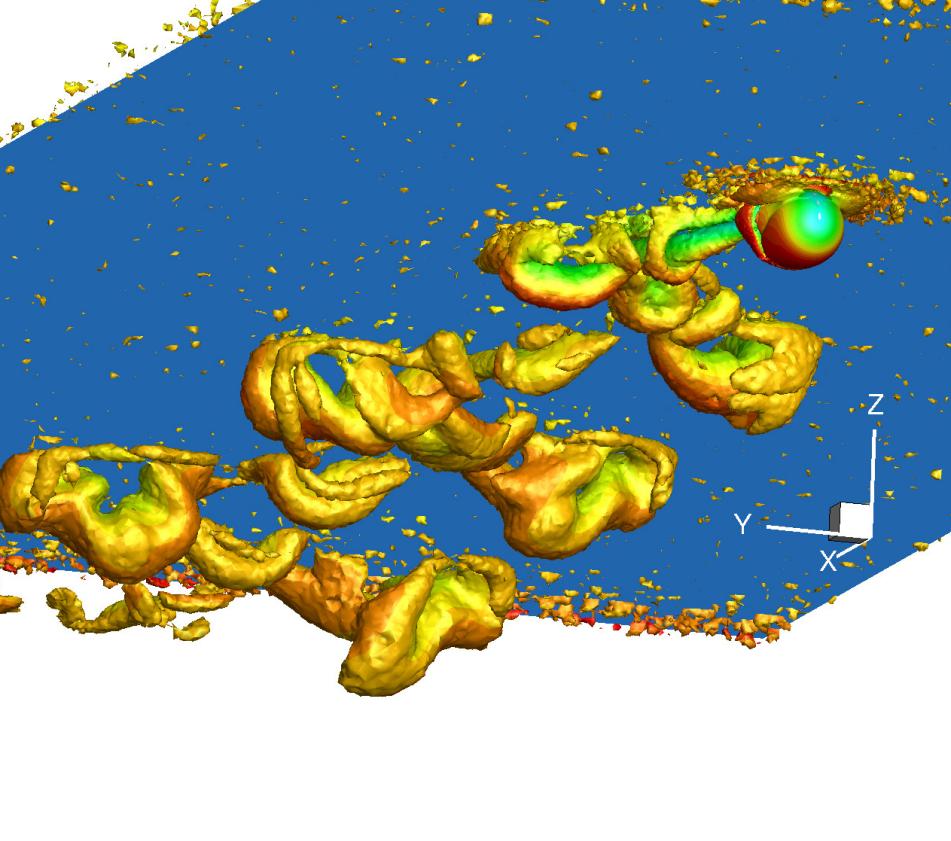}
		\caption{}
	\end{subfigure}%
	\begin{subfigure}[b]{0.3\textwidth}
		\adjincludegraphics[scale=0.14,trim={0.0\width} {0.15\width} {0.02\width} {0.1\width},clip]{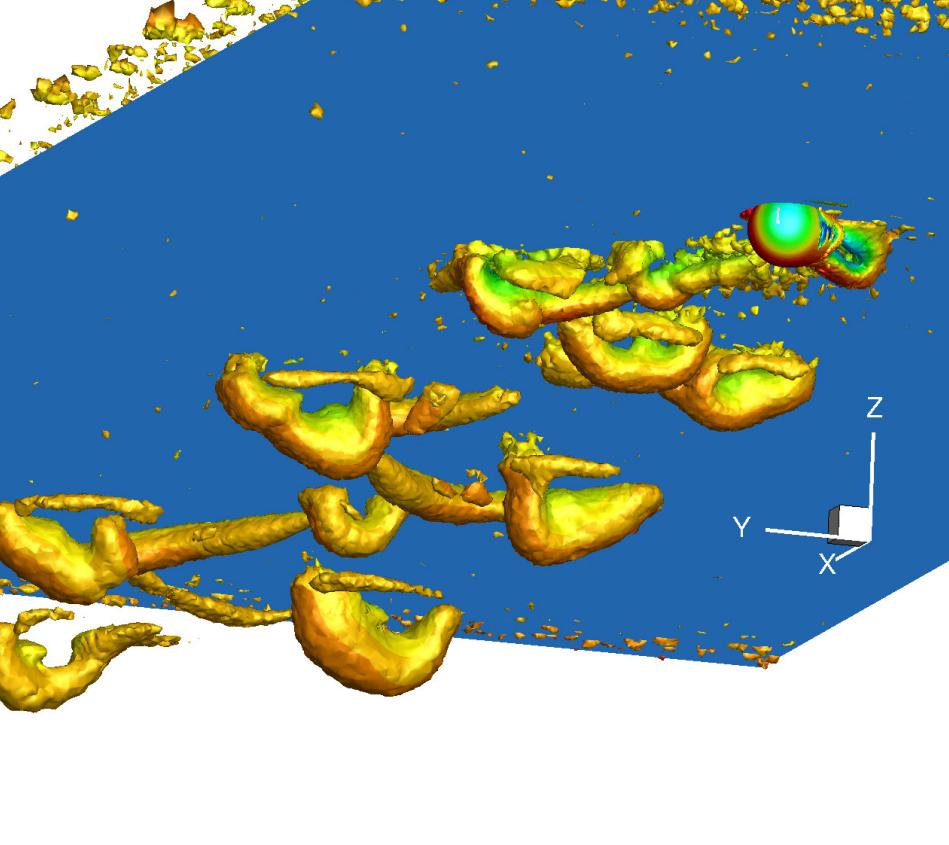}
		\caption{}
	\end{subfigure}%
		\begin{subfigure}[b]{0.1\textwidth}
			\adjincludegraphics[scale=0.13,trim={0.4\width} {0.05\width} {0.3\width} {0.05\width},clip]{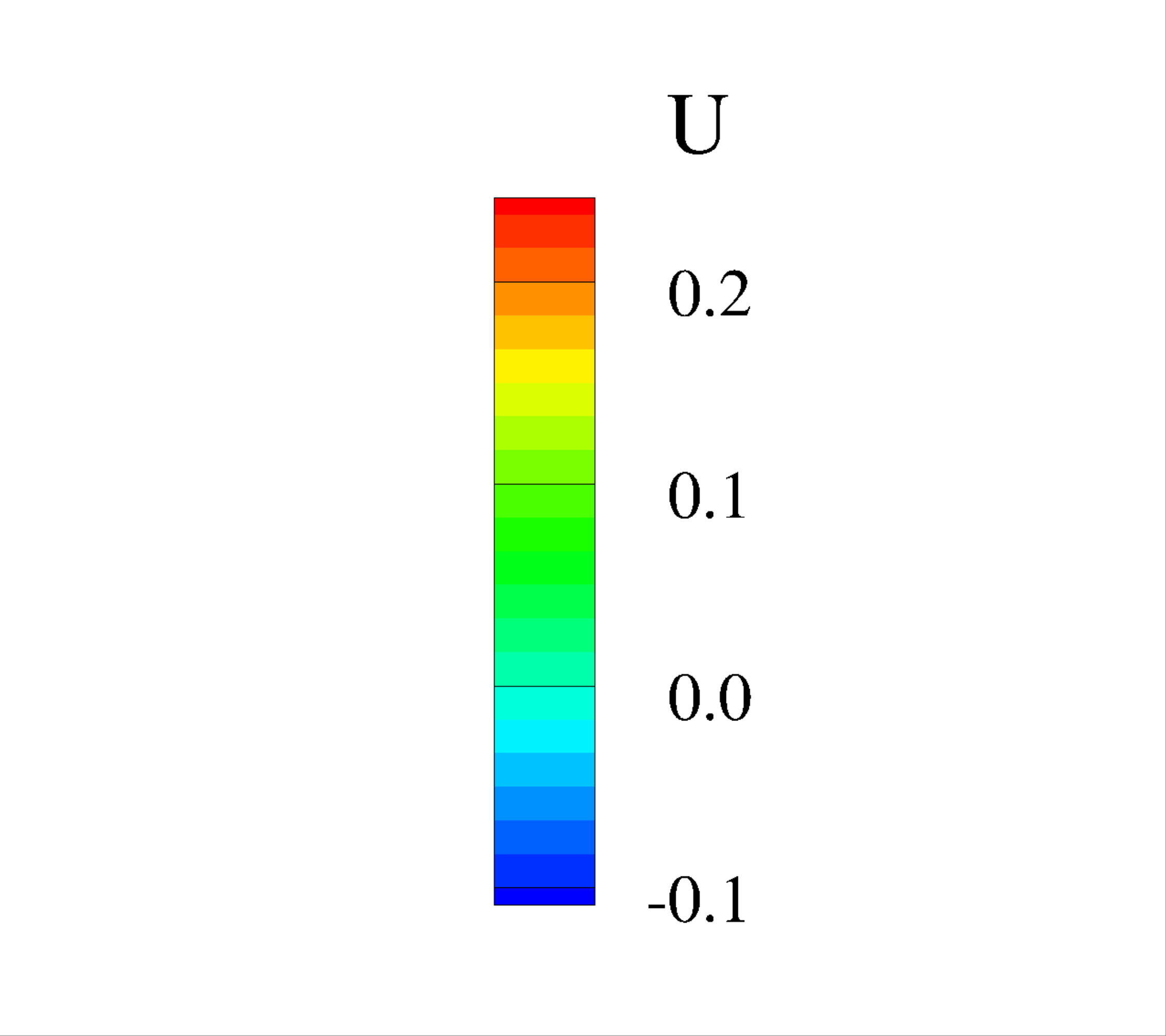}
			\caption*{}
			\caption*{}
		\end{subfigure}
	\caption{Underwater view of the iso-surface wake structures formed behind the 1-DOF sphere at $U^*=10$ and $Re=15\,700$ at stationary state: (a) fully submerged with the immersion ratio of $h^*=1$, (b) the top of the sphere touches the free surface at $h^*=0$ and (c) piercing the free surface with the immersion ratio of $h^*=-0.25$. Iso-surfaces are plotted by the Q-criterion ($Q=2$).}	
	\label{FS_Q_Crit}	
\end{figure}

Fig. \ref{xVor_FS_Ur10_Cycle} shows the streamwise $x$-vorticity contour plots for one complete oscillation cycle for the piercing case at $h^*=-0.25$ at $U^*=10$. The stretched vortex patterns consist of both clockwise and anti-clockwise vorticity loops. When the sphere moves from one side to the other side, the vorticity changes sign accordingly (Fig. \ref{xVor_FS_Ur10_Cycle} (d) and (h) show a clear contrast). In Fig. \ref{xVor_FS_Ur10_Cycle} (d), where the sphere is at its mean position during half-stroke, the blue vortex is trapped with the red vortex loop. On the other hand, when the sphere moves across the opposite side at its mean position, Fig. \ref{xVor_FS_Ur10_Cycle} (h), the vorticity changes sign with the red vortex now trapped with the blue vortex loop. This confirms the existence of the hairpin loops that form from the opposite sides of the sphere and are shed into the downstream wake. The stretched vortex formation structures for the piercing sphere case were also observed in the experiments in \cite{sareen2018}. However, during each stroke, only one sign vortex loop was captured by the PIV in the cross-plane and therefore, no specific wake mode was identified for the piercing sphere case. 
To further analyze the three-dimensional vortical structures, we employ a vortex-identification based on the $Q$-criterion as discussed earlier. Fig. \ref{FS_Q_Crit} shows the $Q$-criterion based vortical structures for the sphere at $h^*=1$, $h^*=0$ and $h^*=-0.25$. The iso-surfaces of quantity $Q$ are shown at a constant positive value and the contour surfaces are colored by the streamwise velocity. The existence of hairpin wake loops for all cases, even when the sphere is piercing the free surface, is observed in our numerical analysis. We find that the vortex loop patterns that were observed for the submerged case, Fig. \ref{FS_Q_Crit} (a), are slightly stretched to elliptical loops for the cases at $h^*=0$ , Fig. \ref{FS_Q_Crit} (b), and the piercing case at $h^*=-0.25$, Fig. \ref{FS_Q_Crit} (c).

\begin{figure}[htbp!]
	\centering
	\begin{subfigure}[b]{0.5\textwidth}
		\centering
		\adjincludegraphics[scale=0.3,trim={0\width} {0.25\width} {0\width} {0.22\width},clip]{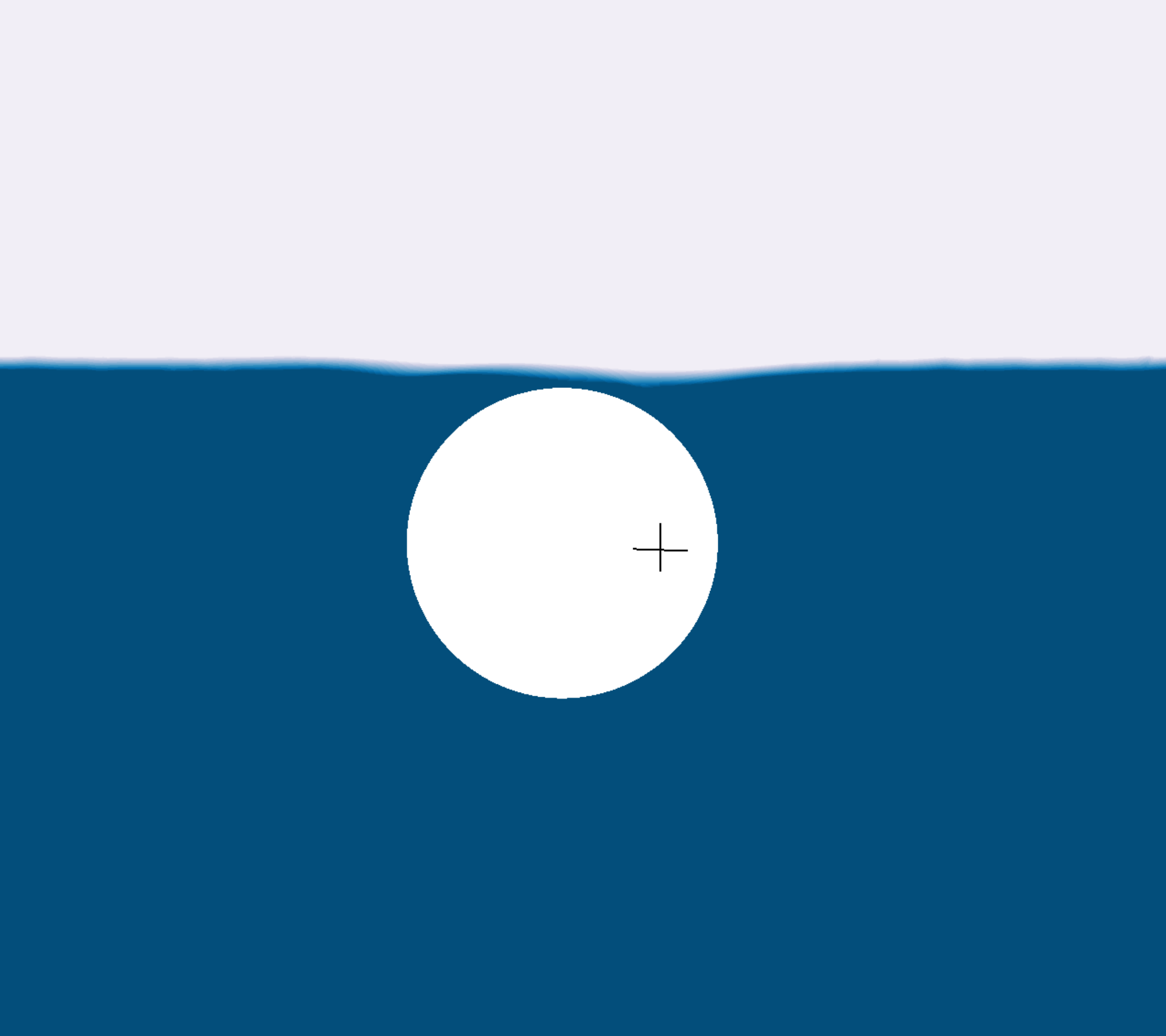}
		\caption*{$(a)$}
	\end{subfigure}%
	\begin{subfigure}[b]{0.4\textwidth}
		\centering
		\adjincludegraphics[scale=0.3,trim={0\width} {0.25\width} {0\width} {0.22\width},clip]{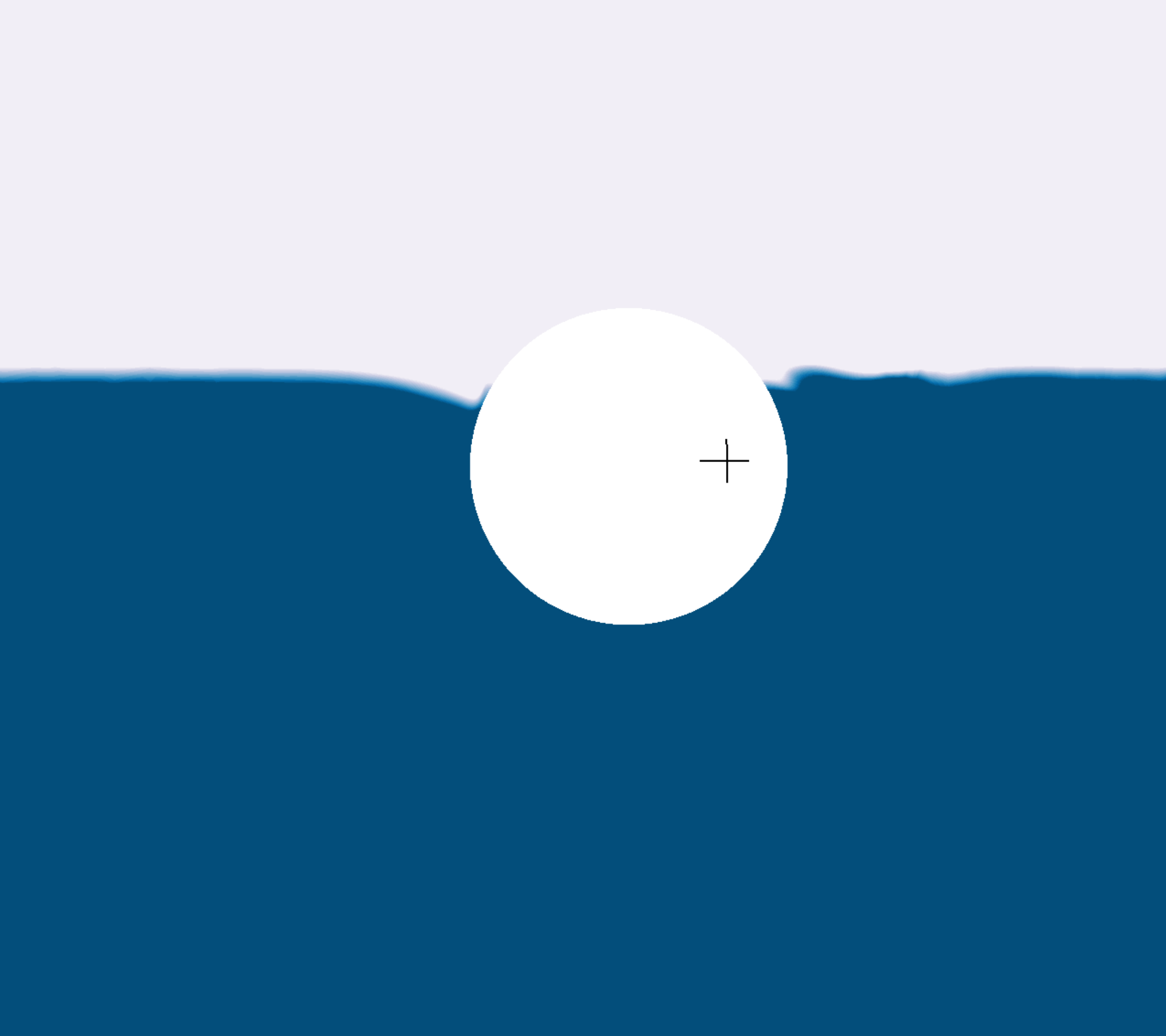}
		\caption*{$(b)$}
	\end{subfigure}%
	\begin{subfigure}[b]{0.2\textwidth}
		\centering
		\adjincludegraphics[scale=0.35,trim={0.65\width} {0.5\width} {0.15\width} {0.0\width},clip]{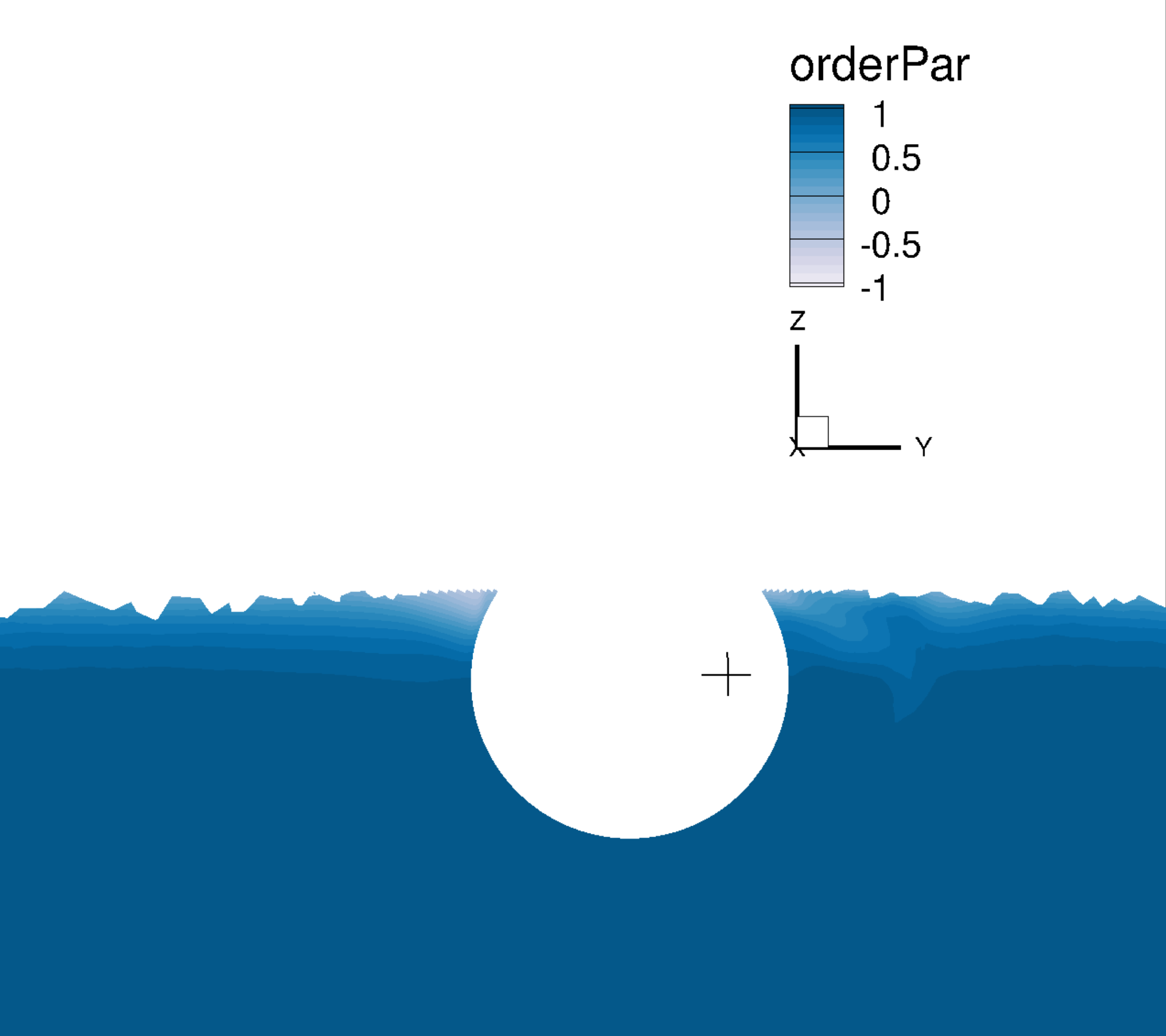}
		\caption*{}
	\end{subfigure}%
	
	\caption{The contour plot of the phase-field order parameter $\phi$ to quantify the free-surface deformation at $\frac{1}{4}D$ downstream on the $Y$-$Z$ plane: (a) submerged sphere at $h^*=0$, and (b) piercing sphere case at $h^*=-0.25$. }
	\label{FSd_hS} 
\end{figure}

\begin{figure}[htbp!]
	\centering
	\begin{subfigure}[b]{0.5\textwidth}
		\centering
		\adjincludegraphics[scale=0.3,trim={0\width} {0.2\width} {0\width} {0.225\width},clip]{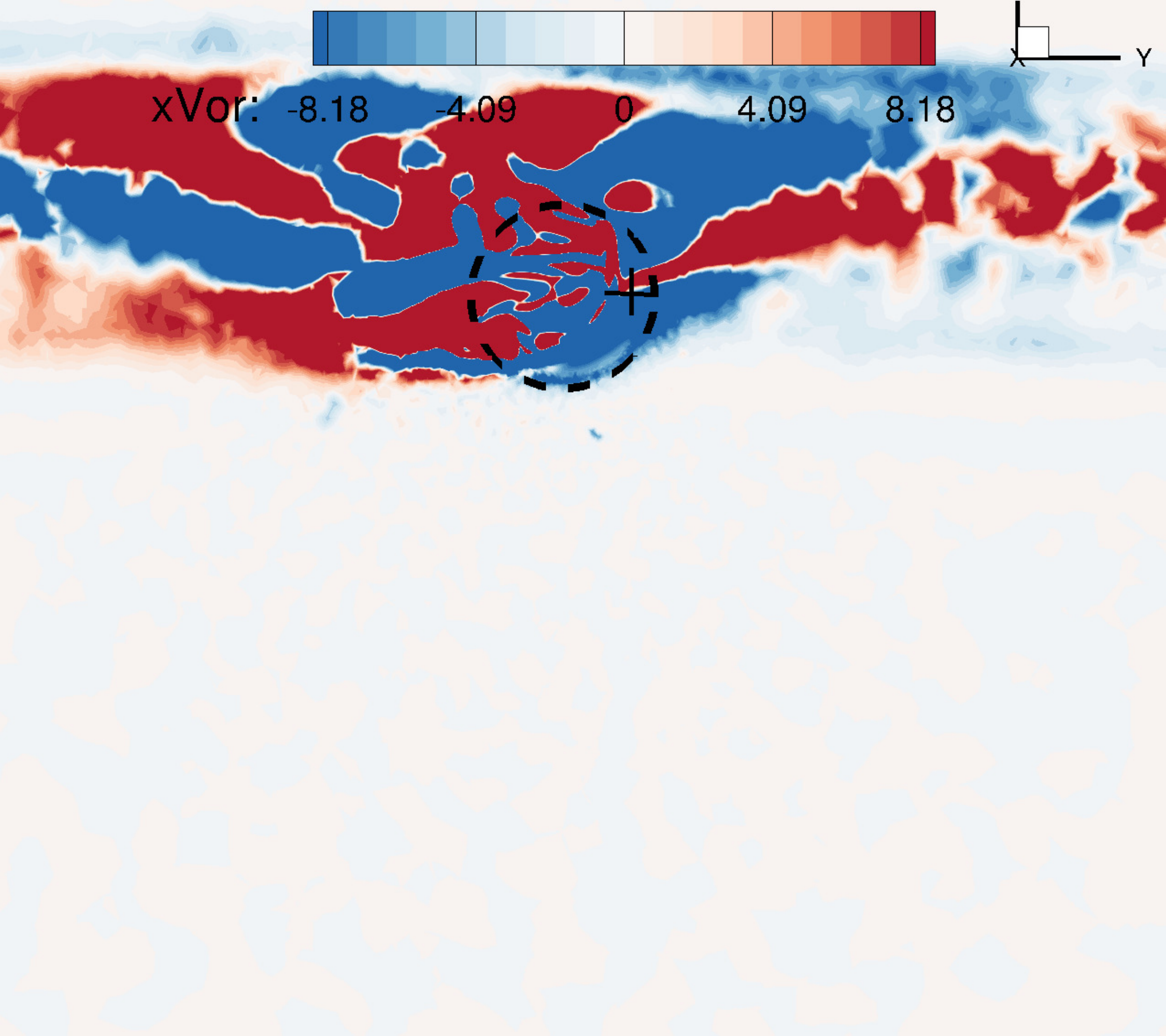}
		\caption*{$(a$ - $1)$}
	\end{subfigure}%
	\begin{subfigure}[b]{0.5\textwidth}
		\centering
		\adjincludegraphics[scale=0.3,trim={0\width} {0.2\width} {0\width} {0.225\width},clip]{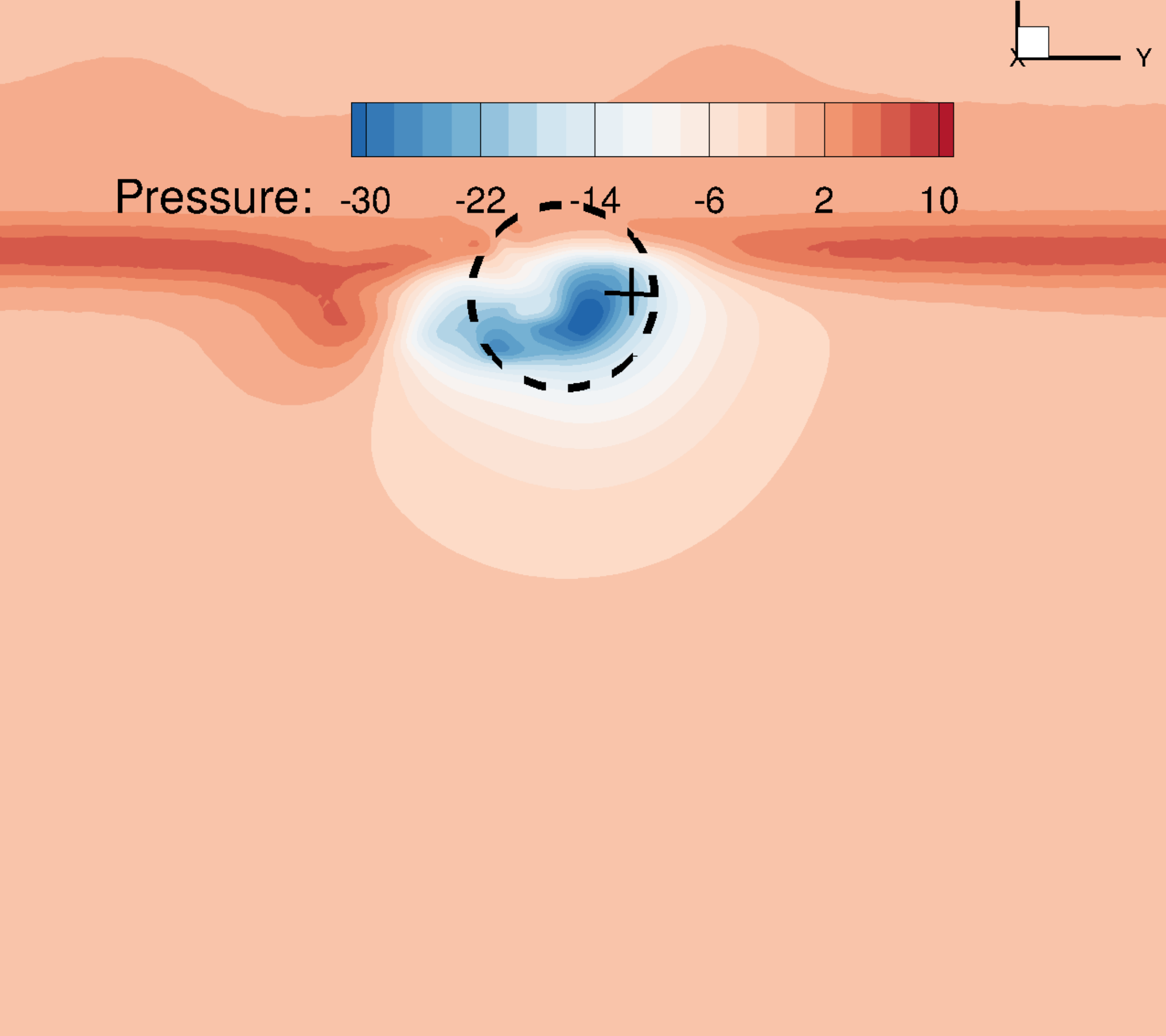}
		\caption*{$(a$ - $2)$}
	\end{subfigure}
	
	\begin{subfigure}[b]{0.5\textwidth}
		\centering
		\adjincludegraphics[scale=0.3,trim={0\width} {0.2\width} {0\width} {0.21\width},clip]{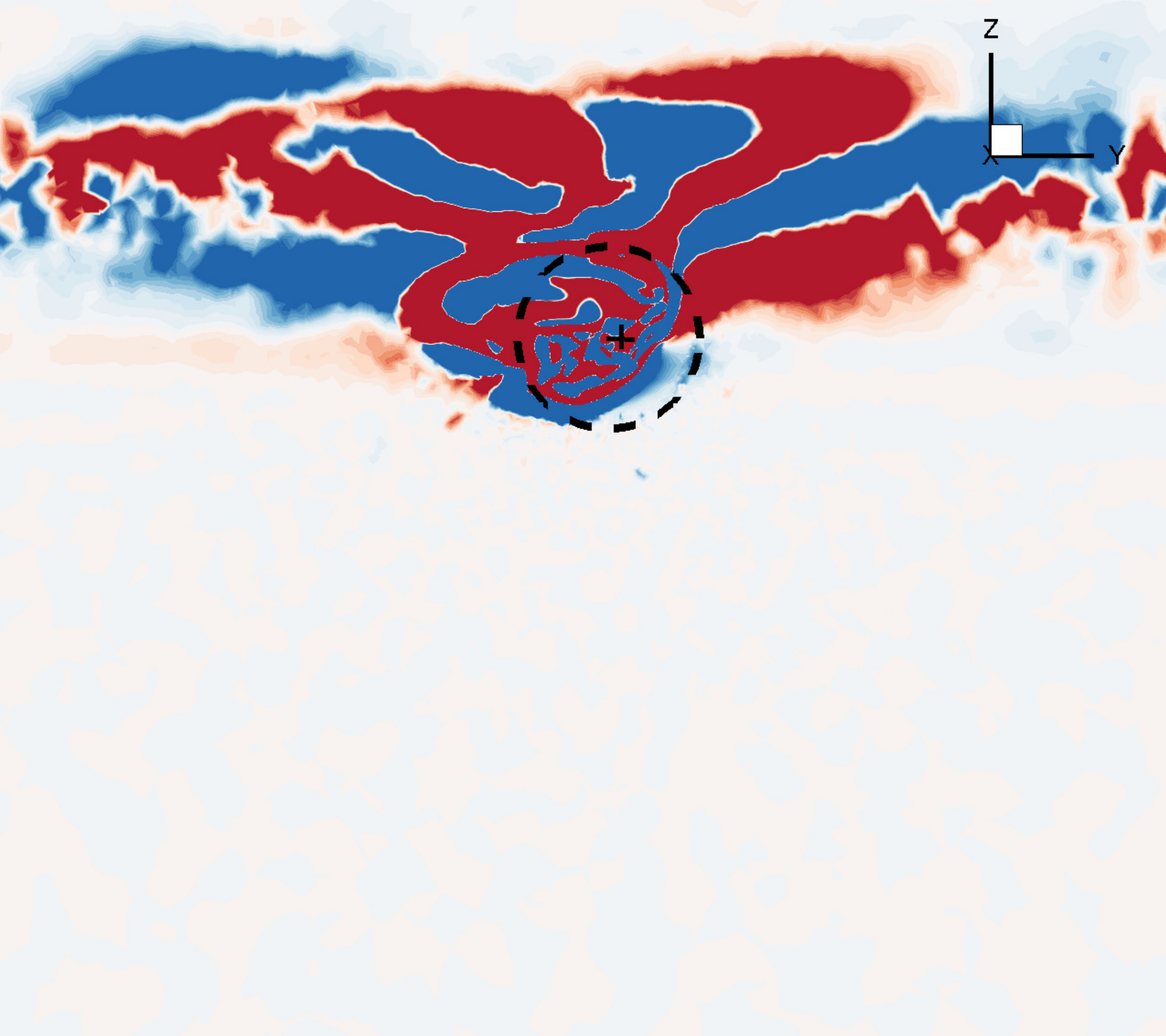}
		\caption*{$(b$ - $1)$}
	\end{subfigure}%
	\begin{subfigure}[b]{0.5\textwidth}
		\centering
		\adjincludegraphics[scale=0.3,trim={0\width} {0.2\width} {0\width} {0.21\width},clip]{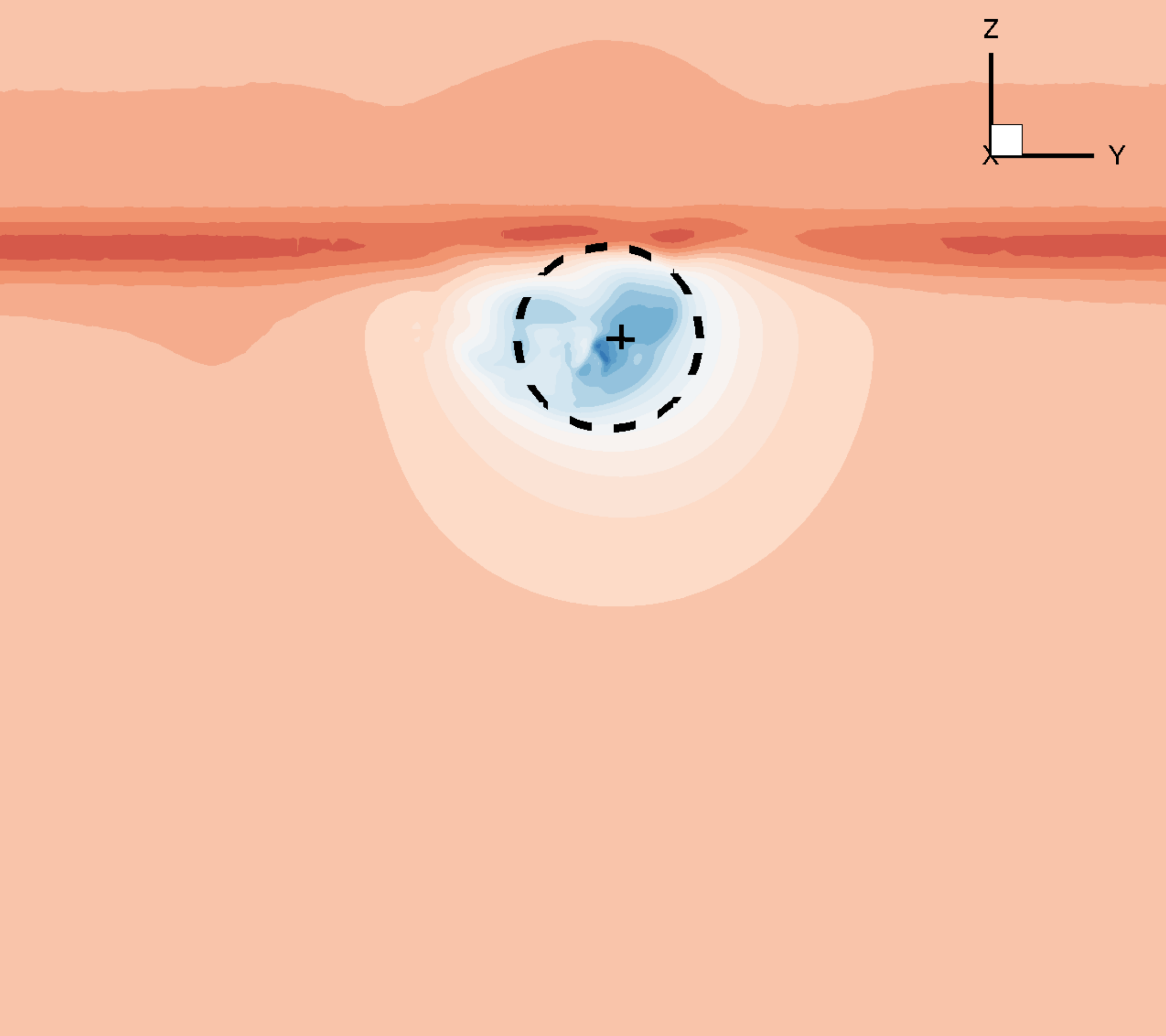}
		\caption*{$(b$ - $2)$}
	\end{subfigure}
	
	\begin{subfigure}[b]{0.5\textwidth}
		\centering
		\adjincludegraphics[scale=0.3,trim={0\width} {0.2\width} {0\width} {0.225\width},clip]{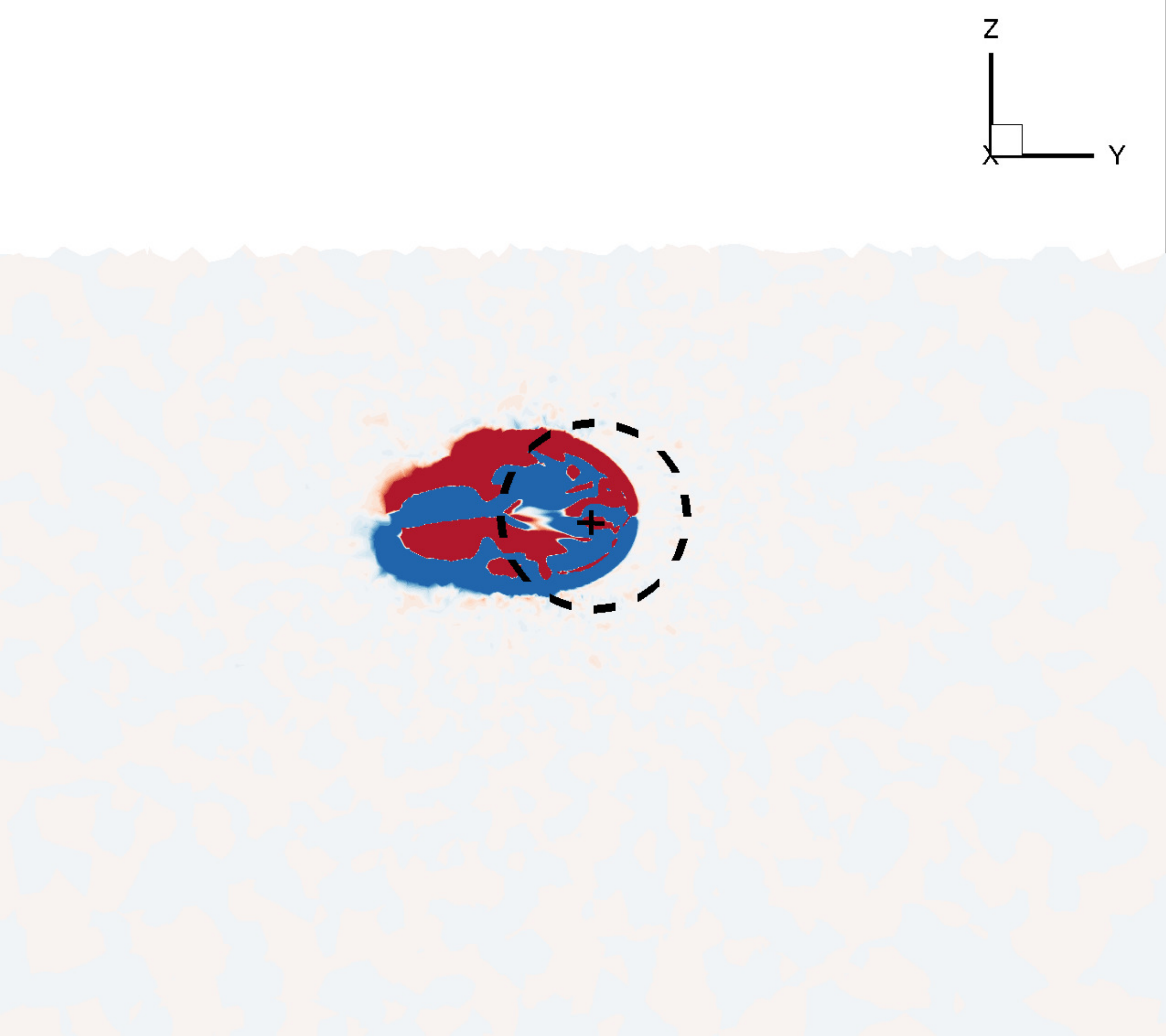}
		\caption*{$(c$ - $1)$}
	\end{subfigure}%
	\begin{subfigure}[b]{0.5\textwidth}
		\centering
		\adjincludegraphics[scale=0.3,trim={0\width} {0.2\width} {0\width} {0.225\width},clip]{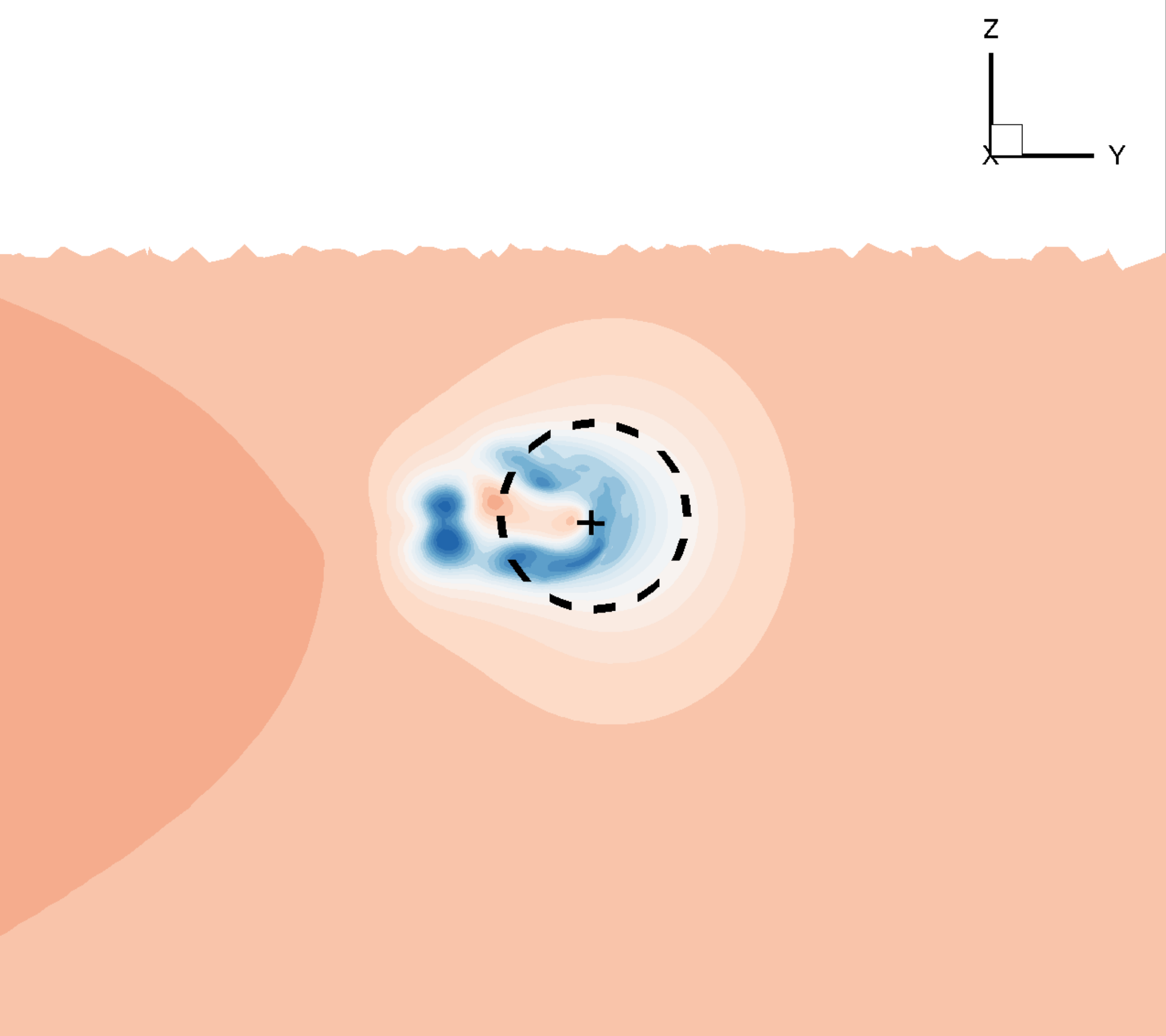}
		\caption*{$(c$ - $2)$}
	\end{subfigure}
	
	\begin{subfigure}[b]{0.5\textwidth}
		\centering
		\adjincludegraphics[scale=0.4,trim={0.05\width} {0.7\width} {0.16\width} {0.0\width},clip]{Photos/38.pdf}
		\caption*{}
	\end{subfigure}%
	\begin{subfigure}[b]{0.5\textwidth}
		\centering
		\adjincludegraphics[scale=0.4,trim={0.05\width} {0.7\width} {0.01\width} {0.0\width},clip]{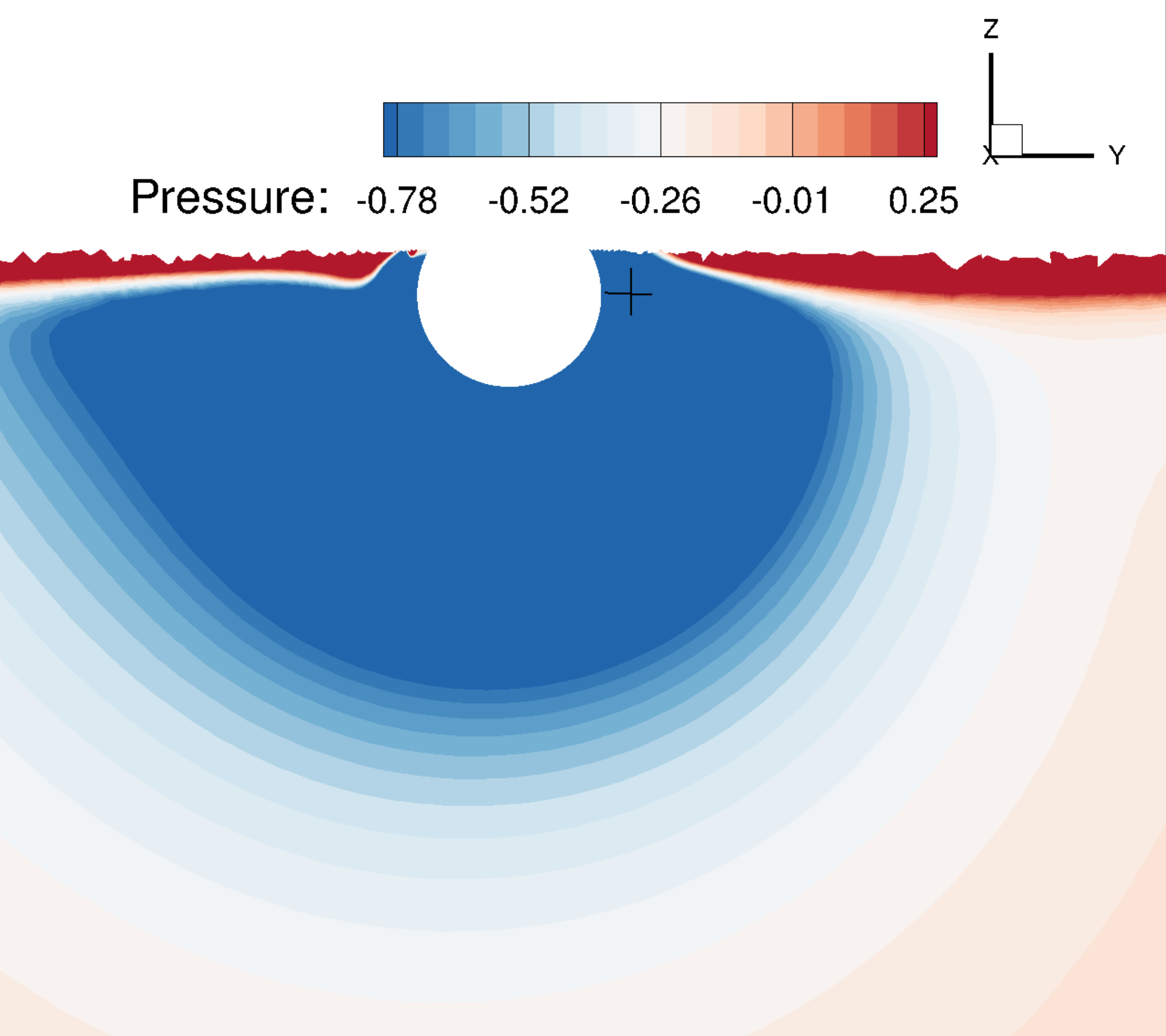}
		\caption*{}
	\end{subfigure}
	
	\caption{Streamwise $x$-vorticity $(\omega_{x}D/U\in[-3.3,3.3])$ and pressure distribution contours $(p/\rho U^2\in[-0.78,0.25])$ plotted at $0.5D$ downstream on the $Y$-$Z$ plane at $U^*=10$, for the sphere at (a) $h^*=-0.25$, (b) $h^*=0$, and (c) $h^*=1$.}
	\label{xVor_Pressure_hS} 
\end{figure}

\begin{figure}[htbp!]
	\centering
	
	\begin{subfigure}[b]{1\textwidth}
		\centering
		\hspace{0cm}
		\adjincludegraphics[scale=0.2,trim={0\width} {0.0\width} {0\width} {0.0\width},clip]{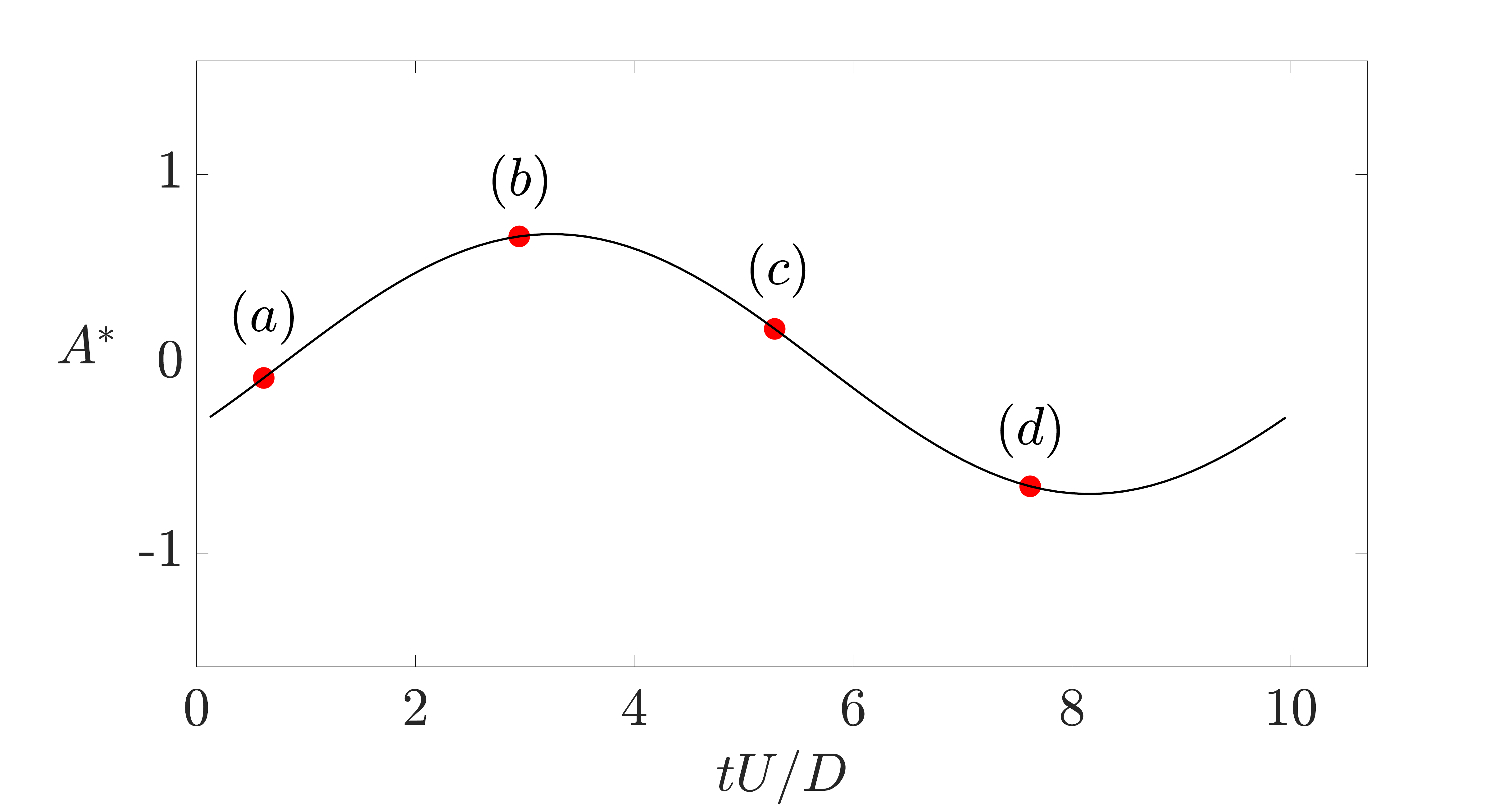}
	\end{subfigure}
	
	\begin{subfigure}[b]{0.5\textwidth}
		\centering
		\adjincludegraphics[scale=0.3,trim={0\width} {0.3\width} {0\width} {0.21\width},clip]{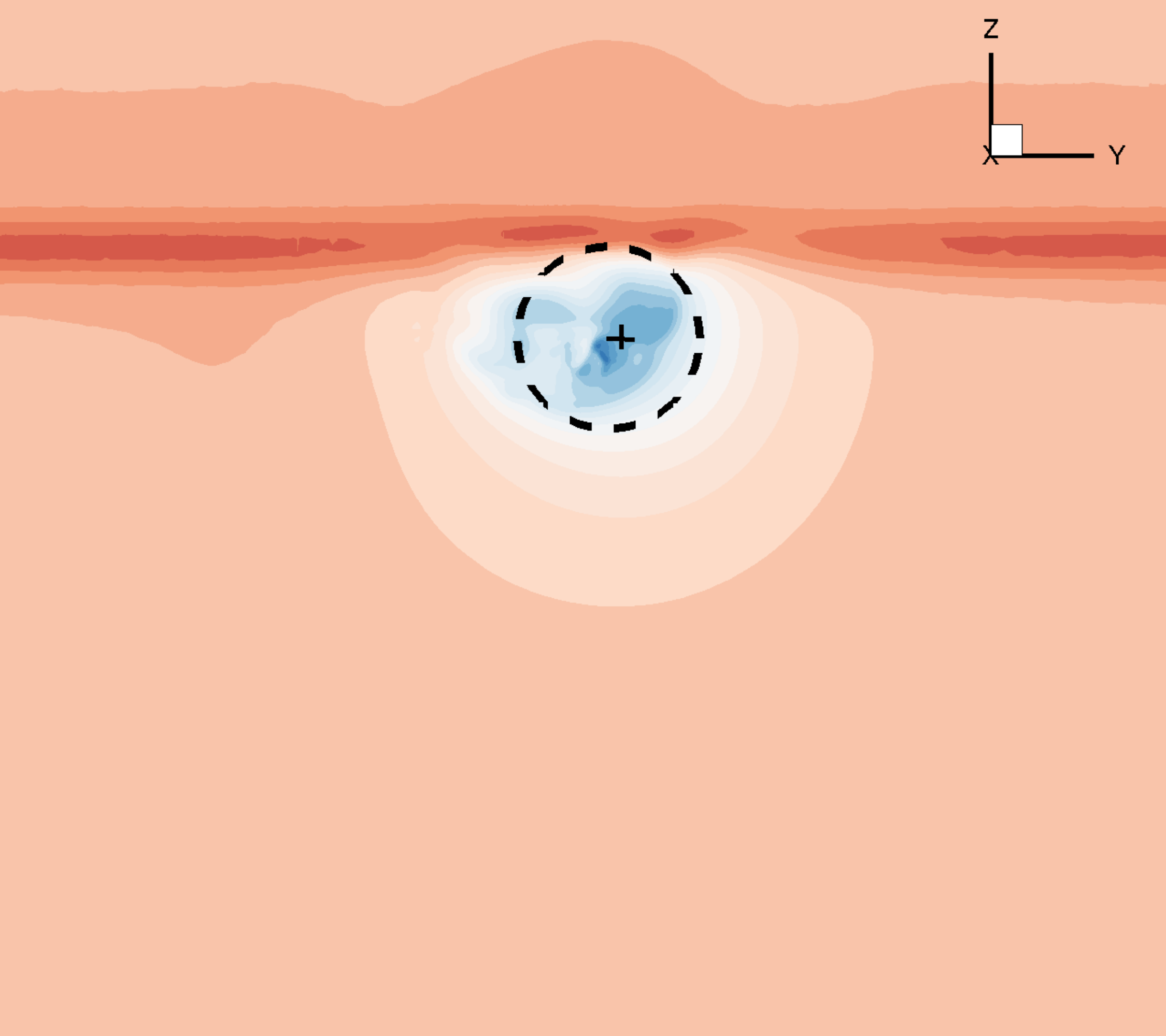}
		\caption*{$(a$ - $1)$}
	\end{subfigure}%
	\begin{subfigure}[b]{0.5\textwidth}
		\centering
		\adjincludegraphics[scale=0.3,trim={0\width} {0.3\width} {0\width} {0.21\width},clip]{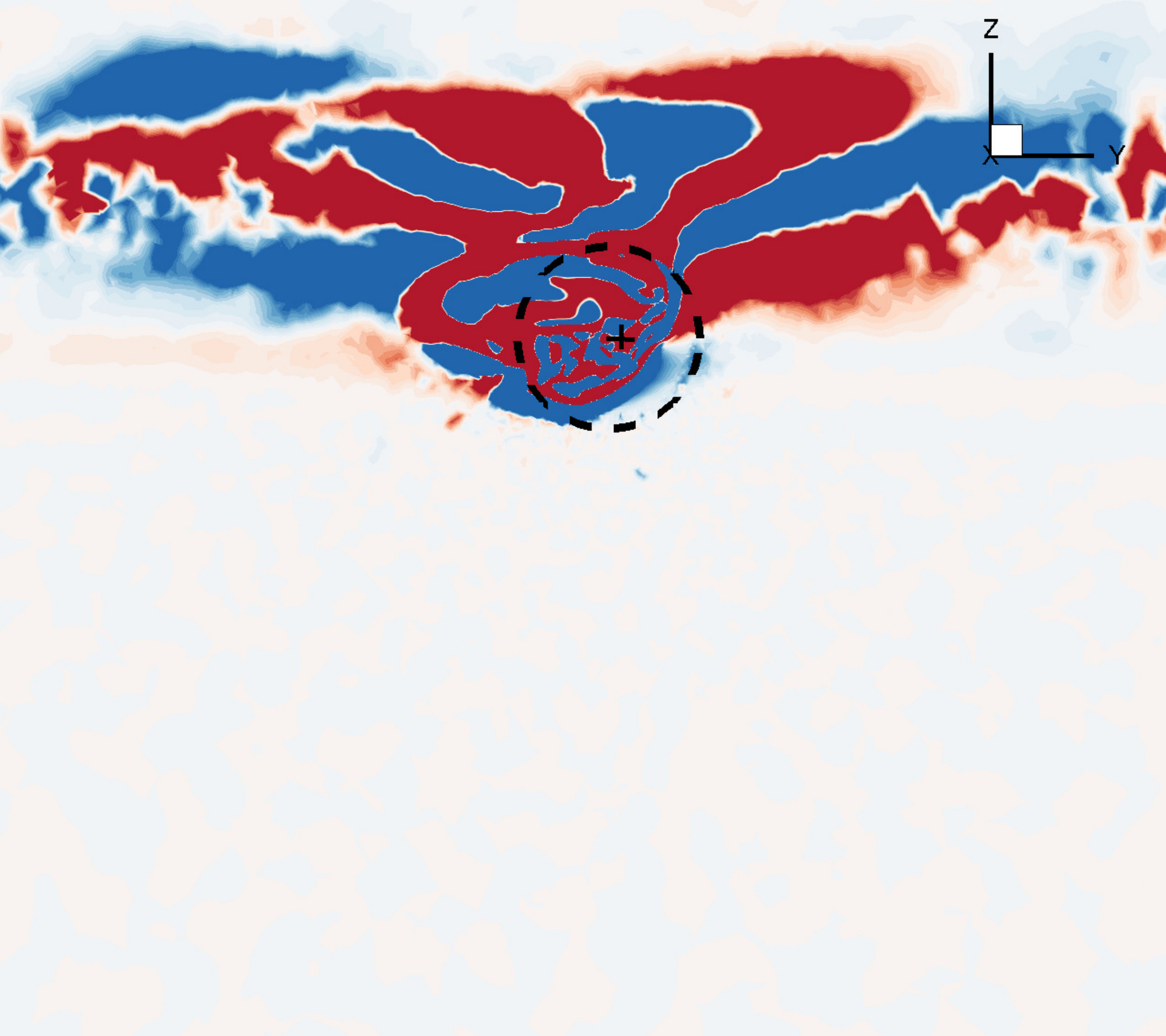}
		\caption*{$(a$ - $2)$}
	\end{subfigure}
	
	\begin{subfigure}[b]{0.5\textwidth}
		\centering
		\adjincludegraphics[scale=0.3,trim={0\width} {0.3\width} {0\width} {0.21\width},clip]{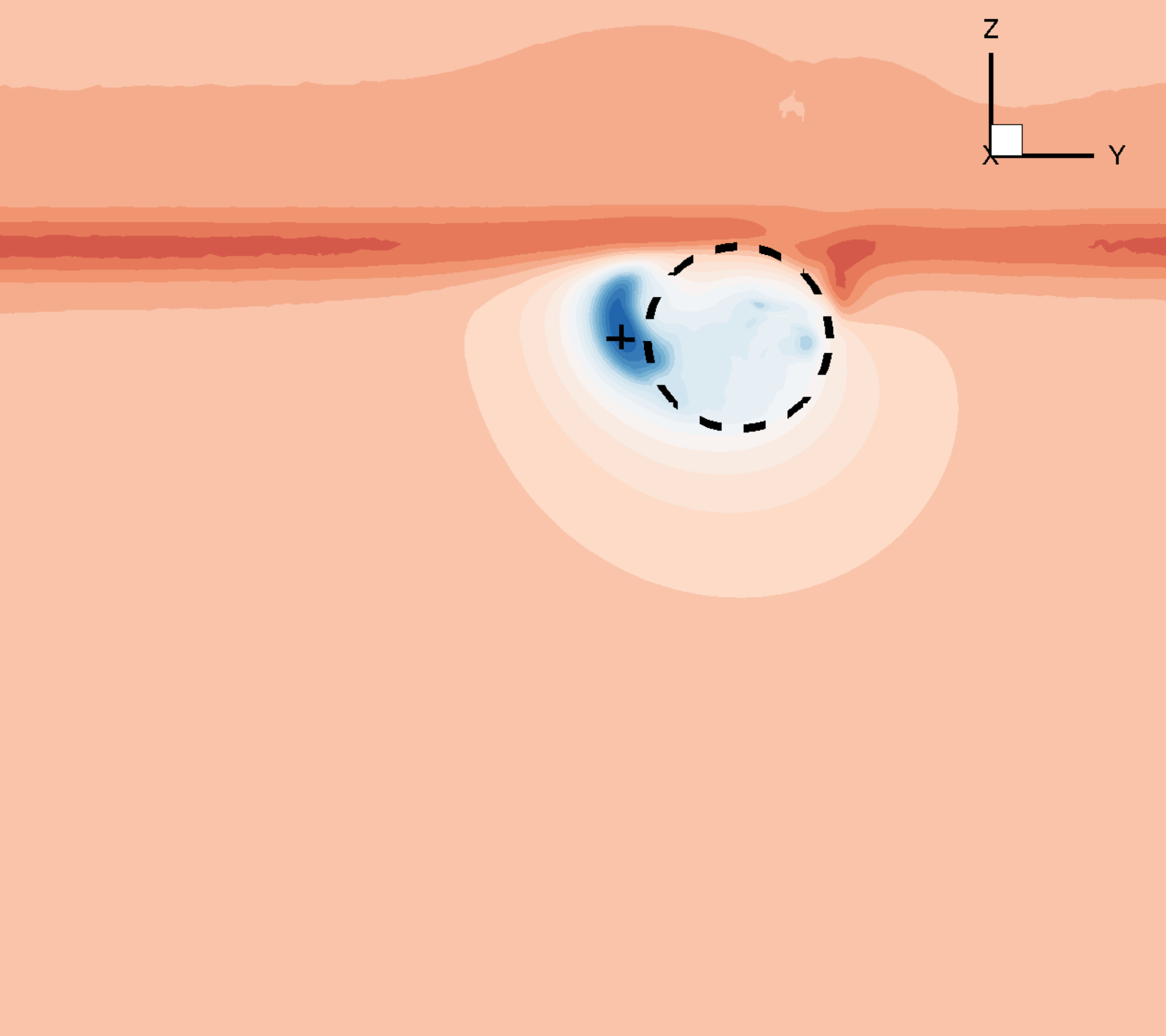}
		\caption*{$(b$ - $1)$}
	\end{subfigure}%
	\begin{subfigure}[b]{0.5\textwidth}
		\centering
		\adjincludegraphics[scale=0.3,trim={0\width} {0.3\width} {0\width} {0.21\width},clip]{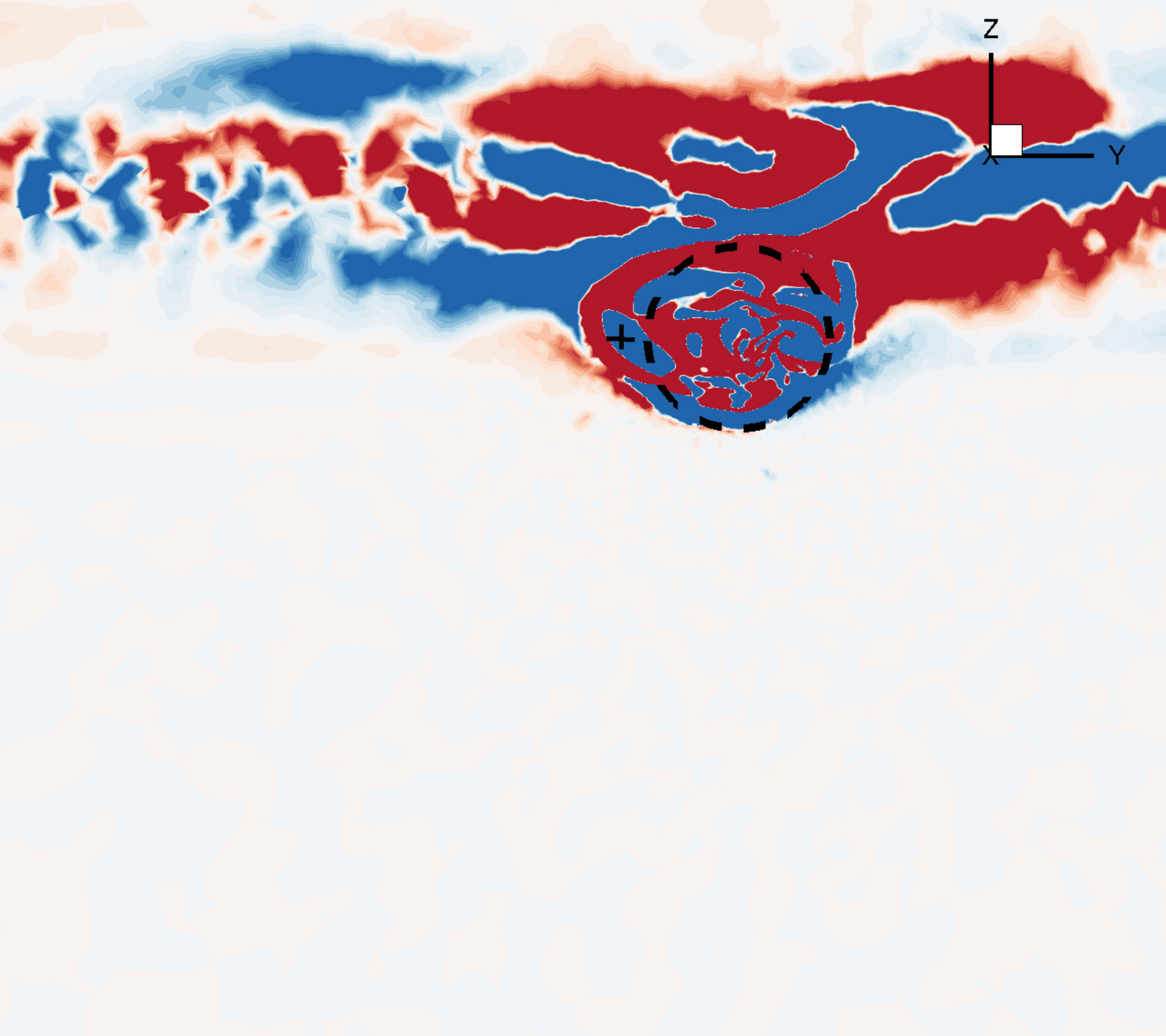}
		\caption*{$(b$ - $2)$}
	\end{subfigure}
	
	\begin{subfigure}[b]{0.5\textwidth}
		\centering
		\adjincludegraphics[scale=0.3,trim={0\width} {0.3\width} {0\width} {0.21\width},clip]{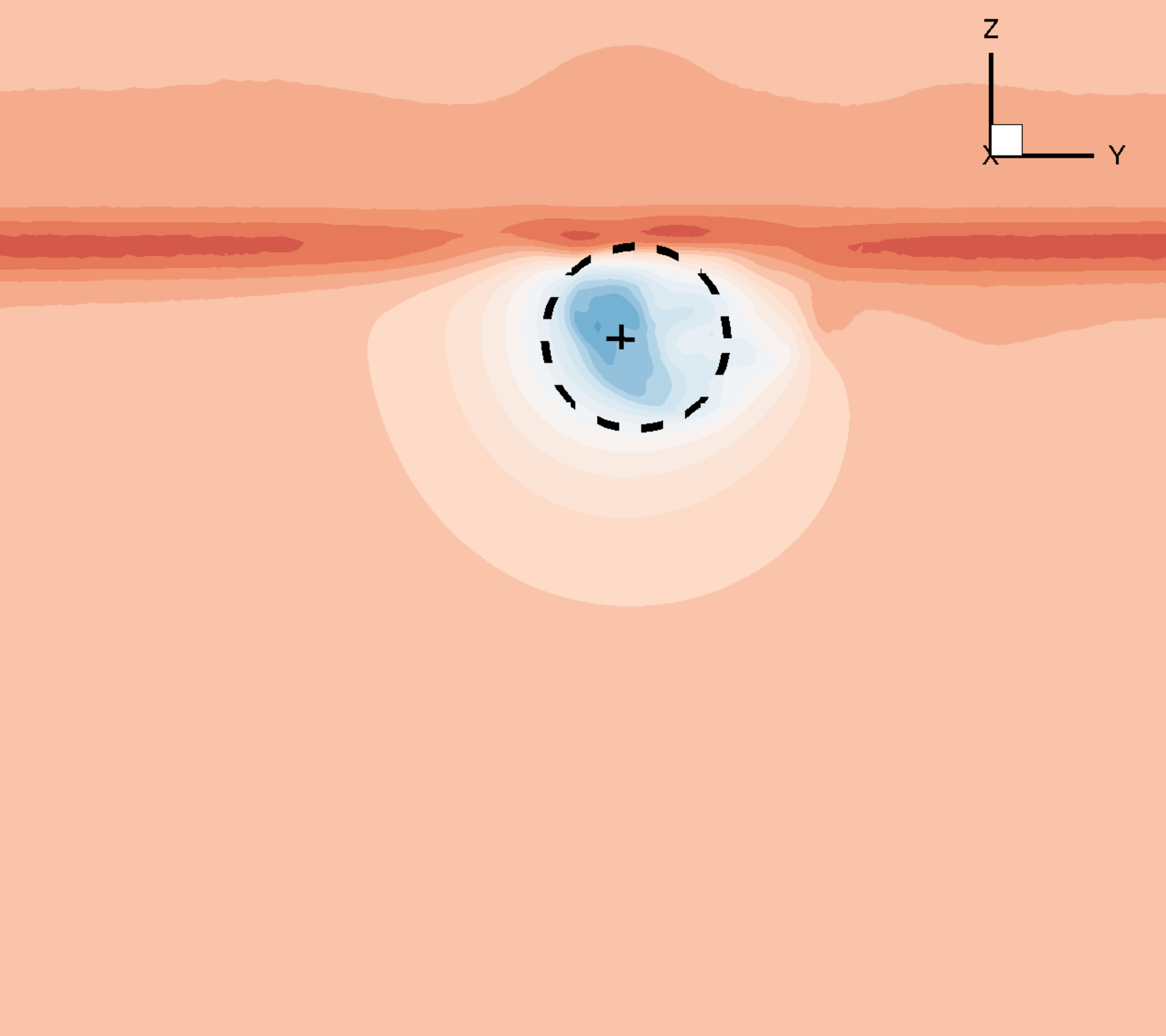}
		\caption*{$(c$ - $1)$}
	\end{subfigure}%
	\begin{subfigure}[b]{0.5\textwidth}
		\centering
		\adjincludegraphics[scale=0.3,trim={0\width} {0.3\width} {0\width} {0.21\width},clip]{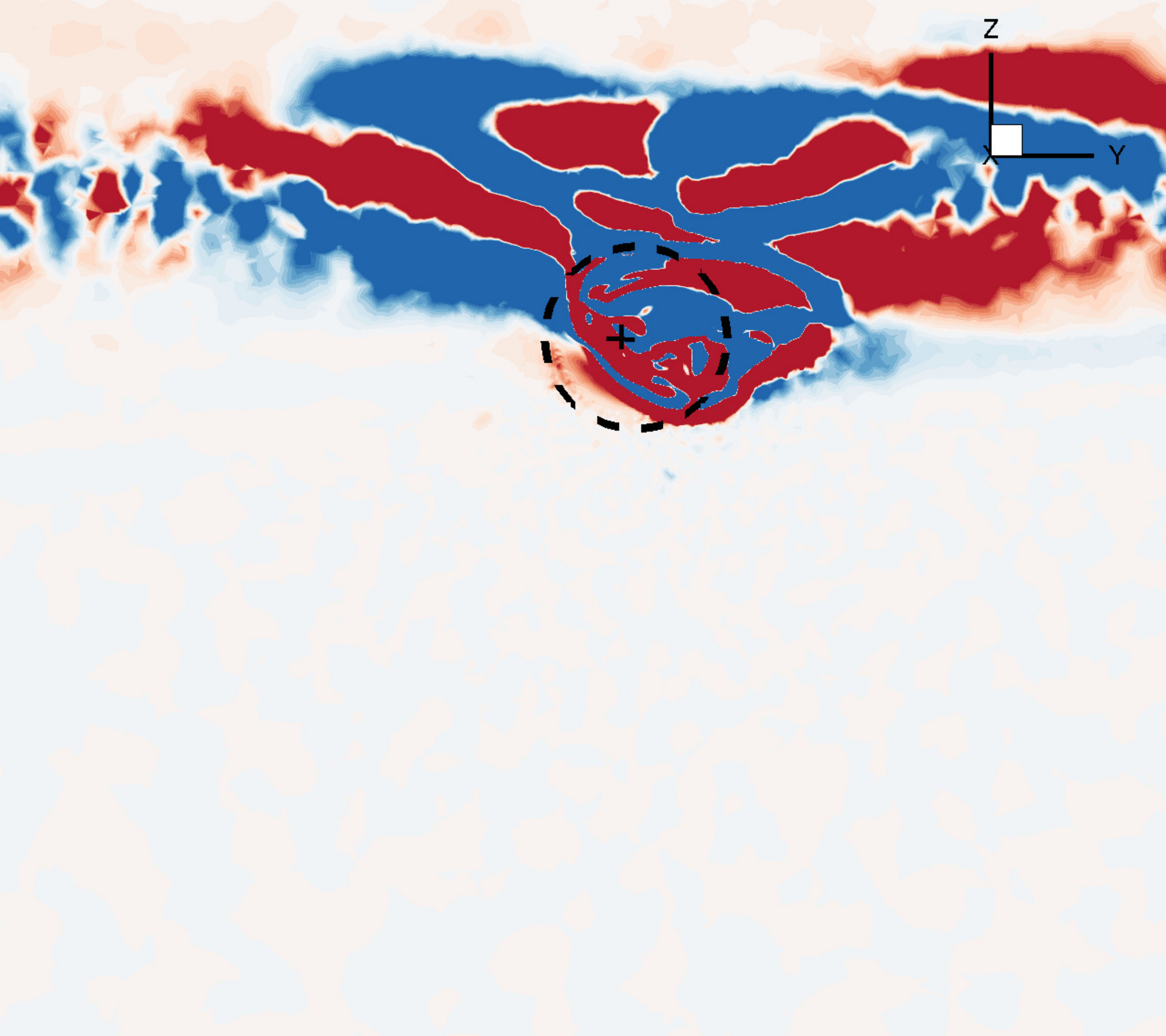}
		\caption*{$(c$ - $2)$}
	\end{subfigure}
	
	\begin{subfigure}[b]{0.5\textwidth}
		\centering
		\adjincludegraphics[scale=0.3,trim={0\width} {0.3\width} {0\width} {0.21\width},clip]{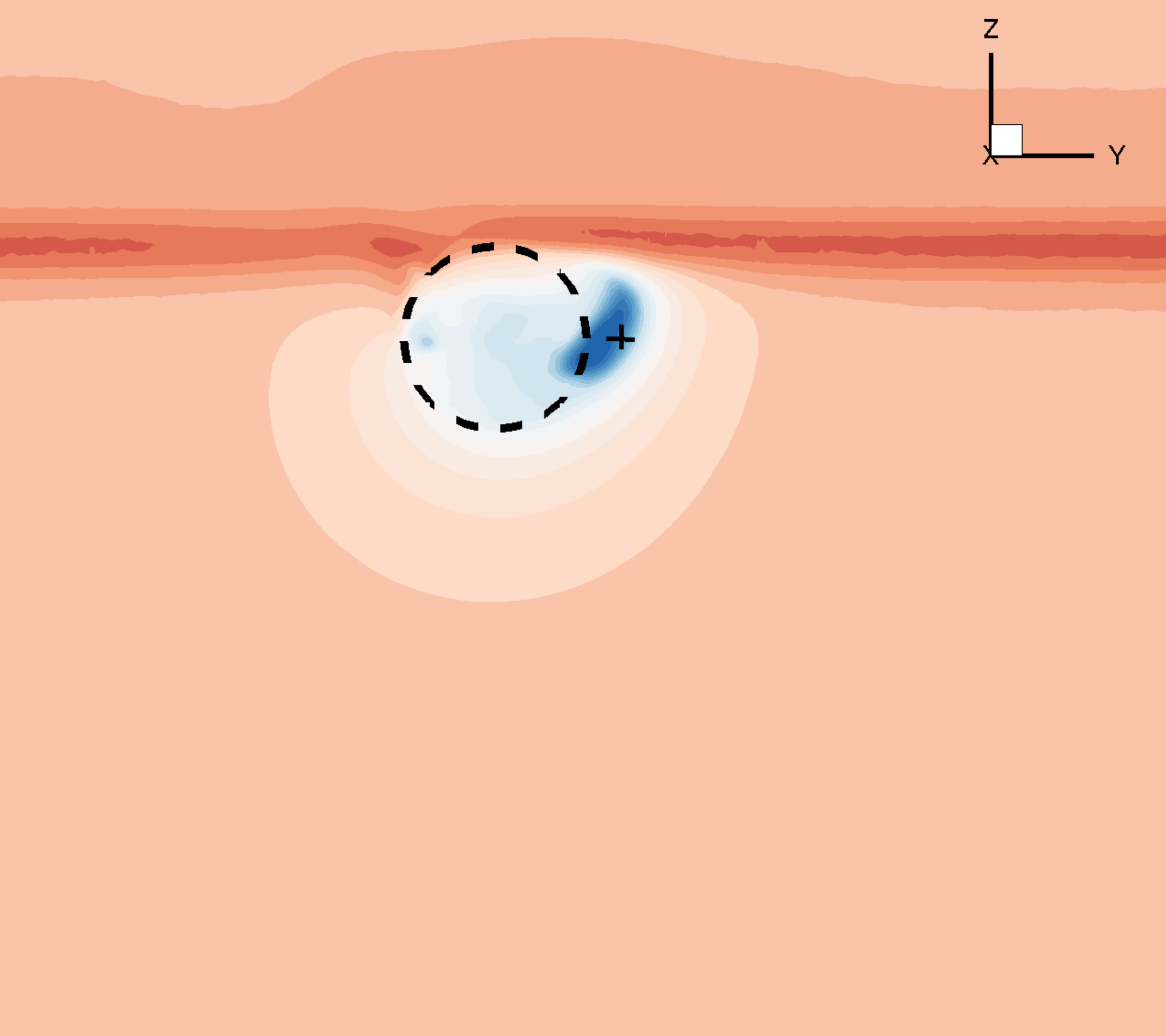}
		\caption*{$(d$ - $1)$}
	\end{subfigure}%
	\begin{subfigure}[b]{0.5\textwidth}
		\centering
		\adjincludegraphics[scale=0.3,trim={0\width} {0.3\width} {0\width} {0.21\width},clip]{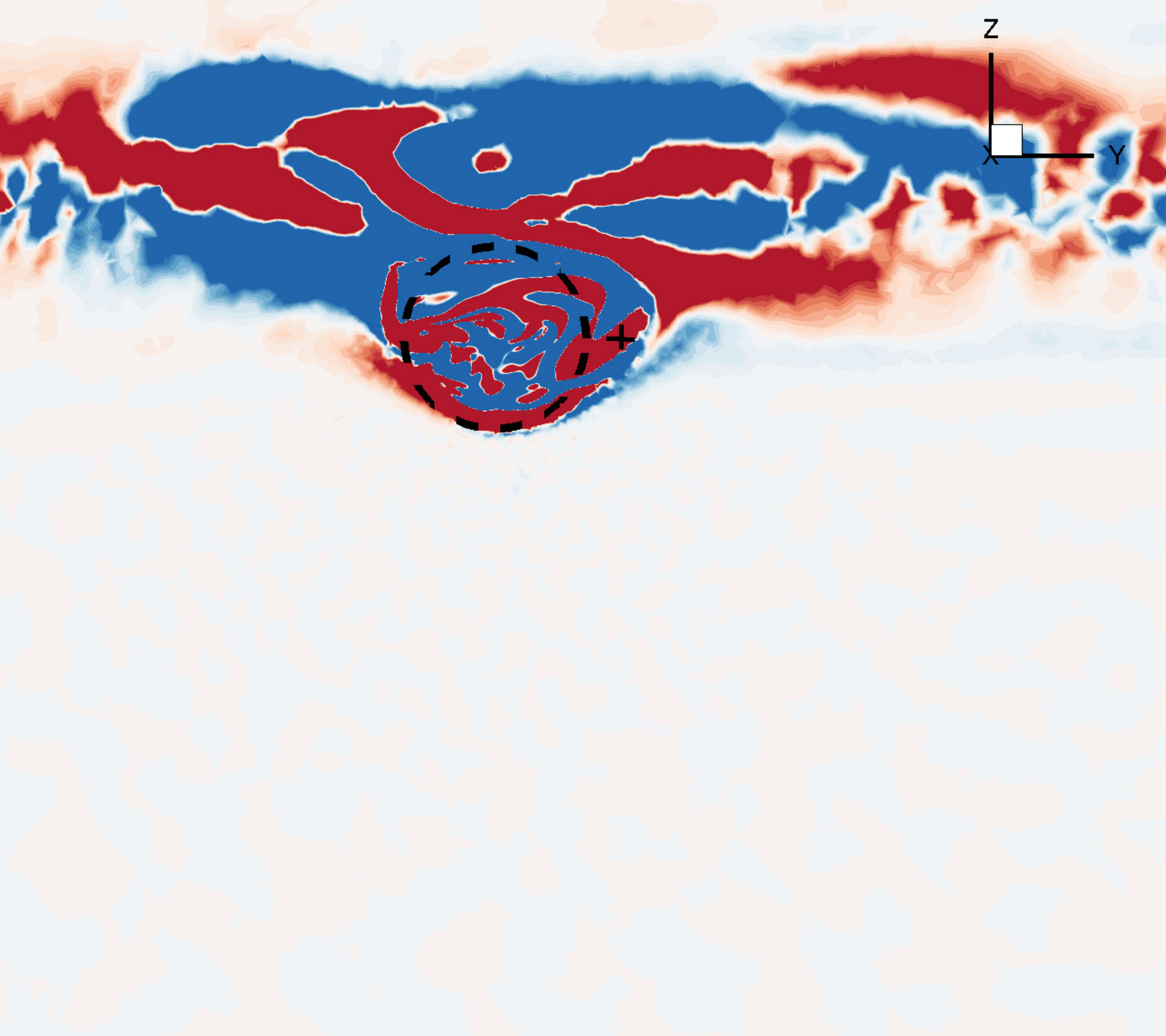}
		\caption*{$(d$ - $2)$}
	\end{subfigure}
	
	\begin{subfigure}[b]{0.5\textwidth}
		\centering
		\adjincludegraphics[scale=0.4,trim={0.05\width} {0.7\width} {0.16\width} {0.0\width},clip]{Photos/78.pdf}
		\caption*{}
	\end{subfigure}%
	\begin{subfigure}[b]{0.5\textwidth}
		\centering
		\adjincludegraphics[scale=0.4,trim={0.05\width} {0.7\width} {0.01\width} {0.0\width},clip]{Photos/38.pdf}
		\caption*{}
	\end{subfigure}
	
	\caption{Evolution of streamwise $x$-vorticity and pressure distribution plotted at $0.5D$ downstream at $U^*=10$ for the submerged sphere at $h^*=0$. Top of  the sphere touches the free surface and one complete oscillation period is considered.}
	\label{hS0_xVor_Pressure} 
\end{figure}

\begin{figure}[htbp!]
	\centering
	
	\begin{subfigure}[b]{1\textwidth}
		\centering
		\hspace{0cm}
		\adjincludegraphics[scale=0.2,trim={0\width} {0.0\width} {0\width} {0.0\width},clip]{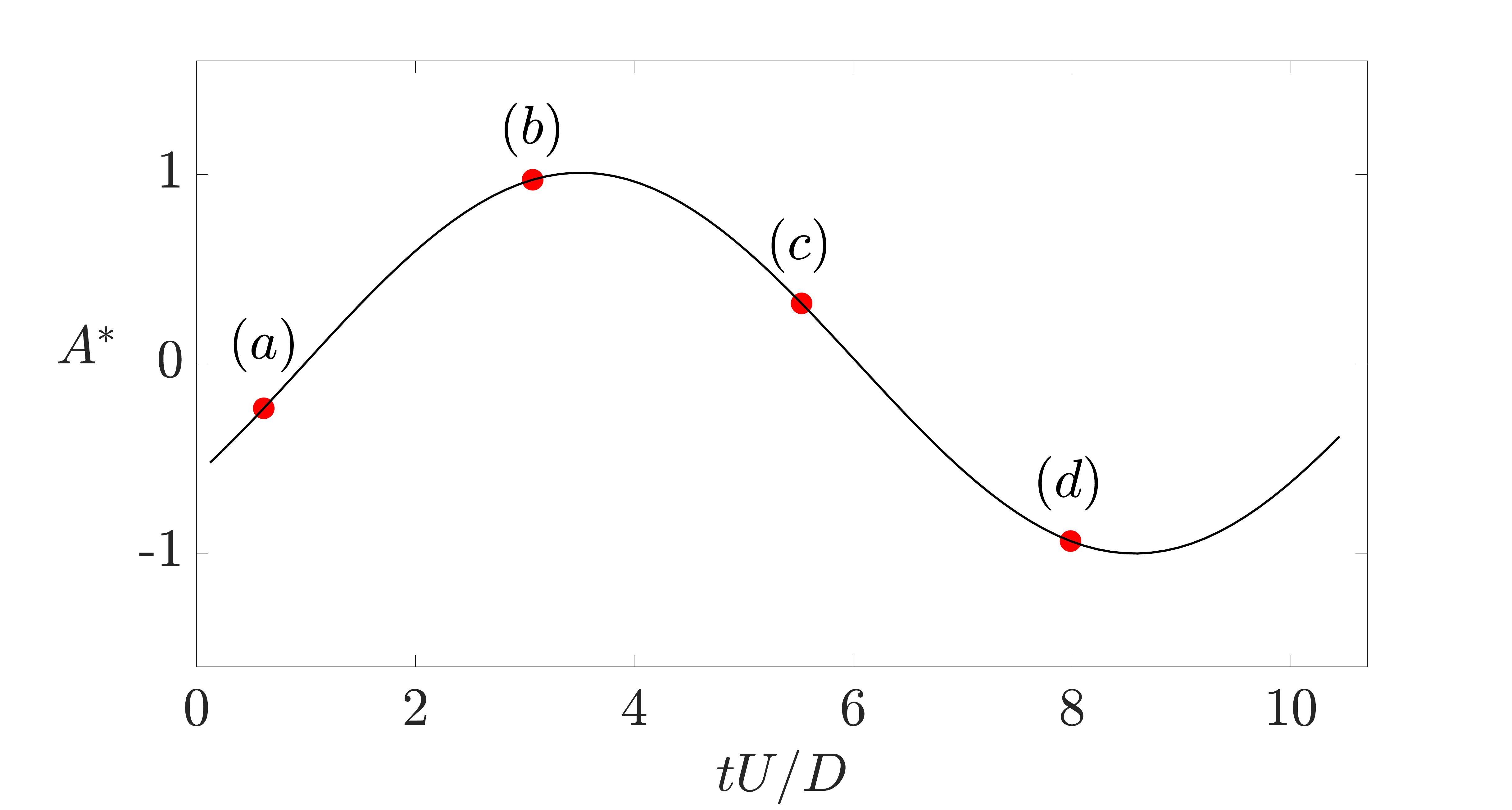}
	\end{subfigure}
	
	\begin{subfigure}[b]{0.5\textwidth}
		\centering
		\adjincludegraphics[scale=0.3,trim={0\width} {0.3\width} {0\width} {0.22\width},clip]{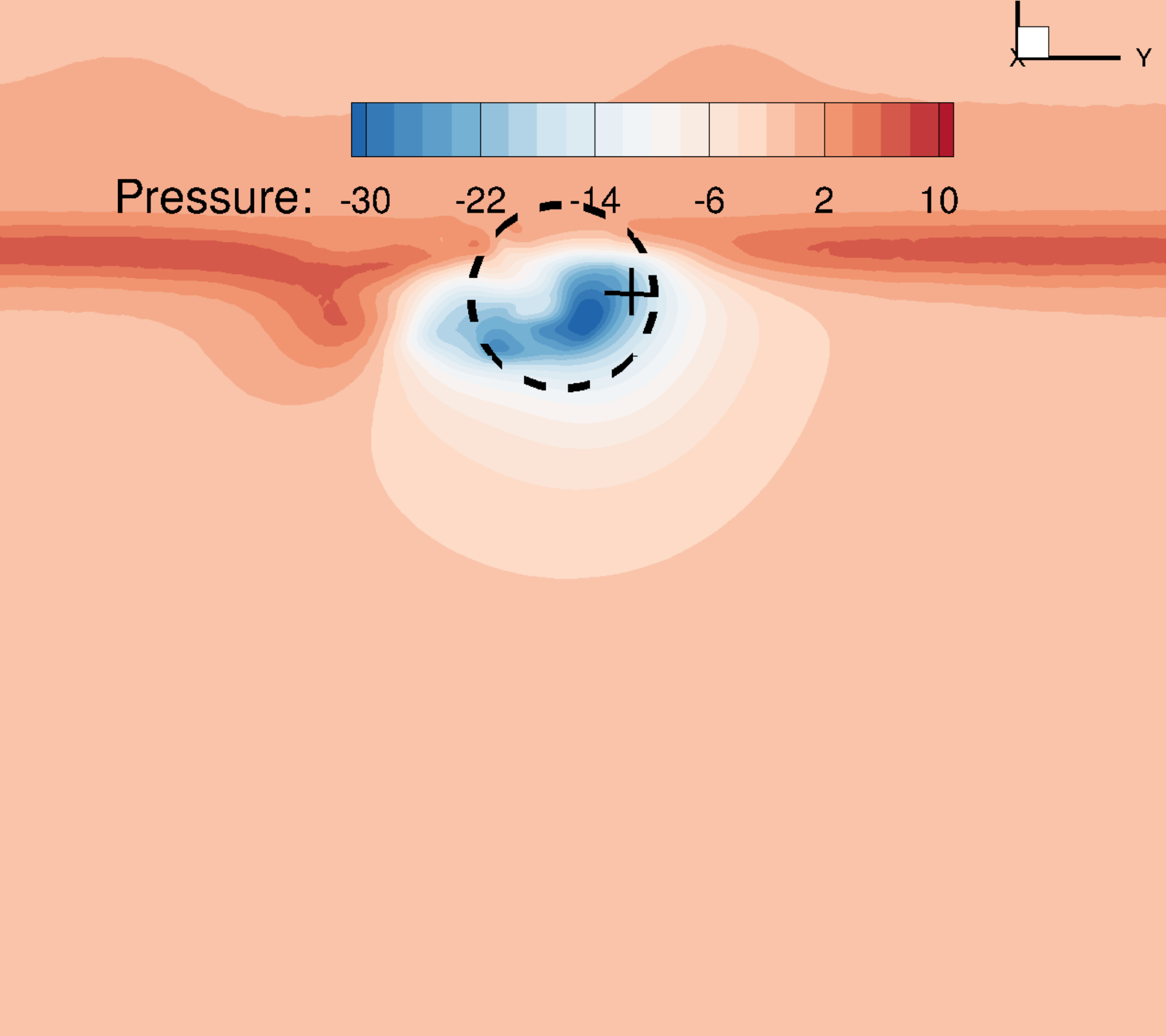}
		\caption*{$(a$ - $1)$}
	\end{subfigure}%
	\begin{subfigure}[b]{0.5\textwidth}
		\centering
		\adjincludegraphics[scale=0.3,trim={0\width} {0.3\width} {0\width} {0.22\width},clip]{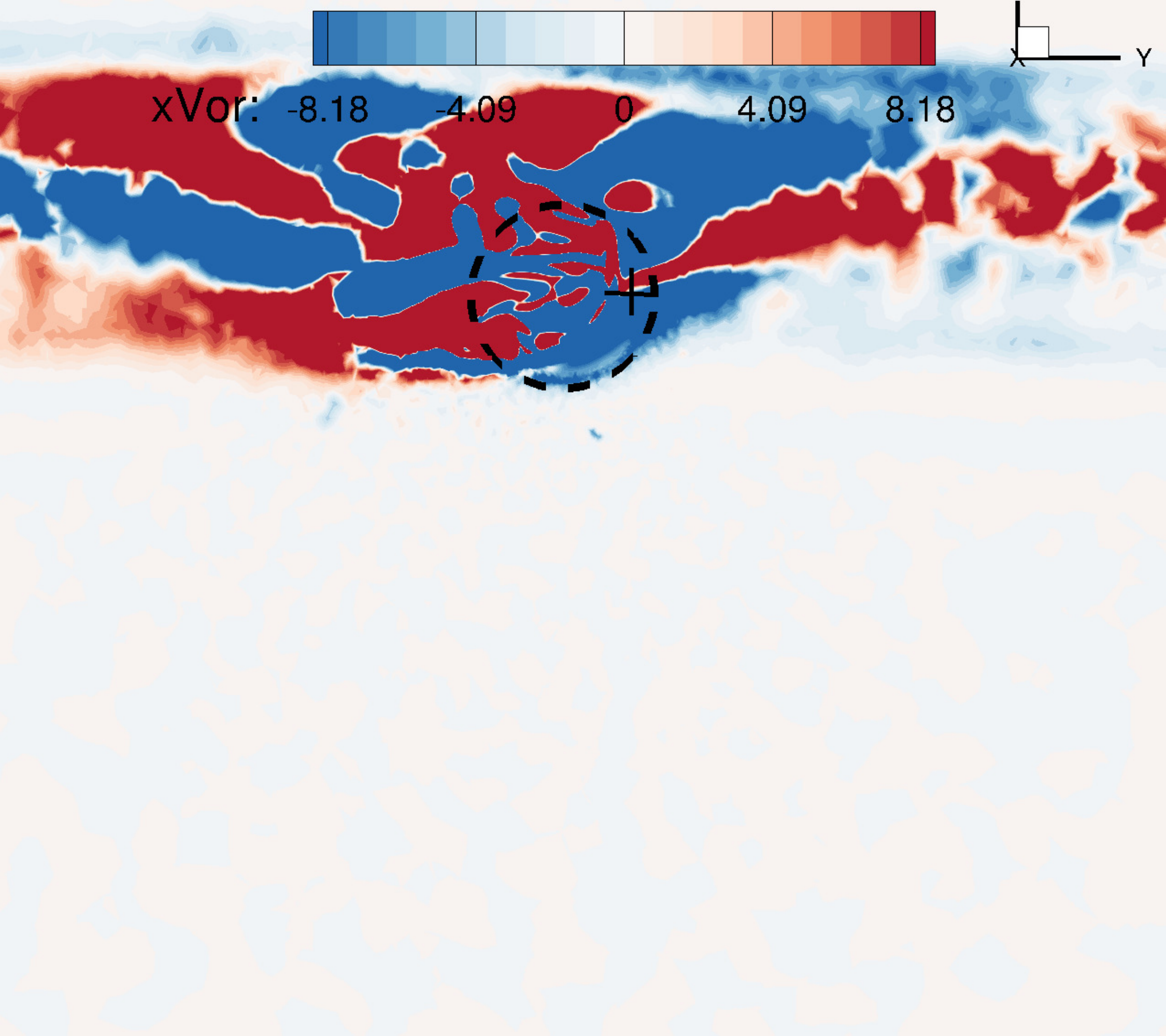}
		\caption*{$(a$ - $2)$}
	\end{subfigure}
	
	\begin{subfigure}[b]{0.5\textwidth}
		\centering
		\adjincludegraphics[scale=0.3,trim={0\width} {0.3\width} {0\width} {0.22\width},clip]{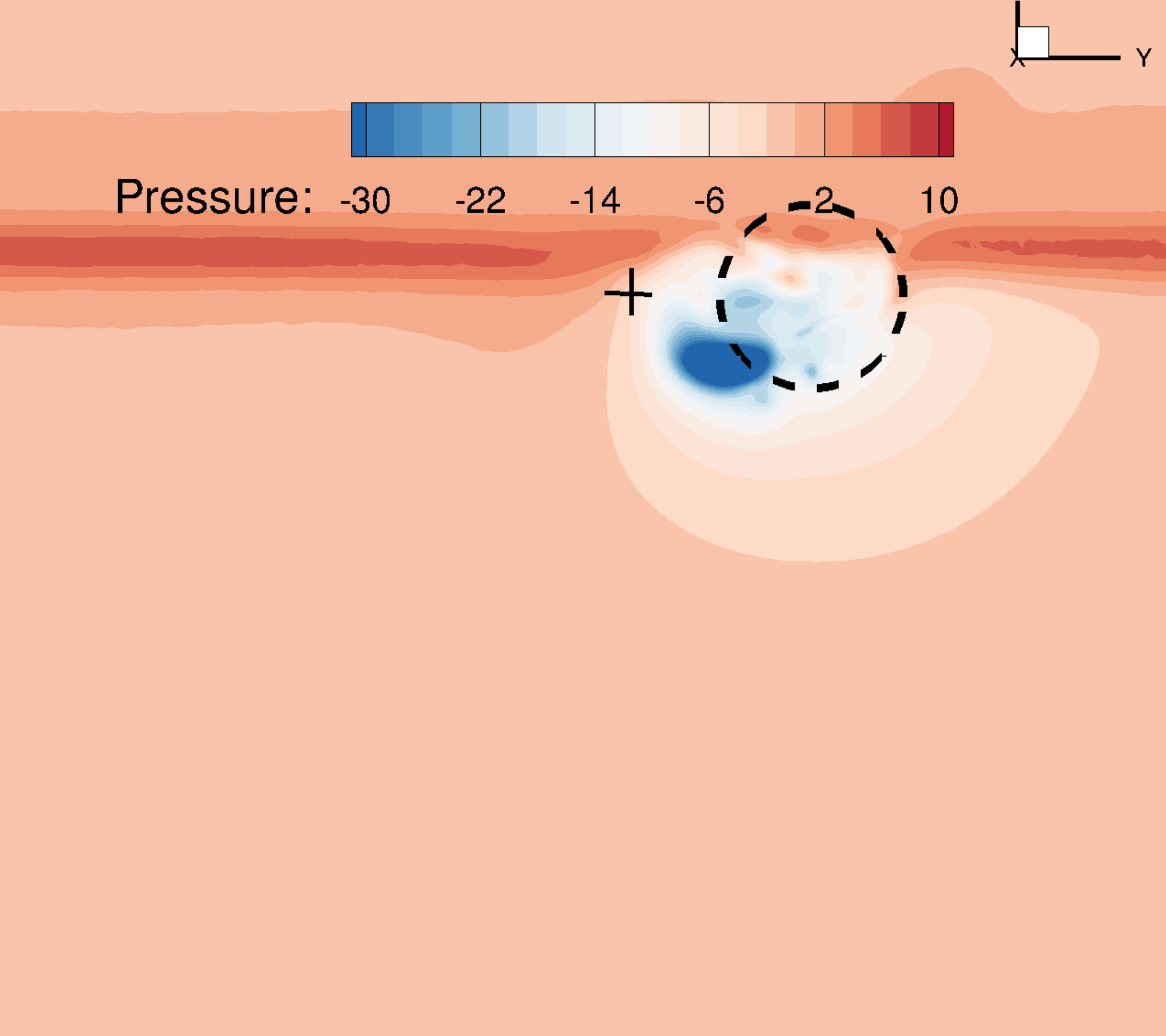}
		\caption*{$(b$ - $1)$}
	\end{subfigure}%
	\begin{subfigure}[b]{0.5\textwidth}
		\centering
		\adjincludegraphics[scale=0.3,trim={0\width} {0.3\width} {0\width} {0.22\width},clip]{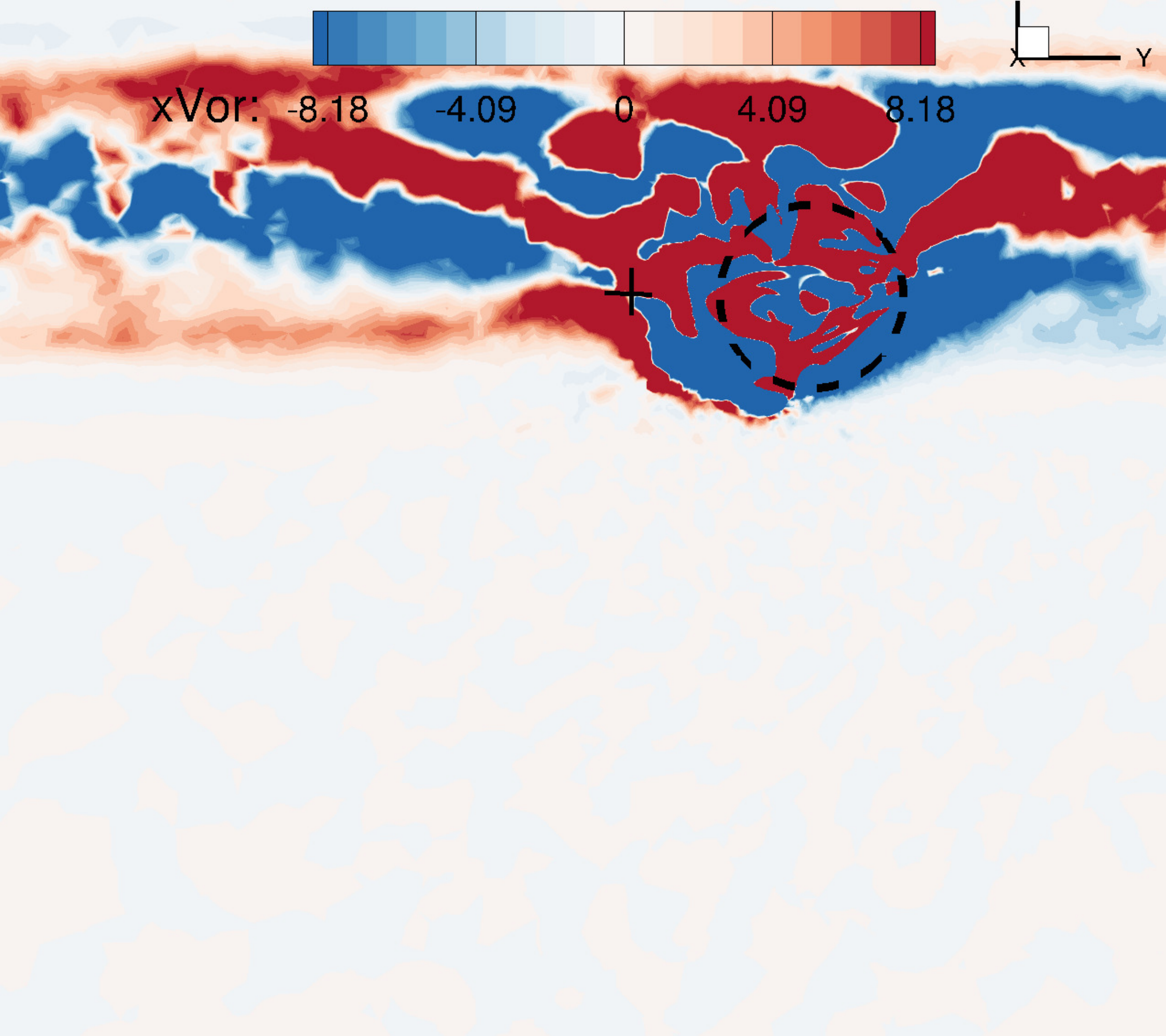}
		\caption*{$(b$ - $2)$}
	\end{subfigure}
	
	\begin{subfigure}[b]{0.5\textwidth}
		\centering
		\adjincludegraphics[scale=0.3,trim={0\width} {0.3\width} {0\width} {0.22\width},clip]{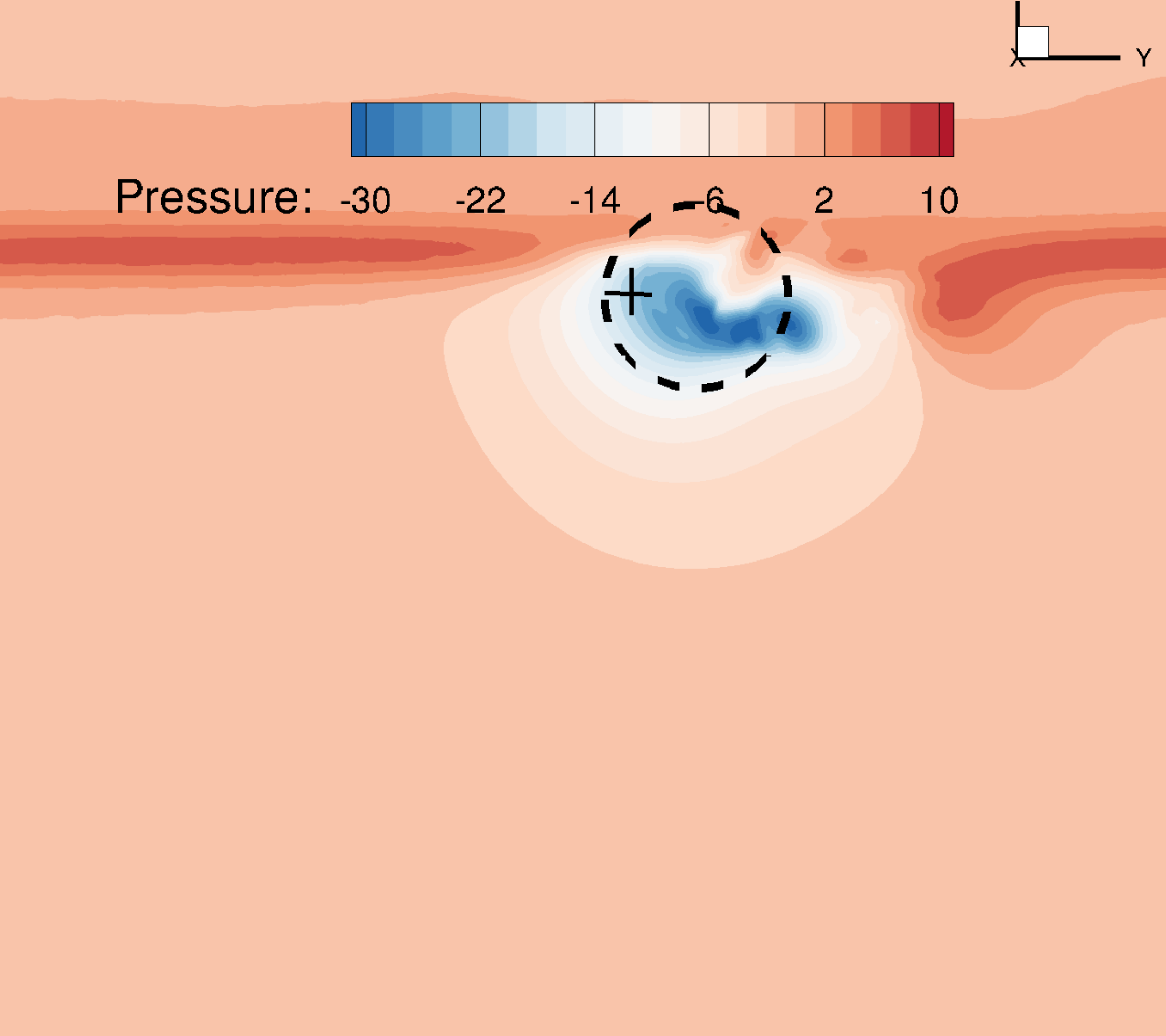}
		\caption*{$(c$ - $1)$}
	\end{subfigure}%
	\begin{subfigure}[b]{0.5\textwidth}
		\centering
		\adjincludegraphics[scale=0.3,trim={0\width} {0.3\width} {0\width} {0.22\width},clip]{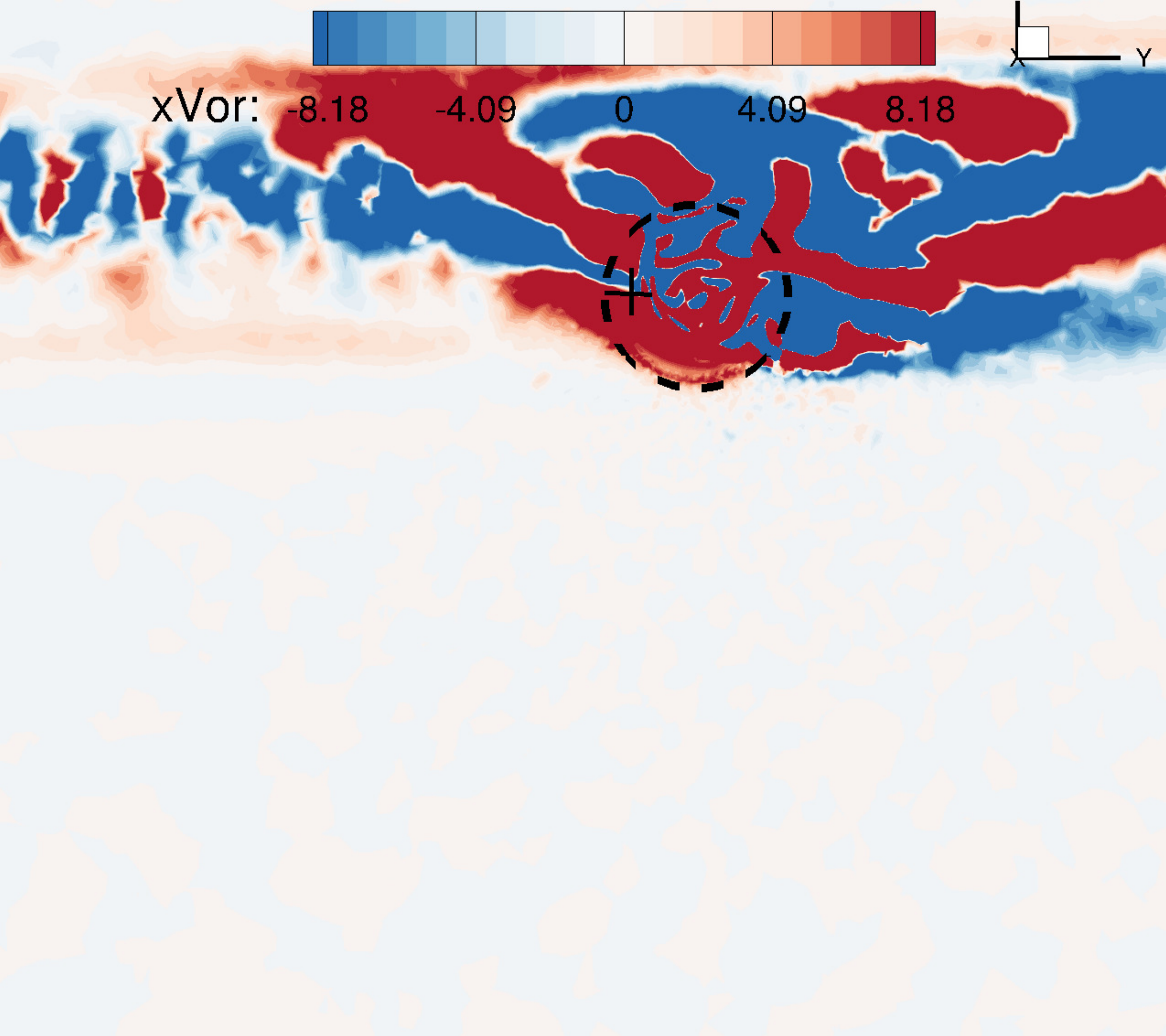}
		\caption*{$(c$ - $2)$}
	\end{subfigure}
	
	\begin{subfigure}[b]{0.5\textwidth}
		\centering
		\adjincludegraphics[scale=0.3,trim={0\width} {0.3\width} {0\width} {0.22\width},clip]{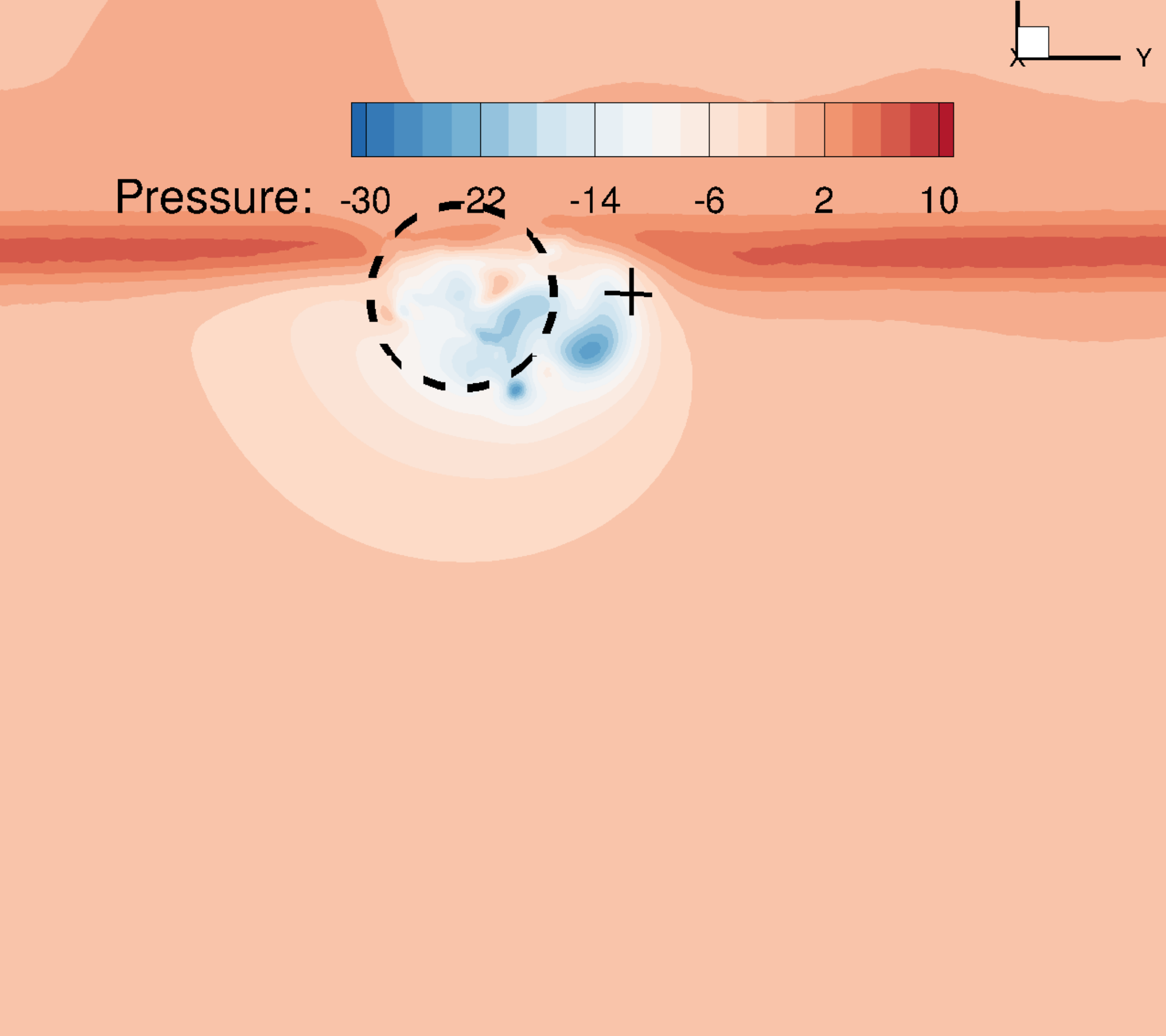}
		\caption*{$(d$ - $1)$}
	\end{subfigure}%
	\begin{subfigure}[b]{0.5\textwidth}
		\centering
		\adjincludegraphics[scale=0.3,trim={0\width} {0.3\width} {0\width} {0.22\width},clip]{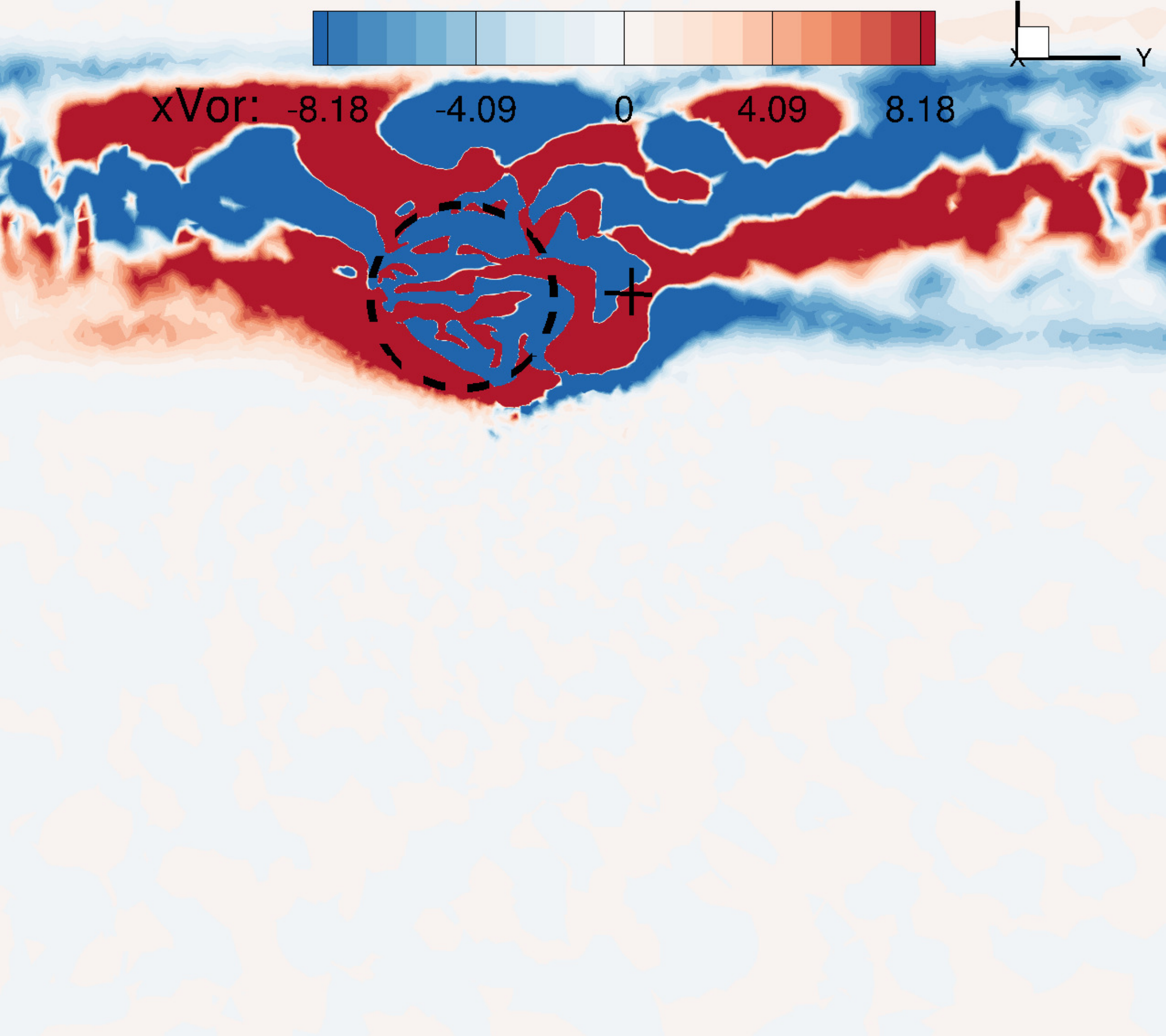}
		\caption*{$(d$ - $2)$}
	\end{subfigure}
	
	\begin{subfigure}[b]{0.5\textwidth}
		\centering
		\adjincludegraphics[scale=0.4,trim={0.05\width} {0.7\width} {0.16\width} {0.0\width},clip]{Photos/78.pdf}
		\caption*{}
	\end{subfigure}%
	\begin{subfigure}[b]{0.5\textwidth}
		\centering
		\adjincludegraphics[scale=0.4,trim={0.05\width} {0.7\width} {0.01\width} {0.0\width},clip]{Photos/38.pdf}
		\caption*{}
	\end{subfigure}
	
	\caption{Evolution of streamwise $x$-vorticity and pressure distribution plotted at $0.5D$ downstream at $U^*=10$ for the piercing sphere at $h^*=-0.25$ for one complete oscillation period.}
	\label{hS-0.25_xVor_Pressure} 
\end{figure}

Fig. \ref{FSd_hS} compares the free-surface deformation for the submerged case at $h^*=0$ and the piercing case at $h^*=-0.25$ at the mean position. It can be seen that the surface deformation for the piercing case is significantly larger compared to the submerged case, where a part of the sphere lies above the water line and causes a noticeable deformation. To understand the vortex dynamics and the force exerted on the sphere, we plot the vorticity formation and the pressure distribution on the cross-flow plane located at $0.5D$ downstream from the center of the sphere for the submerged and the piercing cases. The plots provide an overview of different types of wake behaviors due to the existence of the free surface. 
Fig. \ref{xVor_Pressure_hS} shows the change in the vorticity field and the pressure distribution at different immersion ratios for the sphere at the mean position. As can be seen in Fig. \ref{xVor_Pressure_hS} (a - 1) and (b - 1), the vorticity for the piercing sphere at $h^*=-0.25$, and the vorticity for the submerged sphere when the top of the sphere touches the free surface at $h^*=0$, is significantly different to that of flow past sphere with no free-surface effect at $h^*=1$, Fig. \ref{xVor_Pressure_hS} (c - 1). At $h^*=1$, the $x$-vorticity pattern has feasible symmetry about the horizontal center-line when there is no effect of the free surface. When the sphere comes closer to the free surface at $h^*=0$, the free-surface affects the vortex dynamics substantially. The induced surface distortion due to the sphere motion in the proximity of the free-surface boundary, causes large opposite sign stretched vortex loops at the top region, shown in Fig. \ref{xVor_Pressure_hS} (b - 1). The vorticity generated due to the free surface makes the wake asymmetric about the horizontal center-line. In Fig. \ref{xVor_Pressure_hS} (a - 1), vorticity pattern for the piercing case at $h^*=-0.25$ is remarkably different from the former case at $h^*=0$, where the surface distortion is considerably larger as 25\% of the sphere lies above the waterline. In Fig. \ref{xVor_Pressure_hS} (a - 2), the corresponding pressure distribution for the piercing case shows the high-pressure region on the left side of the sphere due to the induced surface curvature. To further analyze this behavior, we plot the evolution of the vorticity and the pressure distribution for one complete oscillation period for the sphere close to the free surface at $h^*=0$ and the piercing case at $h^*=-0.25$ in Figs. \ref{hS0_xVor_Pressure} and \ref{hS-0.25_xVor_Pressure}, respectively.

\begin{figure}[htbp!]
	\centering
	
	\adjincludegraphics[scale=0.3,trim={0\width} {0.0\width} {0\width} {0.0\width},clip]{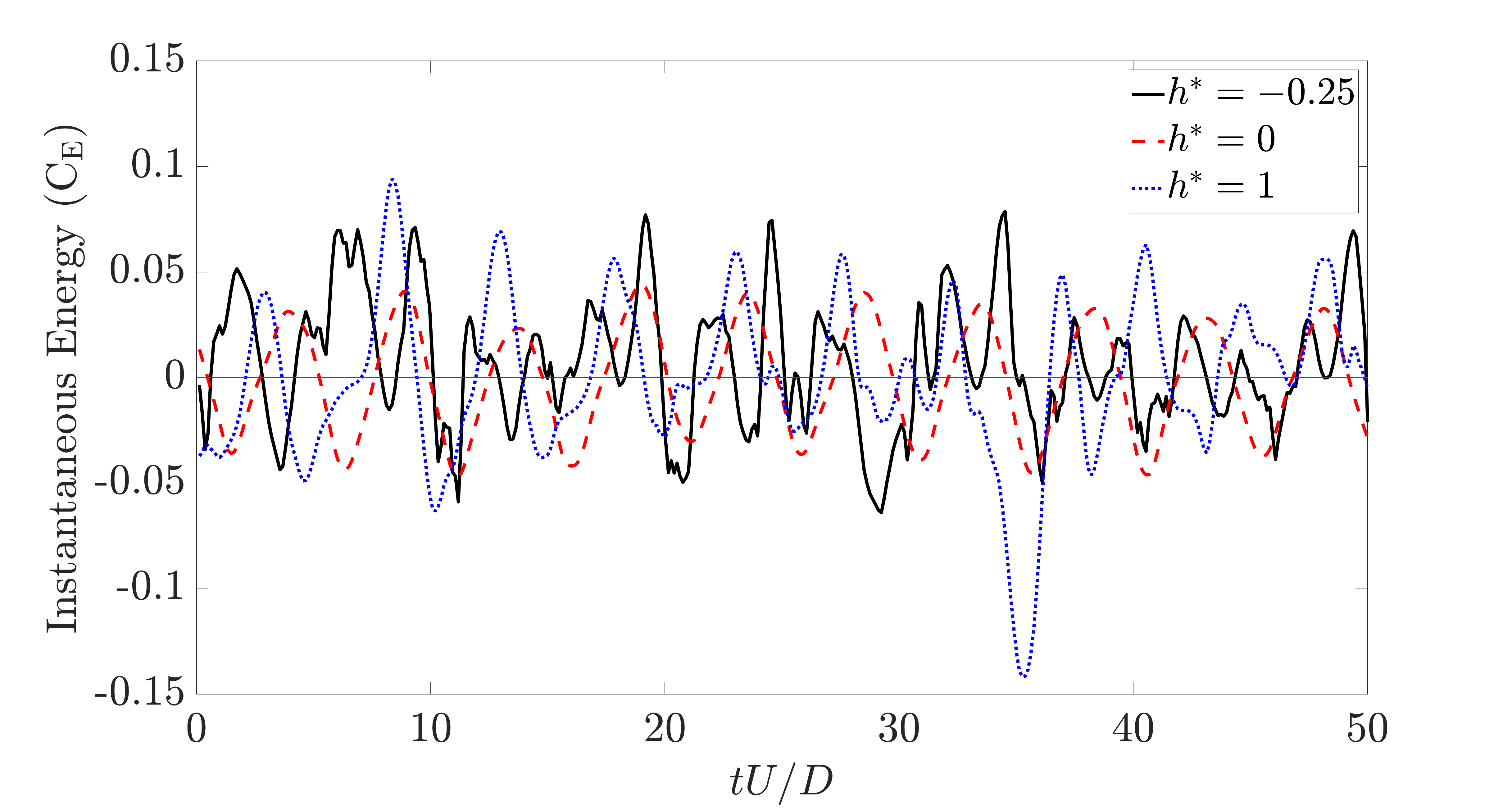}
	
	\caption{Temporal variation of instantaneous energy transfer $C_E$ for the piercing sphere case at $h^*=-0.25$ and the submerged sphere cases at $h^*=0$ and $h^*=1$ at lock-in state with $U^* = 10$, $Re = 15\,700$, and $m^*\zeta = 0.017$. }
	
	\label{Power_hS} 
\end{figure}

In Fig. \ref{hS0_xVor_Pressure}, for the evolution of the vorticity at $h^*=0$, throughout the whole oscillation cycle, a flux of vorticity due to the induced surface distortion appears on the top region. This secondary vorticity flux causes diffusion of the vorticity that is induced due to VIV. The free-surface diffusive vorticity flux acts as a sink of energy and leads to the reduction in the corresponding transverse force acting on the spherical body and hence lower amplitude response. In Fig. \ref{hS-0.25_xVor_Pressure}, for the evolution of the vorticity at $h^*=-0.25$, the surface distortion is considerably larger and consequently strong flux of vorticity is supplied to the wake due to the induced surface distortion. This strong vorticity at the top, induce opposite sign vortex loops immediately below and cross-annihilated small scale vortex structures. This phenomenon alters the vorticity pattern and the synchronization of the vortex shedding. This extra vorticity generation acts as a source of energy supplied to the wake of the piercing sphere and makes the associated vorticity stronger. This leads to a larger transverse force acting on the body and therefore large amplitude oscillations. 
Fig \ref{Power_hS} compares the variation of the instantaneous energy transfer $C_E$ for the piercing case at $h^*=-0.25$ and the submerged cases at $h^*=0$ and $h^*=1$. It is quantified that the non-dimensional time-averaged quantity of the energy transfer ($E$) over each period of motion $T$ for the piercing sphere case at $h^*=-0.25$ is significantly more than the submerged cases at $h^*=0$ and $h^*=1$. This energy transfer sustains the large amplitude oscillations for the piercing sphere case more than all the submerged cases. It can be deduced that the extra vorticity generation at the free surface for the piercing sphere has a significant impact on the synchronization of the vortex shedding and the energy transfer.

\subsection{Effect of mass ratio}
It is known that the non-dimensional parameter mass ratio $m^*$ has a strong influence on the flow-induced vibration. When the sphere pierces the free surface, this ratio varies significantly due to the rapid decrease in the mass of the displaced fluid ($m_d$). However, identifying the effect of the mass ratio on the FIV response for the piercing cases while the immersion ratio is changing cannot be clearly explained  because the geometry of the submerged portion of sphere changes. Here in this subsection, we aim to understand the effect of mass ratio on the FIV characteristics of the sphere piercing the free surface at the fixed immersion ratio $h^*=-0.25$ and zero damping ratio. 

\begin{figure}[htbp!]

	\begin{subfigure}[b]{0.5\textwidth}
			\centering
		\adjincludegraphics[scale=0.17,trim={0\width} {0\width} {0\width} {0.0\width},clip]{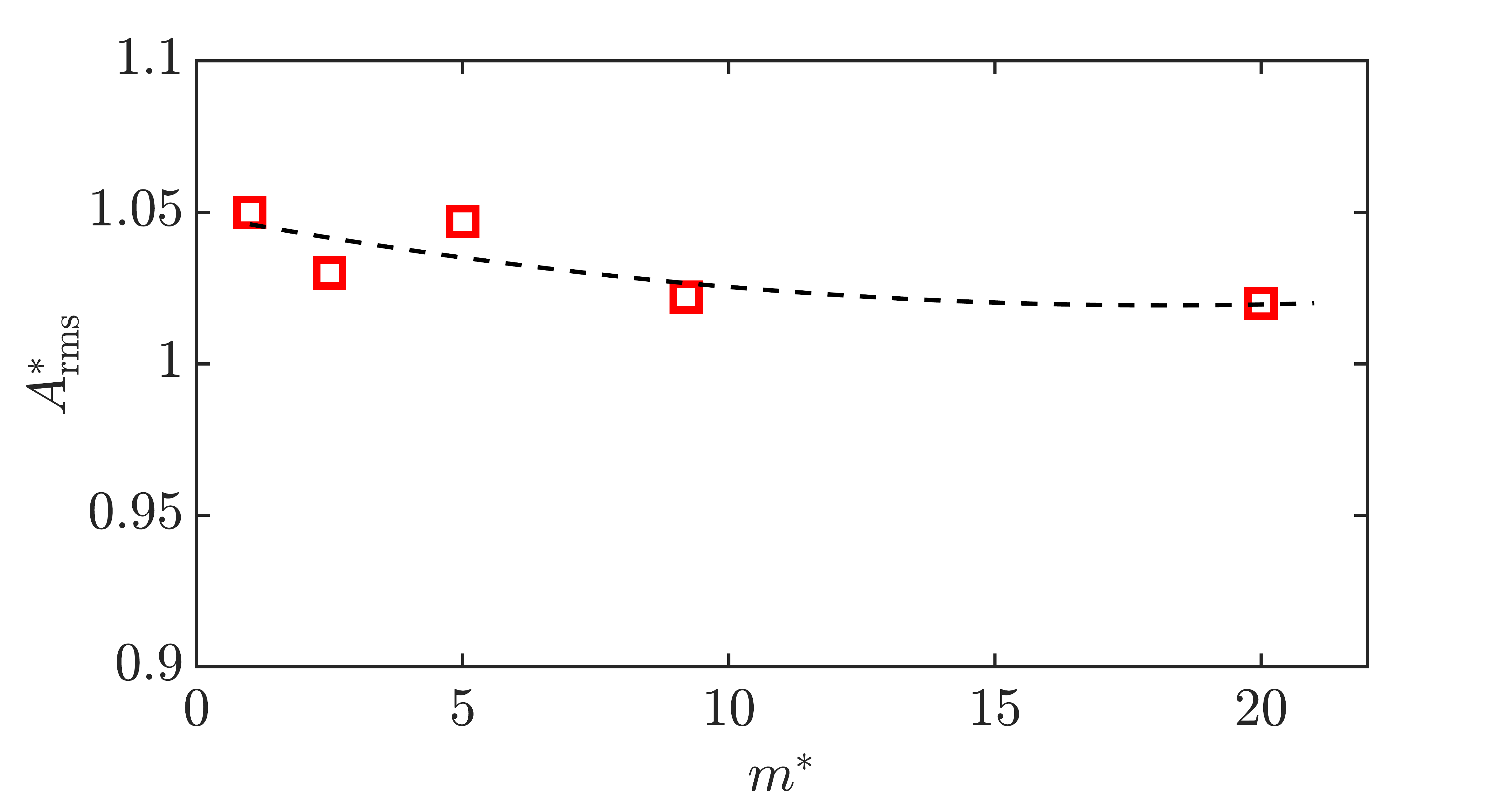}
		\caption{}
	\end{subfigure}%
	\begin{subfigure}[b]{0.5\textwidth}
		\centering
		\adjincludegraphics[scale=0.17,trim={0\width} {0\width} {0\width} {0.0\width},clip]{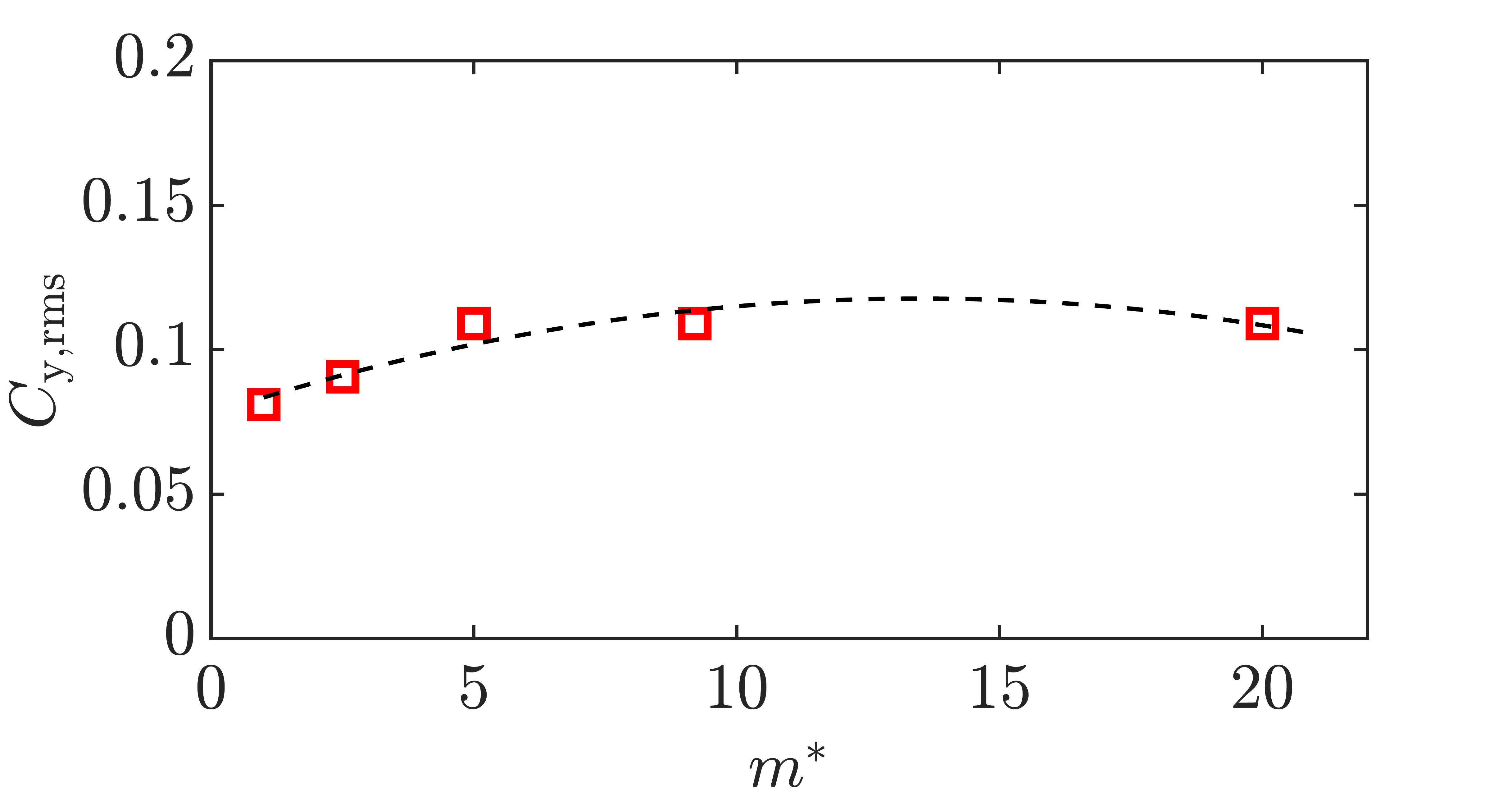}
		\caption{}
	\end{subfigure}
	
	\vspace{2mm}
	
	\begin{subfigure}[b]{0.5\textwidth}
			\centering
		\adjincludegraphics[scale=0.17,trim={0\width} {0\width} {0\width} {0.0\width},clip]{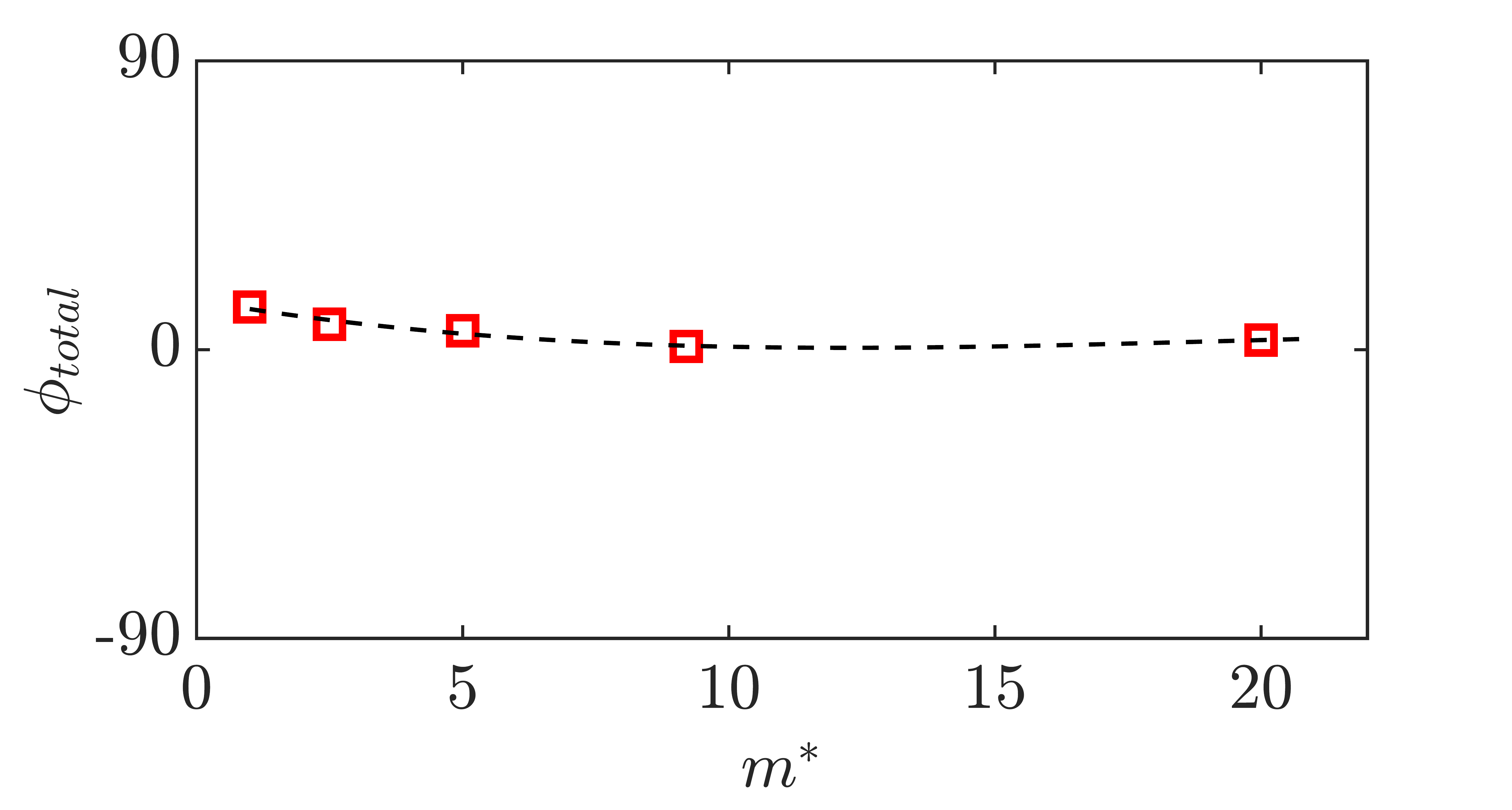}
		\caption{}
	\end{subfigure}%
	\begin{subfigure}[b]{0.5\textwidth}
		\centering
		\adjincludegraphics[scale=0.17,trim={0\width} {0\width} {0\width} {0.0\width},clip]{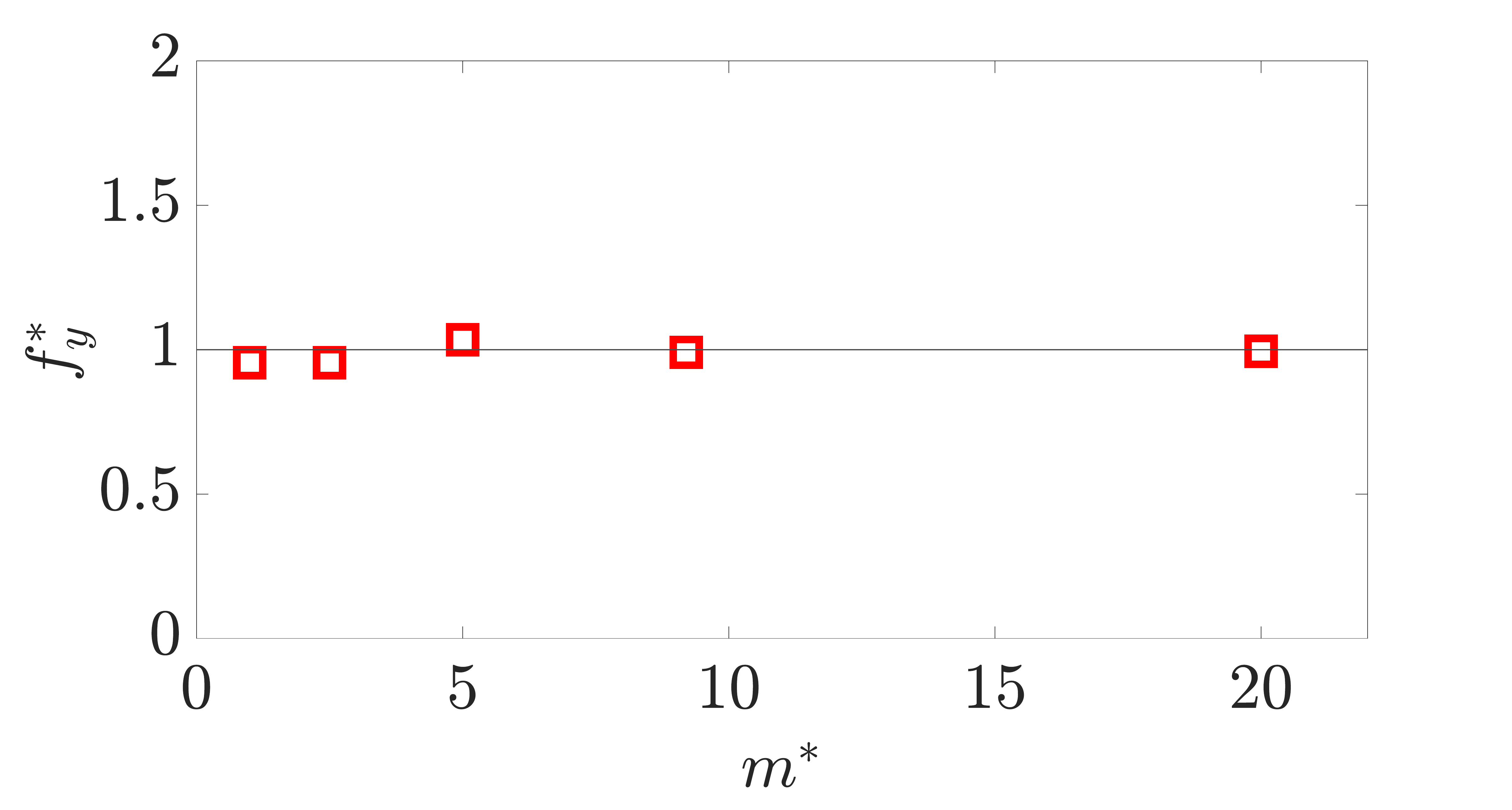}
		\caption{}
	\end{subfigure}

	\caption{Variation of  VIV response parameters as a function of the mass ratio for the sphere piercing the free surface at $h^*=-0.25$, $U^*=10$ and $Re=15\,700$: (a) r.m.s. amplitude response, (b) r.m.s. normalized transverse force, (c) total phase difference and, (d) normalized cross-flow frequency. The normalized cross-flow frequency is defined as $f^*_y=f_y/f_n$, where $f_y$ is the frequency of the oscillations and $f_n$ is the natural frequency of the system.}
	\label{MS_Trend}		
\end{figure}

Fig. \ref{MS_Trend} (a,b) shows the variation of the amplitude response $A^*$ and the normalized transverse force $C_y$, with a range of mass ratio $m^*\in[1,20]$ at identical Reynolds number $Re=15\,700$ and the reduced velocity $U^*=10$. The results show that a small variation of the mass ratio does not have a significant effect on the amplitude response. The variation of the amplitude response with the mass ratio is found to be less than $3 \%$, consistent with previous experimental and numerical investigations for VIV of low-mass-damped fully submerged spheres \cite{govardhan2005vortex, rajamuni2018transverse}. The experimental study in \cite{sareen2018}, considered two different mass ratios at $h^*=-0.25$ where the considerable effect of the mass ratio was reported. In their work, the r.m.s. amplitude response has a noticeable reduction by increasing the mass ratio. This difference may be due to different parameter setup for the mass-damping parameter in the experiments and zero-damping in our numerical simulations in this subsection. Periodic and large amplitude vibrations are observed over a range of mass ratio $m^*\in[1,20]$ in our numerical results.

\begin{figure}[htbp!]
	\centering
	
	\begin{subfigure}[b]{0.5\textwidth}
		\adjincludegraphics[scale=0.28,trim={0.2\width} {0\width} {0.1\width} {0.0\width},clip]{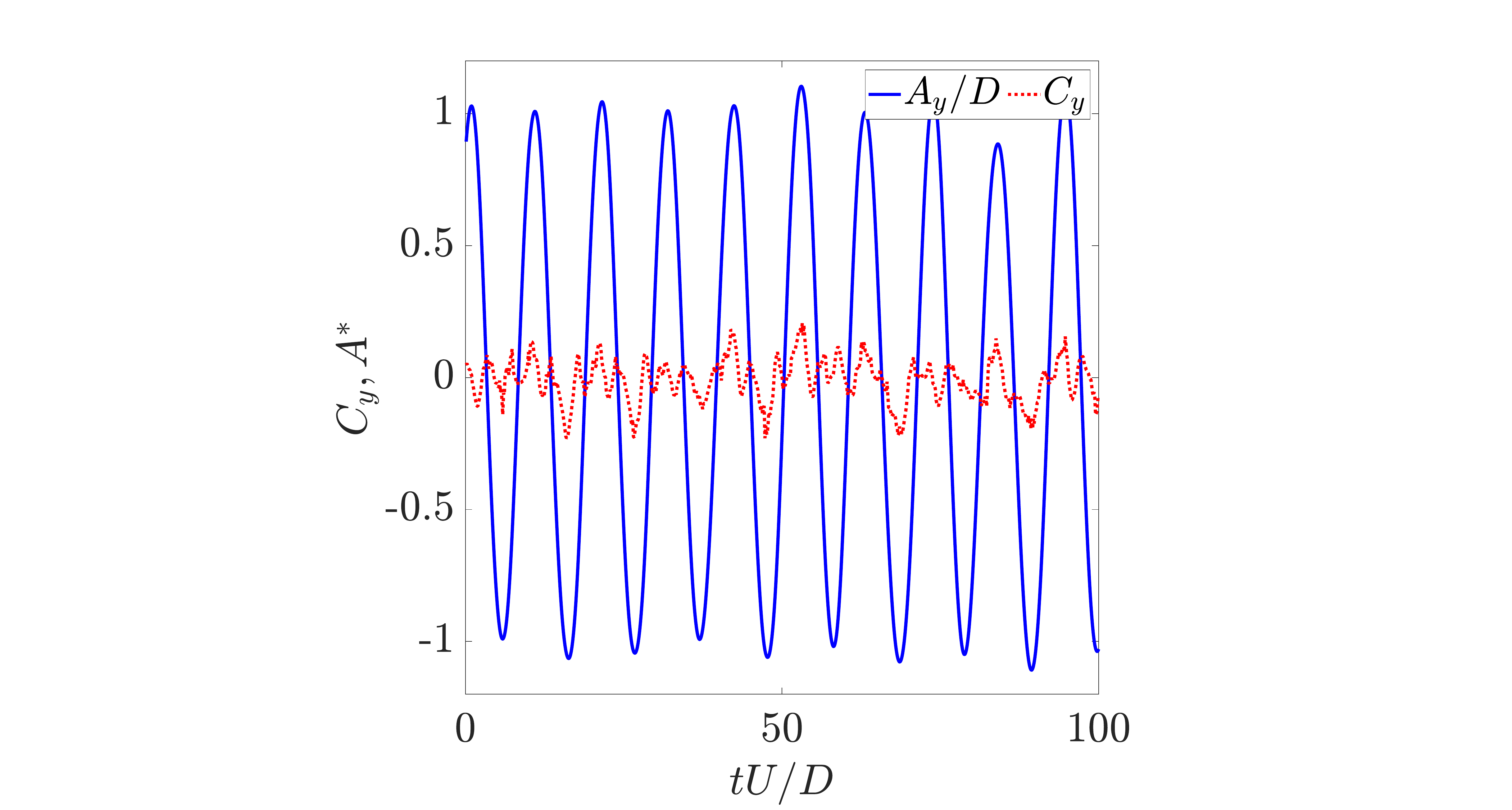}
		\caption*{($a$ - $1$)}
	\end{subfigure}%
	\begin{subfigure}[b]{0.5\textwidth}
		\adjincludegraphics[scale=0.28,trim={0.2\width} {0\width} {0.1\width} {0.0\width},clip]{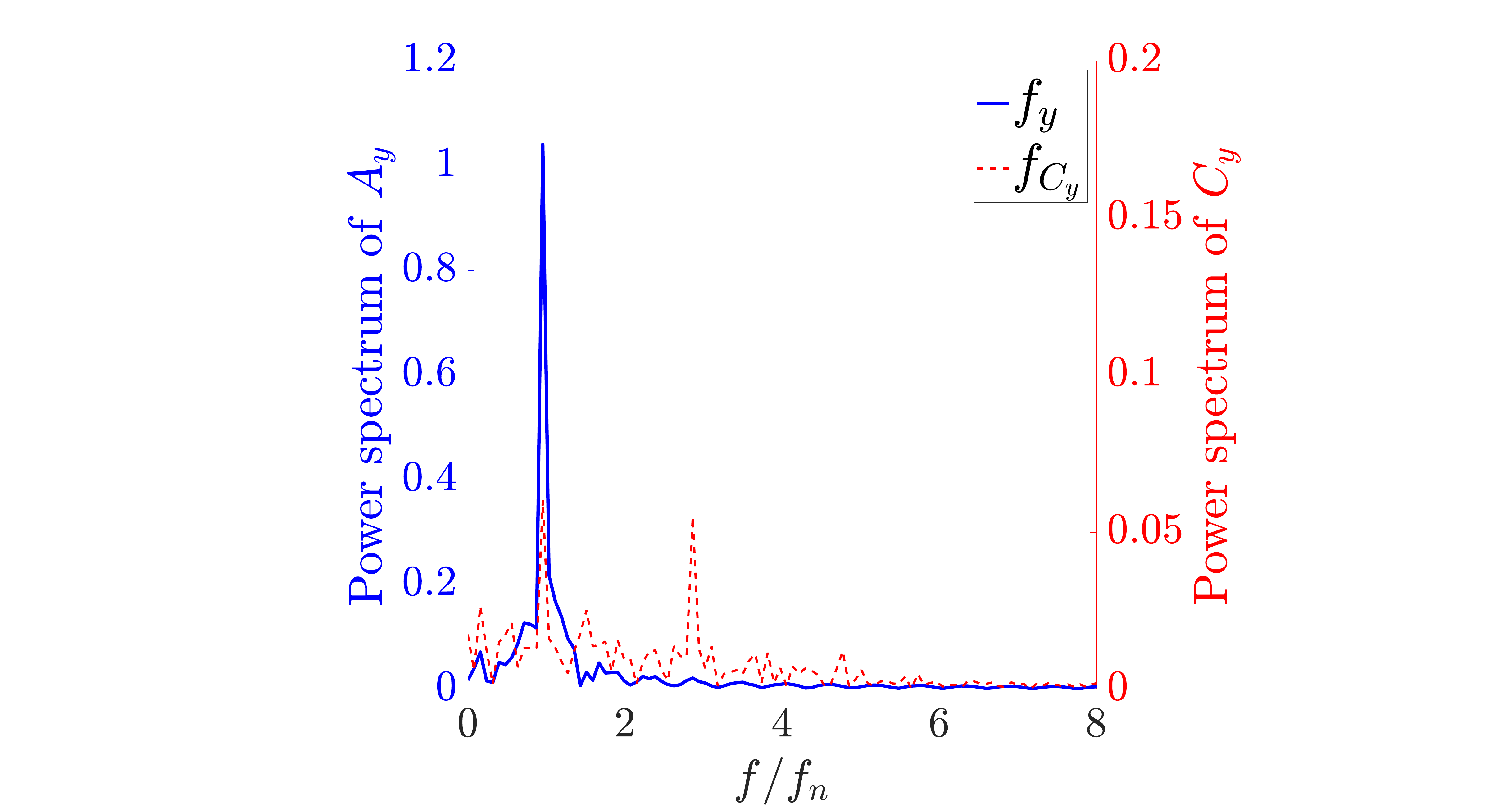}
		\caption*{($a$ - $2$)}
	\end{subfigure}
	
	
	\begin{subfigure}[b]{0.5\textwidth}
		\adjincludegraphics[scale=0.28,trim={0.2\width} {0\width} {0.1\width} {0.0\width},clip]{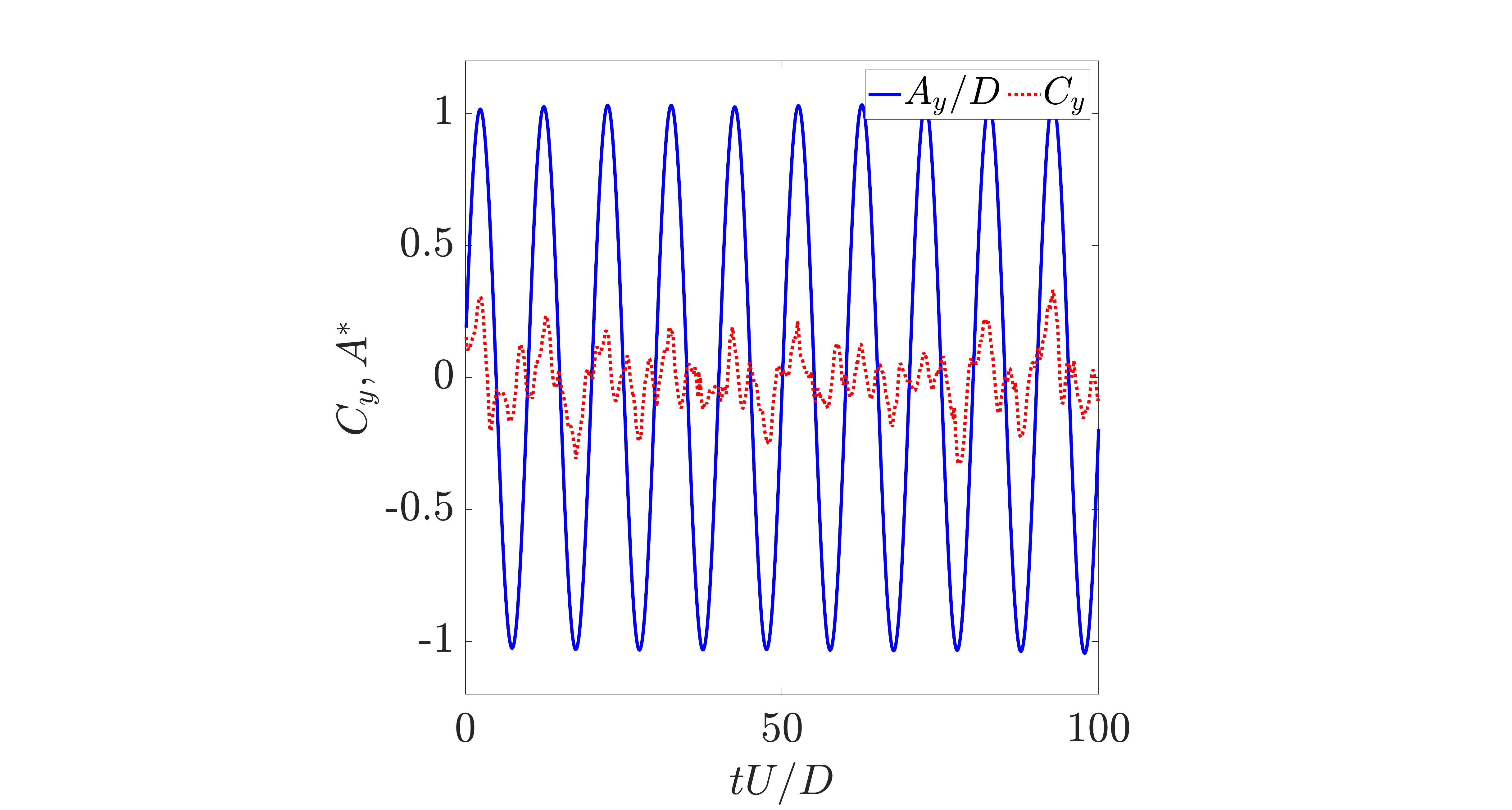}
		\caption*{($b$ - $1$)}
	\end{subfigure}%
	\begin{subfigure}[b]{0.5\textwidth}
		\adjincludegraphics[scale=0.28,trim={0.2\width} {0\width} {0.1\width} {0.0\width},clip]{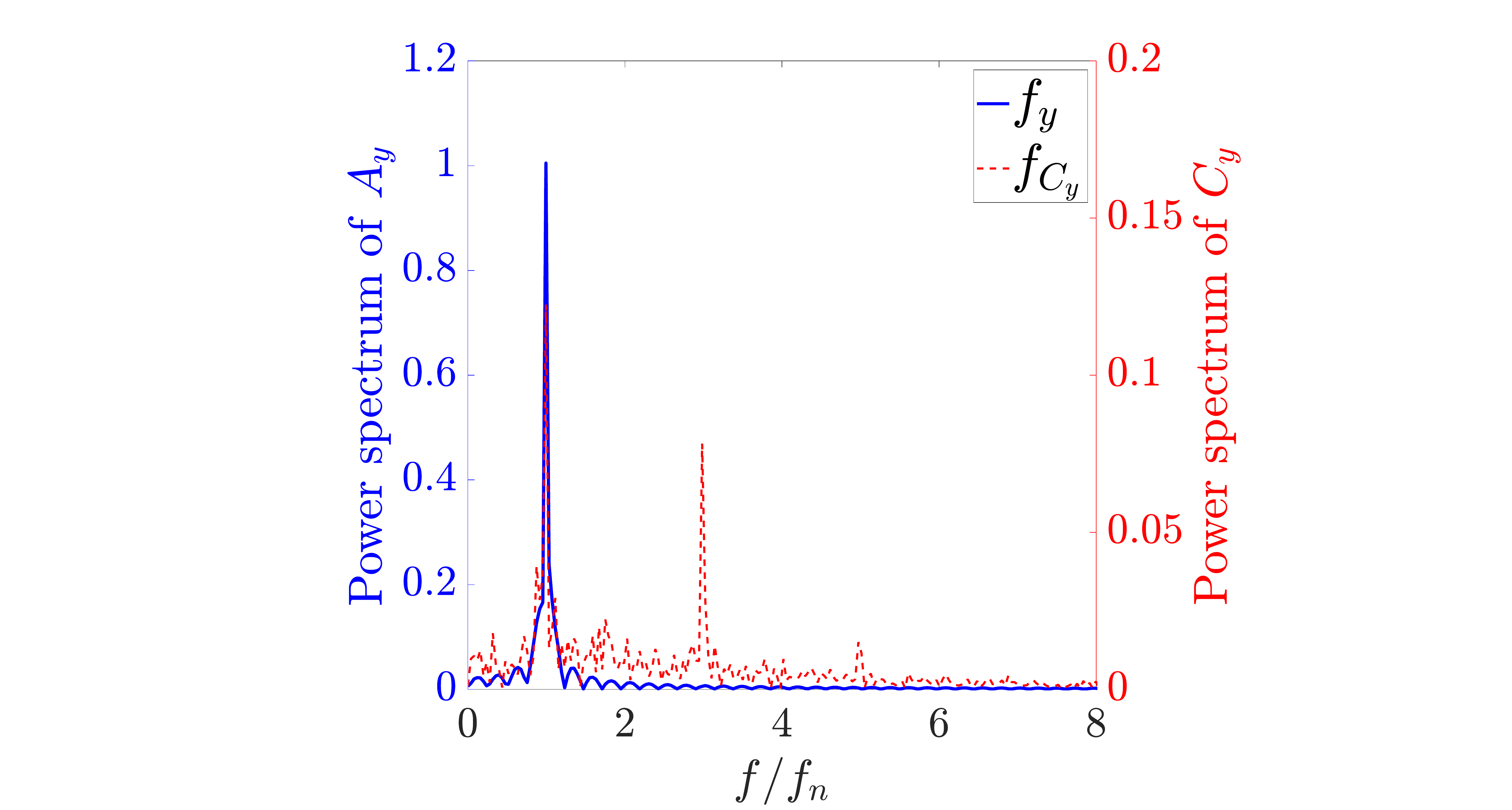}
		\caption*{($b$ - $2$)}
	\end{subfigure}
	
	
	\caption{Time histories of the amplitude response ($A^*$) and the normalized transverse force (${C_y}$)  and their corresponding frequency spectra for the  piercing sphere at $h^*=-0.25$ at two representative mass ratios: (a) $m^*=1$ and (b) $m^*=20$. }
	\label{TH_MS} 
	
\end{figure}

In Fig. \ref{MS_Trend} (c), it is quantified that the vibrations for all the mass ratio range correspond to mode I. The oscillation frequency of the piercing sphere for the entire range of the mass ratio at $U^*=10$ is close to the natural frequency of the system, consistent with the observation for the submerged sphere in \cite{govardhan2005vortex}. 
Fig. \ref{TH_MS} shows the time traces of the amplitude response and the normalized transverse force with their corresponding power spectrum for the piercing case at $h^*=-0.25$ with $m^*=1$ and $m^*=20$. The existence of the third-harmonic behavior of the transverse force is observed for all the mass ratio cases. It can be deduced that the FIV response at $h^*=-0.25$ is relatively insensitive to mass ratio in the range $m^*\in[1,20]$, where the free-surface effect sustains large amplitude oscillations.

\subsection{Effect of Froude Number}
\begin{figure}[htbp!]
	
	\begin{subfigure}[b]{0.5\textwidth}
		\centering
		\adjincludegraphics[scale=0.17,trim={0\width} {0\width} {0\width} {0.0\width},clip]{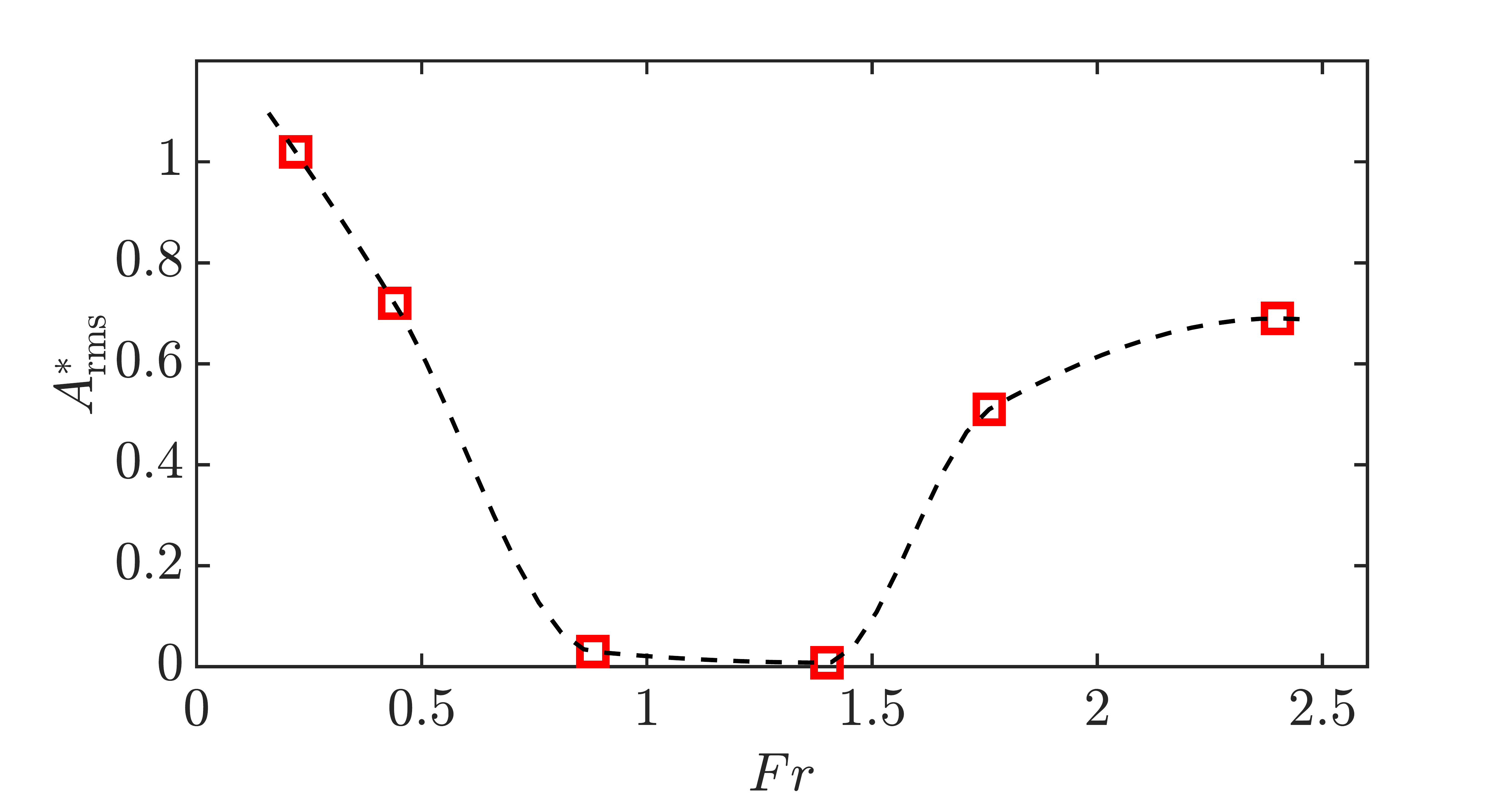}
	\end{subfigure}%
	\begin{subfigure}[b]{0.5\textwidth}
		\centering
		\adjincludegraphics[scale=0.17,trim={0\width} {0\width} {0\width} {0.0\width},clip]{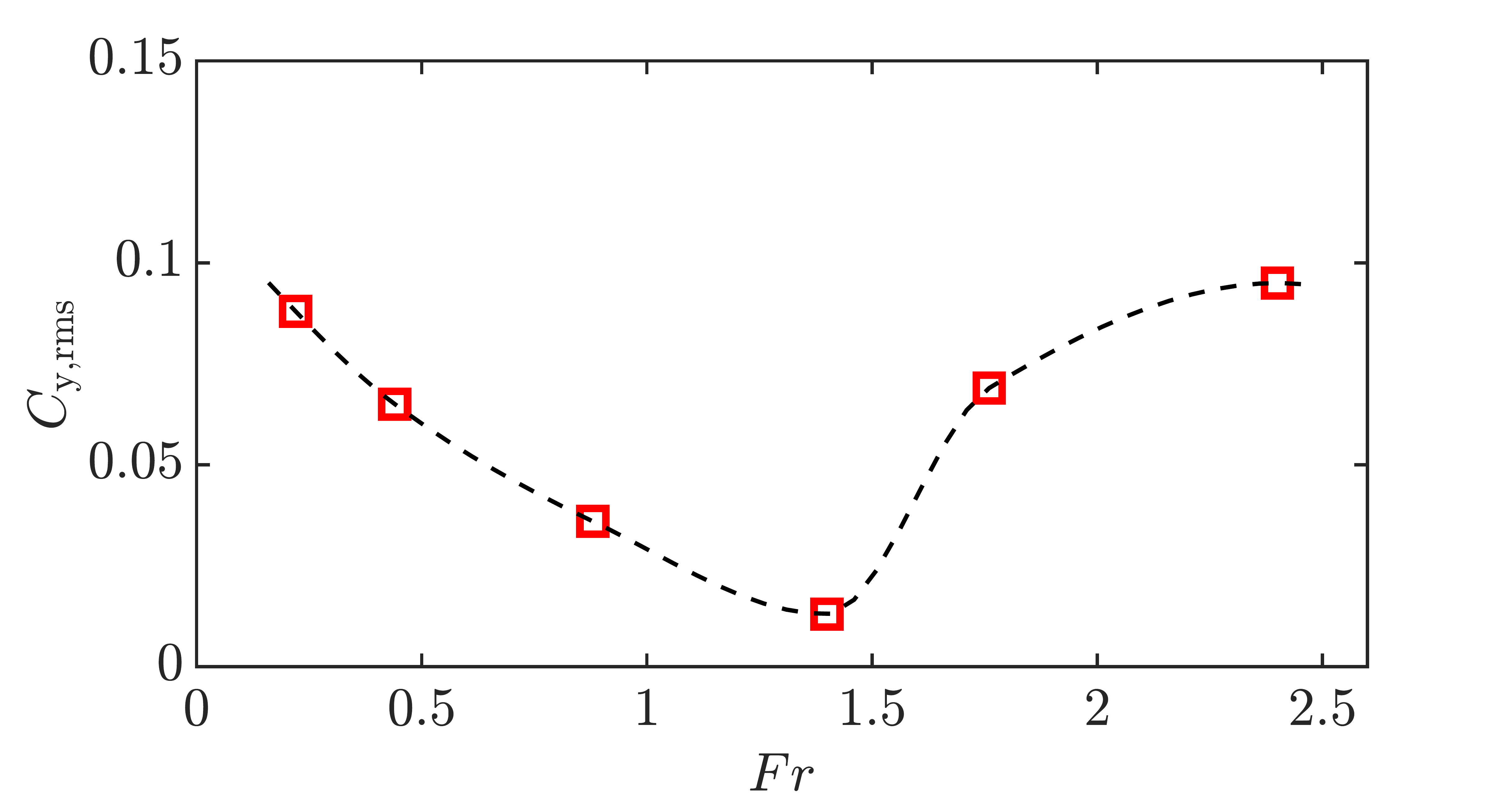}
	\end{subfigure}
	
	\vspace{2mm}
	
	\begin{subfigure}[b]{0.5\textwidth}
		\centering
		\adjincludegraphics[scale=0.17,trim={0\width} {0\width} {0\width} {0.0\width},clip]{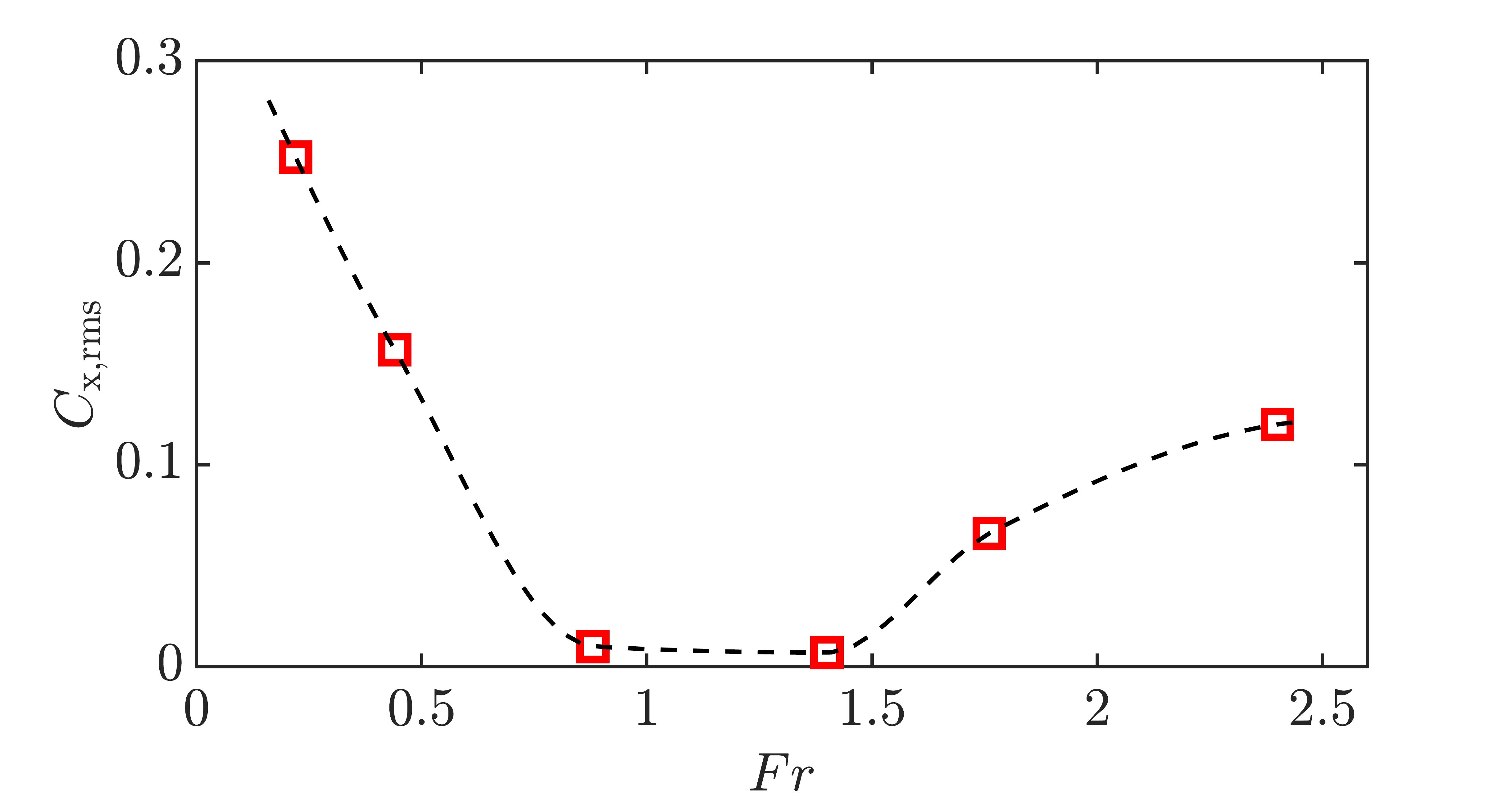}
	\end{subfigure}%
	\begin{subfigure}[b]{0.5\textwidth}
		\centering
		\adjincludegraphics[scale=0.17,trim={0\width} {0\width} {0\width} {0.0\width},clip]{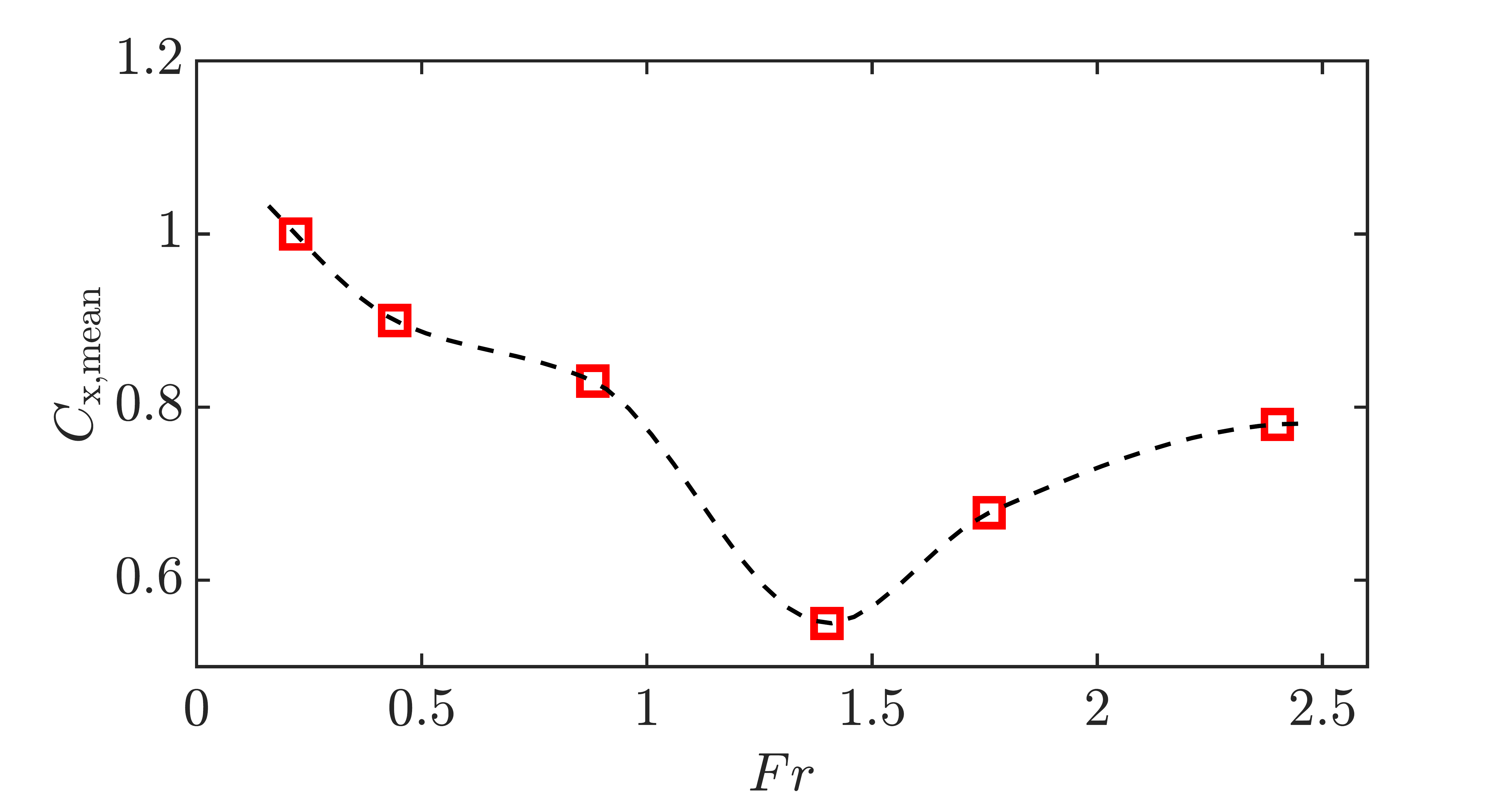}
	\end{subfigure}

	\caption{Variations of r.m.s. amplitude response, r.m.s. normalized force in $y$- and $x$-directions and normalized mean streamwise force as a function of Froude number for the  piercing the  sphere at $h^*=-0.25$, $U^*=10$ and $Re=15\,700$.}
	\label{Fr_Trend}		
\end{figure}

\begin{figure}[htbp!]
	\centering
	
	\begin{subfigure}[b]{0.5\textwidth}
		\adjincludegraphics[scale=0.28,trim={0.2\width} {0\width} {0.1\width} {0.0\width},clip]{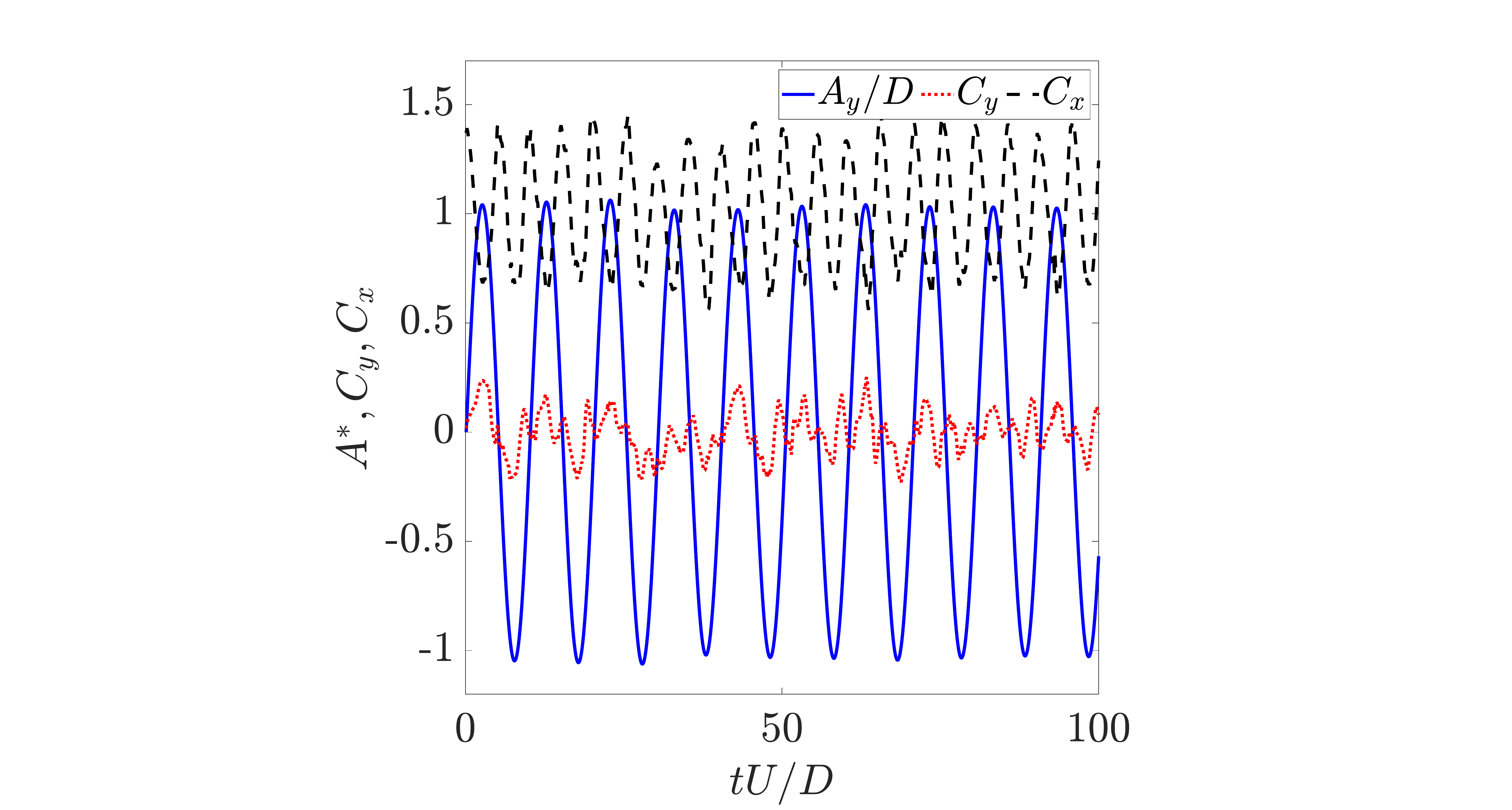}
		\caption*{($a$ - $1$)}
	\end{subfigure}%
	\begin{subfigure}[b]{0.5\textwidth}
		\adjincludegraphics[scale=0.28,trim={0.2\width} {0\width} {0.1\width} {0.0\width},clip]{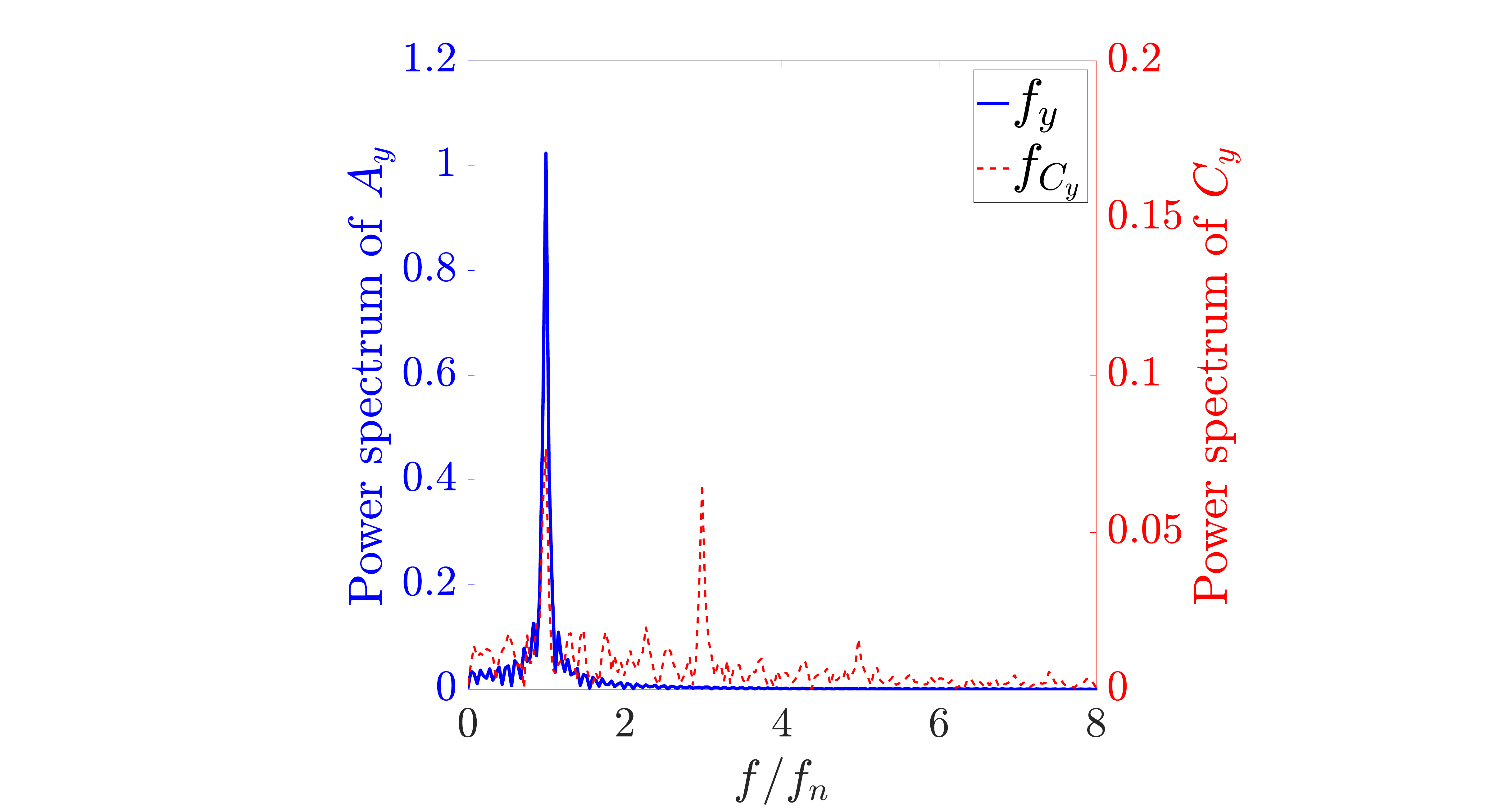}
		\caption*{($a$ - $2$)}
	\end{subfigure}
	
	\begin{subfigure}[b]{0.5\textwidth}
		\adjincludegraphics[scale=0.28,trim={0.2\width} {0\width} {0.1\width} {0.0\width},clip]{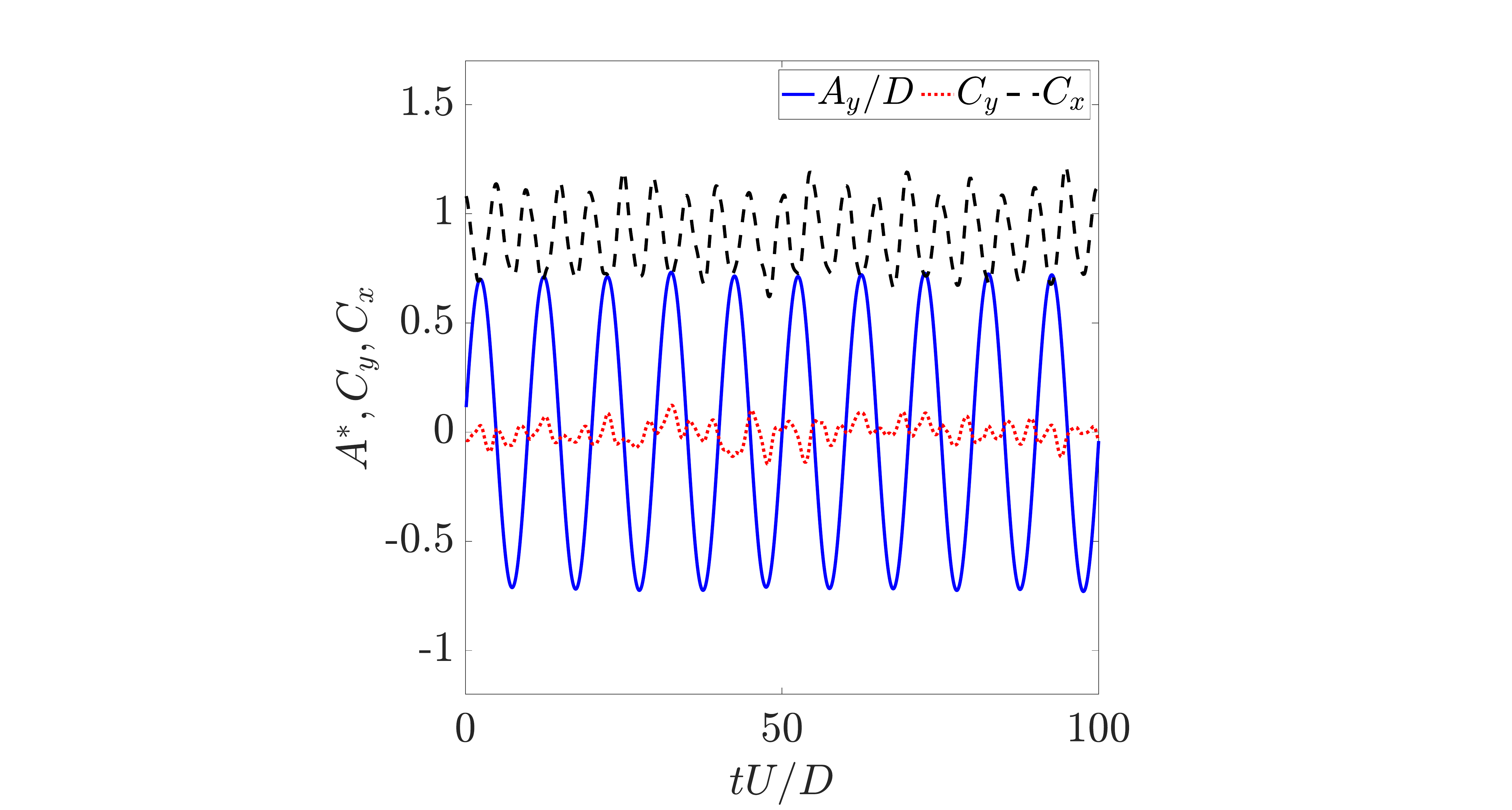}
		\caption*{($b$ - $1$)}
	\end{subfigure}%
	\begin{subfigure}[b]{0.5\textwidth}
		\adjincludegraphics[scale=0.28,trim={0.2\width} {0\width} {0.1\width} {0.0\width},clip]{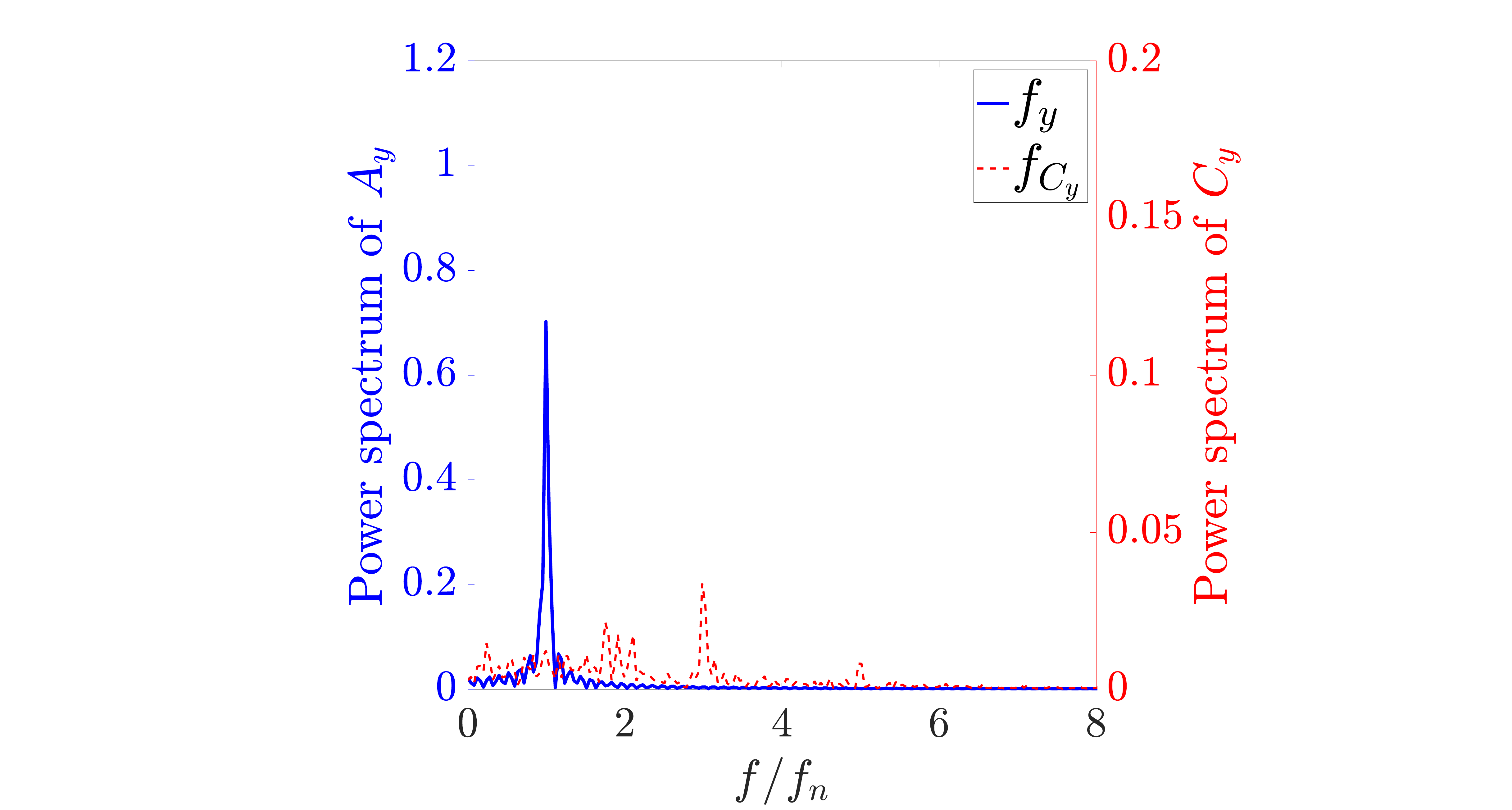}
		\caption*{($b$ - $2$)}
	\end{subfigure}

	\caption{Time histories of the normalized amplitude response ($A^*$) and transverse forces (${C_y}$) and their corresponding frequency spectra for the piercing sphere ($h^*=-0.25$) at two representatives Froude numbers: (a) $Fr=0.22$, and (b) $Fr=0.44$.  Streamwise forces ($C_x$) are also included in ($a$ - $1$) and  ($b$ - $1$) } 
	\label{TH_Fr} 

\end{figure}

To investigate the effect of the free-surface deformation on the transverse VIV response, we next explore the influence of Froude number $Fr$.  All the previous simulations, for validation purpose (Fig. \ref{Validation_Trend}),  are carried out with the range of Froude numbers $Fr\in[0.05,0.45]$ with the experiments performed in \cite{sareen2018}. In the experimental study in \cite{sareen2018}, relatively insensitive effect of the Froude number on the VIV response of the sphere is reported in the range of $Fr\in [0.05,0.45]$ and $U^*\in [3,20]$.  We further investigate the effect of the Froude number on the FIV response at the lock-in state for the piercing sphere case at $h^*=-0.25$, $U^*=10$ and $Re=15\,700$, with the mass ratio $m^*=9.2$. The Froude number is investigated over a range of $Fr\in[0.22,2.4]$  by changing the acceleration due to gravity. 
Fig. \ref{Fr_Trend} shows the variation of the amplitude response $A^*$, the r.m.s. transverse force $C_y$, the r.m.s. and mean streamwise force as a function of Froude number. The results indicate a significant effect of the Froude number on the VIV response. To further analyze the effect of the Froude number, Fig. \ref{TH_Fr} shows the time histories of the amplitude response for the piercing sphere case at $Fr=0.22$ and $Fr=0.44$, where no significant surface deformation is expected \cite{sareen2018}. The r.m.s. amplitude response decreases by about $\sim 30\%$ by doubling the Froude number from $Fr=0.22$ to $Fr=0.44$ at identical reduced velocity $U^*=10$ and Reynolds number $Re=15\,700$.

\begin{figure}[htbp!]
	\centering
	\begin{subfigure}[b]{0.5\textwidth}
		\centering
		\adjincludegraphics[scale=0.3,trim={0\width} {0.3\width} {0\width} {0.22\width},clip]{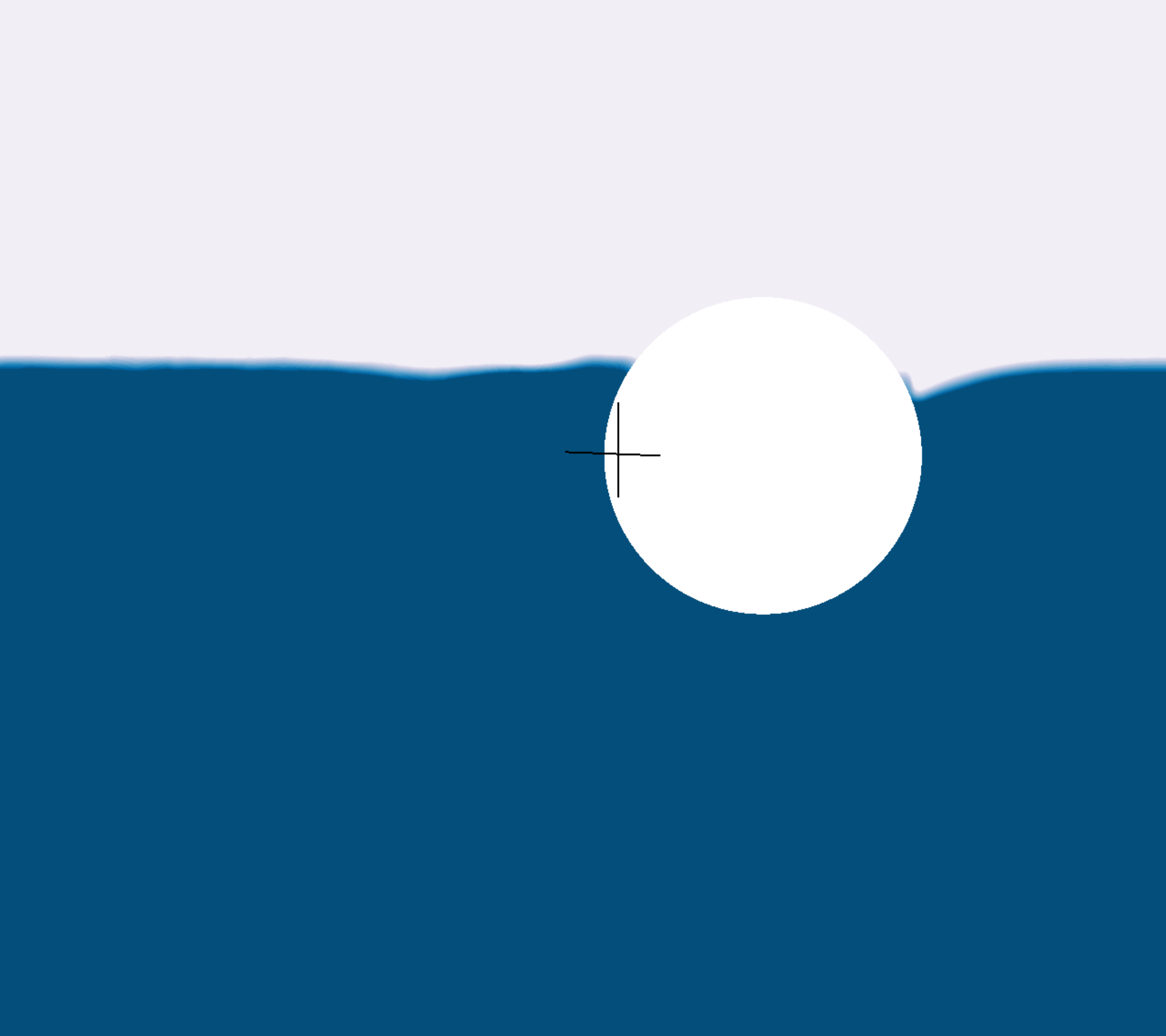}
		\caption*{$(a-1)$}
	\end{subfigure}%
	\begin{subfigure}[b]{0.5\textwidth}
		\centering
		\adjincludegraphics[scale=0.3,trim={0\width} {0.3\width} {0\width} {0.22\width},clip]{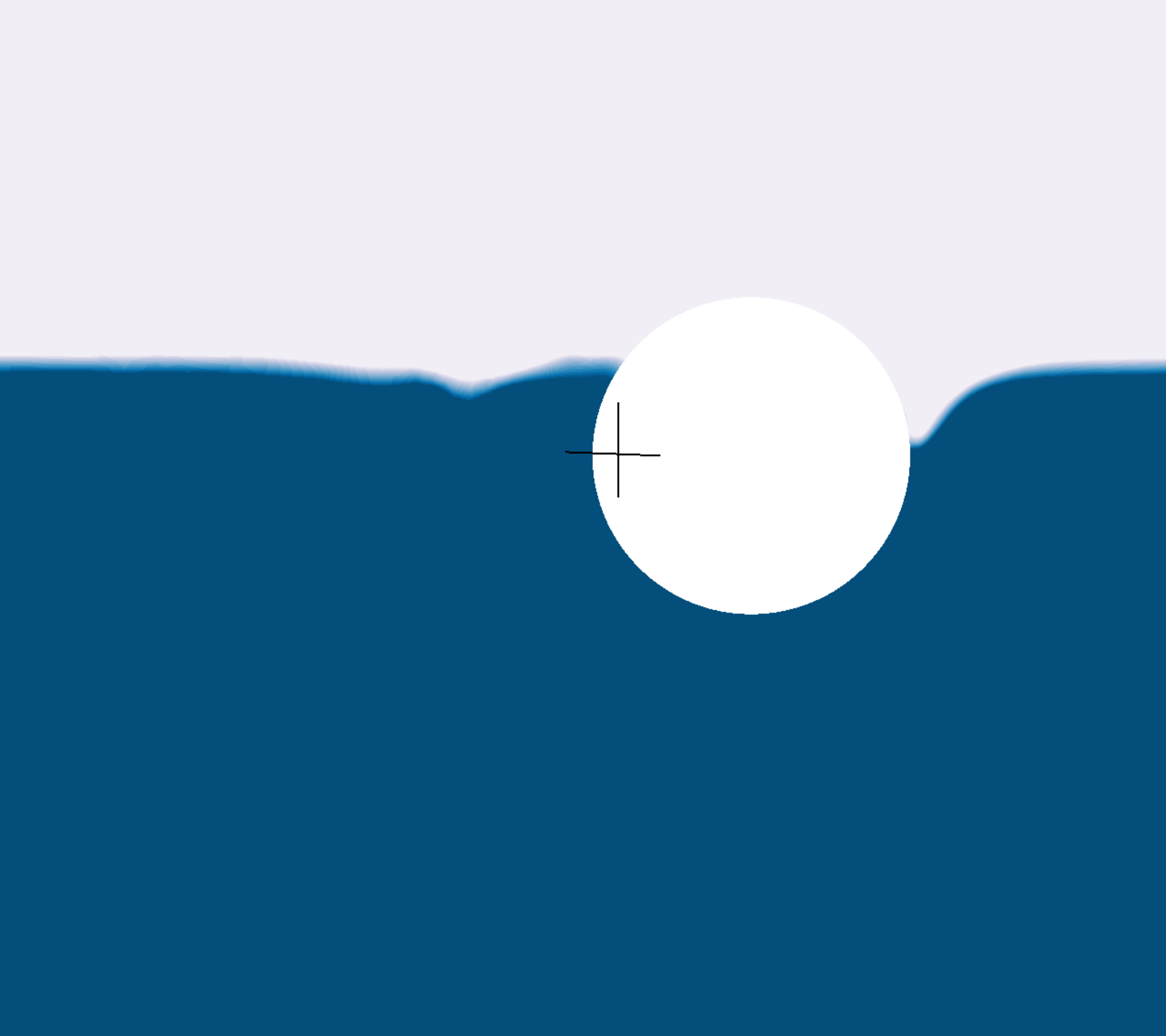}
		\caption*{$(b-1)$}
	\end{subfigure}
	
	\begin{subfigure}[b]{0.5\textwidth}
		\centering
		\adjincludegraphics[scale=0.3,trim={0\width} {0.3\width} {0\width} {0.22\width},clip]{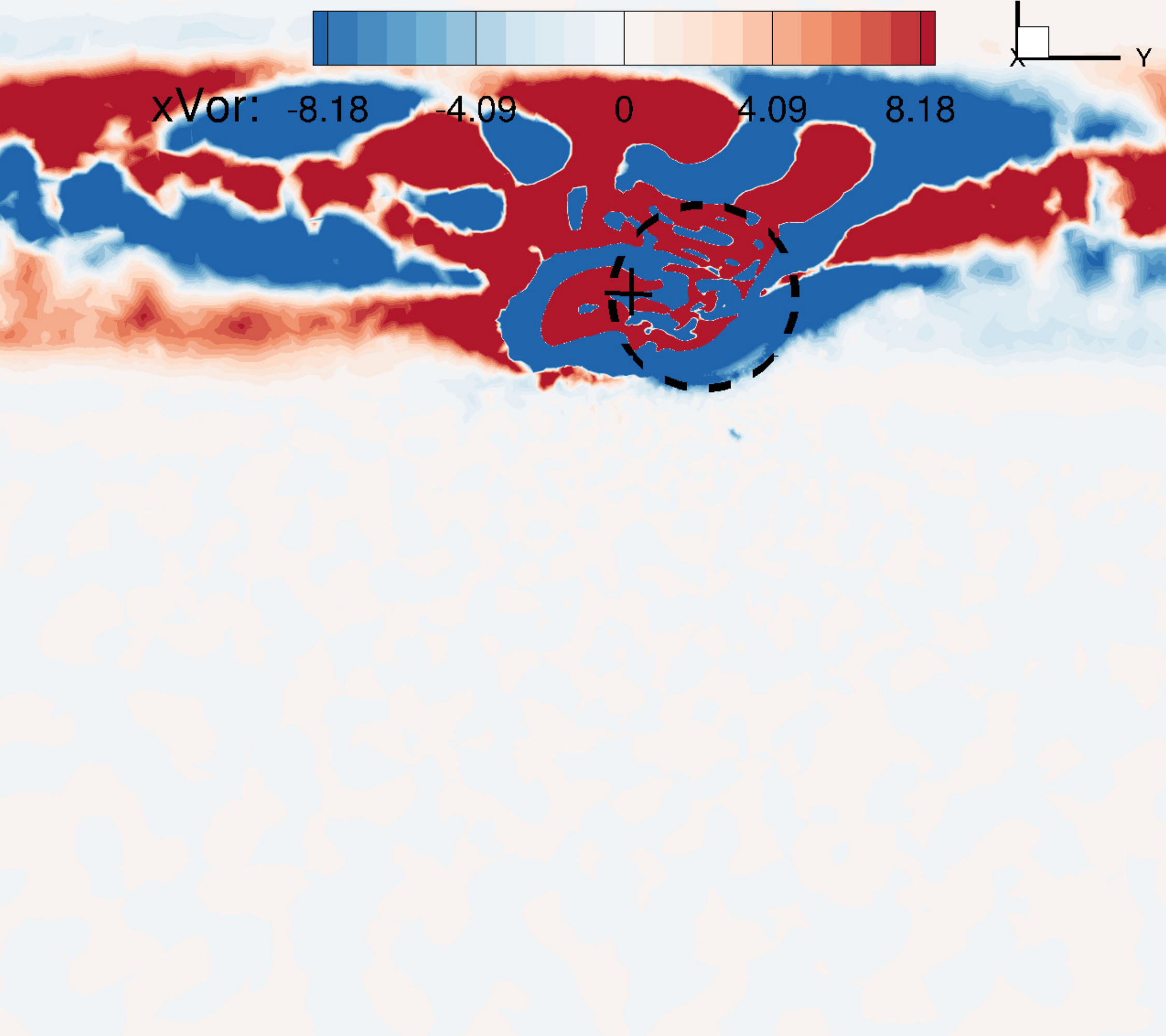}
		\caption*{$(a-2)$}
	\end{subfigure}%
	\begin{subfigure}[b]{0.5\textwidth}
		\centering
		\adjincludegraphics[scale=0.3,trim={0\width} {0.3\width} {0\width} {0.22\width},clip]{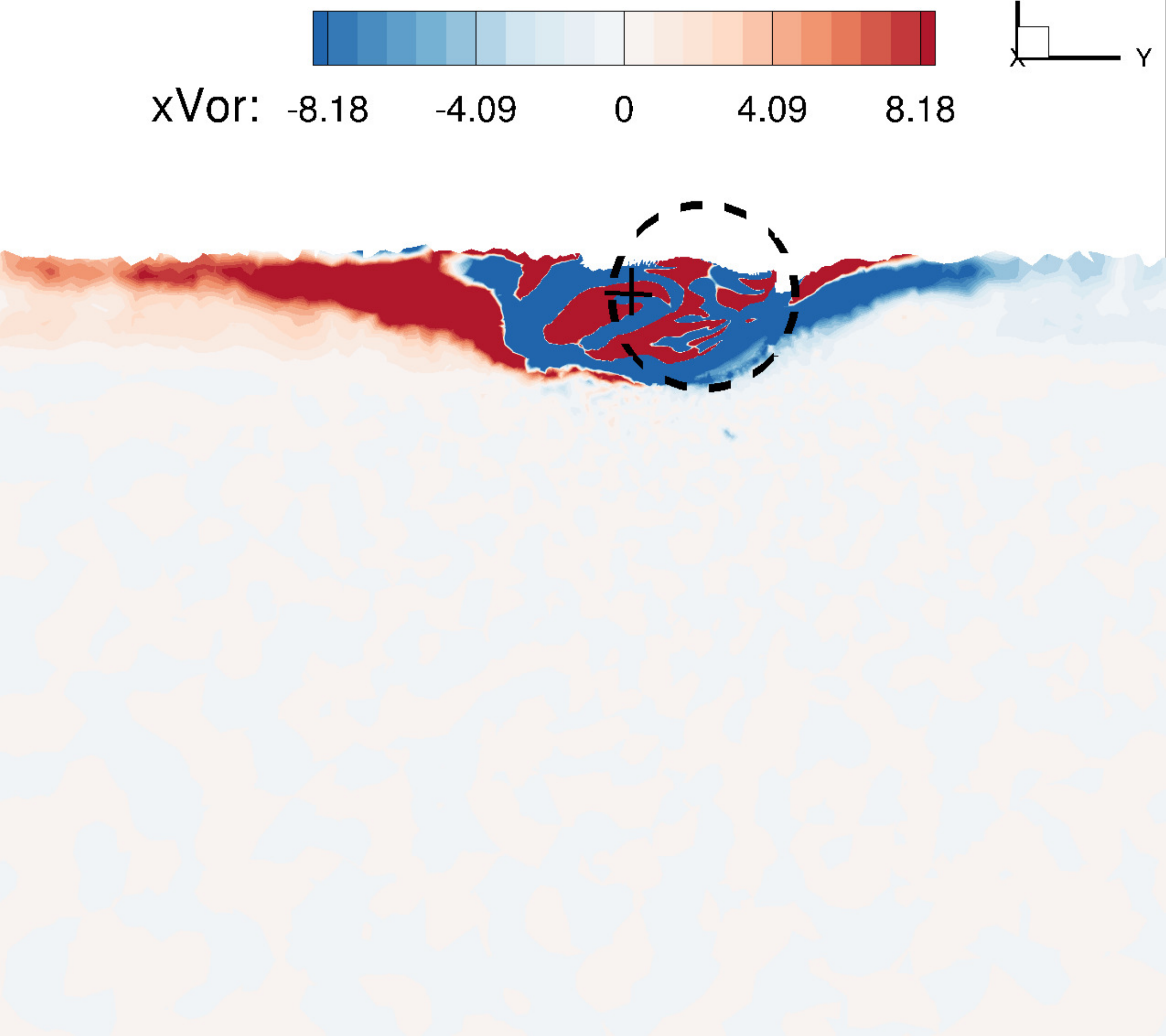}
		\caption*{$(b-2)$}
	\end{subfigure}

	\begin{subfigure}[b]{0.5\textwidth}
		\centering
		\adjincludegraphics[scale=0.3,trim={0\width} {0.3\width} {0\width} {0.22\width},clip]{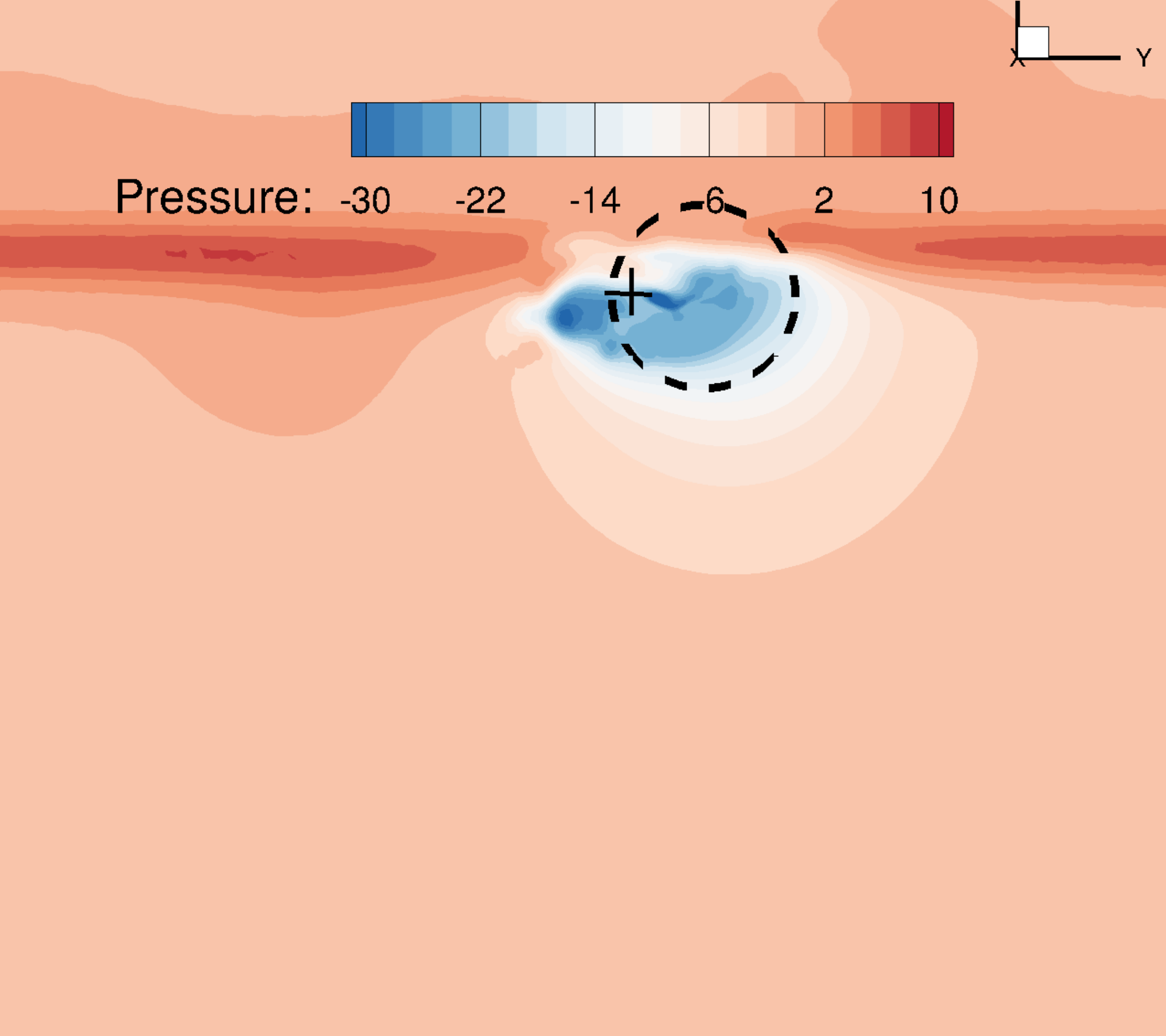}
		\caption*{$(a-3)$}
	\end{subfigure}%
	\begin{subfigure}[b]{0.5\textwidth}
		\centering
		\adjincludegraphics[scale=0.3,trim={0\width} {0.3\width} {0\width} {0.22\width},clip]{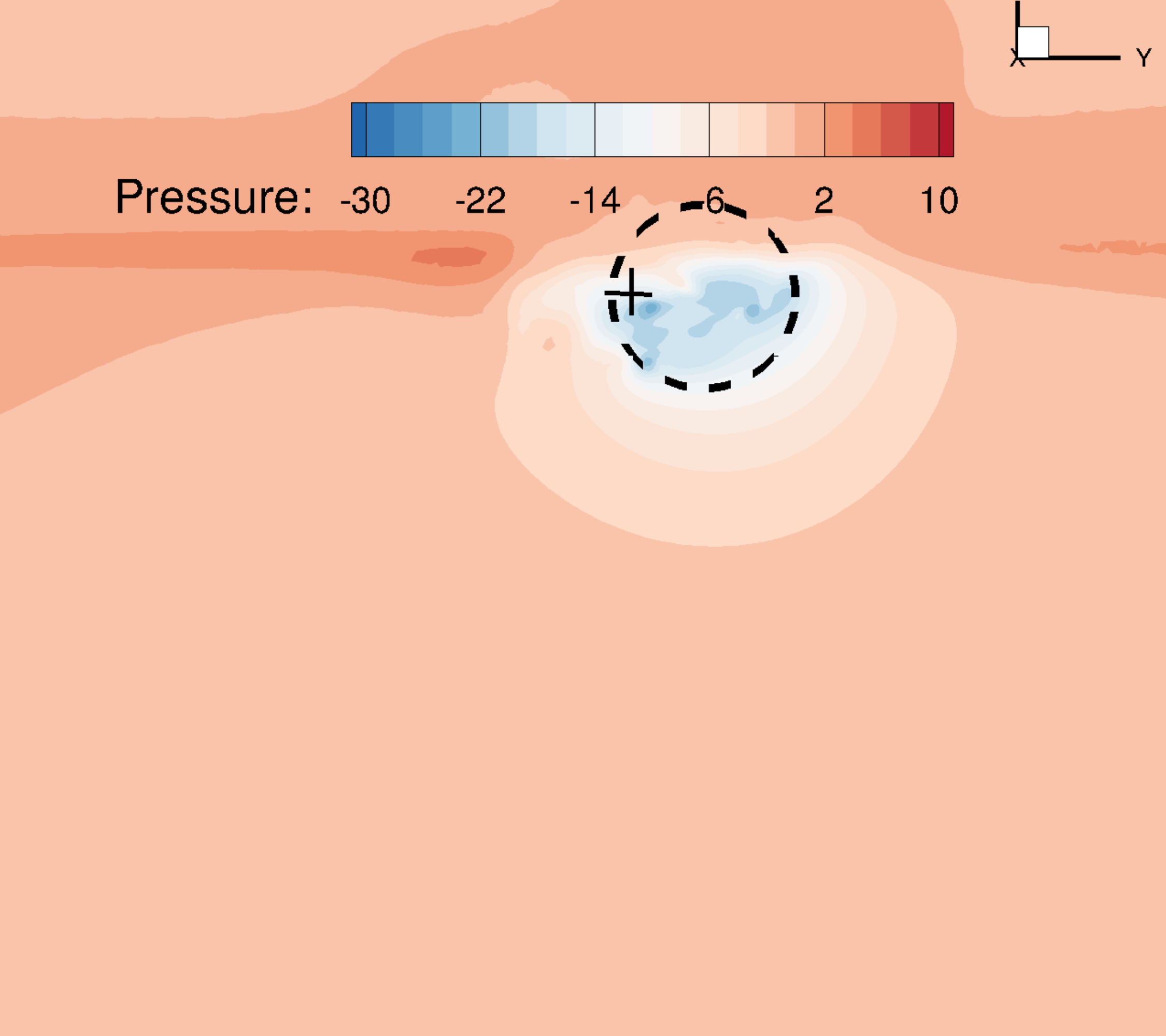}
		\caption*{$(b-3)$}
	\end{subfigure}

	\begin{subfigure}[b]{0.5\textwidth}
		\centering
		\adjincludegraphics[scale=0.4,trim={0.05\width} {0.7\width} {0.16\width} {0.0\width},clip]{Photos/38.pdf}
		\caption*{}
	\end{subfigure}%
	\begin{subfigure}[b]{0.5\textwidth}
		\centering
		\adjincludegraphics[scale=0.4,trim={0.05\width} {0.7\width} {0.01\width} {0.0\width},clip]{Photos/78.pdf}
		\caption*{}
	\end{subfigure}
	
	\vspace{-0.5cm}
	
	\caption{Flow visualizations for the piercing sphere case at $h^*=-0.25$ and $U^*=10$ for two representative Froude numbers (a) $Fr=0.22$ and, (b) $Fr=0.44$: Free-surface deformation quantified with the order parameter ($\phi$) at $\frac{1}{4}D$ downstream (top), normalized streamwise $x$-vorticity (middle) and pressure distribution (bottom) contours plotted at $0.5D$ downstream.} 
	\label{xVor_FS_Ur10_Compare_Fr_2} 
\end{figure}

Fig. \ref{TH_Fr} (b - 2) shows the frequency spectrum at $Fr=0.44$ compared to Fig. \ref{TH_Fr} (a - 2) at $Fr=0.22$. The third harmonic frequency at $Fr=0.44$ is found to be the only dominant force frequency on the sphere. This higher harmonic behavior is expected due to the free-surface effects. Despite the case at $Fr=0.22$, the first harmonic frequency at $Fr=0.44$ has almost disappeared. 
To have a better understanding, Fig. \ref{xVor_FS_Ur10_Compare_Fr_2} shows the surface deformation, the normalized vorticity and the pressure distribution plots at $Fr=0.22$ and $Fr=0.44$. From  Fig. \ref{xVor_FS_Ur10_Compare_Fr_2} (a - 1) and (b - 1),  it can be seen that the surface deformation at higher Froude number $Fr=0.44$ is considerably larger than the case at $Fr=0.22$. Through the vorticity plots in Fig. \ref{xVor_FS_Ur10_Compare_Fr_2}, it is found that the vorticity supplied to the wake by the free surface at $Fr=0.22$ is much stronger than $Fr=0.44$. In Fig. \ref{xVor_FS_Ur10_Compare_Fr_2} (a - 2) for the case at $Fr=0.22$, the strong negative sign vorticity (blue vortex loop) at the top left corner, generated due to the free-surface distortion, has completely disappeared for the case at $Fr=0.44$ as shown in Fig. \ref{xVor_FS_Ur10_Compare_Fr_2} (b - 2). Since the free-surface boundary is allowed to deform due to the stress-free condition, the vorticity at the top region for the higher Froude number case causes a large surface deformation and dissipates the energy. It can be deduced that at higher Froude number $Fr=0.44$, the strength of supplied vorticity due to the free surface is reduced significantly. For $Fr=0.22$,  Fig. \ref{xVor_FS_Ur10_Compare_Fr_2} (a - 3) shows the existence of the high-pressure region on the top left corner due to the induced surface curvature and the extra supplied vorticity. In Fig. \ref{xVor_FS_Ur10_Compare_Fr_2} (b - 3) at $Fr=0.44$, it is found that the pressure level on top left region is decreased substantially compared to the case at $Fr=0.22$.



\begin{figure}[htbp!]
	
	\begin{subfigure}[b]{0.5\textwidth}
		\centering
		\adjincludegraphics[scale=0.21,trim={0.01\width} {0.25\width} {0.02\width} {0.0\width},clip]{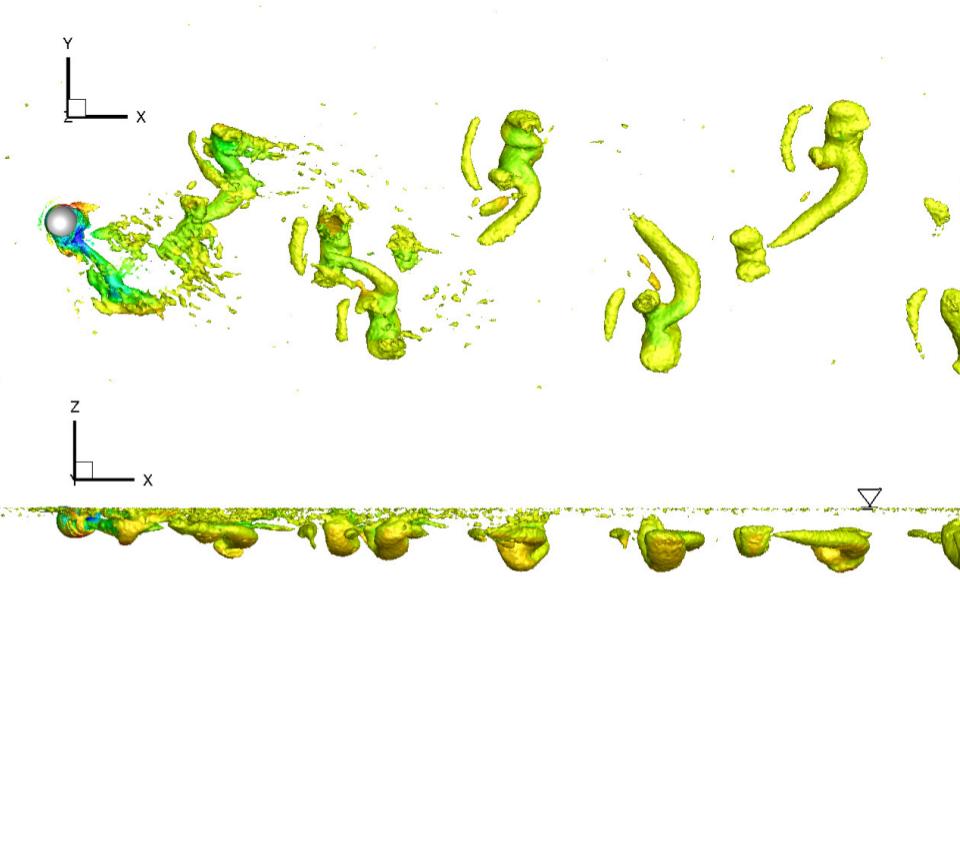}
	\end{subfigure}%
	\begin{subfigure}[b]{0.5\textwidth}
		\centering
		\adjincludegraphics[scale=0.3,trim={0.01\width} {0.25\width} {0.02\width} {0.0\width},clip]{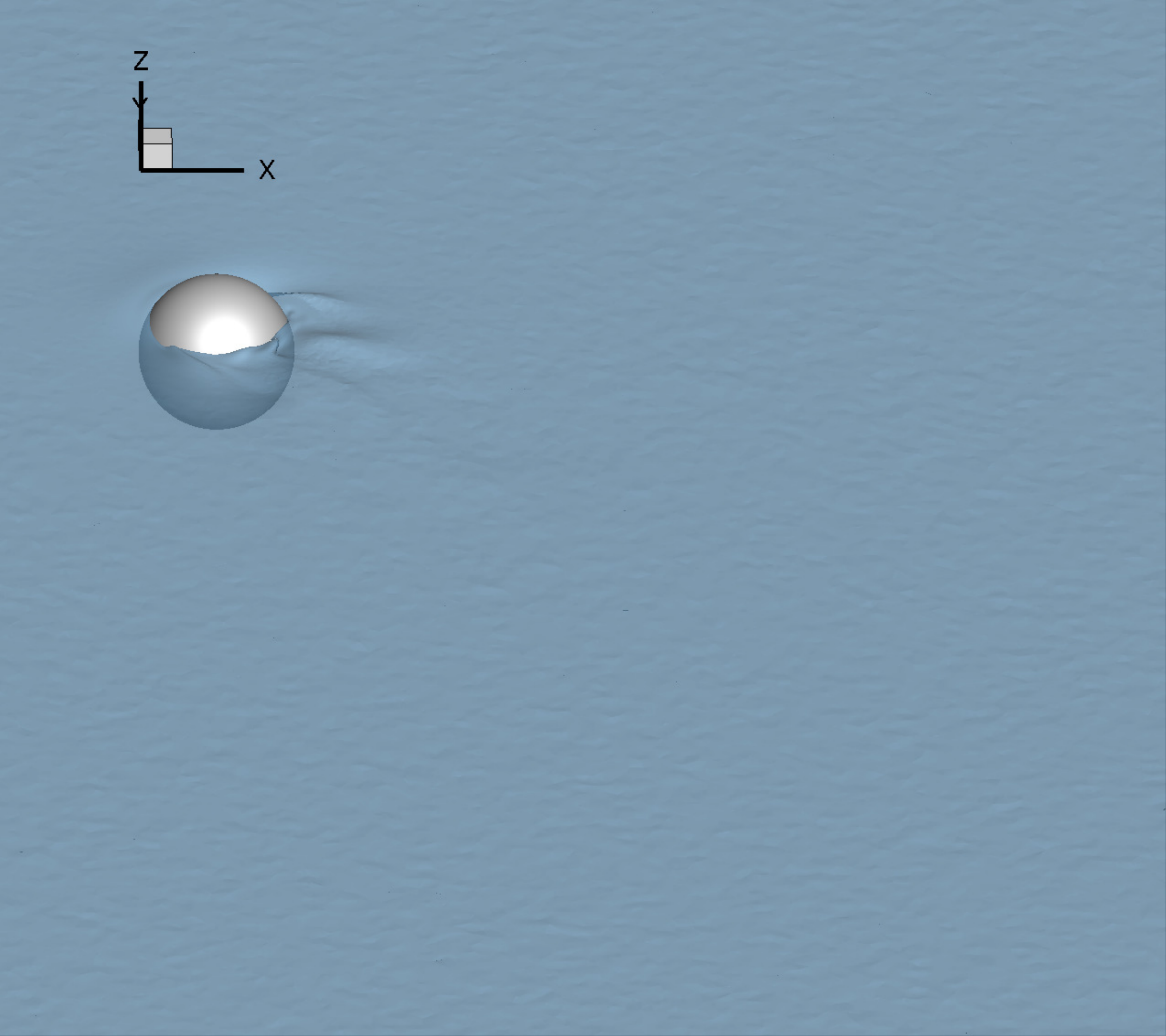}
	\end{subfigure}
	
	\caption*{(a) $Fr=0.22$}

	
	\begin{subfigure}[b]{0.5\textwidth}
		\centering
		\adjincludegraphics[scale=0.21,trim={0.01\width} {0.25\width} {0.02\width} {0.0\width},clip]{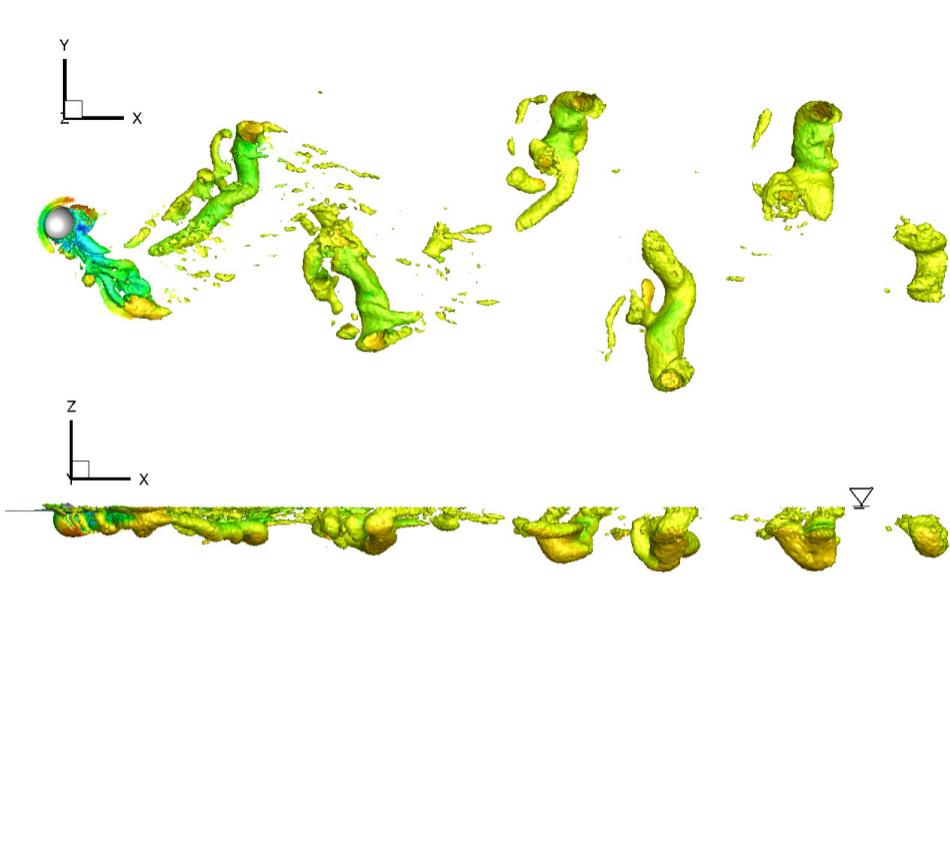}
	\end{subfigure}%
	\begin{subfigure}[b]{0.5\textwidth}
		\centering
		\adjincludegraphics[scale=0.3,trim={0.01\width} {0.25\width} {0.02\width} {0.0\width},clip]{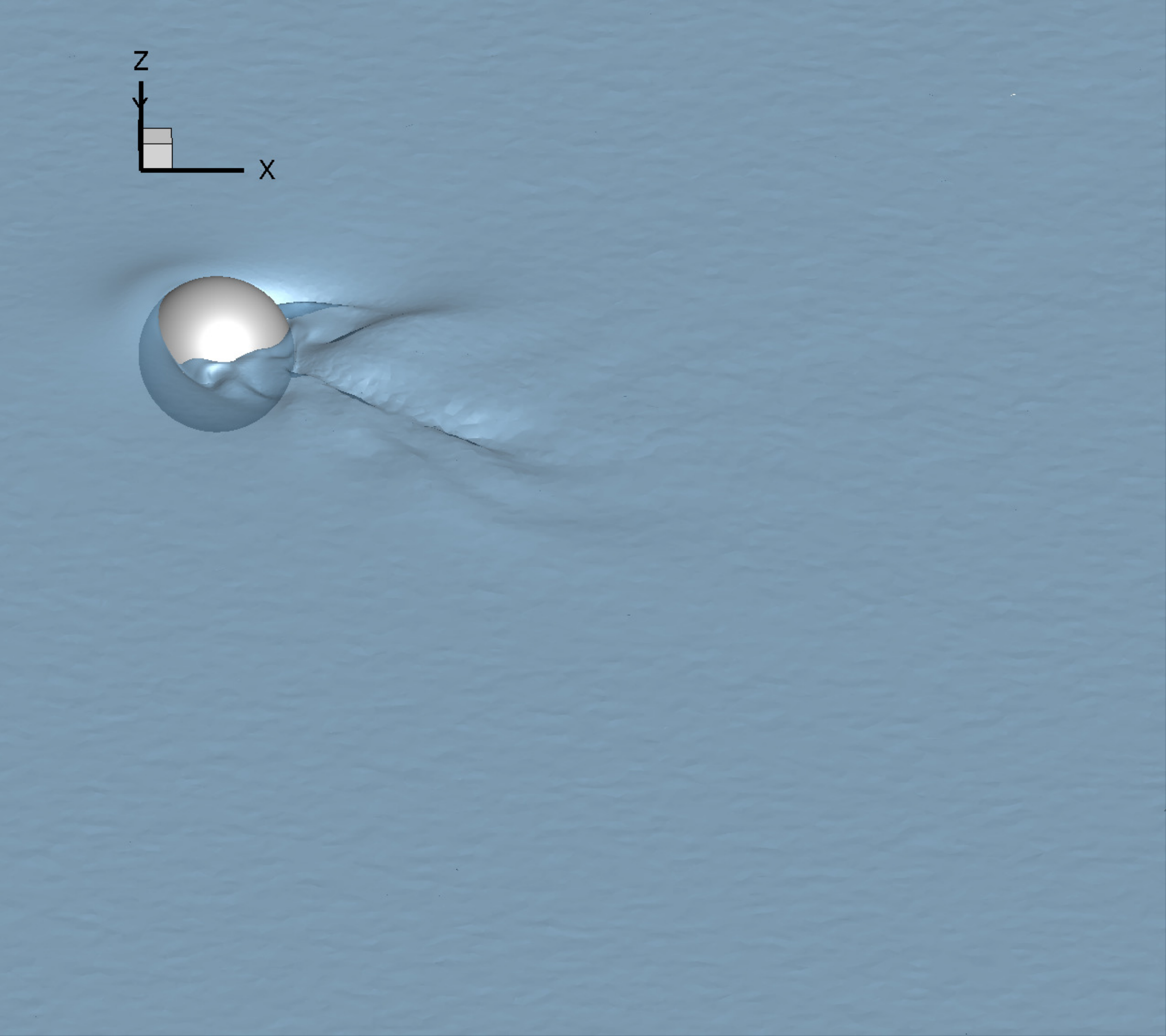}
	\end{subfigure}

	\caption*{(b) $Fr=0.44$}

	
	\begin{subfigure}[b]{0.5\textwidth}
		\centering
		\adjincludegraphics[scale=0.21,trim={0.01\width} {0.25\width} {0.02\width} {0.0\width},clip]{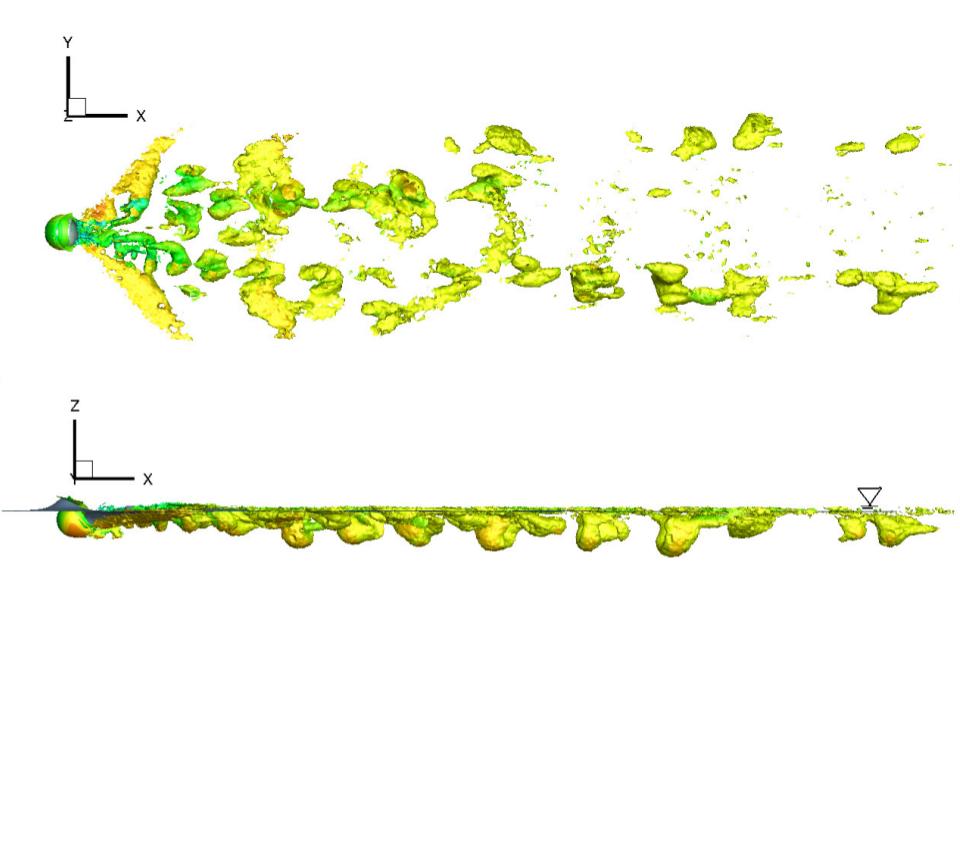}
	\end{subfigure}%
	\begin{subfigure}[b]{0.5\textwidth}
		\centering
		\adjincludegraphics[scale=0.3,trim={0.01\width} {0.25\width} {0.02\width} {0.0\width},clip]{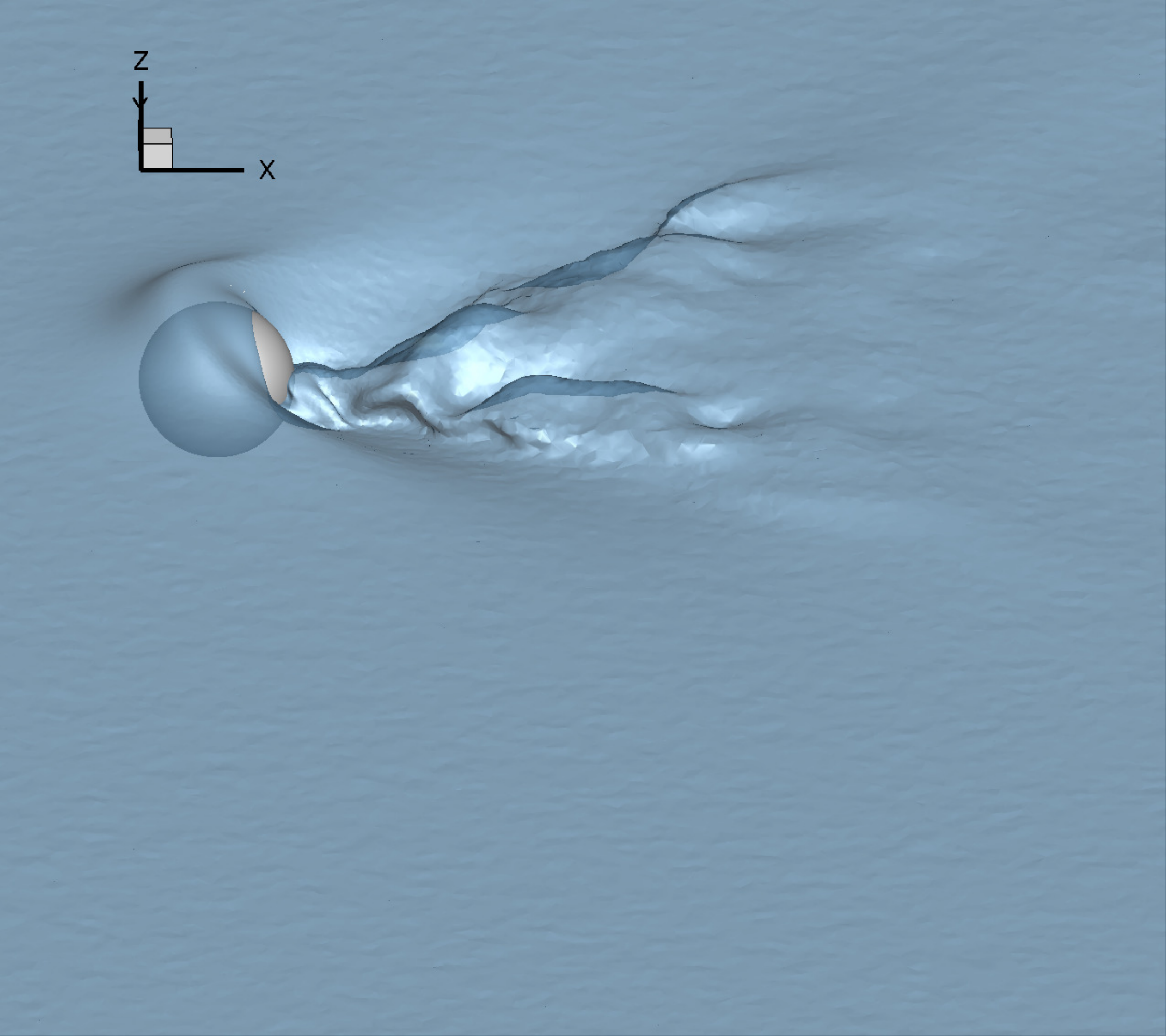}
	\end{subfigure}
	
	\caption*{(c) $Fr=0.88$}
	
	
	\begin{subfigure}[b]{0.5\textwidth}
		\centering
		\adjincludegraphics[scale=0.21,trim={0.01\width} {0.2\width} {0.02\width} {0.0\width},clip]{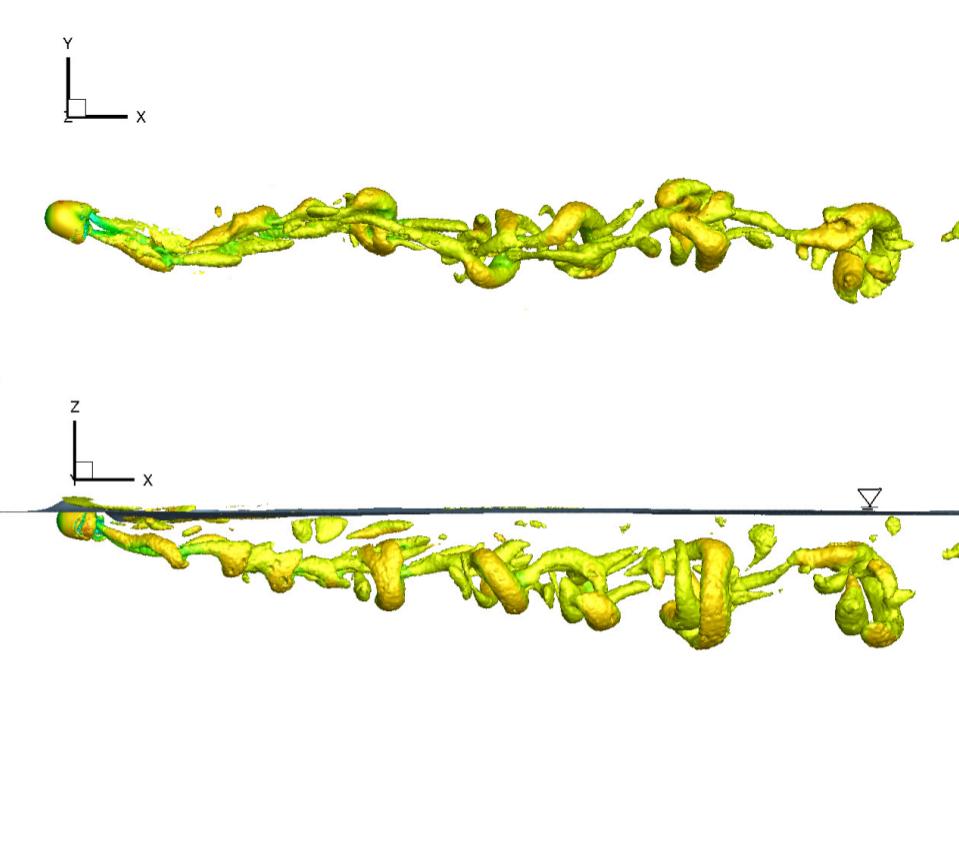}
	\end{subfigure}%
	\begin{subfigure}[b]{0.5\textwidth}
		\centering
		\adjincludegraphics[scale=0.3,trim={0.01\width} {0.2\width} {0.02\width} {0.0\width},clip]{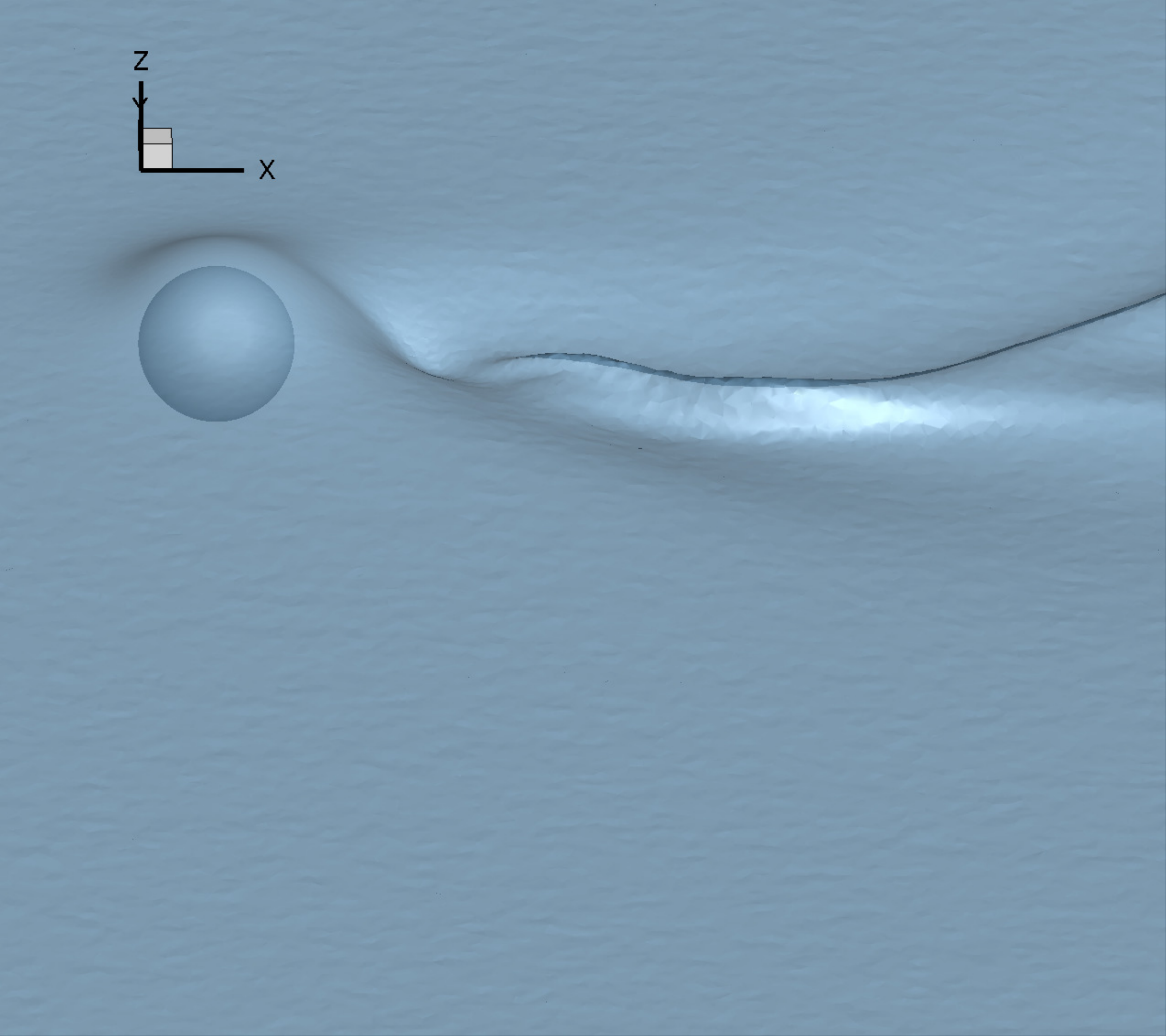}
	\end{subfigure}

	\caption*{(d) $Fr=1.76$}

	\begin{subfigure}[b]{0.5\textwidth}
		\centering
					\hspace{-1.5cm}
					\vspace{-0.2cm}
		\adjincludegraphics[scale=0.22,trim={0.3\width} {0.74\width} {0.16\width} {0.04\width},clip]{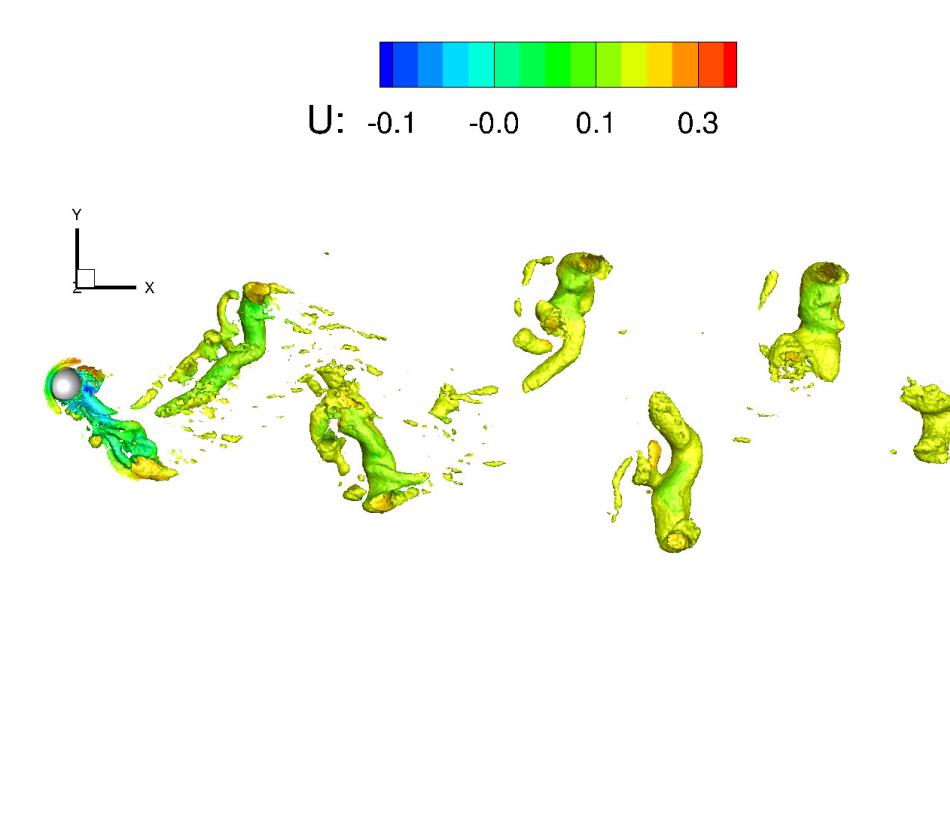}
		\caption*{}
	\end{subfigure}

	\vspace{-0.5cm}

	\caption{Iso-surface of wake structures formed behind the sphere and the free-surface deformation for the piercing sphere at $h^*=-0.25$ for different Froude numbers. Iso-surfaces of the 3D vortical structures are plotted by the Q-criterion ($Q=2$) and the iso-surfaces of the free-surface deformation are plotted by the order parameter ($\phi=0$).}
	\label{Q_Crit_Fr_2}
\end{figure}

Fig. \ref{Q_Crit_Fr_2} shows the three-dimensional wake structures along with the surface deformation at four different Froude numbers. By comparing the cases at $Fr=0.22$ and $Fr=0.44$, where the surface deformation is not substantial, it can be seen that at lower Froude number $Fr=0.22$, the hairpin type structures at near wake region are generated, although the sphere pierces the free surface. The upper vortex loops at downstream flow are detached from the lower loops and lose their strength through diffusion into the free surface. At higher Froude number $Fr=0.44$, the upper vortex loops of the hairpin vortex structures get diffused into the free surface immediately in the near wake region and cause larger surface deformation. The upper vortex loops completely disappear downstream, which results in the reduction of the circulation and corresponding transverse force on the sphere.

\begin{figure}[htbp!]
	\centering
	
	\begin{subfigure}[b]{0.5\textwidth}
		\adjincludegraphics[scale=0.28,trim={0.2\width} {0\width} {0.1\width} {0.0\width},clip]{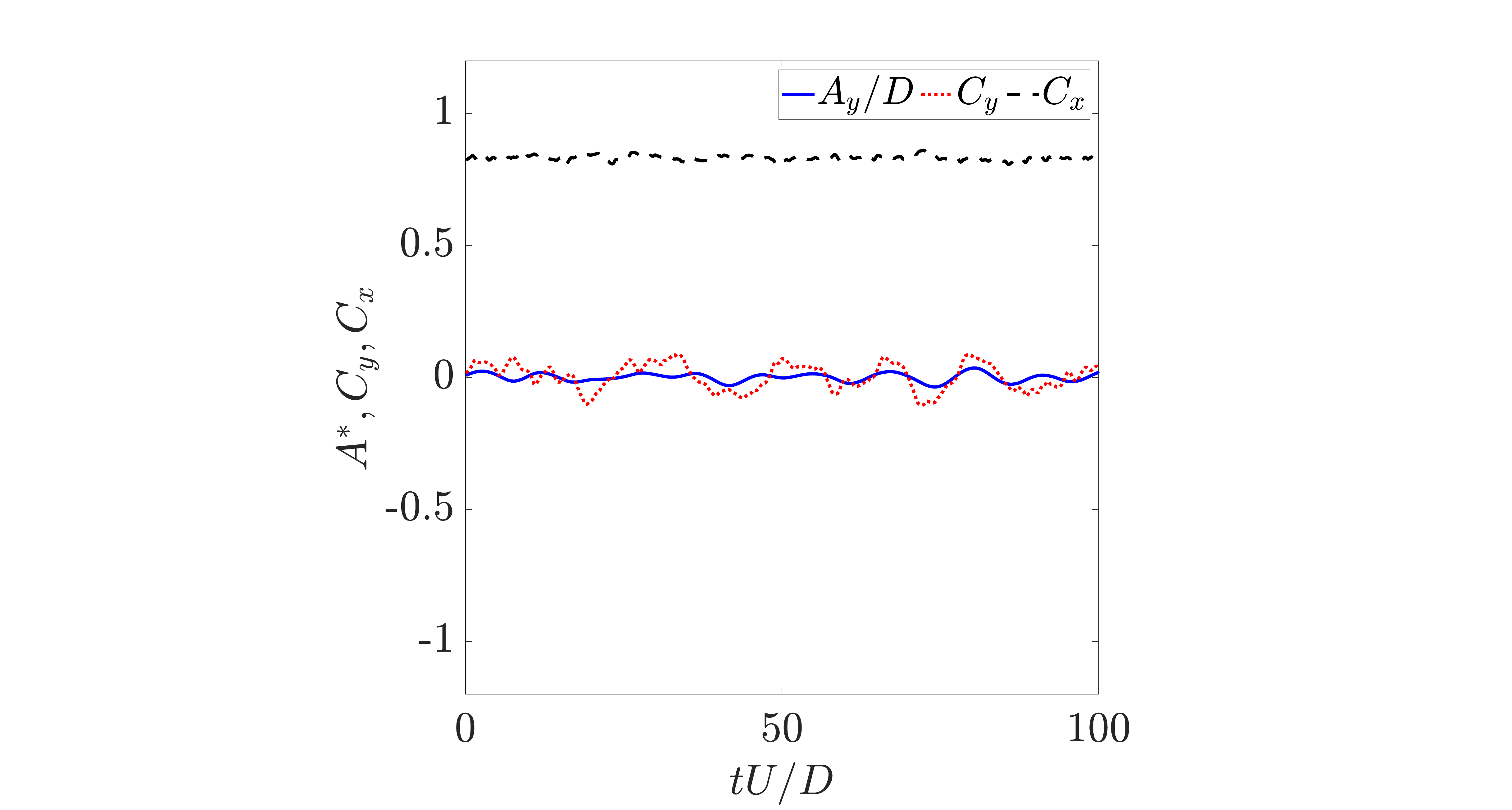}
		\caption*{($a$ - $1$)}
	\end{subfigure}%
	\begin{subfigure}[b]{0.5\textwidth}
		\adjincludegraphics[scale=0.28,trim={0.2\width} {0\width} {0.1\width} {0.0\width},clip]{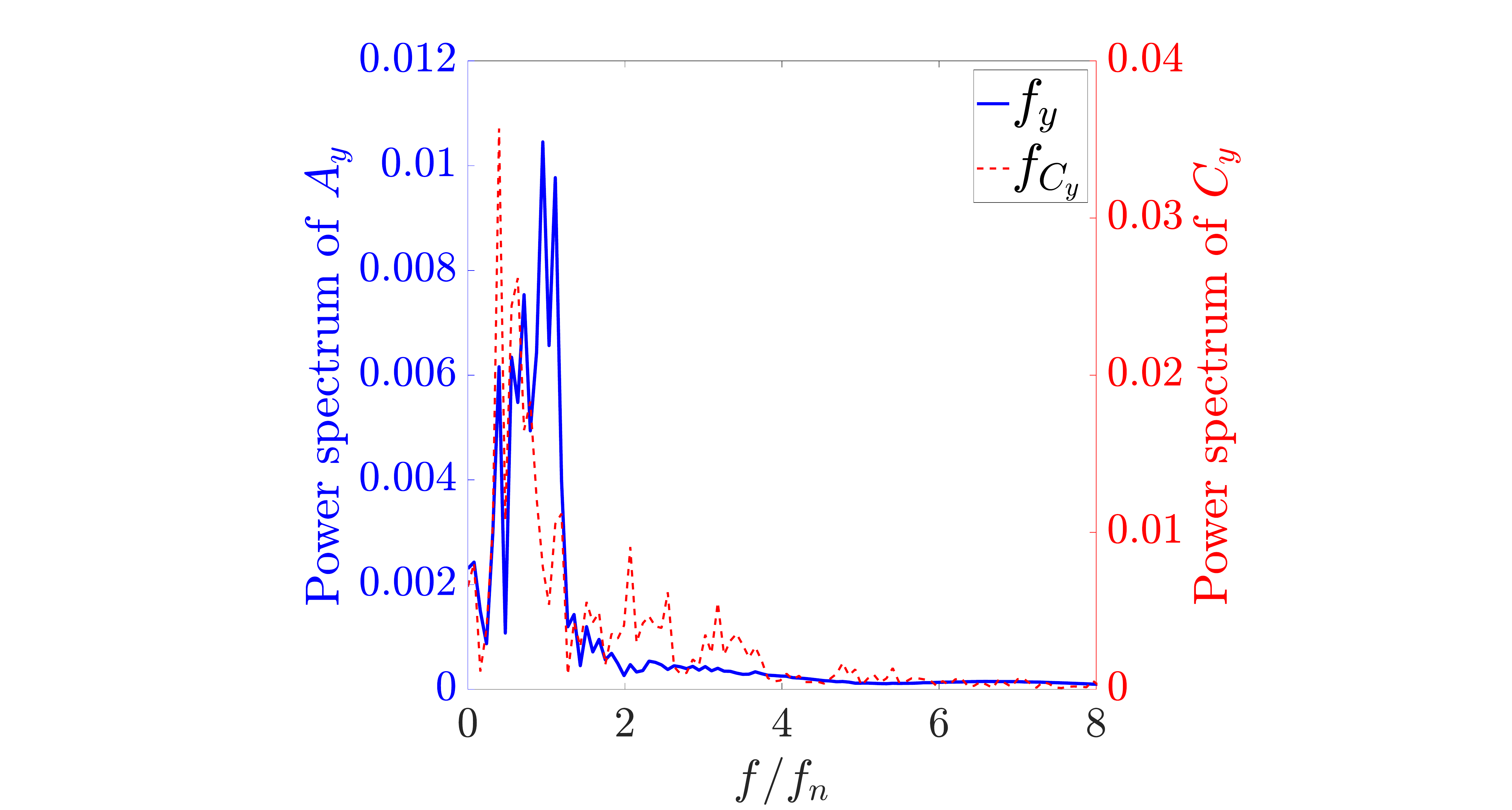}
		\caption*{($a$ - $2$)}
	\end{subfigure}

	\begin{subfigure}[b]{0.5\textwidth}
		\adjincludegraphics[scale=0.27,trim={0.2\width} {0\width} {0.1\width} {0.0\width},clip]{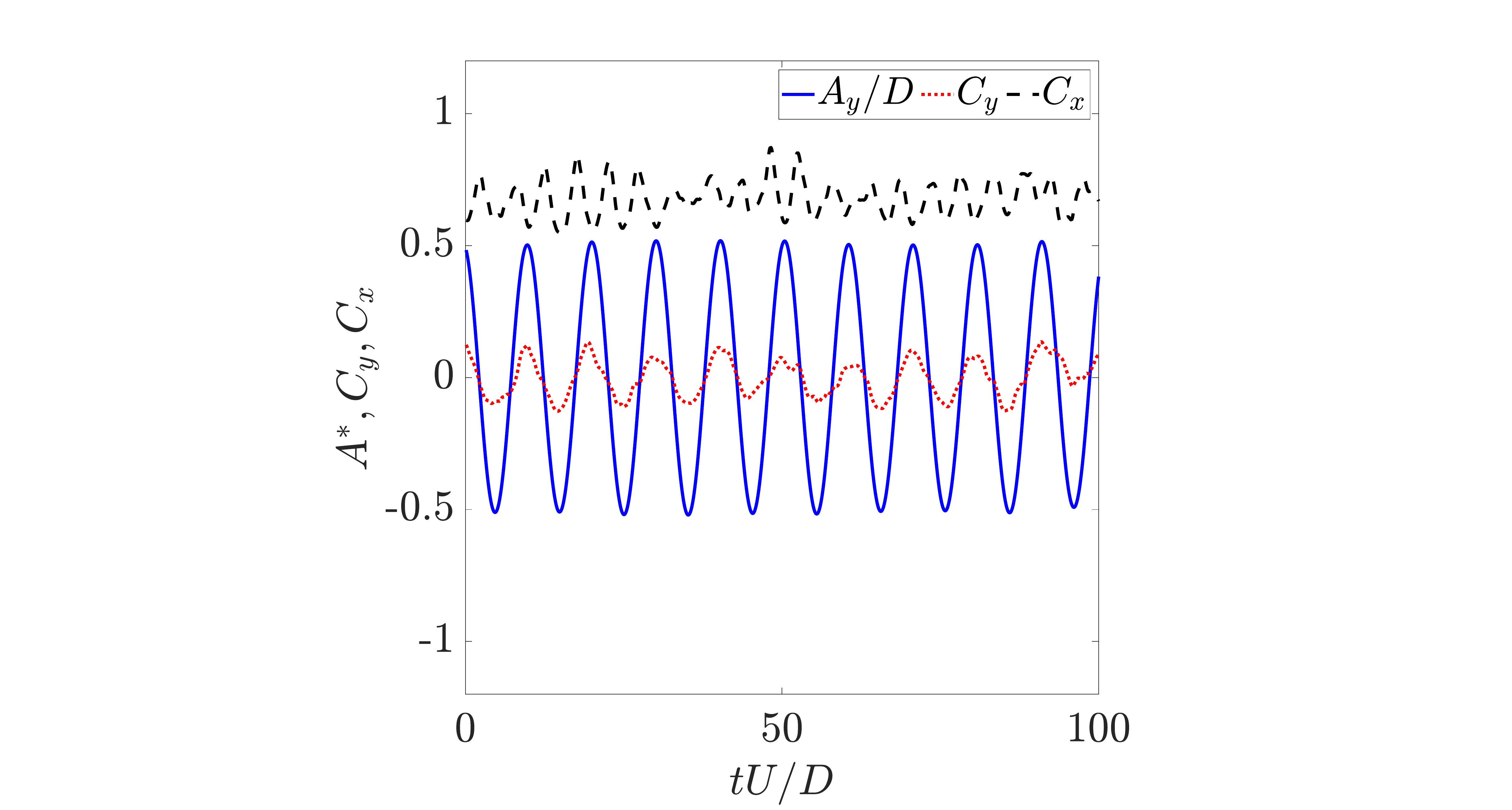}
		\caption*{($c$ - $1$)}
	\end{subfigure}%
	\begin{subfigure}[b]{0.5\textwidth}
		\adjincludegraphics[scale=0.27,trim={0.2\width} {0\width} {0.1\width} {0.0\width},clip]{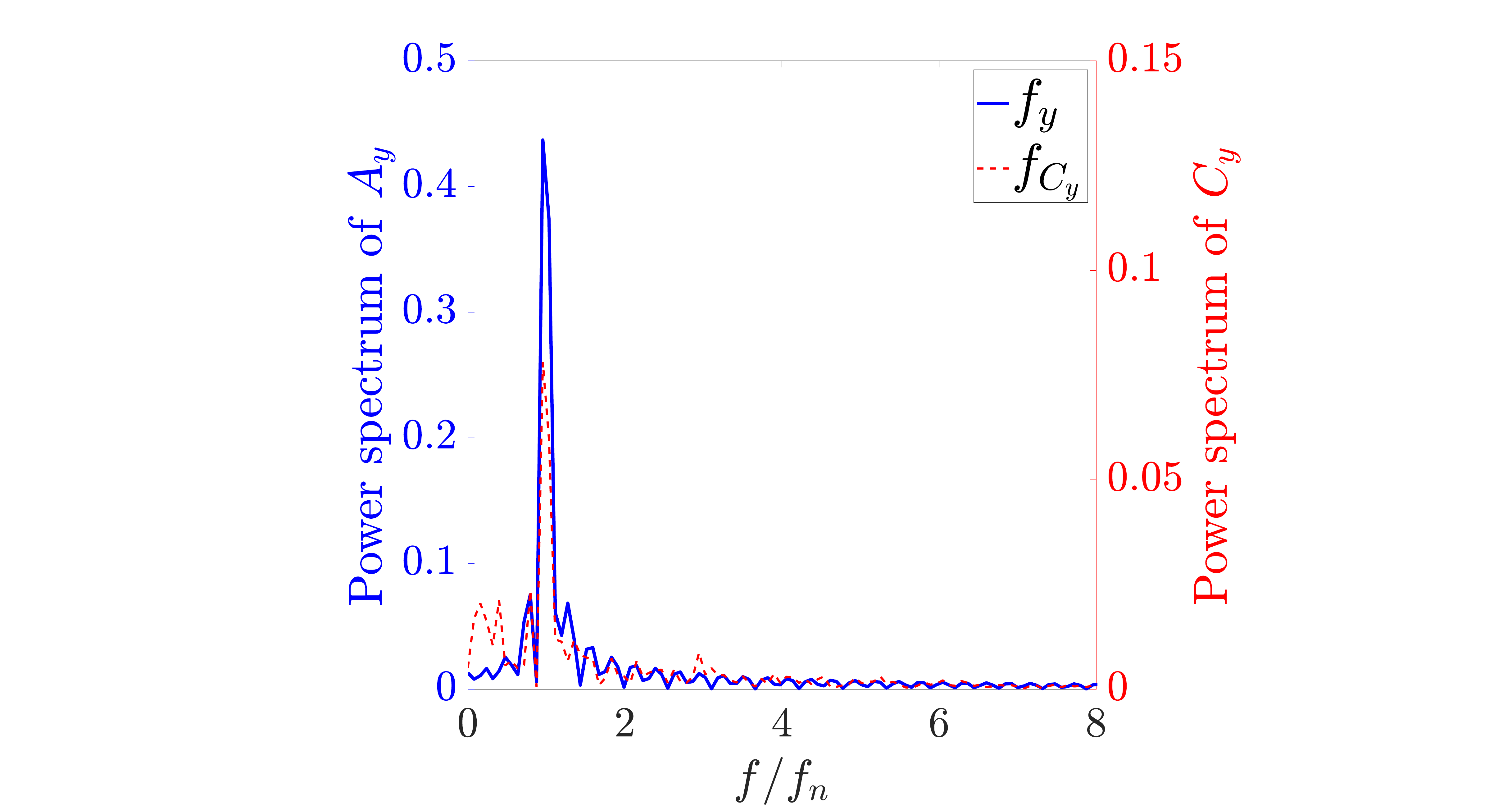}
		\caption*{($c$ - $2$)}
	\end{subfigure}

	\caption{Time histories of the normalized amplitude response ($A^*$) and transverse forces (${C_y}$) and their corresponding frequency spectra for the piercing sphere ($h^*=-0.25$) at two representatives Froude numbers: (a) $Fr=0.88$, and (b) $Fr=1.76$.  Streamwise forces ($C_x$) are also included in ($a$ - $1$) and  ($c$ - $1$).} 
	\label{TH_Frx4x8} 
	
\end{figure}

At higher Froude number cases $Fr\in[0.88,2.4]$, the surface deformation becomes substantial. As can be seen in Fig. \ref{Q_Crit_Fr_2}, for the case at $Fr=0.88$, the large surface deformation covers the front side of the sphere and the back side of the sphere is exposed to air. Therefore, the extreme surface distortion breaks the synchronization of vortex shedding and prevents the formation of the hairpin vortex loops completely. This causes a large reduction in the hydrodynamic transverse force on the sphere and a significant decrease in amplitude response, as shown in Fig \ref{TH_Frx4x8} (a). By further increasing the Froude number to $Fr=1.76$, the free surface covers the entire part of the sphere that is above the undisturbed free-surface level. The hairpin vortex structures shed behind the sphere while the centreline of the wake is directed downwards in the vertical $z$-direction. The transverse hydrodynamic force on the sphere recovers the strength and the amplitude response is increased. The FIV response at $Fr=1.76$ is comparable with the fully submerged cases where the only dominant shedding frequency matches with the oscillation frequency of the sphere, as can be seen in Fig. \ref{TH_Frx4x8} (b).

\begin{figure}[htbp!]

		\begin{subfigure}[b]{1\textwidth}
				\centering
				\adjincludegraphics[scale=0.24,trim={0\width} {0.0\width} {0\width} {0.0\width},clip]{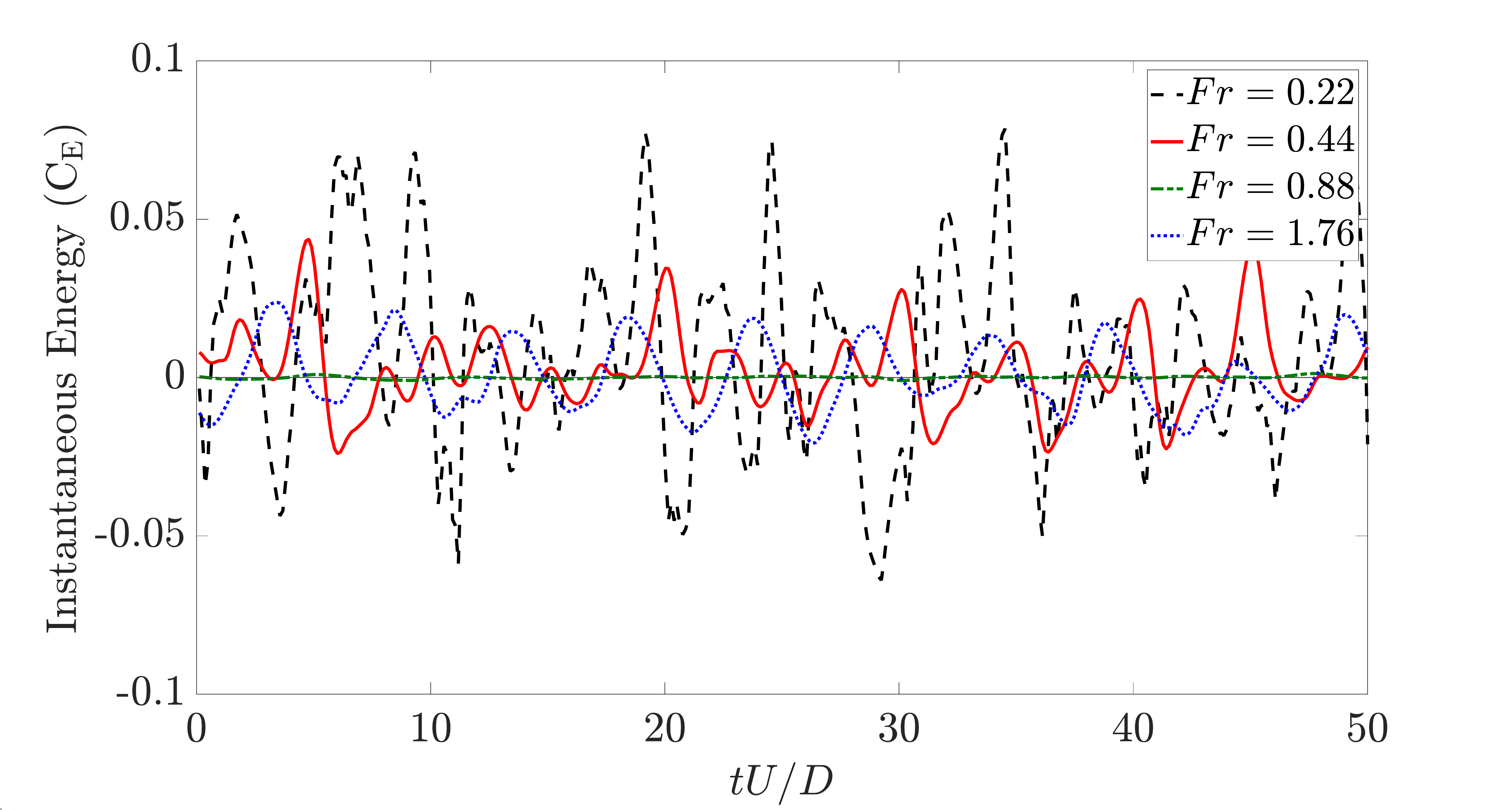}
			\caption{}
		\end{subfigure}

		\begin{subfigure}[b]{1\textwidth}
				\centering
				\adjincludegraphics[scale=0.24,trim={0\width} {0.0\width} {0\width} {0.0\width},clip]{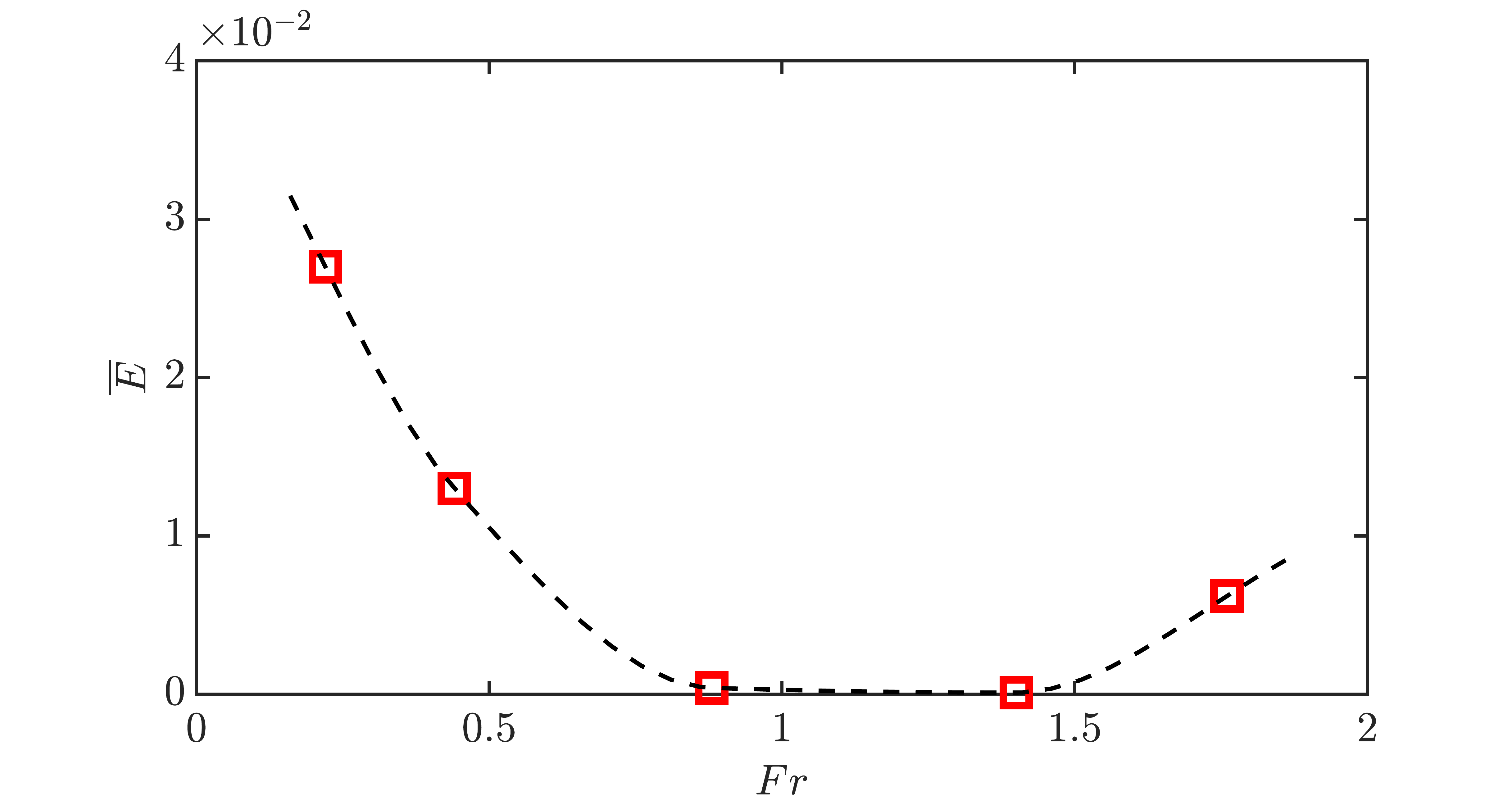}
			\caption{}
		\end{subfigure}
	
	\caption{Dependence of energy transfer on Froude number for the piercing sphere at $h^*=-0.25$, $U^* = 10$, and $Re = 15\,700$:  (a) temporal variation of instantaneous energy transfer $C_E$, and (b)  mean energy transfer over 10 oscillation periods. } 
	
	\label{Power_Fr} 
\end{figure}
The energy transfer between the flow and the oscillating sphere for different Froude numbers is characterized by the time-dependent energy coefficient ($C_E$). Fig. \ref{Power_Fr} (a) compares the variation of the instantaneous energy transfer $C_E$ for the piercing case. The non-dimensional time-averaged quantity of energy transfer (${E}$) is extracted for 10 oscillation cycles when the system reaches a stationary state. $\overline{E}$ denotes the mean quantity of the non-dimensional energy transfer over the 10 cycles, which is shown in Fig. \ref{Power_Fr} (b). It is found that by increasing the Froude number in the range $Fr\in[0.22,0.88]$, the amount of net energy transfer per oscillation cycle decreases. The reduction in the net energy transfer leads to a significant reduction of the amplitude response. By further increase in the Froude number for $Fr\ge1.4$, the amount of net energy transfer per cycle is increased as the free-surface deformation becomes substantial, covering the entire sphere surface at $h^*=-0.25$. 
In summary, we find that the large-amplitude VIV response is strongly sensitive to the Froude number at the range of $Fr\in [0.22,2.4]$. For the piercing sphere case, by increasing the Froude number in the lower range of $Fr\in [0.22,0.44]$, wherein the surface deformation is not substantial, the near-surface vorticity diffuses into the deformable free surface which in turn leads to significant energy dissipation and reduction in the amplitude response. At higher Froude number range $Fr\in [0.8,2.4]$, the surface deformation becomes substantial and the free surface covers the entire part of the sphere surface above the undisturbed free-surface level for $Fr\ge1.4$, altering the wake dynamics and increasing the VIV amplitude.
%


\section{Concluding remarks}
\label{Conclusion}
A numerical study has been performed to investigate the effect of the free surface on the FIV response of a transversely vibrating sphere in the proximity to a free surface.
We employed the recently developed three-dimensional fluid-structure-free-surface interaction solver to explore the FIV response of fully and partially submerged sphere configurations.
To begin, we first examined the VIV phenomenon and the wake modes of a fully submerged freely vibrating sphere in a wide range of Reynolds number $Re\in[300,30\,000]$ at the lock-in state.
We found that the sphere begins to move along a linear trajectory with hairpin vortex-shedding mode, eventually transforming into a circular trajectory with the spiral mode in its stationary state for $Re \in [2\,000,6\,000]$. 
By examining the mode transitions and the motion trajectories we found that the mode transition is strongly sensitive to Reynolds number. We observed that the motion trajectories at higher Reynolds number range $Re\in [12\,000,30\,000]$ show a chaotic response with a combination of linear motions and circular-type motions at the periodic state, where the vortex structure modes transform frequently from the hairpin mode to the spiral mode and \textit{vice versa}.

The FIV response of a sphere in the proximity of the free surface was investigated at three stages of immersion ratios at $h^*=1$, $h^*=0$ and $h^*=-0.25$ at the lock-in regime. 
Successful validation of the sphere by considering the effect of free surface has been established for the first time through quantitative and qualitative comparison with the experiments. 
The important findings of the paper can be summarized as follows: 
\begin{itemize}
\item For the fully submerged cases, the amplitude response of the sphere vibration when it touches the free surface at $h^*=0$ is decreased by $\sim20\%$ compared to the case at $h^*=1$. The vorticity plot in the cross-plane $0.5D$ downstream at $h^*=0$ revealed a diffusion of the vorticity flux due to induced free-surface distortion on the top of the sphere. The free surface changes the vorticity structure significantly and causes the vorticity pattern to become asymmetric along the horizontal plane. The free surface acts as a sink of energy which leads to the reduction in the transverse force and the amplitude response of the elastically mounted sphere.
\item The amplitude response for the piercing case at $h^*=-0.25$ is increased dramatically with the maximum peak-to-peak amplitude of $\sim2D$, larger than all the submerged cases studied. We observed that the free-surface distortion for the piercing case is considerably larger due to the complex interaction of the free surface with the piercing sphere geometry and the sphere wake. It was found that a strong flux of vorticity is supplied due to the piercing sphere/free-surface interaction. The existence of third-harmonic in the transverse force is related to the extra free-surface vorticity flux.  Increased streamwise vorticity gives rise to a relatively larger transverse force to the piercing sphere at $h^*=-0.25$, resulting in a relatively greater positive energy transfer per cycle to sustain the large-amplitude vibrations.
\item The effect of the mass ratio on the amplitude response for the piercing sphere case at $h^*=-0.25$ was studied over a range of $m^*\in[1,20]$ at the lock-in state.
The FIV response was found to be relatively insensitive to the mass ratio $m^*$ in the range studied, although increasing the mass ratio led to a slight reduction in the peak amplitude.
The existence of the third-harmonic behavior of the transverse force was observed for all the mass ratios.
\item Lastly, it is found that the FIV response is strongly sensitive to the Froude number for the piercing sphere case at $h^*=-0.25$ over a range of $Fr\in[0.22,2.4]$ at the lock-in state.
For the Froude number range $Fr\in [0.22,0.44]$, where surface deformation was not substantial, we observed that the amplitude response is decreased by $\sim30\%$ at $Fr=0.44$ compared to the case at $Fr=0.22$. It was found that at $Fr=0.44$, the strength of vorticity flux due to the free-surface distortion is considerably lower than the case at $Fr=0.22$. 
At higher Froude number range $Fr\in [0.8,2.4]$, the surface deformation becomes substantial. At $Fr=0.88$, the large surface deformation covers the front side of the sphere and the backside of the sphere is exposed to air. The extreme surface distortion breaks the synchronization of vortex shedding and prevents the formation of the hairpin vortex loops.
For $Fr\ge1.4$, the free surface covers the entire part of the sphere surface above the undisturbed free-surface level, altering the wake dynamics and the FIV response.
Hairpin vortex structures form behind the sphere with the centreline of the wake slightly shifted downwards in the vertical $z$-direction.
This results in an increase of the transverse hydrodynamic force and the amplitude response of the sphere.
Further research is required to systematically decompose the vorticity contributions of the sphere/free-surface and the wake/free-surface interactions for a broader range of physical parameters.

\end{itemize}

 


 \section*{Acknowledgment}
	The authors would like to acknowledge the Natural Sciences and Engineering Research Council of Canada (NSERC) for the funding. This research was supported in part through computational resources and services provided by Advanced Research Computing at the University of British Columbia.

\bibliography{A} 
\bibliographystyle{unsrt}

\end{document}